\documentclass{aa}  

\usepackage{graphicx}
\usepackage{txfonts}

\newcommand{\be}{\begin{equation}}
\newcommand{\ee}{\end{equation}}

\usepackage[nice]{nicefrac}

\usepackage[normalem]{ulem}

\begin{document} 

    \title{Probing the radial acceleration relation and the strong equivalence principle with the Coma cluster ultra-diffuse galaxies
    }
    \titlerunning{Radial acceleration relation and strong equivalence principle in Coma UDGs}
\authorrunning{Freundlich et al.}
    
   %\subtitle{XXX}

   \author{Jonathan Freundlich
          \inst{1}%\fnmsep\thanks{XX}
          \and
          Benoit Famaey
          \inst{1}
          \and
          Pierre-Antoine Oria
          \inst{1}
          \and
          Michal B\'ilek
          \inst{1,2}
          \and
          Oliver M\"uller 
          \inst{1}
          \and
          Rodrigo Ibata
          \inst{1}
          }

   \institute{Universit\'e de Strasbourg, CNRS UMR 7550, Observatoire astronomique de Strasbourg, 67000 Strasbourg, France\\
   \email{jonathan.freundlich@astro.unistra.fr}
   \and
    European Southern Observatory, Karl-Schwarzschild-Str. 2, D-85748 Garching, Germany 
   }

   \date{Received XXX; accepted XXX}
 
  \abstract
  {The tight radial acceleration relation (RAR) obeyed by rotationally supported disk galaxies is one of the most successful a priori predictions of the modified Newtonian dynamics (MOND) paradigm on galaxy scales. Another important consequence of MOND as a classical modification of gravity is that the strong equivalence principle (SEP) -- which requires the dynamics of a small, free-falling, self-gravitating system not to depend on the external gravitational field in which it is embedded -- should be broken. Multiple tentative detections of this so-called external field effect (EFE) of MOND have been made in the past, but the systems that should be most sensitive to it are galaxies with low internal gravitational accelerations residing in galaxy clusters within a strong external field. Here, we show that ultra-diffuse galaxies (UDGs) in the Coma cluster do lie on the RAR, and that their velocity dispersion profiles are in full agreement with isolated MOND predictions, especially when including some degree of radial anisotropy. However, including a breaking of the SEP via the EFE seriously deteriorates this agreement. We discuss various possibilities to explain this within the context of MOND, including a combination of tidal heating and higher baryonic masses. We also speculate that our results could mean that the EFE is screened in cluster UDGs. The fact that this would happen precisely within galaxy clusters, where classical MOND fails, could be especially relevant to the nature of the residual MOND missing mass in clusters of galaxies.}


   \maketitle
%
%-------------------------------------------------------------------

\section{Introduction}
\label{section:introduction}

Alternatives to general relativity (GR) have been the subject of active research over the past decades \citep[e.g.][and references therein]{Clifton2012}, the reason being that GR alone cannot explain the observed dynamics on scales ranging from galaxies to the largest scales in the observable Universe without additional degrees of freedom. These are denoted dark matter (DM) and dark energy, and their actual nature is probably the most pressing issue for modern physics. 
A great deal of research into modified gravity concentrates on the dark energy question, which is related to the cosmological constant problem; however, it has also been suggested since the early 1980s \citep{Milgrom1983b,Milgrom1983c, Milgrom1983a,Milgrom1984, Bekenstein1984} that the phenomena attributed to DM might also be, at least partly, related to new gravitational degrees of freedom rather than to new particles in the matter sector as generally assumed. 

%\subsection{MOND}

This approach, known as modified Newtonian dynamics \citep[MOND; see e.g.][for reviews]{Sanders2002,Famaey2012,Milgrom2014} postulates that the gravitational acceleration $g$ approaches $\sqrt{g_N a_0}$ when the Newtonian gravitational acceleration $g_N$ falls below a characteristic acceleration scale $a_0\approx 10^{-10}~\rm m \, s^{-2}$ but remains Newtonian above this threshold. This allows one to directly predict the dynamics of galaxies from their baryonic mass distribution alone.
This empirical modification of the gravitational law was initially proposed \citep{Milgrom1983b,Milgrom1983c, Milgrom1983a} to solve the missing mass problem in the high surface brightness galaxies known at the time, and especially their asymptotically flat circular velocity curves \citep[e.g.,][]{Bosma1978, Rubin1978, Faber1979}. It is particularly intriguing that this simple recipe has survived almost 40 years of scrutiny at galactic scales, as it has been able to predict the dynamics of a wide variety of galaxies \citep[e.g. ][]{Begeman1991, Sanders1996, McGaugh1998,Sanders1998,deBlok1998,Sanders2007,Gentile2007a,Swaters2010,Gentile2011, Famaey2012,Milgrom2012,McGaugh2013a,McGaugh2013b,Sanders2019}, including low surface brightness and dwarf galaxies where internal accelerations can be well below $a_0,$ such that the MOND acceleration should a priori deviate significantly from the Newtonian acceleration. This was a core prediction of the original MOND papers and one of its most intriguing successes. MOND can also provide possible answers to various other puzzles in galaxy dynamics, such as the prevalence of bulgeless discs \citep{Combes2014} and fast bars \citep{Tiret2007,Tiret2008,Roshan2021}, or the detailed kinematics of polar ring galaxies \citep{Lughausen2013}. Generally speaking, there now appears to be a clear and direct connection between the baryonic mass distribution and the rotation curve in most disc galaxies, known as the radial acceleration relation \citep[RAR;][]{McGaugh2016,Lelli2017}, and this empirical relation is actually indistinguishable from the original MOND prescription \citep{Li2018}. Evaluating whether the RAR holds for all types of galaxies and in all environments is thus of high importance when assessing the viability of MOND as an alternative to particle DM in galaxies.

It is important to note that it was also originally predicted that the non-linearity of the MOND acceleration would typically lead to a violation of the strong equivalence principle of GR, according to which the internal dynamics of a self-gravitating system embedded in a constant gravitational field should not depend on the external field strength. Within MOND, systems embedded in an external field stronger than their internal one should experience an `external field effect' \citep[EFE;][]{Milgrom1983a,Bekenstein1984,Famaey2012,McGaugh2013a, McGaugh2013b, Milgrom2014, Wu2015, Haghi2019} whose consequence is notably that the deviations from Newtonian dynamics are suppressed if the external field is strong enough, and in particular if it is larger than $a_0$. Its influence can be important for the stability and secular evolution of galaxies even when it is weak \citep{Banik2020}, and it can create interesting features such as asymmetric tidal tails of globular clusters \citep{Thomas2018}.
The EFE is for instance an observational necessity to allow the dynamics of wide binary stars to remain consistent with MOND (\citealp{Pittordis2019}, \citealp{Banik2019}, although see also \citealp{Hernandez2021}).
Because of this EFE, a rotationally supported (pressure supported) system in isolation is expected to have a higher rotational velocity (velocity dispersion) than the same system around a massive host \citep[e.g.][]{Wu2007,Gentile2007c,McGaugh2013a,McGaugh2013b,Pawlowski2014,Pawlowski2015,McGaugh2016_crater, Hees2016, Haghi2016, Muller2019a, Chae2020, Chae2021}.
In particular, the latter should not follow the RAR, which is contrary to the case of the more isolated systems that should lie on the RAR. This breaking of the strong equivalence principle should be an unmistakable sign of MOND, and it is therefore important to test it for galaxies with internal gravitational accelerations lower than the external field in which they are embedded: this is the focus of the present work.

The need for DM within GR is of course not limited to galaxies. Expanding MOND predictions to the cosmological regime requires a relativistic framework for the paradigm. In order to retain the success of the standard $\Lambda$CDM cosmological model on large scales, some hybrid models have been proposed, for instance, where GR is retained but gravity is effectively modified in galaxies through some exotic properties of DM itself, such as in dipolar DM \citep{Blanchet2009,Bernard2015,Blanchet2015} or superfluid DM \citep{Khoury2015, Berezhiani2015,Berezhiani2018,Berezhiani2019}. More traditional relativistic MOND theories rely on a multi-field framework (typically with a scalar and a vector field in addition to the metric), as originally proposed by \citet{Bekenstein2004}, but adapted to pass the most recent constraints from gravitational waves \citep{Skordis2019}. It has recently been shown, as a proof of concept, how the angular power spectrum of the cosmic microwave background (CMB) could be reproduced in such a framework \citep{Skordis2020}; the scalar field, which gives rise to the MOND behaviour in the quasi-static limit, also plays the role of DM in the time-dependent cosmological regime, thereby providing an analogue to cosmological DM for the CMB. 
However, the real challenge for such an approach, and for MOND in general, is to explain the mass discrepancy in galaxy clusters. It has indeed long been known that applying the MOND recipe to galaxy clusters yields a residual missing  mass problem in these  objects~\citep[e.g.,][]{Sanders1999,Sanders2003,Pointecouteau2005,Natarajan2008,Angus2008a}. This is essentially because, contrary to the case of galaxies, there is observationally a need for  DM even where the observed acceleration is larger than $a_0$, meaning that the  MOND prescription is not enough to explain the observed discrepancy. In the  central parts of clusters, the ratio of MOND dynamical mass to  
observed baryonic mass can reach a value of 10. This cluster missing mass problem extends to giant ellipticals residing at the centre of clusters \citep{Bilek2019a}. It is also clear that this residual missing mass must be collisionless \citep{Clowe2006,Angus2007}, and it has hence been proposed that it could be made of cold, dense molecular gas clouds \citep{Milgrom2008} or some form of hot dark matter (HDM) such as sterile neutrinos, which would not condense on galaxy scales \citep{Angus2010,Haslbauer2020}. In such cases, the residual missing mass should be an important gravitational source contributing to the EFE acting on galaxies residing in clusters.
On the other hand, if the residual MOND missing mass problem would itself be a gravitational phenomenon, it would then not necessarily contribute to the EFE as
a source. Therefore, studying the dynamics of galaxies residing in galaxy clusters, and in particular whether the EFE can be detected there, should provide powerful constraints for relativistic model-building in the MOND context and also illuminate our understanding of scaling relations with environment in the cold dark matter (CDM) paradigm. Galaxies with a very low internal gravity, hence ultra-diffuse ones, are best suited for such a study.

%\subsection{Ultra-Diffuse Galaxies}

Ultra-diffuse galaxies (UDGs) are low central surface brightness ($\rm \mu_{g,0}>24~mag.arcsec^{-2}$) objects with optical luminosities typical of dwarf galaxies ($L\sim 10^7-10^8 L_\odot$) but effective radii comparable to that of the Milky Way ($r_{\rm  eff}> 1.5\rm ~kpc$). Such galaxies have been identified for several decades \citep[e.g.][]{Fosbury1978,Sandage1984, Impey1988, Karachentsev2000} but have recently undergone a revival of interest as deep imaging observations revealed their ubiquity both in groups and clusters \citep[e.g. ][]{VanDokkum2015,vanDokkum2015b,Janowiecki2015,Koda2015,Mihos2015, Mihos2017,Munoz2015,Martinez-Delgado2016, vanderBurg2016, Yagi2016,Merritt2016, Janssens2017,Shi2017,  Venhola2017,Muller2018, Zaritsky2019,Habas2020}  and in the field \citep[e.g.][]{Leisman2017,Roman2017,Prole2019}.
Different scenarios have been proposed to explain their formation, invoking them as either
(i) failed Milky Way-like galaxies that lost their gas after forming their first stars \citep{VanDokkum2015, vanDokkum2015b, vanDokkum2016, Yozin2015, Peng2016, Lim2018, Pandya2018,Villaume2021}, possibly due to denser group and cluster environments; 
(ii) the high-spin tail of the dwarf galaxy population \citep{Amorisco2016, Rong2017,Shi2017}; 
(iii) tidal debris from mergers or tidally disrupted dwarfs \citep{Beasley2016b,Greco2018,Toloba2018, Jiang2019,Carleton2019}; or 
(iv) galaxies whose spatial extent results from dynamical heating due to stellar feedback \citep{DiCintio2017,Chan2018, Jiang2019, Freundlich2020a,Freundlich2020b}.
Within the  CDM model of structure formation, their particularly low surface brightness allows us to probe the galaxy-halo connection in the low-mass regime, with an ongoing debate regarding their DM content \citep{Beasley2016,Peng2016,vanDokkum2016, VanDokkum2017, VanDokkum2018a, VanDokkum2019a,Martin2018,Lim2018,Wasserman2018,Nusser2018,Nusser2019,Hayashi2018,DuttaChowdhury2019, DuttaChowdhury2020, Laporte2019, Danieli2019,Emsellem2019,Fensch2019,Trujillo2019, Haslbauer2019,Gannon2020, Gannon2021, Muller2021}. 
The possible diversity of their DM and gas content \citep{Janowiecki2015,Papastergis2017,Leisman2017,Shi2017,Karunakaran2020} may indicate that more than one of the formation scenarios coexist \citep[cf. also][]{Toloba2018,Ruiz-Lara2018,Ferre-Mateu2018, Chan2018, Jiang2019, Forbes2020, Sales2020}.

%\subsection{UDGs in MOND}

Ultra-diffuse galaxies in clusters provide a testing ground for MOND and the EFE given the singularly low internal accelerations stemming from their low surface brightness and the strong external field.
The small velocity dispersion observed in the two group UDGs NGC 1052-DF2 and NGC 1052-DF4, inferring dynamical masses close to their stellar masses, was initially interpreted as a challenge for MOND \citep{VanDokkum2018a, VanDokkum2019a}. 
Indeed, the dynamical effect attributed to DM in the CDM model, and to a modification of the gravitational law within MOND in isolation, would be absent. 
However, taking the EFE into account removes or significantly lessens the tension \citep{Famaey2018, Kroupa2018, Muller2019a, Haghi2019}. 
On the other hand, the large velocity dispersion of the Coma cluster UDG DF44 \citep{vanDokkum2016, vanDokkum2019b} and its relative agreement with the isolated MOND prediction without EFE has been used to place constraints on its distance from the cluster centre within MOND, or on a potential need for an additional baryonic mass \citep{Bilek2019,Haghi2019b}. 
Different approaches have been used to take the EFE into account in this context:
\cite{Kroupa2018} and \cite{Haghi2019} used fitting functions for the one-dimensional line-of-sight velocity dispersion in an external field stemming from MOND $N$-body simulations by \cite{Haghi2009};
\cite{Famaey2018} and \cite{Muller2019a} used the
 one-dimensional analytical expression for the acceleration field in the presence of an external field from \cite{Famaey2012}, Eq. (59), together with the \cite{Wolf2010} relation for the line-of-sight velocity dispersion. 
\cite{Bilek2019} and \cite{Haghi2019b} did not quantitatively assess the EFE on the velocity dispersion for DF44, since the data required as little of it as possible. 
We propose examining the question of the EFE in this galaxy more quantitatively and expanding the study to a larger sample of UDGs.

%\subsection{Present work}

In this article, we compare observed stellar velocity dispersion measurements of 11 Coma cluster UDGs with MOND modified gravity predictions with and without EFE. 
We also propose a new analytic formula in quasi-linear MOND \citep[QUMOND,][]{Milgrom2010}, assuming a spherical self-gravitating system embedded in a constant external field, which is tested against numerical computations and compared with existing formulae.
Throughout this work, we adopt a distance of 100 Mpc, a distance modulus $(m-M)_0=35$, and central coordinates R.A.: 12:59:48.75 and Dec.: +27:58:50.9 (J2000)  \citep[e.g.,][]{Carter2008,Alabi2020} for the Coma cluster.
In Section \ref{section:observations}, we present the sample of Coma cluster UDGs with stellar velocity dispersion measurements. 
In Section \ref{section:isolated}, we model their dynamics within MOND in isolation. 
In Section \ref{section:EFE}, we present our new QUMOND formula for the EFE and model these UDGs including this effect. We discuss the implications of the confrontation between the MOND predictions and the observations in Section \ref{section:discussion}, and we conclude in Section \ref{section:conclusion}.

%-------------------------------------------------------------------

\begin{figure*}
\centering
\includegraphics[width=0.85\textwidth,trim={3cm 3cm 3cm 3cm},clip]{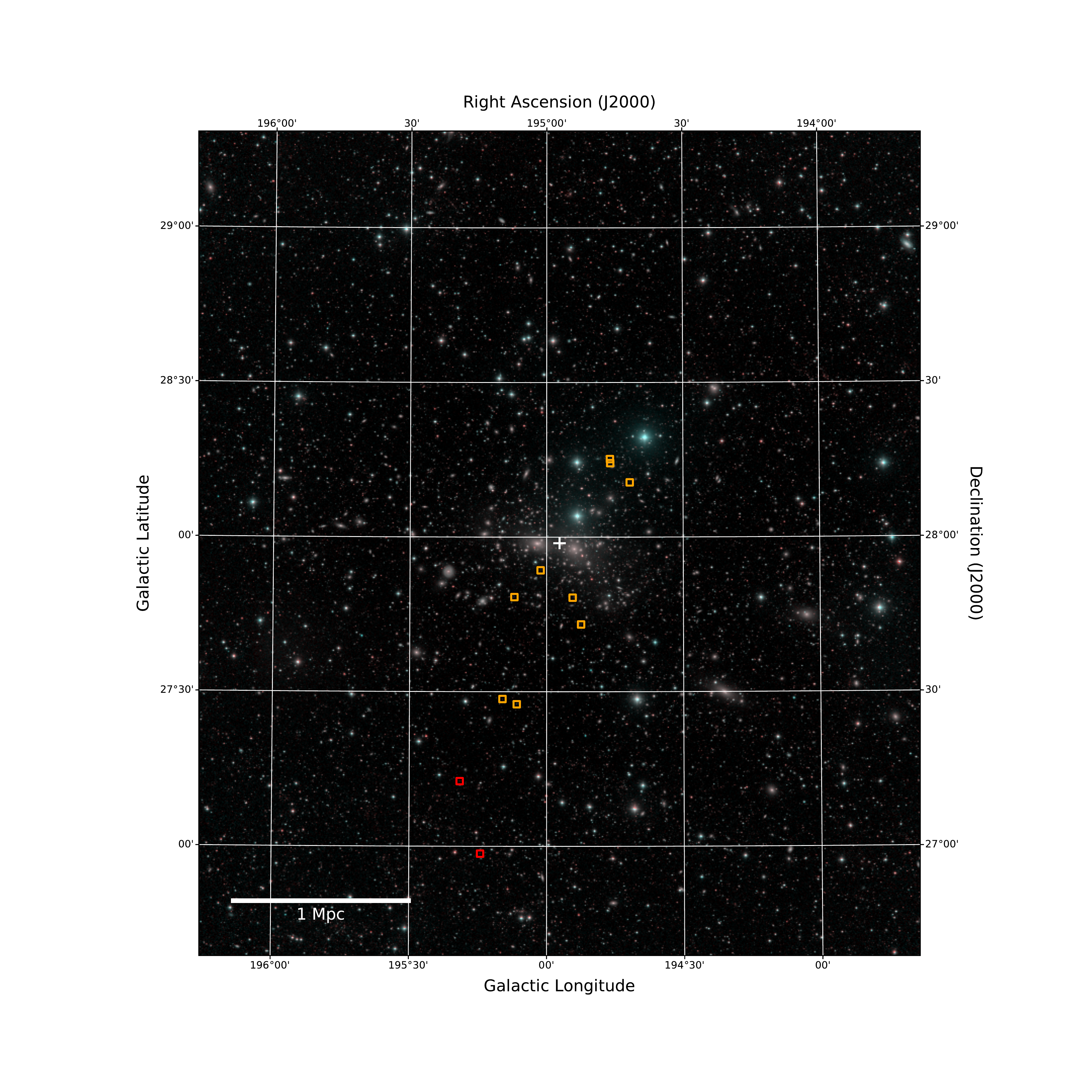}
\caption{Spatial distribution of the Coma cluster UDGs modelled in this article projected on a colour image of the cluster created from the Dragonfly $g$ and $r$ images \protect\citep{Abraham2014,VanDokkum2015}. 
The adopted centre of the Coma cluster is indicated by the white cross. 
DF44 and DFX1, which are the furthest away from the centre, are indicated in red, the \protect\cite{Chilingarian2019} UDGs are displayed in orange. 
}
\label{fig:coma}%
\end{figure*}

\section{A sample of Coma cluster UDGs}
\label{section:observations}

The Coma cluster is one of the nearest and most-studied rich and dense cluster environments. 
\cite{VanDokkum2015} notably coined the term ultra-diffuse galaxies when they uncovered its large population of low surface brightness objects with the Dragonfly Telephoto Array \citep{Abraham2014}, and the first spatially resolved kinematics of a UDG was obtained for a Coma cluster UDG: DF44 \citep{vanDokkum2019b}.

\subsection{DF44}

The velocity and velocity dispersion profiles of DF44 were obtained with Keck Cosmic Web Imager \citep[KCWI,][]{Morrissey2012, Morrissey2018}, an integral field unit (IFU) spectrograph on the Keck II telescope optimised for low surface brightness objects. These profiles were studied in the context of Newtonian gravity assuming CDM by \cite{vanDokkum2019b}, in the context of MOND by \cite{Bilek2019} and \cite{Haghi2019b}. \cite{Wasserman2019} further considered Newtonian gravity with fuzzy dark matter, an exotic form of DM \citep[cf. e.g. ][]{Hu2000}. 
The relatively high velocity dispersion profile, with a luminosity-weighted $\sigma_{\rm eff}=33_{-3}^{+2}~\rm km s^{-1}$ at the half-light radius, is interpreted within the CDM model as indicative of a galaxy dominated by DM even at its centre \citep{vanDokkum2019b}.
\cite{Bilek2019} and \cite{Haghi2019b} showed that this velocity dispersion profile fits well with the MOND prediction in isolation. These two studies used the \citet{VanDokkum2015, vanDokkum2016,VanDokkum2017,vanDokkum2019b} $I$-band luminosity $L_I=3.0\pm 0.6 \times 10^8~\rm L_\odot$ and projected half-light radius $R_e=4.7~\rm kpc$ of DF44 to infer the gravitational acceleration within MOND and, from it, the line-of-sight velocity dispersion profile through Jeans modelling. \cite{Bilek2019} assumed a spherical S\'ersic profile given the projected S\'ersic index $n=0.94$, fixed the stellar mass-to-light ratio to 1.3, and considered different anisotropy parameters. \cite{Haghi2019b} assumed a spherical Plummer profile and left both the dynamical mass-to-light ratio and the anisotropy parameter free. 
\cite{Bilek2019} interpreted the agreement between the velocity dispersion profile with the MOND prediction in isolation as an indication that DF44 is either far from the cluster centre (the projected distance is 1.8 Mpc), undergoing disruption, or embedded in a dark baryonic halo. It was also noted that some degree of tangential anisotropy allowed a better fit of the outer data points, which is also true in the CDM context.

\subsection{Ten additional UDGs}
\label{section:10_additionnal}

While \cite{VanDokkum2015} initially reported 47 UDGs in the Coma cluster using the Dragonfly Telephoto Array, \cite{Koda2015} and \cite{Yagi2016} increased the number to 854 using Subaru archival data, and \cite{Alabi2020} discovered a further 29 of them.
Amongst all Coma cluster UDGs, about 30 have been spectroscopically confirmed \citep{Kadowaki2017, Alabi2018, Ruiz-Lara2018, Chilingarian2019}. Most lack stellar velocity and velocity dispersion measurements, which are particularly challenging to obtain given their low surface brightness, such that internal dynamics and DM content within the CDM model are generally poorly constrained. 
\cite{Chilingarian2019} obtained spatially resolved stellar velocity profiles and velocity dispersions for nine UDGs in the Coma cluster, seven of them being taken from the \cite{Yagi2016} catalogue, using the Binospec spectrograph on the 6.5mm MMT telescope \citep{Fabricant2019}.
They modelled them within the CDM model, obtaining DM fractions between 50 to 90\% within the half-light radius. We note that amongst the nine UDGs for which \cite{Chilingarian2019} obtained spatially resolved velocity profiles, only three have spatially resolved velocity dispersion profiles. 
Together with DF44, this thus yields ten Coma cluster UDGs with spatially resolved stellar velocity profiles and velocity dispersion measurements, four of them having spatially resolved velocity dispersion profiles. 
We note that one galaxy amongst them (J130005.40+275333.0) displays a $20\rm ~km~s^{-1}$ rotation, and three others (J125848.94+281037.1, J125904.06+281422.4, and J130028.34+274820.5) show hints of rotation ($<10~\rm km~s^{-1}$), always at least twice as low as the typical level of the line-of-sight velocity dispersion.

To this sample, we further add DFX1, a UDG whose central velocity dispersion is reported in \citet{VanDokkum2017} together with its $I$-band Sérsic index, half-light radius, and magnitude. We do not include UDGs for which the velocity dispersion has been measured only from the globular cluster population, notably as some disagreement between stellar and globular cluster measurements has been reported (cf. for example \citealp{Muller2020} versus \citealp{Forbes2021}).

\subsection{Sample properties}

\newcommand\T{\rule{0pt}{2.6ex}}       % Top strut
\newcommand\B{\rule[-1.2ex]{0pt}{0pt}} % Bottom strut

\begin{table*}
\caption{Sample UDGs and their properties}             
\label{table:sample}      
\centering          
\begin{tabular}{llllllll}     
\hline\hline       
Name & $d$ $\rm [kpc]$ & $d_{\rm mean}$ $\rm [kpc]$ & $L$ $\rm [10^8~L_\odot]$ & $R_e$ $\rm [kpc]$& $n$ & $\rm M/L$ $\rm [M_\odot/L_\odot]$& $\sigma_{\rm eff}$ $\rm [km s^{-1}]$ \T\\ 
(1) & (2) & (3) & (4) & (5) & (6) & (7) & (8) \B\\
\hline
DF44$^a$ & 1809.0 & 2336.2 & 3.0 & 4.7 & 0.85 & 1.3$^\dagger$ & $33 \pm 3$\T\\
DFX1$^a$ & 1457.1 & 1994.9 & 1.9 & 3.6 & 0.90 & 1.3$^\dagger$ & $30\pm 7^\dagger $ \B\\
\hline
J125848.94+281037.1$^b$ & 523.9 & 1129.3 & 0.38 & 1.5 & 0.95 & $1.32\pm 0.16$ & $18\pm13$ \T\\
J125904.06+281422.4$^b$ & 534.9 & 1138.7 & 0.43 & 1.4 & 0.76 & $1.48\pm 0.40$ & $25\pm7$\\
J125904.20+281507.7$^b$ & 553.1 & 1154.4 & 0.46 & 1.6 & 1.36 & $1.41\pm 0.11$ & $17\pm10$ \\
J125929.89+274303.0$^b$ & 475.3 & 1088.0 & 1.2  & 2.9 & 1.21 & $1.14\pm 0.25$ & $21\pm7$\\
J125937.23+274815.2$^b$ & 317.0 & 960.2 & 0.40 & 1.8 & 1.81 & $1.45\pm 0.12$ & $22\pm8$\\
J130005.40+275333.0$^b$ & 187.6 & 867.8 & 0.82 & 1.8 & 0.99 & $0.73\pm 0.11$ & $37\pm6$ \\
J130026.26+272735.2$^b$ & 940.9 & 1504.6 & 3.6  & 3.7 & 1$^\dagger$ & $0.37\pm 0.02$ & $19\pm5$\\
J130028.34+274820.5$^b$ & 397.7 & 1023.9 & 0.71 & 2.1 & 1.37 & $0.79\pm 0.12$ & $23\pm7$\\
J130038.63+272835.3$^b$ & 937.0 & 1501.0 & 1.5  & 1.9 & 1$^\dagger$ & $0.41\pm 0.03$ & $27\pm5$\B\\
\hline 
\end{tabular}
\begin{minipage}{1\linewidth}
        \T \textbf{Notes.} 
        (1) Name of the UDG, as reported by (a) \citet{VanDokkum2015,vanDokkum2016,VanDokkum2017,vanDokkum2019b}, (b)  \citet{Chilingarian2019}. 
        (2) Projected distance from the centre of the Coma cluster. 
        (3) Average distance from the centre inferred from the Einasto distribution from \citet[][cf. Section \protect\ref{section:coma}]{vanderBurg2016}. 
        (4) Stellar luminosity, in the $I$-band for DF44 and DFX1, and in the $R$-band for the others. 
        (5) (6) Projected half-light radius and Sérsic index from $I$- or $R$-band fits. A Sérsic index $n=1$ is assumed for two UDGs identified by \citet[][marked by a dagger]{Chilingarian2019}. 
        (7) Stellar mass-to-light ratio in  $I$- or $R$-band, taken to be 1.3 for the first two UDGs (marked by a dagger) and derived from fitting the spectrum with single stellar population models by \citet{Chilingarian2019} for the others.
        (8) Effective stellar velocity dispersion within the de-projected half-light radius or, when not available, central stellar velocity dispersion (marked by a dagger). 
\end{minipage}
\end{table*}

In the present article, we model the 11 Coma cluster UDGs with velocity dispersion measurements within MOND, namely DF44 and the ten UDGs mentioned above. DF44 was already modelled within MOND by \cite{Bilek2019} and \cite{Haghi2019}, but the EFE was not quantitatively assessed, which justifies its inclusion in our current study. 

Figure~\ref{fig:coma} shows the Coma cluster as seen by Dragonfly in the $r$ and $g$ bands together with the positions of the UDGs at stake. 
For DF44 and DFX1, we retain the $I$-band luminosity, projected half-light radius, and S\'ersic index from \citet{VanDokkum2015, vanDokkum2016, VanDokkum2017}. \citet{vanDokkum2019b} obtained stellar mass-to-light ($\rm M/L$) ratios in the 1-1.5 range. We assume here $\rm M/L=1.3$ as  \cite{Bilek2019}. 
For the \cite{Chilingarian2019} UDGs, we used the best-fit $R$-band luminosities, projected half-light radii, S\'ersic indices from the \cite{Yagi2016} catalogue when available, and the stellar $\rm M/L$ ratios obtained by \cite{Chilingarian2019} by fitting the full Binospec spectrum with simple stellar population models. We did not leave the stellar mass-to-light ratio as a free parameter, so our results are direct predictions, not fits.
For the two UDGs identified by \cite{Chilingarian2019} that are not in the \cite{Yagi2016} catalogue, we used the $R$-band luminosities and projected half-light radii obtained by \cite{Chilingarian2019}, and we assumed S\'ersic indices $n=1$. 
We note that while \cite{vanDokkum2019b} assumed a distance of 100 Mpc to the Coma cluster, \cite{Chilingarian2019} used 99 Mpc, but we neglected the 1\% difference. 
Table~\ref{table:sample} summarises the properties of the sample UDGs, including the luminosity-weighted effective stellar velocity dispersion $\sigma_{\rm eff}$. 

The sample distribution of radial velocities (which can be found in \citealp{vanDokkum2015b}, \citealp{VanDokkum2017}, and \citealp{Chilingarian2019}) is centred at $6654\rm ~km~s^{-1}$ with a dispersion of $922~\rm km~s^{-1}$. This is in line with typical Coma cluster galaxies \citep[e.g.][with galaxies centred at $6853\rm ~km~s^{-1}$ and a dispersion of $1082~\rm km~s^{-1}$]{Colless1996}, such that there is a priori no reason to believe the sample kinematics are biased.

%-------------------------------------------------------------------

\section{MOND models in isolation}
\label{section:isolated}

\subsection{Radial acceleration relation}
\label{section:RAR}

Since the radial acceleration relation \citep[RAR;][]{McGaugh2016, Milgrom2016, Lelli2017,Li2018} between the radial dynamical acceleration at the half-light radius, inferred from the velocity dispersion, and the radial gravitational acceleration predicted by the observed baryonic distribution within this radius, is a core prediction of MOND in isolation, we first check whether the Coma cluster UDGs do fall on this relation or not.
Within MOND, the tight scatter of this relation linking the dynamical and baryonic masses of galaxies over a wide range of galaxy types, sizes, and morphologies is indeed a consequence of the modification of the gravitational law.

\subsubsection{Observed versus inferred acceleration}
\label{section:method_rar}

From the effective stellar velocity dispersion $\sigma_{\rm eff}$ within the de-projected half-light radius $r_{1/2}\approx (4/3) R_e$, where $R_e$ is the half-light radius on the sky, we derived the Newtonian dynamical mass within $r_{1/2}$ using the \cite{Wolf2010} formula: 
\be
M_{1/2, \rm W10}\approx 4 G^{-1} \sigma_{\rm eff}^2 R_e
\ee
(cf. also \citealt{Kretschmer2021} for a discussion on the proportionality factor based on cosmological zoom-in simulations), and we estimated the corresponding `observed' gravitational acceleration:
\be
g_{\rm obs} = \frac{G M_{1/2, \rm W10}}{r_{1/2}^2} \approx \frac{9}{4} \frac{\sigma_{\rm eff}^2}{R_e}. 
\ee
This gravitational acceleration can be compared to that due to the baryons stemming from the stellar distribution $M(r)$:
\be
g_{\rm bar} = GM(r_{1/2})/r_{1/2}^2.
\ee
We assume that the uncertainties on $g_{\rm obs}$ are dominated by those on $\sigma_{\rm eff}$, i.e., $dg_{\rm obs}/g_{\rm obs}=2d\sigma_{\rm eff}/\sigma_{\rm eff}$, and that the uncertainties on $g_{\rm bar}$ are a combination of those on the $\rm M/L$ ratio and a conservative 30\% uncertainty on the luminosity. 

We note that this approach is obviously less precise than the detailed Jeans modelling that we performed afterwards, but it gives a first indication as to whether the Coma UDGs agree with the isolated MOND predictions or not.

\subsubsection{UDG spherical mass model}
\label{section:method_spherical}

To describe the three-dimensional stellar mass distribution of each UDG, we assumed a uniform stellar mass-to-light $\rm M/L$ ratio and de-projected the two-dimensional Sérsic light profile using the semi-analytical approximation proposed by \cite{LimaNeto1999}: 
\be
\label{eq:rho}
\rho(r) = \rho_0 \left(\frac{r}{a}\right)^p \exp \left[ -\left(\frac{r}{a}\right)^{1/n} \right], 
\ee
where $n$ is the Sérsic index, $a=R_e \exp[0.1789-0.6950\times n-n \ln n]$, and $p= 1- 0.6097/n +0.05563/n^2$. These expressions are given in their Eqs. (13) and (19), and in Appendix A of \cite{Marquez2000} for the updated value of the $1/n^2$ term of $p$, which was already used in \cite{Bilek2019}. 
The two-dimensional Sérsic density profile indeed lacks an actual analytical de-projection in three dimensions, and the \cite{LimaNeto1999} approximation is accurate within 5\% for $n$ between 0.5 and 10 and radii between $10^{-2}$ and $10^3 R_e$. 
The corresponding enclosed mass profile is 
\be
\label{eq:M}
M(r) = 4\pi \rho_0 n a^3 \gamma\left(3n-np,\left(\frac{r}{a}\right)^{1/n}\right), 
\ee
where $\gamma(s,x)=\int_0^x t^{s-1}e^{-t} dt$ is the incomplete gamma function, such that $\rho_0 = M_0/[4\pi n a^3 \Gamma(3n-np)]$, given the total mass $M_0$.

\subsubsection{Resulting RAR}

\begin{figure*}
\centering
\includegraphics[width=0.6\linewidth,trim={0.2cm 0.6cm 0cm 0.4cm},clip]{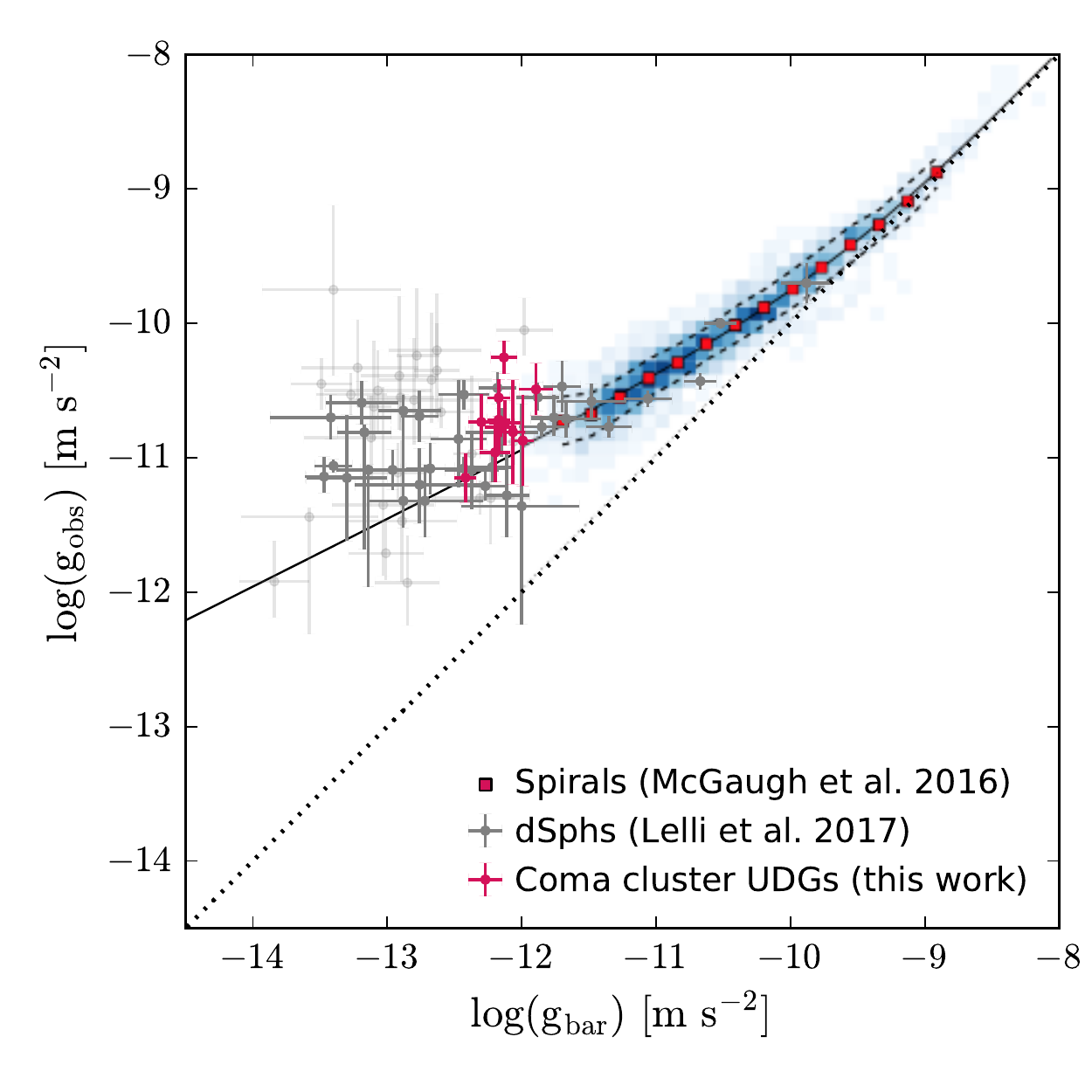}
\caption{
Radial acceleration relation for the Coma cluster UDGs (magenta error bars) together with that obtained for SPARC local spiral galaxies by \citet[][blue colour scale, red squares, and dashed lines for the binned mean and standard deviation]{McGaugh2016} and for dSphs of the Local Group \protect\citep[][light grey error bars for their least reliable data and dark grey error bars for their better quality sample]{Lelli2017}. The plain black line shows the fit using Eq.~(\ref{eq:RAR}) obtained by \protect\citet{McGaugh2016} on the SPARC sample of spiral galaxies.
The Coma cluster UDGs fall within the scatter obtained for the dSphs and in reasonable agreement with the empirical function of \protect\citet{McGaugh2016}, which reproduces MOND phenomenology. 
}
\label{fig:rar}%
\end{figure*}

Table~\ref{table:RAR} indicates the values of $g_{\rm obs}$ and $g_{\rm bar}$ for the sample UDGs; Fig.~\ref{fig:rar} displays them on the radial acceleration relation obtained by \cite{McGaugh2016} for local spiral galaxies from the Spitzer Photometry and Accurate Rotation Curves (SPARC) sample \citep{Lelli2016_SPARC} and for local dwarf spheroidals (dSphs) by \cite{Lelli2017}, including satellites of the Milky Way and of Andromeda. 
As for the latter, the UDG data points are generally above but in reasonable agreement with the following empirical fitting function: 
\be
\label{eq:RAR}
g_{\rm obs} = \frac{g_{\rm bar}}{1-e^{-\sqrt{g_{\rm bar}/a_0}}}
,\ee
with $a_0=1.20 \times 10^{-10}~\rm m s^{-2}$ obtained for the local spirals and shown as a plain black line \citep{McGaugh2008, McGaugh2016, Lelli2017}.
This function reproduces the MOND phenomenology in isolation: $g_{\rm obs}\approx g_{\rm bar}$ when $g_{\rm bar}\gg a_0$ and $g_{\rm obs}\approx \sqrt{g_{\rm bar} a_0}$ when $g_{\rm bar}\ll a_0$. 
The fact that the Coma cluster UDGs fall on the same radial acceleration relation as spirals and dSphs not only shows that their dynamical and baryonic masses are similarly correlated (cf. also \citealt{McGaugh2000}, \citealt{McGaugh2005, McGaugh2012}, \citealt{Lelli2016, Lelli2019}, and \citealt{McGaugh2021} on the baryonic Tully-Fisher relation), but it also hints towards the UDGs being well described by MOND in isolation. We now turn to a more detailed Jeans modelling of the velocity dispersion profiles to confirm this rough first hint.

\begin{table}
\caption{Observed and baryonic accelerations at the half-light radius for the radial acceleration relation.}             
\label{table:RAR}      
\centering          
\begin{tabular}{lll}     
\hline\hline       
Name & $\log(g_{\rm bar}) ~\rm [m s^{-2}]$ & $\log(g_{\rm obs})~ \rm [m s^{-2}]$ \T\B\\ 
\hline
DF44 & $-12.16\pm0.12$ & $-10.77\pm0.07$ \T\\
DFX1 & $-12.12\pm 0.12$ & $-10.74\pm 0.17$ \B\\
\hline
J125846.94+281037.1 & $-12.06 \pm 0.09$ & $-10.81 \pm 0.39$\T\\
J125904.06+281422.4 & $-11.89 \pm 0.13$ & $-10.49 \pm 0.19$\\
J125904.20+281507.7 & $-11.99 \pm 0.08$ & $-10.87 \pm 0.34$\\
J125929.89+274303.0 & $-12.20 \pm 0.11$ & $-10.96 \pm 0.22$\\
J125937.23+274815.2 & $-12.17 \pm 0.09$ & $-10.72 \pm 0.24$\\
J130005.40+275333.0 & $-12.13 \pm 0.10$ & $-10.25 \pm 0.12$\\
J130026.26+272735.2 & $-12.42 \pm 0.08$ & $-11.15 \pm 0.18$\\
J130028.34+274820.5 & $-12.30 \pm 0.10$ & $-10.73 \pm 0.21$\\
J130038.63+272835.3 & $-12.17 \pm 0.08$ & $-10.55 \pm 0.14$\B\\
\hline 
\end{tabular}
\end{table}

\subsection{MOND velocity dispersion in isolation}
\label{section:method_isolated}

We compare the observed velocity dispersion measurements with the profiles expected within MOND, assuming that the galaxies are sufficiently far away from the centre of the Coma cluster and from other galaxies to neglect the EFE, that they can be approximated as spherical, and that $\rm M/L$ is uniform for each galaxy. We followed similar steps to those of \cite{Bilek2019} to determine the expected velocity dispersion profiles in isolation, which are summarised below. 
We highlight that the predicted profiles are obtained without any free fitting parameters; they depend only on the stellar distribution, approximated by a Sérsic sphere, the fixed $\rm M/L$, and the assumed anisotropy. The parameters describing the stellar distribution and $\rm M/L$ (fixed at its value from the literature) are indicated in Table~\ref{table:sample}. We neglect any gaseous component.

\subsubsection{Line-of-sight velocity dispersion}
\label{section:method_isolated_los}

The velocity dispersion of a spherical collisionless system in equilibrium is related to the gravitational acceleration $g$ through the Jeans equation \citep[][Eq. 4.215]{BinneyTremaine2008}:
\be \label{eq:Jeans}
\frac{1}{\rho}\frac{d(\rho \sigma_r^2)}{dr} +\frac{2\beta}{r}\sigma_r^2 = g,
\ee
where $\rho$ is the density, $\sigma_r$ the radial velocity dispersion, and $\beta=1-\sigma_t^2/\sigma_r^2$ the anisotropy parameter, with $\sigma_t$ being the tangential velocity dispersion. 
The line-of-sight velocity dispersion $\sigma_{\rm los}$, which is the observable quantity, can be expressed as a function of $\sigma_r$ by projecting the velocity ellipsoid along the line of sight, with 
\be
\label{eq:sigma_los}
\frac{1}{2} \Sigma(R) \sigma_{\rm los}^2 (R) = \int_R^{+\infty} \frac{\rho \sigma_r^2 r dr}{\sqrt{r^2-R^2}} - R^2 \int_R^{+\infty} \frac{\beta \rho \sigma_r^2 dr}{r\sqrt{r^2-R^2}}
\ee
\citep{Binney1982}, where $\Sigma$ is the surface density. Given $g$, it is thus possible to recover $\sigma_{\rm los}$ by injecting Eq.~(\ref{eq:Jeans}) into Eq.~(\ref{eq:sigma_los}). The appendix of \cite{Mamon2005} provides formulas expressing $\sigma_{\rm los}$ as a single integral for specific anisotropy profiles, including the case of a uniform $\beta$, on which we focused in this work.

\subsubsection{MOND phenomenology}
\label{section:method_isolated_mond}

\begin{figure}
\centering
\includegraphics[width=1\linewidth,trim={1.8cm 0cm 3.75cm 0cm},clip]{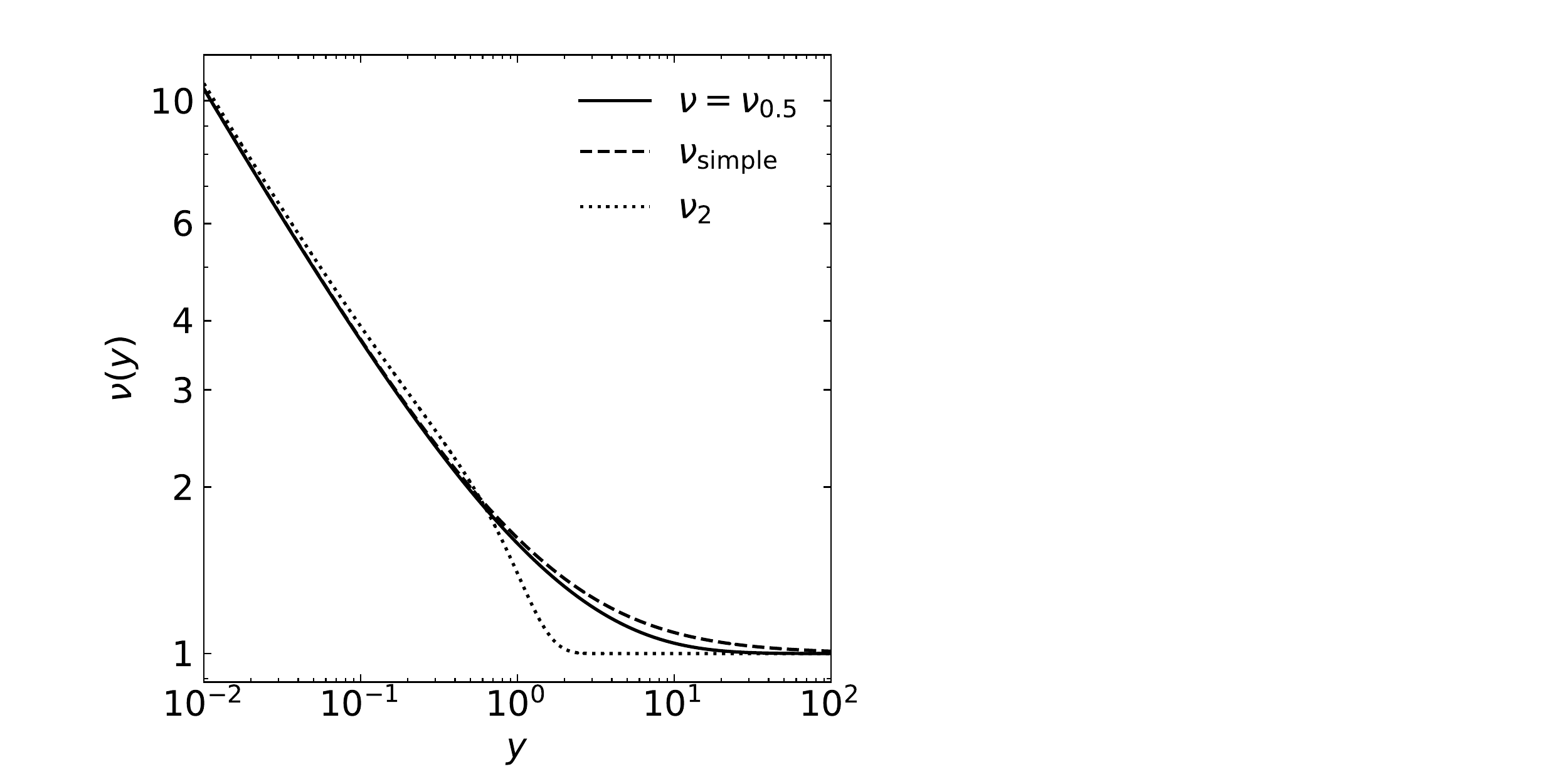}
\caption{Comparison between the three MOND interpolating functions considered in this work, namely $\nu$ from Eq.~(\ref{eq:nu}), the simple interpolating function from Eq.~(\ref{eq:nu_simple}), and $\nu_2$ from Eq.~(\ref{eq:nu_alpha}) with $\alpha=2$. The fiducial $\nu$ is very similar to the simple interpolating function. 
}
\label{fig:nu_functions}%
\end{figure}

In spherical symmetry, MOND modified gravity reduces to the following relation, sometimes dubbed Milgrom's relation: 
\be
\label{eq:MOND_nu}
\vec{g} = \nu\left(\frac{g_N}{a_0}\right) \vec{g_N}
,\ee
where $\nu$ is an interpolating function such that $\nu(y)\rightarrow 1$ for $y\gg 1$ and $\nu(y) \rightarrow y^{-1/2}$ for $y\ll 1,$ and $g_N$ is the Newtonian gravitational field deduced from the baryonic mass distribution \citep[e.g.][]{Famaey2012}. Although $y\nu(y)$ should be monotonically increasing for $g$ and $g_N$ to be equivocally determined, MOND allows some freedom in the shape of the interpolating function \citep{Milgrom1983a}. 
In the following, we adopt the function
\be
\label{eq:nu}
\nu(y) = \frac{1}{1-e^{-\sqrt{y}}} 
,\ee
corresponding to Eq.~(\ref{eq:RAR}), which  seems to be favoured by galaxy rotation curves \citep{Famaey2012}. 
However, \cite{Hees2016} noted that this function is excluded in the strong acceleration regime by Solar System data. Therefore, we also considered the more general family of functions \citep[e.g.][]{Hees2016}:
\be
\label{eq:nu_alpha}
\nu_\alpha(y) = \left(1-e^{-y^\alpha}\right)^{-1/2\alpha} + \left(1-1/2\alpha \right) e^{-y^\alpha} ,
\ee
for which $\alpha=0.5$ gives the previously adopted interpolating function, and for which $\alpha=2$ safely passes the Solar System constraints. We also consider the so-called simple interpolating function
\be
\label{eq:nu_simple}
\nu_{\rm simple}(y) = \frac{1+(1+4y^{-1})^{1/2}}{2}, 
\ee
which yields good fits in the intermediate and weak gravity regime of galaxies \citep{Famaey2005,Zhao2006, Sanders2007, Gentile2011}. We note that the adopted  interpolating function (Eq.~(\ref{eq:nu}) or Eq.~(\ref{eq:nu_alpha}) with $\alpha=0.5$) is very similar to the simple interpolating function in the intermediate and weak gravity regime, as can be seen in Fig.~\ref{fig:nu_functions}.

As in Section~\ref{section:RAR}, we assumed a fixed and uniform $\rm M/L$ and de-projected the two-dimensional Sérsic light profile using the semi-analytical approximation proposed by \cite{LimaNeto1999} into the spherical three-dimensional mass profile $M(r)$ given by Eq.~(\ref{eq:M}). 
Assuming the UDG to be isolated, the Newtonian acceleration is simply $g_N(r)=GM(r)/r^2$, from which we can derive the MOND acceleration $g$ using Eq.~(\ref{eq:MOND_nu}) and the line-of-sight velocity dispersion using the \cite{Mamon2005} formulas.

\begin{figure*}
\centering
\includegraphics[width=0.46\textwidth,trim={0cm 0cm 0cm 0cm},clip]{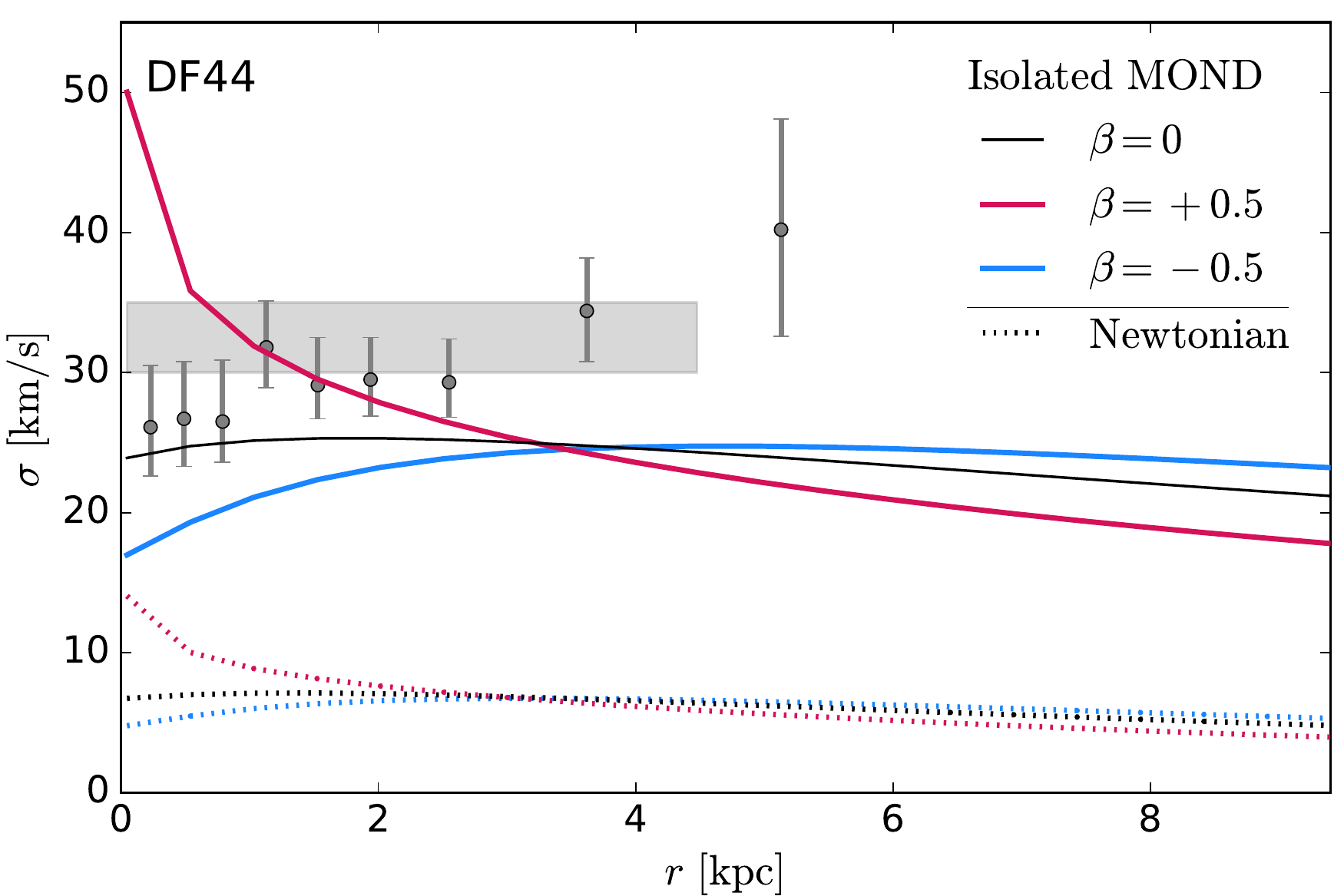}
\hfill
\includegraphics[width=0.46\textwidth,trim={0cm 0cm 0cm 0cm},clip]{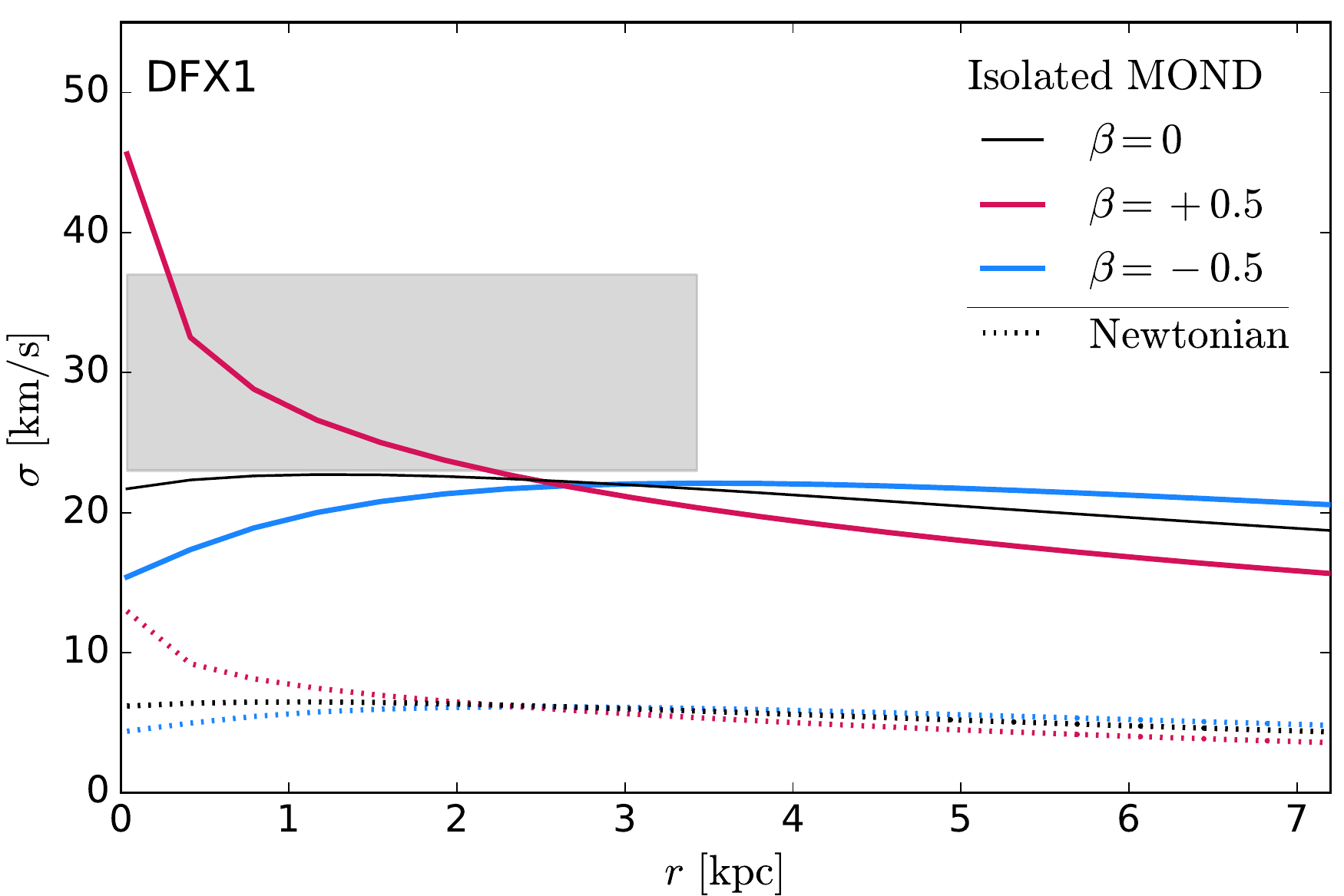}
\includegraphics[width=0.46\textwidth,trim={0cm 0cm 0cm 0cm},clip]{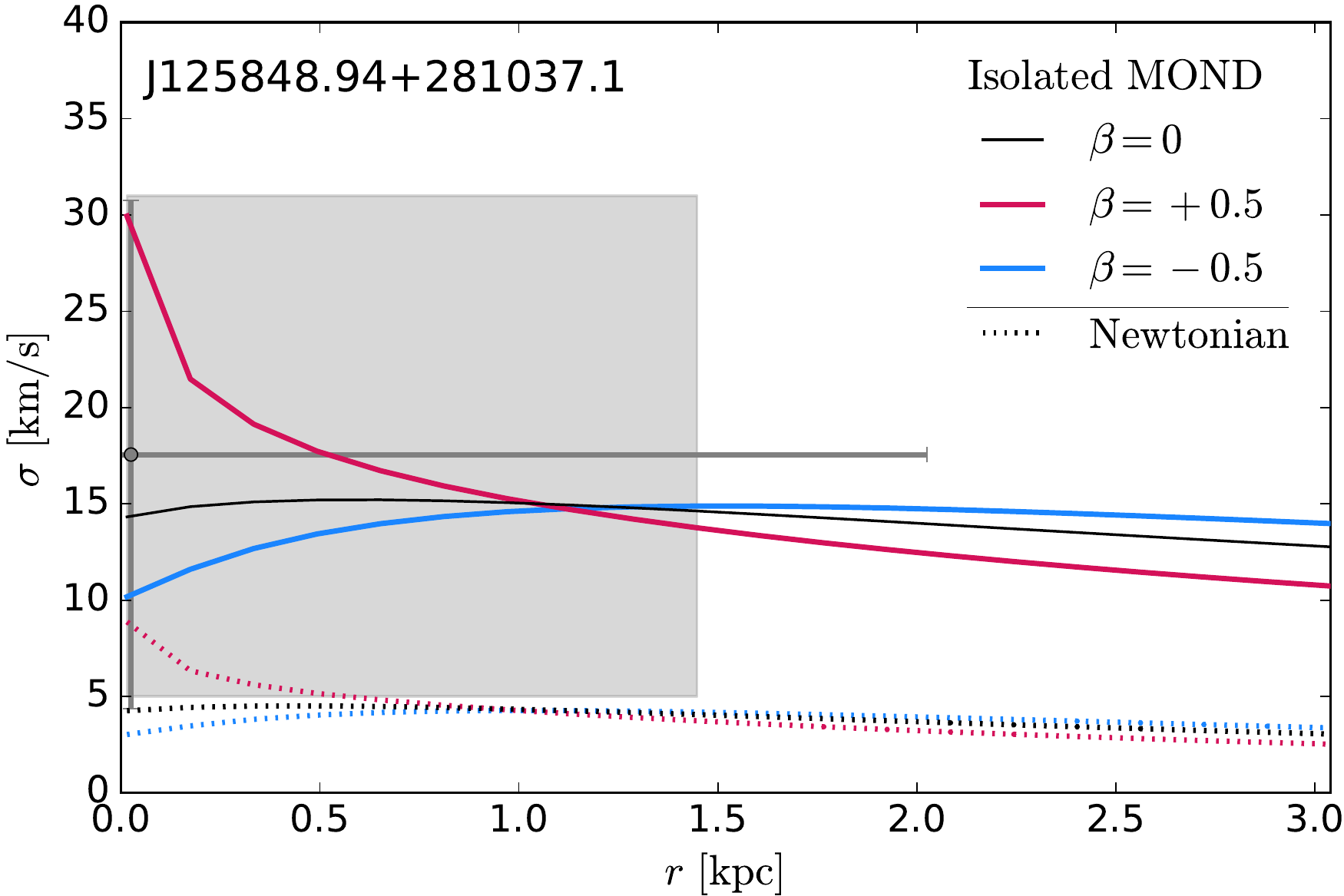}
\hfill
\includegraphics[width=0.46\textwidth,trim={0cm 0cm 0cm 0cm},clip]{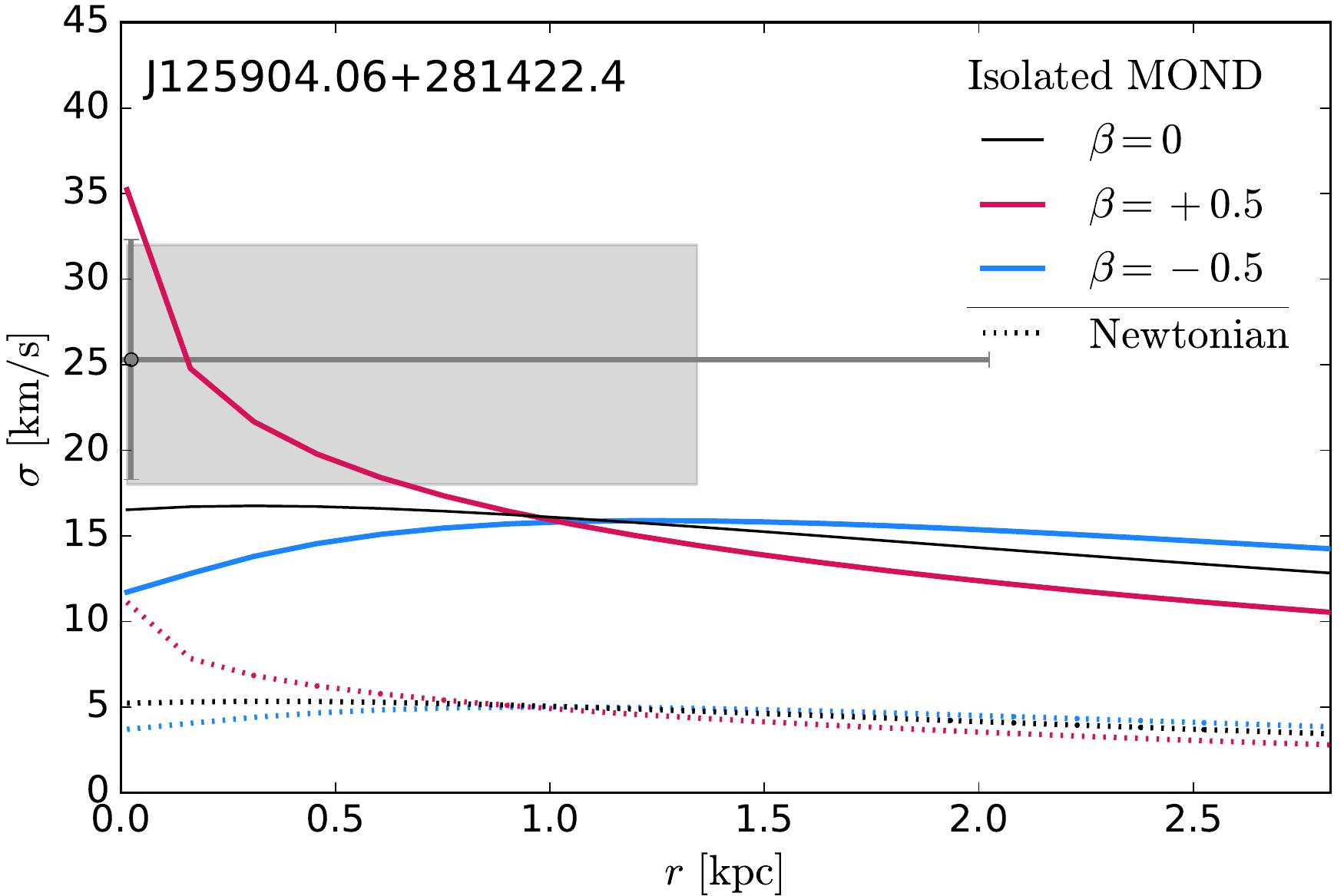}
\includegraphics[width=0.46\textwidth,trim={0cm 0cm 0cm 0cm},clip]{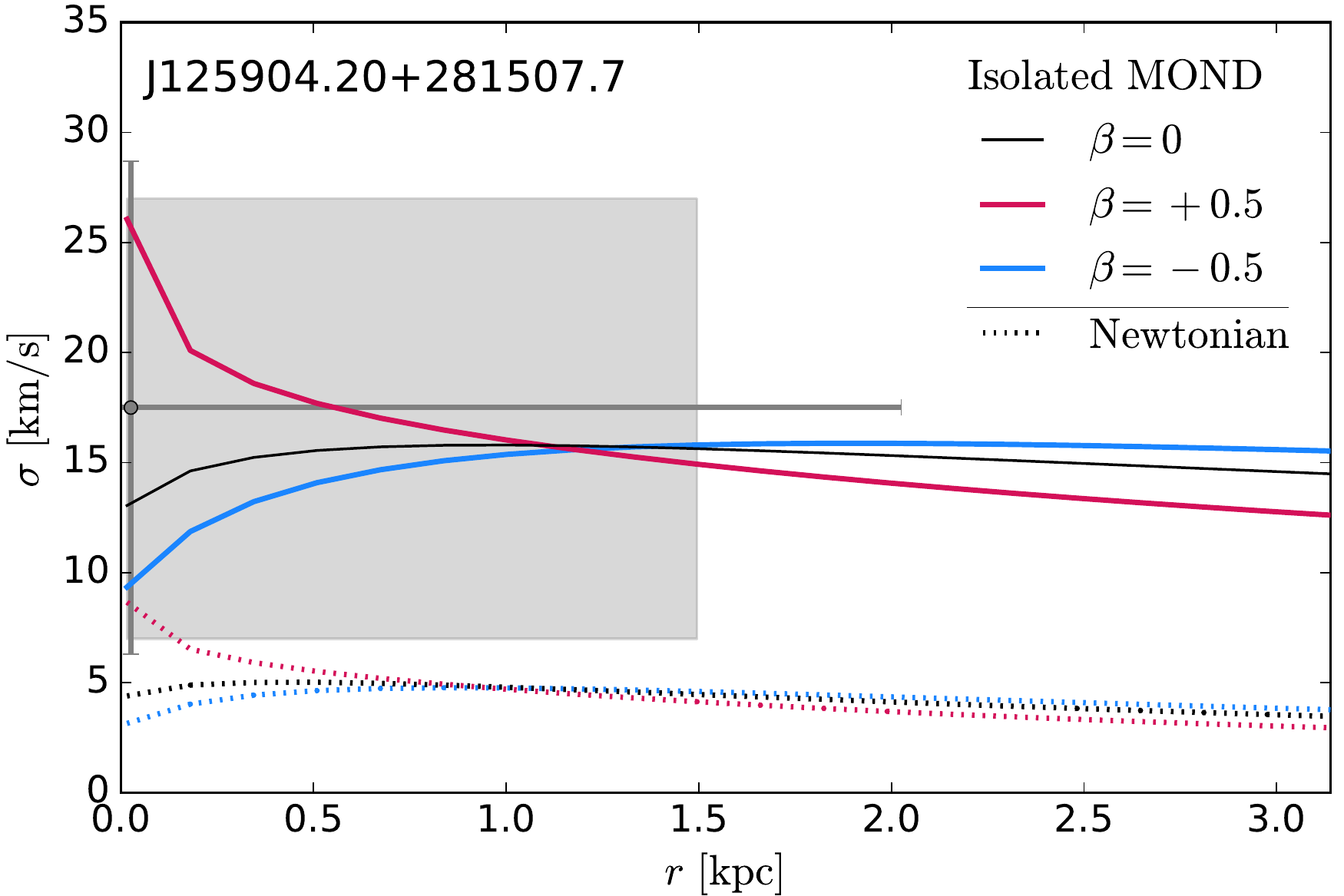}
\hfill
\includegraphics[width=0.46\textwidth,trim={0cm 0cm 0cm 0cm},clip]{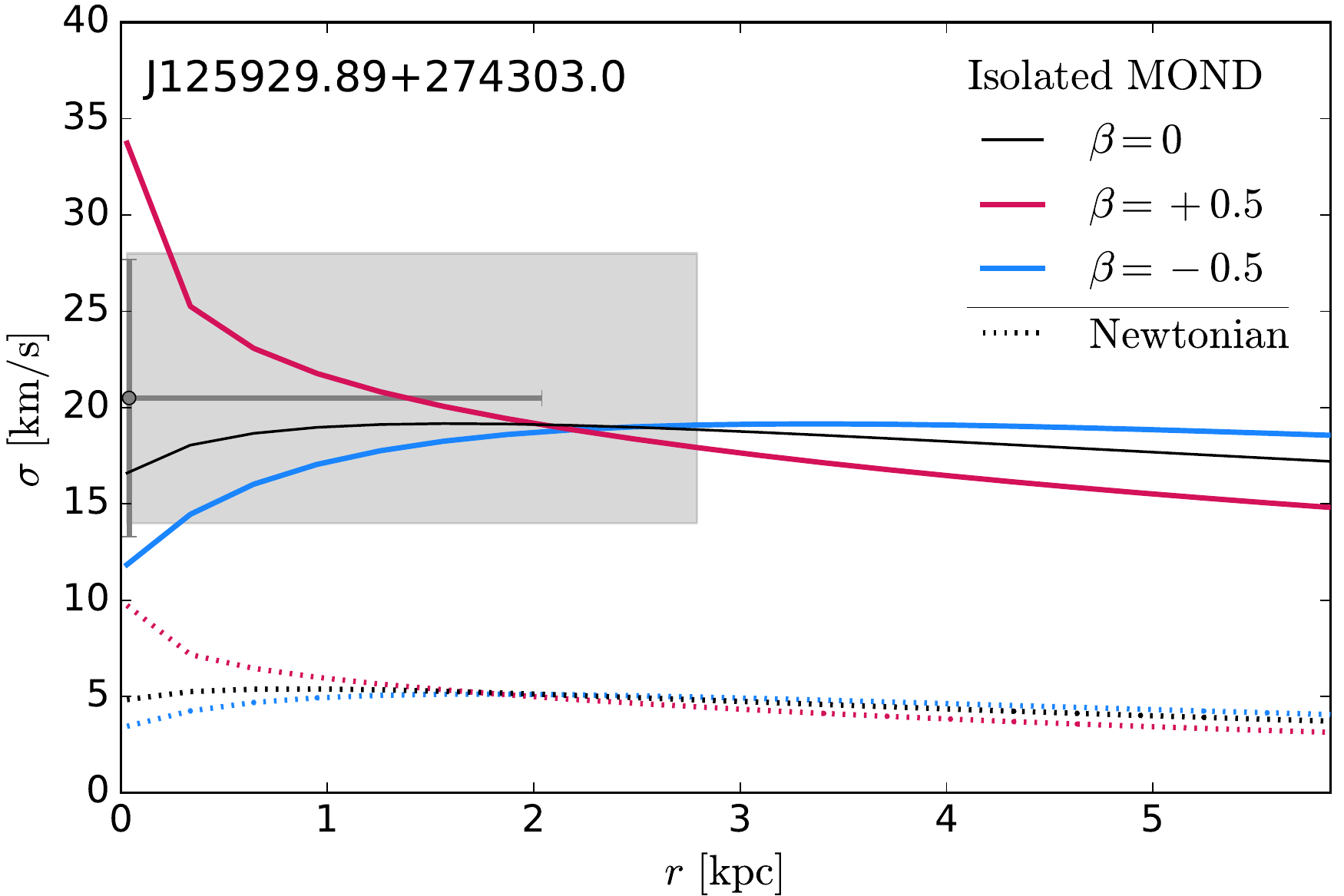}
\includegraphics[width=0.46\textwidth,trim={0cm 0cm 0cm 0cm},clip]{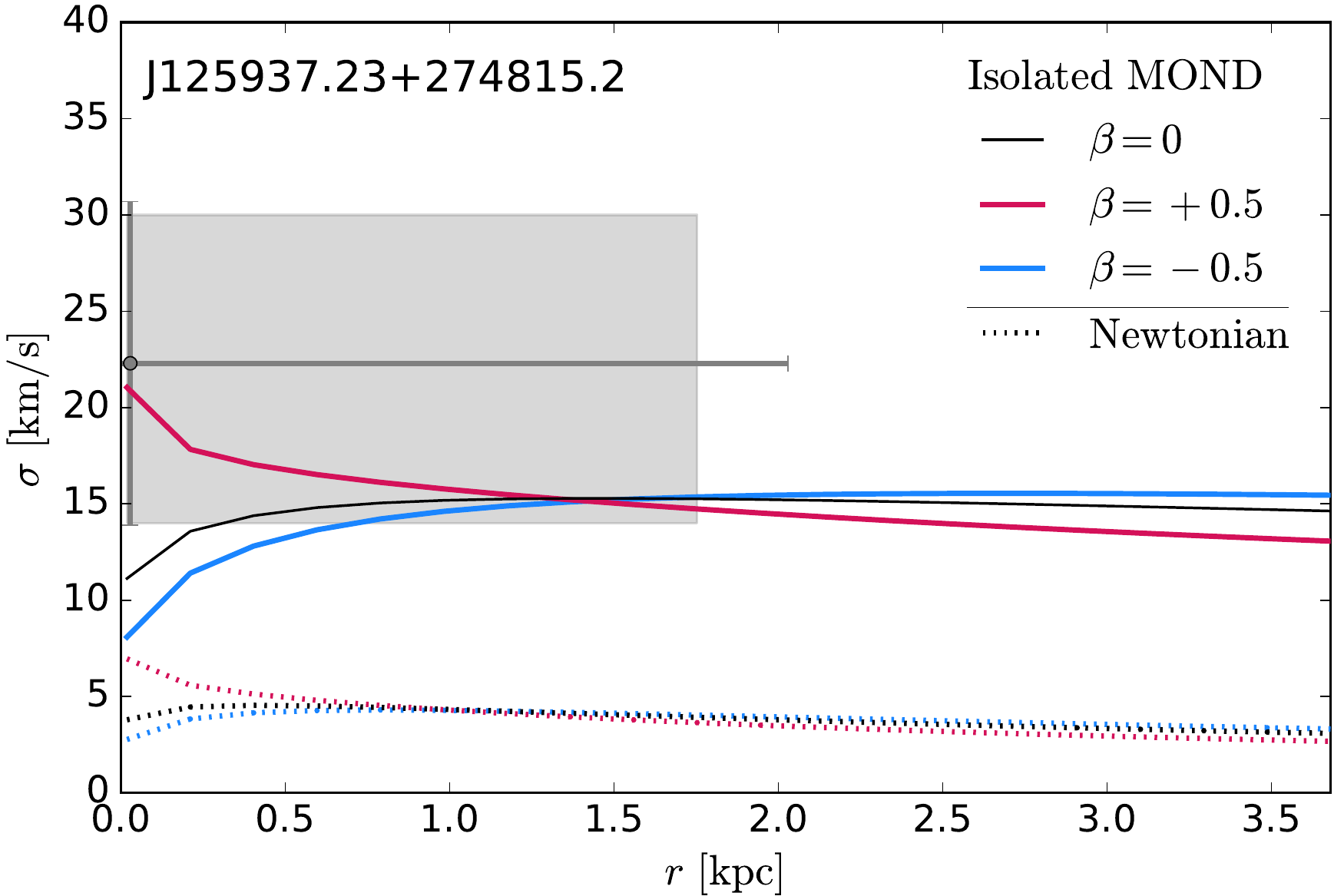}
\hfill
\includegraphics[width=0.46\textwidth,trim={0cm 0cm 0cm 0cm},clip]{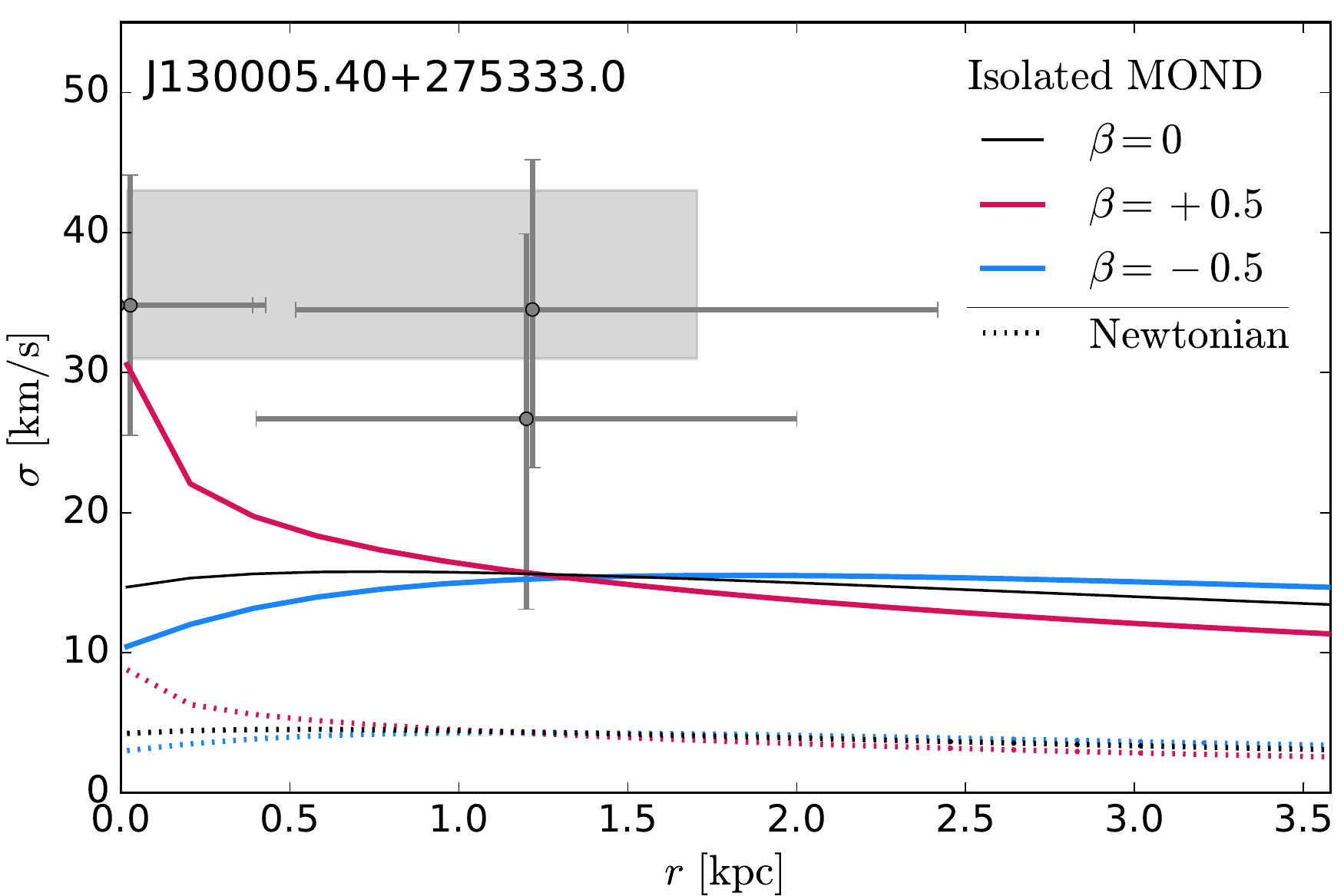}
\caption{
Comparison between the measured line-of-sight velocity dispersion of the sample UDGs (grey error bars and/or shaded area), the MOND model in isolation (plain lines), and the Newtonian model (dotted lines). Black corresponds to a uniform anisotropy parameter $\beta=0$ (isotropic), magenta to $\beta=+0.5$ (radial), and blue to $\beta=-0.5$ (tangential). 
Data points stem from \citet{vanDokkum2019b} for DF44, \citet{VanDokkum2017} for DFX1, and from \citet{Chilingarian2019} for the other UDGs. The shaded area corresponds to the effective velocity dispersion or, when not available (in the case of DFX1), central stellar velocity dispersion.
}
\label{fig:isoloated}
\end{figure*}

\setcounter{figure}{3}
\begin{figure*}
\centering
\includegraphics[width=0.46\textwidth,trim={0cm 0cm 0cm 0cm},clip]{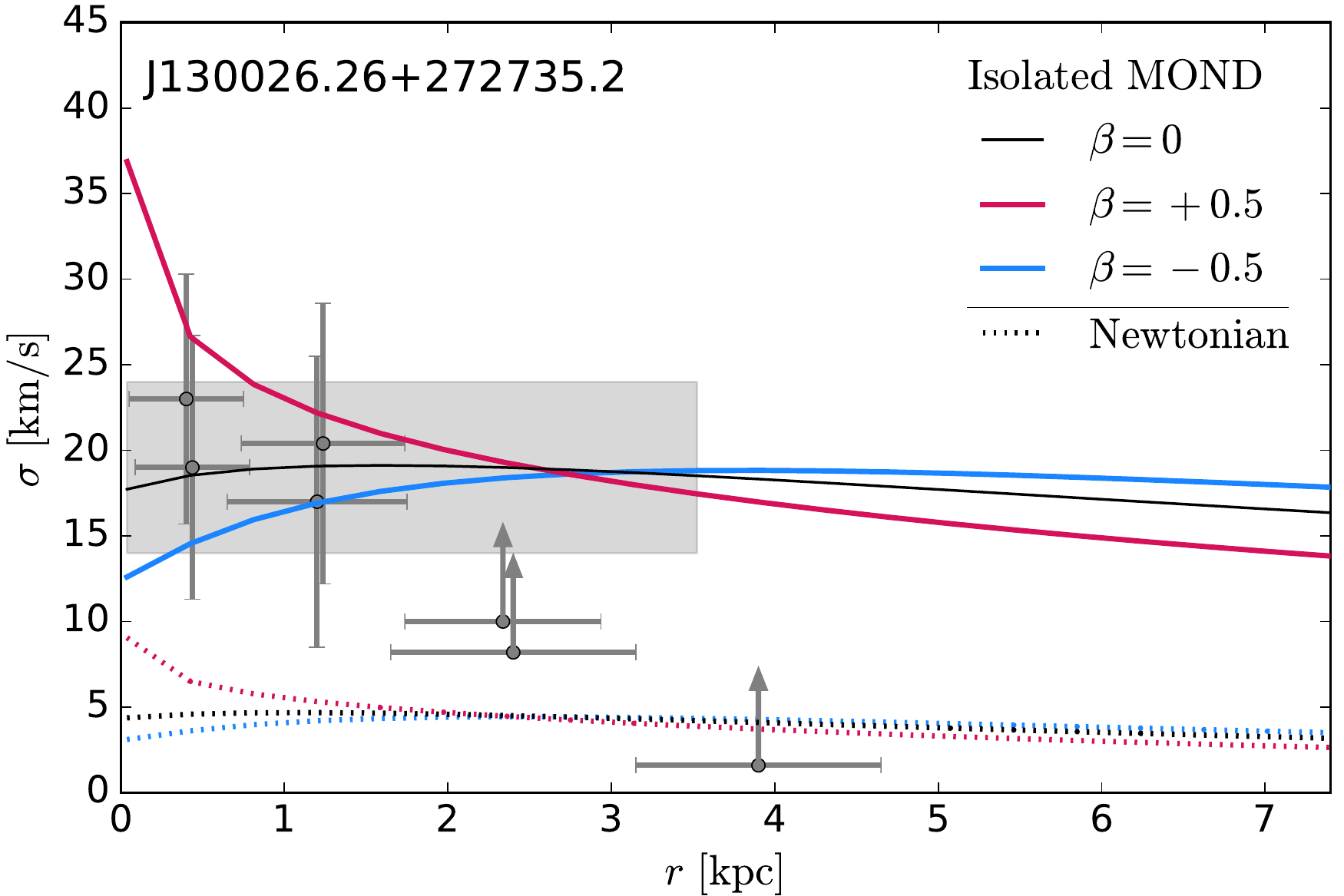}
\hfill
\includegraphics[width=0.46\textwidth,trim={0cm 0cm 0cm 0cm},clip]{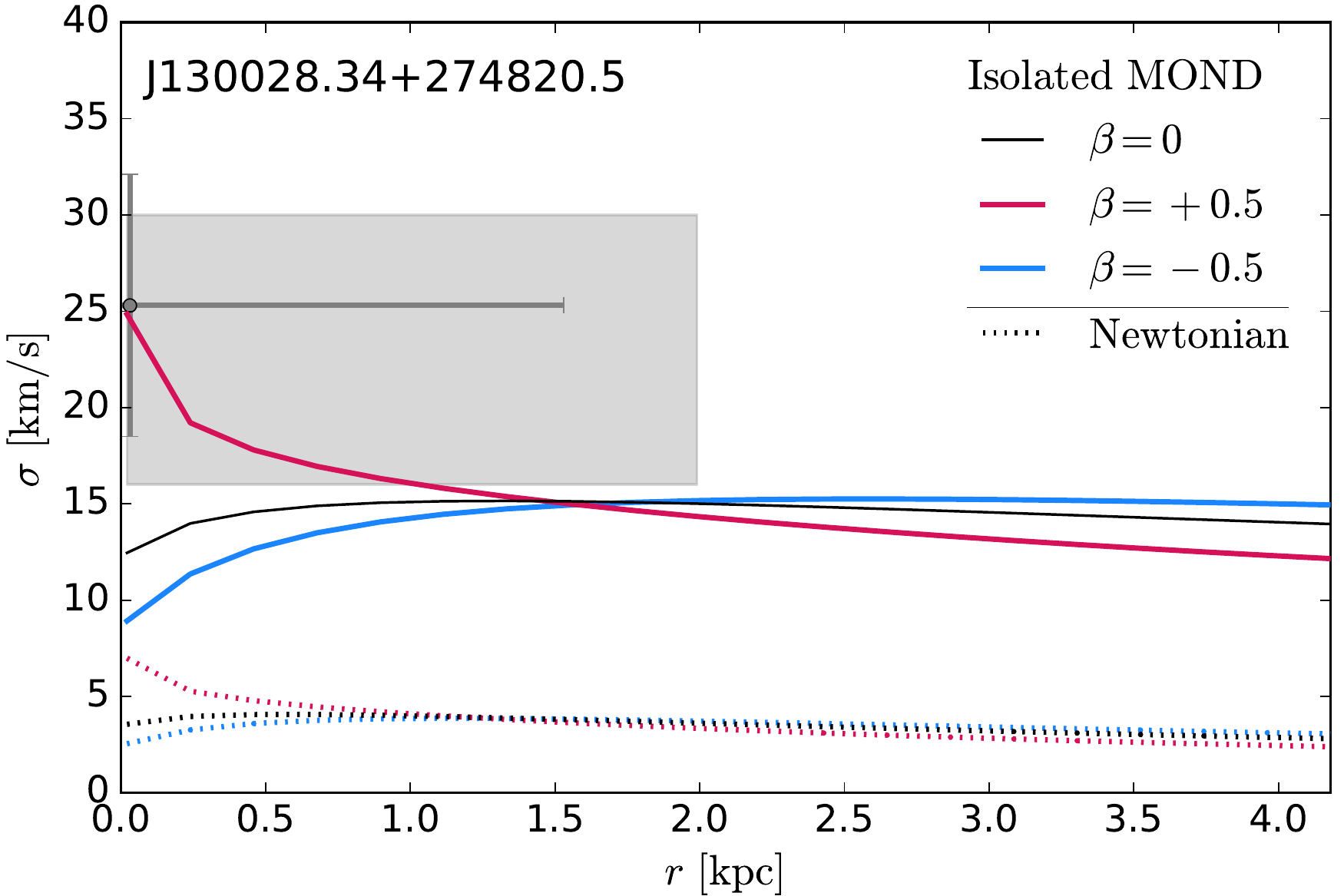}
\includegraphics[width=0.46\textwidth,trim={0cm 0cm 0cm 0cm},clip]{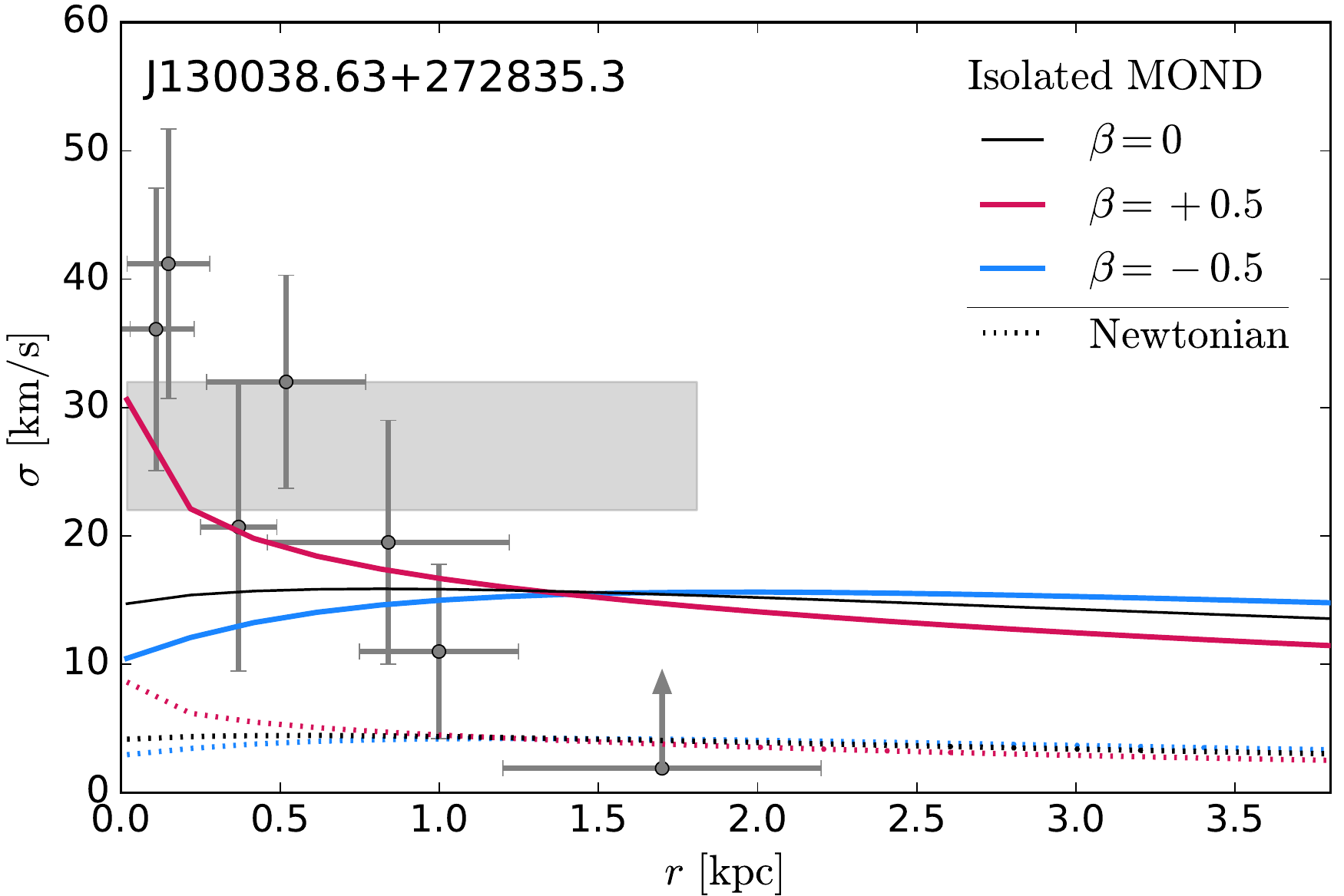}
\hfill \hspace{1cm}
\caption{
Continued. 
}
\label{fig:isolated2}%
\end{figure*}

\subsubsection{Resulting velocity dispersion profiles}
\label{section:isolated_profiles}

Figure~\ref{fig:isoloated} compares the observed velocity dispersion measurements with the MOND predictions in isolation given the stellar distribution, as outlined in the previous sections. For each UDG, we consider uniform anisotropy parameters $\beta=0$ (isotropic, in black), $\beta=+0.5$ (radially biased, in magenta), and $\beta=-0.5$ (tangentially biased, in blue). We also indicate the Newtonian prediction given the stellar distribution without dark matter (dotted lines). 
The interpolating function adopted was that of Eq.~(\ref{eq:nu}), namely $\nu_{0.5}$ in the terminology of Eq.~(\ref{eq:nu_alpha}), but we also considered $\nu_2$ and the simple interpolating function of Eq.~(\ref{eq:nu_simple}), yielding similar results because these galaxies are all in the weak-acceleration regime where all interpolating functions are similar.  
Given that there is no free parameter involved on each curve, the figure shows a remarkably good agreement between the observed velocity dispersion measurements and the isolated MOND models. The only parameter that has been changed from one curve to the other is the anisotropy, and we note that the agreement is even better for a slight radial anisotropy, $\beta=+0.5$. 
Such a radial anisotropy may inform us about the formation of these galaxies in MOND (cf. Section~\ref{section:survivor}). 
As expected, the Newtonian models without dark matter largely underestimate the velocity dispersions, for example by more than 3$\sigma$ in the case of DF44, as was already reported by \citet{Bilek2019}. 
We noted that four galaxies display some -- or a hint of -- rotation (cf. Section~\ref{section:10_additionnal}). Taking this into account in detail would largely complicate the present analysis, but the magnitude of the correction would typically only be of the order of $1~\rm km~s^{-1}$ or less. In other words, given the level of rotation hinted at in these four galaxies, the same galaxies without rotation would typically see an increase of their velocity dispersion data by about $1~\rm km~s^{-1}$, which would obviously not change our conclusions.

%-------------------------------------------------------------------

\section{MOND models with external field effect}
\label{section:EFE}

\subsection{Mass and UDG distributions of the Coma cluster}
\label{section:coma}

\begin{figure}
\centering
\includegraphics[width=1\linewidth,trim={1.6cm 0cm 3.55cm 0cm},clip]{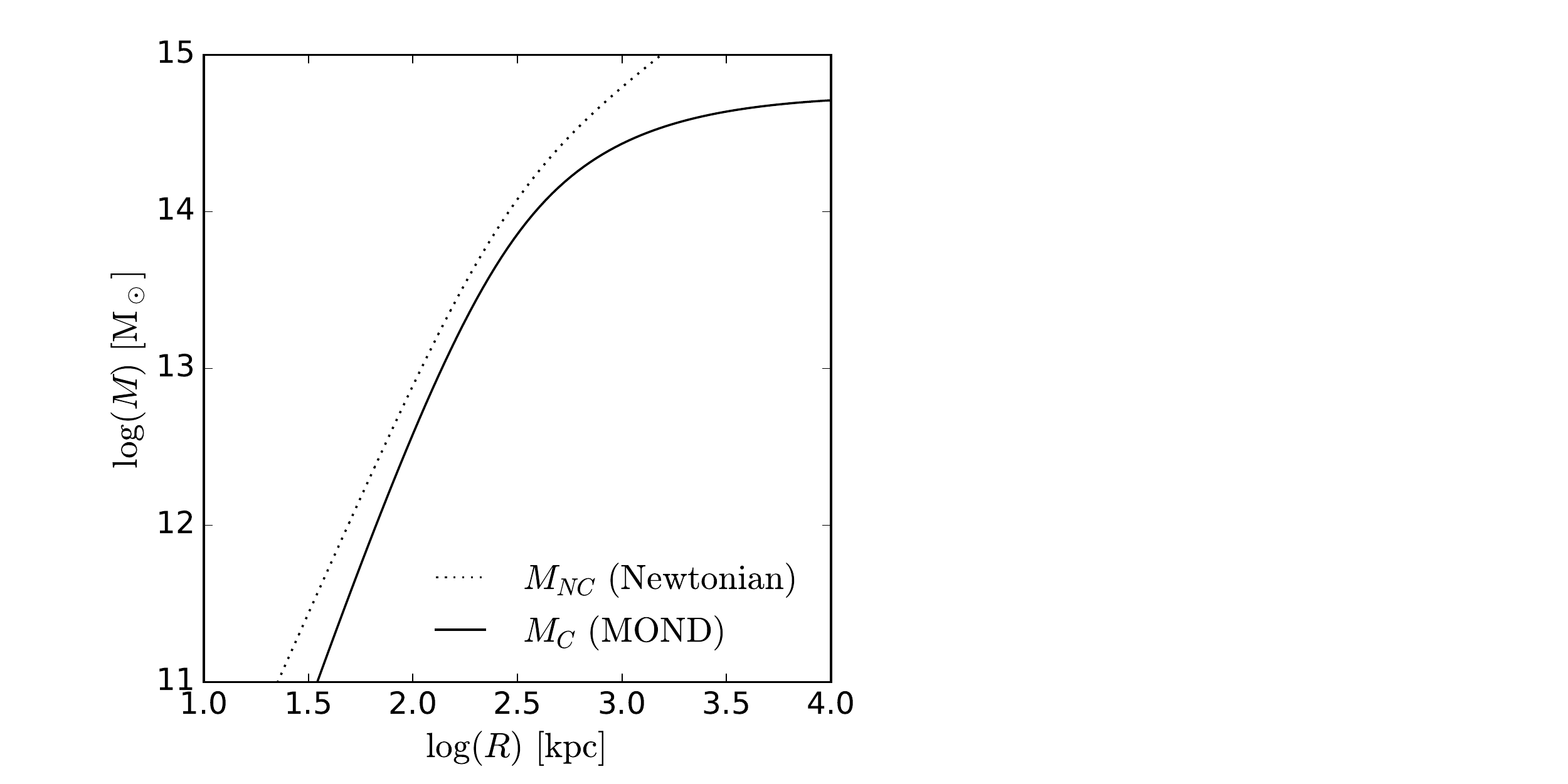}
\caption{
Coma cluster cumulative dynamical mass derived by \protect\cite{Sanders2003} from the hydrostatic equilibrium of the X-ray-emitting hot gas in Newtonian dynamics (dotted black line) and in MOND (plain black line). While the Newtonian dynamical mass ($M_{NC}$) continues to increase above $\rm 1~Mpc$, the MOND mass ($M_C$) converges towards $5.6\times 10^{14}~\rm M_\odot$. 
}
\label{fig:coma_mass}%
\end{figure}

To assess the MOND velocity dispersion with EFE for the Coma cluster UDGs, one first needs to estimate the external gravitational field they inhabit, stemming from the cluster mass distribution. We derived the mass profile $M_C(R)$ of the Coma cluster within MOND as in \cite{Sanders2003} from the hydrostatic equilibrium of the X-ray-emitting hot gas, assumed to be isothermal. 
Using the simple interpolation function (Eq.~(\ref{eq:nu_simple})) instead of that used by \cite{Sanders2003} in their Eq.~(8), since Fig.~\ref{fig:nu_functions} shows how closely the simple interpolation function follows our fiducial interpolation function (Eq.~(\ref{eq:nu})), the mass profile yields 
\be
\label{eq:MC}
M_{C}(R) = \frac{a_C}{a_C+a_0} M_{NC}, 
\ee
where
\be
\label{eq:aC}
a_{C}(R) = - \frac{kT_C}{\mu m_p} \frac{1}{R} 
 \frac{{\rm d} \ln \rho_{C} }{{\rm d} \ln R}
\ee
is the gravitational acceleration, 
\be
\label{eq:MNC}
M_{NC}(R) =  - \frac{kT_C}{G \mu m_p} R \frac{{\rm d} \ln \rho_{C} }{{\rm d} \ln R}
\ee
is the Newtonian mass profile,
and the intracluster X-ray-emitting hot gas density is described by a $\beta$-model:\be
\label{eq:rho_c}
\rho_{C}(R) =\rho_{C0} \left[ 1+\left(\frac{R}{r_C}\right)^2\right]^{-1.5\beta_C}
\ee
with an exponent $\beta_C=0.71$, a temperature $kT_C=8.6~\rm keV$, a mean atomic weight $\mu=0.61$, a characteristic radius $r_C=276~\rm kpc$, a central density $\rho_{C0} = 9.0 \times 10^4~\rm M_\odot~\! kpc^{-3}$, and $m_p$ being the proton mass \citep{Reiprich2001}.
Fig.~\ref{fig:coma_mass} shows the corresponding mass profile. 
The external Newtonian acceleration at a distance $R$ from the centre of the Coma cluster is $g_{Ne}(R)=-G M_C(R)/R^2$. 

UDGs are observed at given projected distances, but their actual distances from the cluster centre may be significantly larger. \cite{vanderBurg2016} shows that the projected radial distribution of UDGs in clusters is generally steeper than NFW at the outskirts, flatter than NFW towards the centre, and on average well fit with an \cite{Einasto1965} profile: 
\be
\rho_{\rm Ein}(R) = \rho_{-2} \exp \left\{ -\frac{2}{\alpha_{\rm Ein}} \left[\left(\frac{R}{r_{-2}}\right)^{\alpha_{\rm Ein}}-1 \right]\right\}
,\ee
with concentration $c_{\rm Ein}\equiv R_{200}/r_{-2}=1.83$ and exponent $\alpha_{\rm Ein}=0.92$. Assuming that Coma cluster UDGs follow this distribution enables one to retrieve the most probable distance and the average distance of a UDG from the cluster centre given its projected distance. The probability for an object at projected distance $d$ to be between $R$ and $R+dR$ given the three-dimensional distribution $\rho_{\rm Ein}$ is
\be
dp \propto \frac{\rho_{\rm Ein}(R)}{\sqrt{1-\nicefrac{d^2}{R^2}}} dR, 
\ee 
from which the mode and the average distance $d_{\rm mean} = \int_{d}^{+\infty} R dp$ can be derived. We note that the average distance $d_{\rm mean}$ is always larger than the mode, both being larger than the projected distance $d$. The resulting average distances are indicated in Table~\ref{table:sample}. 
We used $R_{200}=2.94\rm ~Mpc$ \citep{Kubo2007}. 

\begin{figure}
\centering
\includegraphics[width=1\linewidth,trim={2.2cm 0cm 3.cm 0cm},clip]{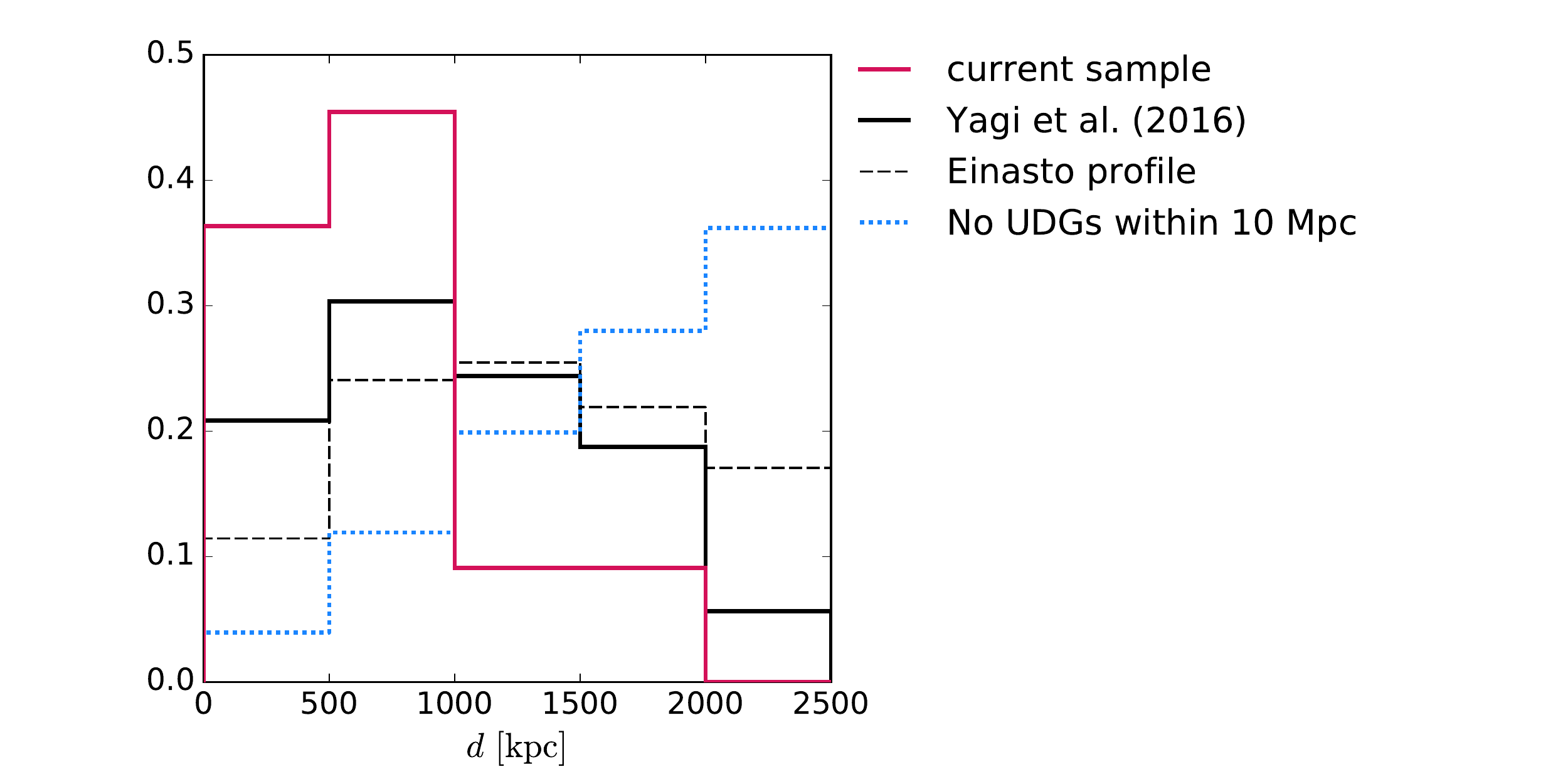}
\caption{
Distribution of the projected distance of the sample UDGs (plain red) and of the parent \protect\citet{Yagi2016} Coma cluster UDG catalogue (plain black) compared to the average Einasto profile derived by \protect\citet[][dashed black]{vanderBurg2016} and a distribution with no UDGs within 10 Mpc (dotted blue). A constant UDG density within 10 Mpc would closely follow the latter, given the x-axis range. The  \protect\citet{vanderBurg2016} Einasto distribution provides a reasonable fit to the Coma cluster UDGs; a distribution without UDGs within 10 Mpc or with a constant UDG density are  disfavoured. 
}
\label{fig:distance}%
\end{figure}

In Fig.~\ref{fig:distance}, we show the projected distance distribution of the sample UDGs (magenta) together with that of the parent Coma cluster UDG population from the \cite{Yagi2016} catalogue (black) and that corresponding to the \cite{vanderBurg2016} Einasto distribution (dashed black), which was obtained for a sample of eight nearby galaxy clusters excluding Coma. The Einasto distribution provides a reasonable fit to the \cite{Yagi2016} distribution, hence validating our estimate of the average distance $d_{\rm mean}$. For comparison, the figure also shows the distribution that would result from having no UDGs below 10 Mpc and a UDG density decreasing as $1/r^3$ beyond (dotted blue). The increase of this latter curve is dominated by the increase of the area spanned by the linearly spaced radial bins. A uniform distribution within 10 Mpc would thus yield a similar curve in the regime shown in the plot ($d<2.5\rm ~Mpc$). The discrepancy between the two black curves and the dotted blue curve together with the \cite{vanderBurg2016} Einasto distribution generally advocates for the UDG distribution to peak towards the centre rather than being pushed beyond 10 Mpc or remaining constant.

\subsection{An analytic formula for the EFE in QUMOND}
\label{section:formula}

In Newtonian dynamics, the internal dynamics of a self-gravitating system embedded in a constant gravitational field does not depend on the external field strength because of linearity. This is not the case in MOND, where non-Newtonian effects appear where the absolute values of the internal acceleration $g$ and of the external acceleration  $g_e$ are both less than $a_0$. 
In MOND, a system with $g_e<g<a_0$ presents a deep MOND behaviour, a system with $g<a_0<g_e$ is Newtonian, and a system with $g<g_e<a_0$ is Newtonian with a re-normalised gravitational constant, depending on $g_e$.  
\citet[][Eqs. (59) and (60),]{Famaey2012} provide a formula expressing the MOND internal gravitational field $g$ as a function of the internal Newtonian field $g_N$ and the external field $g_e$ (or its Newtonian counterpart $g_{Ne}$) in one dimension, i.e., when $g$ and $g_e$ are aligned, namely, for the case of QUMOND:
\begin{equation}
\label{eq:g_parallel}
g_\parallel=\nu\left(\frac{g_N+g_{Ne}}{a_0}\right) g_N  +\left[ \nu\left(\frac{g_N+g_{Ne}}{a_0}\right) -\nu\left(\frac{g_{Ne}}{a_0}\right)\right] g_{Ne}. 
\end{equation}
It turns out that simulations within the \citet{Bekenstein1984} version of MOND tend to agree well with this estimate \citep{Haghi2019}, which yields a rather strong effect of the external field on the internal dynamics. However, in the QUMOND version \citep{Milgrom2010}, this one-dimensional formula would  a priori appear to be an upper limit on the impact of the EFE.

In the context of QUMOND, we hereafter propose a simple approximate analytical formula for spherical systems by averaging the typical effect over the sphere. We validate the formula using a numerical Poisson solver in QUMOND in Appendix~\ref{section:EFE_numerical}. 
We consider a spherical system, characterised by a internal gravitational field $\vec{g}$ embedded in a constant external field $\vec{g_e}$. 
Following Eq.~(\ref{eq:MOND_nu}) in the QUMOND formalism, the external field can be expressed in terms of its Newtonian counterpart $\vec{g_{Ne}}$ through
\be
\vec{g_e} = \nu\left(\frac{g_{Ne}}{a_0}\right) \vec{g_{Ne}}, 
\ee
while the total field yields
\be
\vec{g} + \vec{g_e} = \nu\left(\frac{|\vec{g_{N}}+\vec{g_{Ne}}|}{a_0}\right) \left( \vec{g_{N}} + \vec{g_{Ne}}\right), 
\ee
where $\vec{g_{N}}$ is the Newtonian internal field. 
Hence, the internal gravitational field
\be
\label{eq:gvec}
\vec{g} =  \nu\left(\frac{|\vec{g_{N}}+\vec{g_{Ne}}|}{a_0}\right) \vec{g_N} +\left[\nu\left(\frac{|\vec{g_{N}}+\vec{g_{Ne}}|}{a_0}\right)-\nu\left(\frac{g_{Ne}}{a_0}\right) \right] \vec{g_{Ne}}, 
\ee
whose intensity depends on the position on the sphere. 
\begin{figure}
\centering
\includegraphics[width=0.7\linewidth,trim={0cm 0.5cm 0cm 0.5cm},clip]{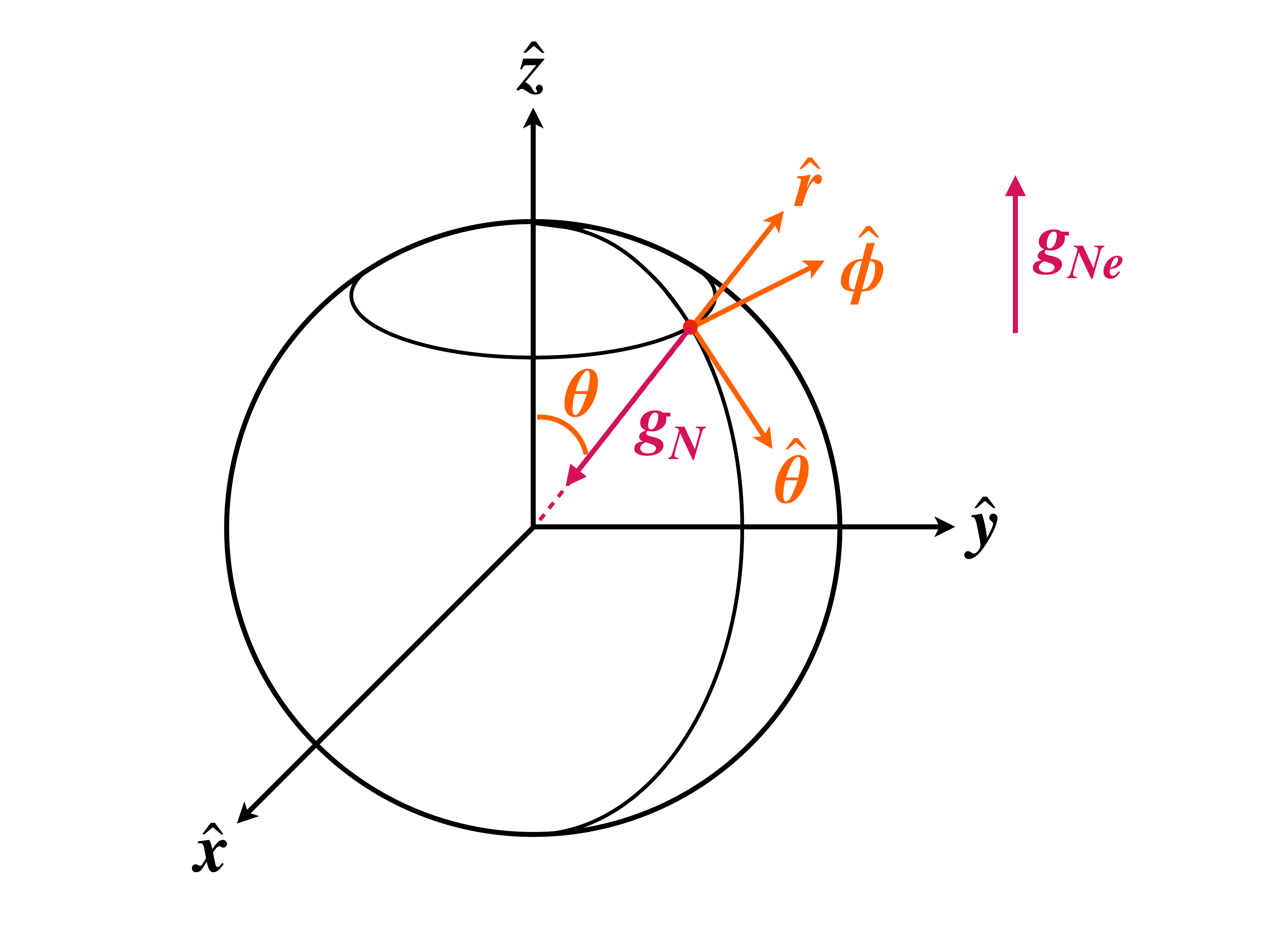}
\caption{
Spherical self-gravitating system embedded in a constant external field.
}
\label{fig:coordinates}%
\end{figure}
Using the standard spherical coordinates $(\vec{\hat{r}},\vec{\hat{\theta}},\vec{\hat{\phi}})$ such that $\vec{g_{N}}=-g_N ~\vec{\hat{r}}$ and $\vec{g_{Ne}}=g_{Ne}\vec{\hat{z}}$ with $\vec{\hat{z}}=\cos\theta~\vec{\hat{r}} -\sin\theta ~\vec{\hat{\theta}}$, as shown in Fig.~\ref{fig:coordinates}, we have
\be
\label{eq:gnorm_cos}
|\vec{g_{N}}+\vec{g_{Ne}}|=\sqrt{g_N^2 + g_{Ne}^2 -2 g_N g_{Ne}\cos \theta}
\ee
and
\begin{multline}
\label{eq:g_vec}
\vec{g} = \left[-\nu\left(\frac{|\vec{g_{N}}\!+\!\vec{g_{Ne}}|}{a_0}\right) g_N \!+\! \left[\nu\left(\frac{|\vec{g_{N}}\!+\!\vec{g_{Ne}}|}{a_0}\right)-\nu\left(\frac{g_{Ne}}{a_0}\right) \right] g_{Ne} \cos \theta \right] \vec{\hat{r}}\\
\!-\!\left[ \nu\left(\frac{|\vec{g_{N}}\!+\!\vec{g_{Ne}}|}{a_0}\right)-\nu\left(\frac{g_{Ne}}{a_0}\right)\right] g_{Ne} \sin \theta ~\vec{\hat{\theta}}. 
\end{multline}
We now average this expression over the sphere of a given radius $r$ and approximate $\langle \nu(x)\rangle$ by $\nu(\langle x\rangle)$. 
Integrating over the sphere yields an average
\be
\label{eq:avnorm}
\langle |\vec{g_{N}}+\vec{g_{Ne}}| \rangle=\frac{|g_N+g_{Ne}|^3-|g_N-g_{Ne}|^3}{6g_N g_{Ne}}. 
\ee
This norm yields $\langle |\vec{g_{N}}+\vec{g_{Ne}}| \rangle = g_N + g_{Ne}^2/3g_N$ when $g_N\geq g_{Ne}$ and $\langle |\vec{g_{N}}+\vec{g_{Ne}}| \rangle = g_{Ne} + g_{N}^2/3g_{Ne}$ when $g_N\leq g_{Ne}$. 
Approximating 
\be
\nu\left(\frac{|\vec{g_{N}}\!+\!\vec{g_{Ne}}|}{a_0}\right) \approx \nu \left(\frac{\langle |\vec{g_{N}}\!+\!\vec{g_{Ne}}|\rangle}{a_0}\right),
\ee
we obtain the average radial and tangential internal gravitational fields 
\be
\label{eq:gr_norm}
\langle g_r \rangle = \nu\left(\frac{\langle |\vec{g_{N}}\!+\!\vec{g_{Ne}}|\rangle}{a_0}\right) g_N
\ee
and
\be
\langle g_\theta \rangle = \frac{\pi}{4} \left[ \nu\left(\frac{\langle |\vec{g_{N}}\!+\!\vec{g_{Ne}}|\rangle}{a_0}\right)-\nu\left(\frac{g_{Ne}}{a_0}\right)\right] g_{Ne}. 
\ee
The expression for $\langle g_r \rangle$ yields 
\be
\label{eq:gr}
\langle g_r \rangle = 
\begin{cases}
\displaystyle \nu\left( \frac{g_N}{a_0}+\frac{g_{Ne}^2}{3g_N a_0}\right) g_N ~~~~~{\rm when}~~ g_N\geq g_{Ne} \\
\displaystyle \nu\left( \frac{g_{Ne}}{a_0}+\frac{g_N^2}{3g_{Ne} a_0}\right) g_N ~~{\rm when}~~ g_N\leq g_{Ne},
\end{cases}
\ee
from which we can retrieve the MOND behaviour when \mbox{$g_{Ne}<g_N<a_0$} and the  Newtonian behaviour with a re-normalised gravitational constant when $g_{N}<g_{Ne}<a_0$.
As already noted, for instance by \cite{Banik2015}, an unusual consequence of the MOND EFE is that the gravitational attraction due to a given mass may not always be directed towards it. 
The expression for $\langle g_r \rangle$ can be injected in the Jeans equation to retrieve the average velocity dispersion at any radius. 

Equations~(\ref{eq:gvec}) and (\ref{eq:gnorm_cos}) with $\cos \theta=0$ further enable us to express the internal gravitational field in the case where $g$ and $g_e$ are perpendicular: 
\begin{equation}
\label{eq:g_perp}
g_\perp= \sqrt{\nu^2\left( \frac{\sqrt{g_N^2+g_{Ne}^2}}{a_0} \right) g_N^2 + \left( \nu\left(\frac{\sqrt{g_N^2+g_{Ne}^2}}{a_0}\right) -\nu\left(\frac{g_{Ne}}{a_0}\right)\right)^2 g_{Ne}^2}. 
\end{equation}
The second term in the square root of Eq.~(\ref{eq:g_perp}) is small compared to the first one, such that
\begin{equation}
\label{eq:g_perp_approx}
g_\perp\approx \nu\left( \frac{\sqrt{g_N^2+g_{Ne}^2}}{a_0} \right) g_N. 
\end{equation}
Given that the form of Eq.~(\ref{eq:gr_norm}), 
$\langle g_r\rangle$ is close to $g_\perp$: the average radial acceleration field approaches the perpendicular one. Thus, either Eq.~(\ref{eq:gr}) or Eq. (\ref{eq:g_perp_approx}) can be used to retrieve the internal acceleration field with EFE in QUMOND.

\begin{figure}
\centering
\includegraphics[width=1\linewidth,trim={0cm 0cm 0cm 0cm},clip]{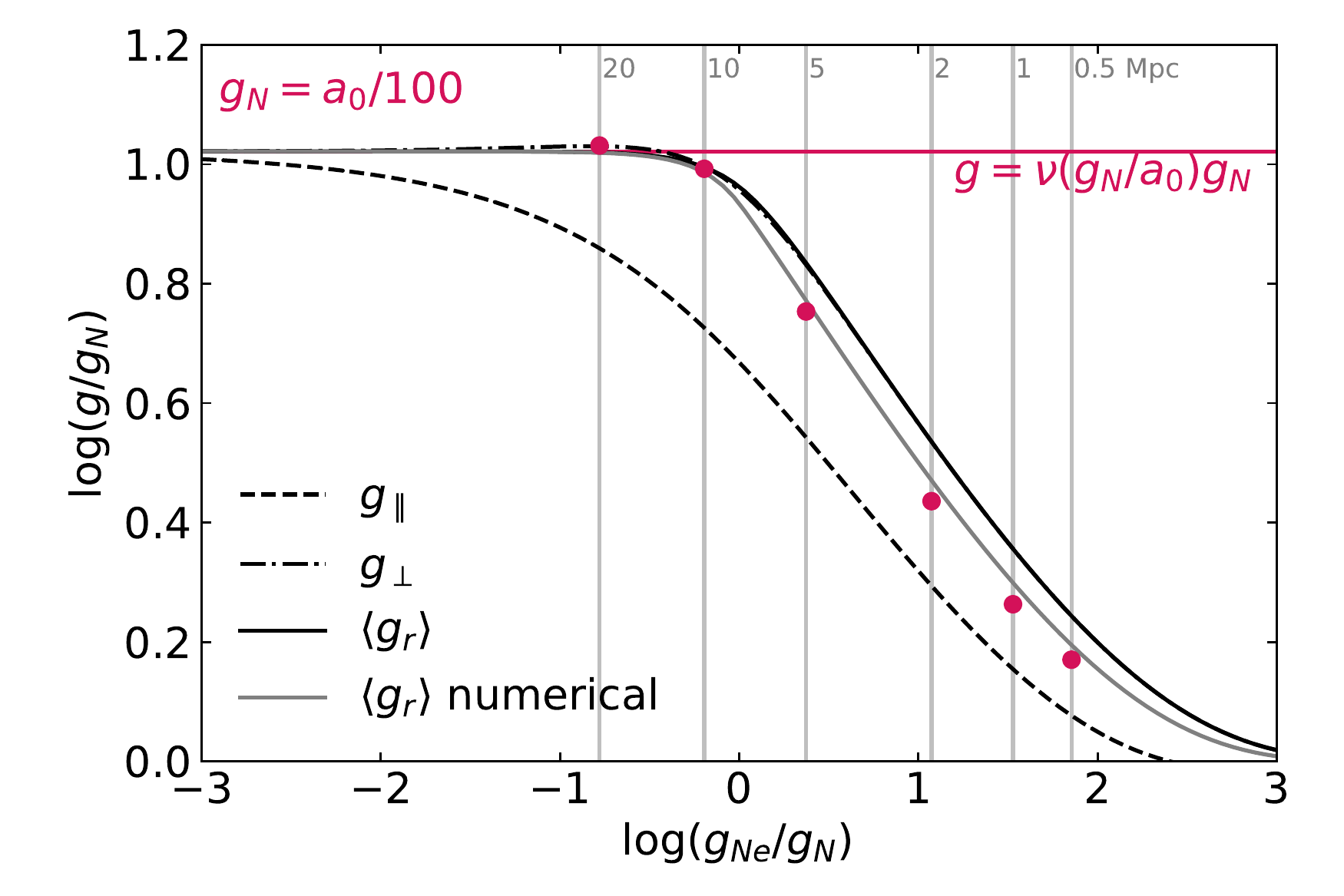}
\caption{MOND gravitational acceleration $g$ stemming from the different analytic formulae as a function of the Newtonian external field $g_{Ne}$, for a given internal Newtonian acceleration $g_N=a_0/100$. The magenta data points correspond to the radial acceleration field where $g_N=a_0/100$ in the QUMOND numerical integrations plotted in Fig.~\ref{fig:EFE_formula_g}, which consider a Plummer sphere at distances $d=$ 0.5, 1, 2, 5, 10, and 20 Mpc (from right to left) from a point mass representing the Coma cluster. We refer the reader to  Fig.~\ref{fig:EFE_formula_g} in the Appendix for the values of $g_{Ne}$ corresponding to each distance $d$ within the Coma cluster. The grey line corresponds to a numerical integration of the average radial acceleration of Eq.~(\ref{eq:g_vec}). }
\label{fig:EFE_formula_ggn}
\end{figure}

Figure~\ref{fig:EFE_formula_ggn} shows the behaviour of the different EFE formulae as a function of the Newtonian external field, for a given typical internal Newtonian acceleration $g_N = a_0/100$ (cf. Fig.~\ref{fig:EFE_formula_g}). 
The different formulas transition from the deep MOND regime ($g=\nu(g_N/a_0) g_N$, highlighted as a magenta horizontal line) to the Newtonian regime ($g=g_N$). 
Since $g_\parallel$ assumes the fields to be aligned in the same direction, it systematically yields a lower gravitational acceleration than $\langle g_r \rangle \approx g_\perp$. Therefore, using our QUMOND formula minimises the EFE compared to $g_\parallel$, which might be closer to estimates in the \citet{Bekenstein1984} formulation \citep{Haghi2019}.

We refer the reader to Appendix~\ref{section:EFE_numerical} for a validation of our formula in the context of a numerical Poisson solver in QUMOND, as developed by \cite{Oria2021}. Points corresponding to the acceleration where $g_N = a_0/100$ in the QUMOND numerical integrations shown in Fig.~\ref{fig:EFE_formula_ggn} (in magenta) make it possible to visualise the success of Eq.~(\ref{eq:gr}) to approximate the QUMOND acceleration with EFE throughout the transition from the deep-MOND, low-$g_{Ne}$ regime to the Newtonian, high-$g_{Ne}$ regime. 
We also compute the mean radial acceleration of Eq.~(\ref{eq:g_vec}) numerically (grey line), i.e., integrating over $\theta$ numerically, and we find that it overlaps perfectly with the data points from the Poisson solver. %
The expressions for $\langle g_r \rangle \approx g_\perp$ only very slightly underestimate the EFE, and thus provide a very good estimate of the acceleration field and of the resulting velocity dispersion in QUMOND. All other formulae will generate a stronger EFE.

\subsection{Application to the Coma cluster UDGs}
\label{section:sigma_EFE}

We applied Eq.~(\ref{eq:gr}) and the \cite{Mamon2005} formulas to derive the velocity dispersion profiles predicted in MOND with EFE for the Coma cluster UDGs of the sample. The internal Newtonian acceleration $g_N$ is obtained as in Section~\ref{section:isolated} from the stellar distribution, assuming a Sérsic sphere, a uniform $\rm M/L$ (assumed to be 1.3 for DF44 and DFX1, measured for the other galaxies), and no gas. 
The external Newtonian acceleration $g_{Ne}$ is derived for fiducial distances from the centre of the Coma cluster using the cluster mass distribution obtained by \cite{Sanders2003} and explained in detail in Section \ref{section:coma}. 

\begin{figure*}
\centering
\includegraphics[width=0.46\textwidth,trim={0cm 0cm 0cm 0cm},clip]{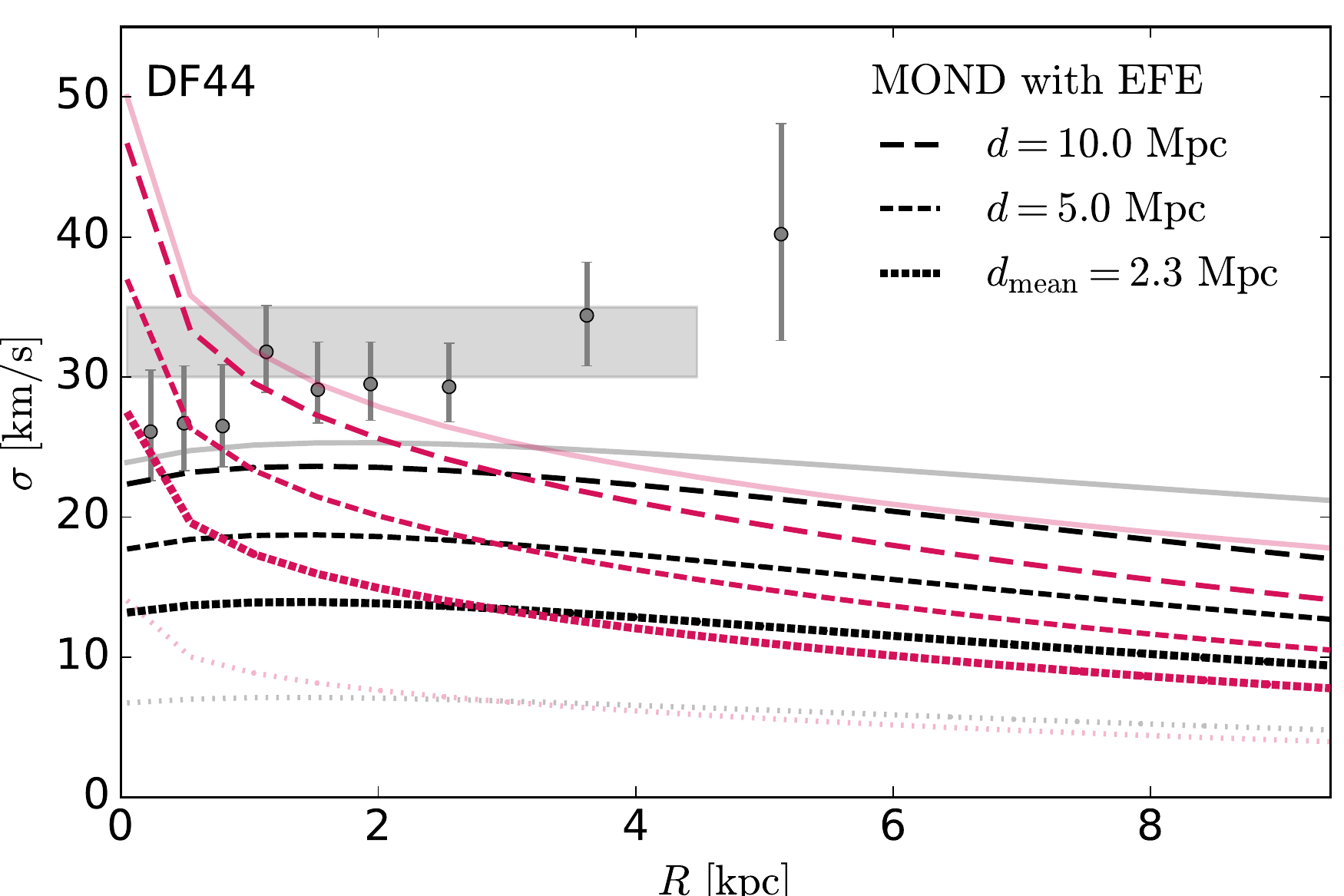}
\hfill
\includegraphics[width=0.46\textwidth,trim={0cm 0cm 0cm 0cm},clip]{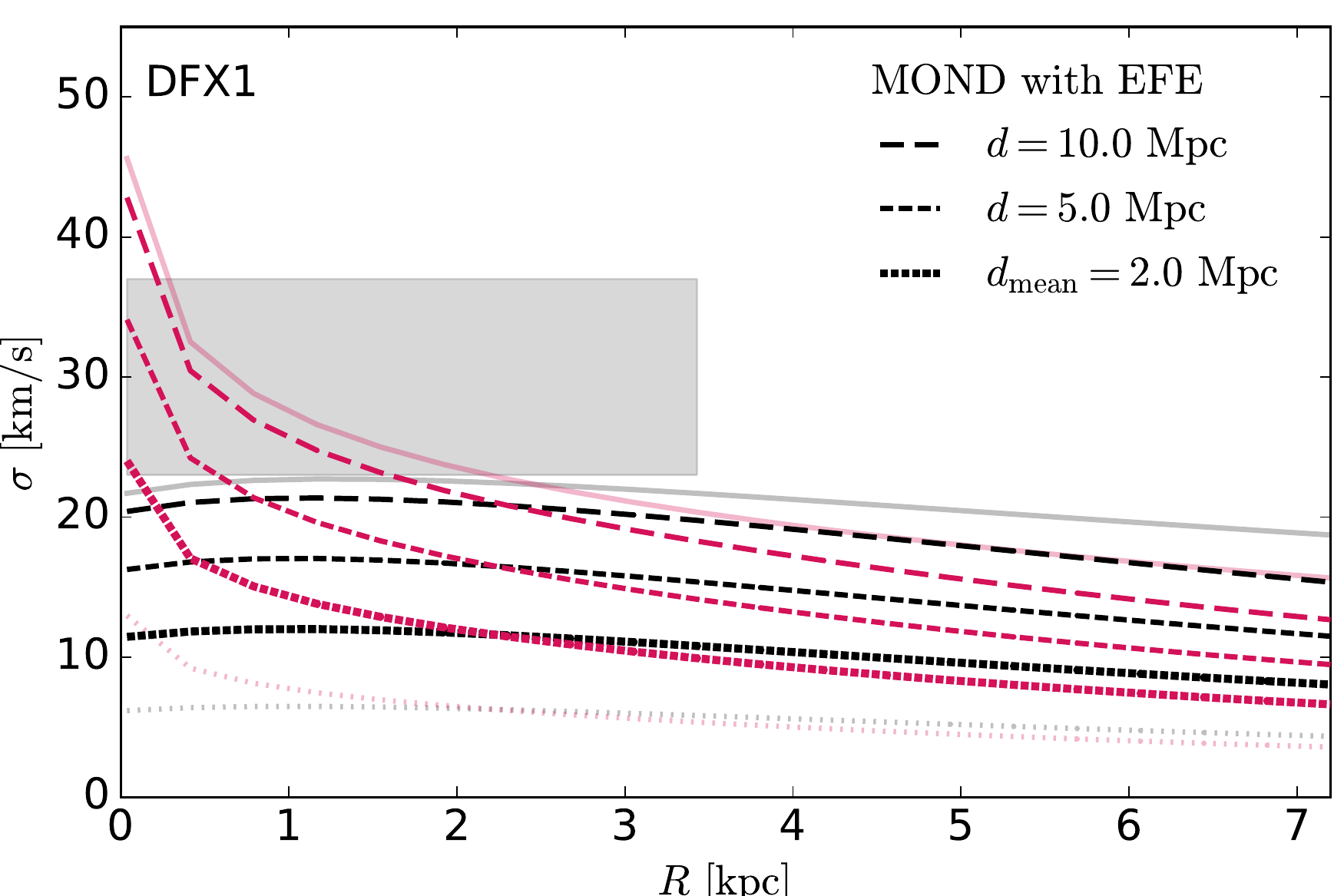}
\includegraphics[width=0.46\textwidth,trim={0cm 0cm 0cm 0cm},clip]{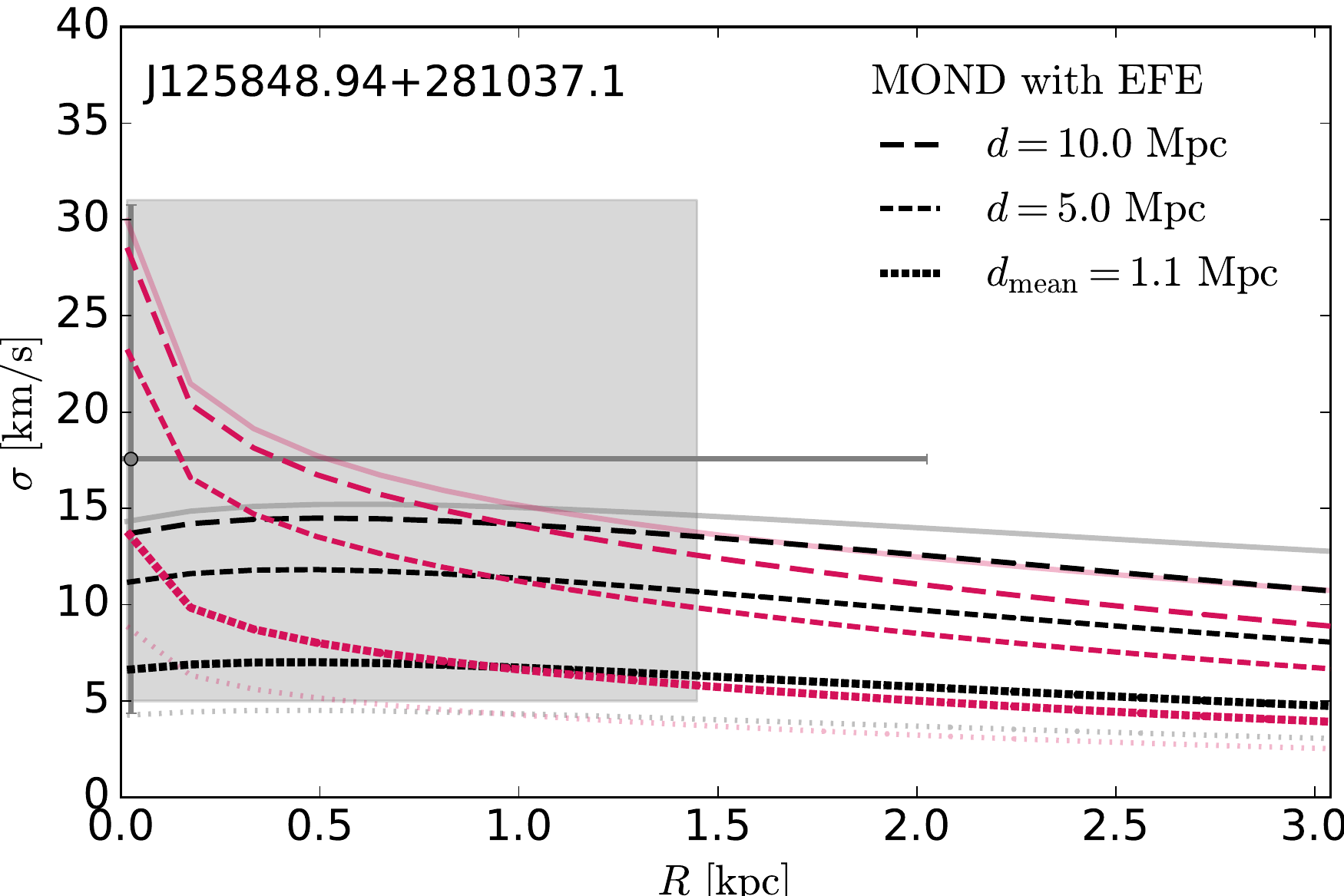}
\hfill
\includegraphics[width=0.46\textwidth,trim={0cm 0cm 0cm 0cm},clip]{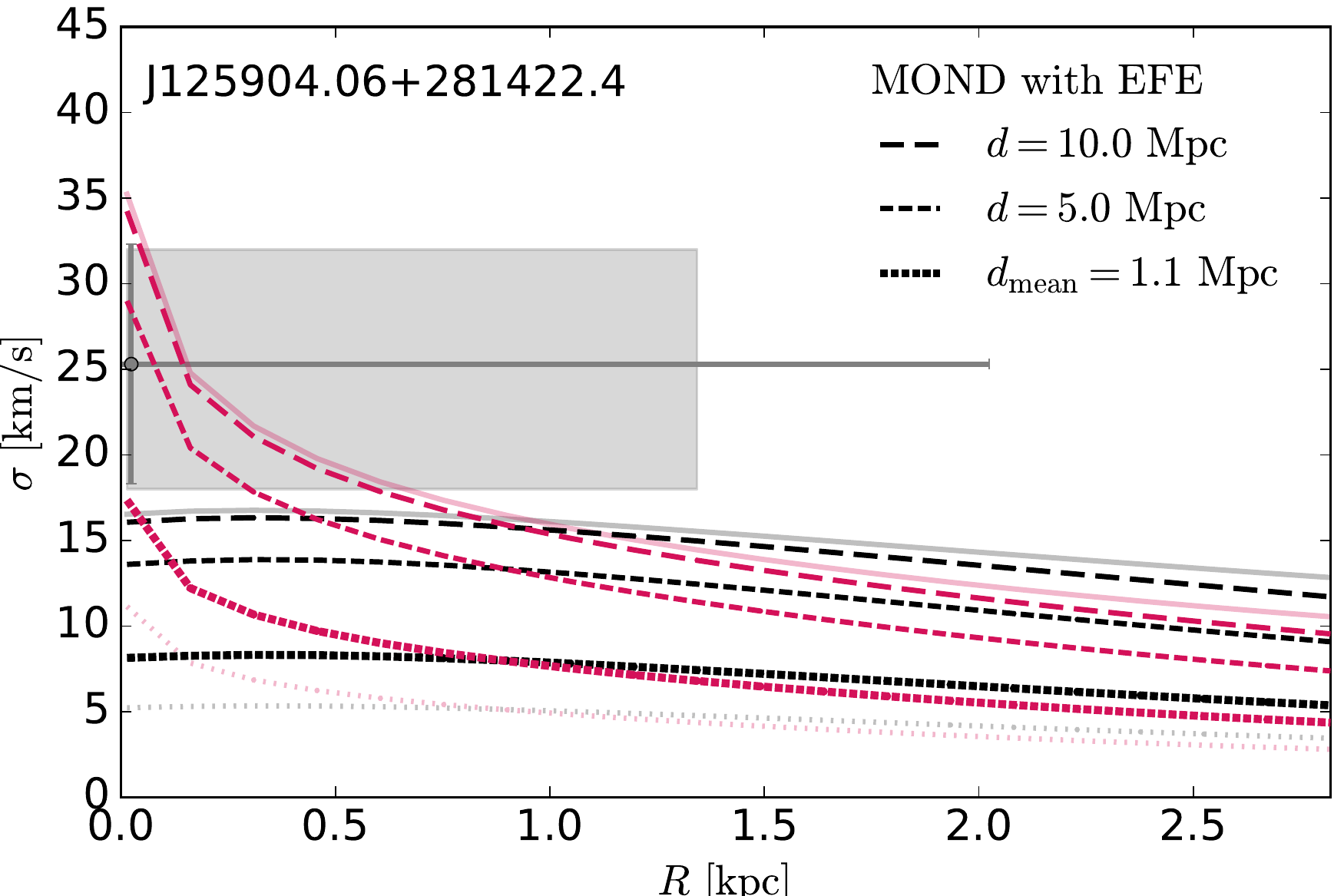}
\includegraphics[width=0.46\textwidth,trim={0cm 0cm 0cm 0cm},clip]{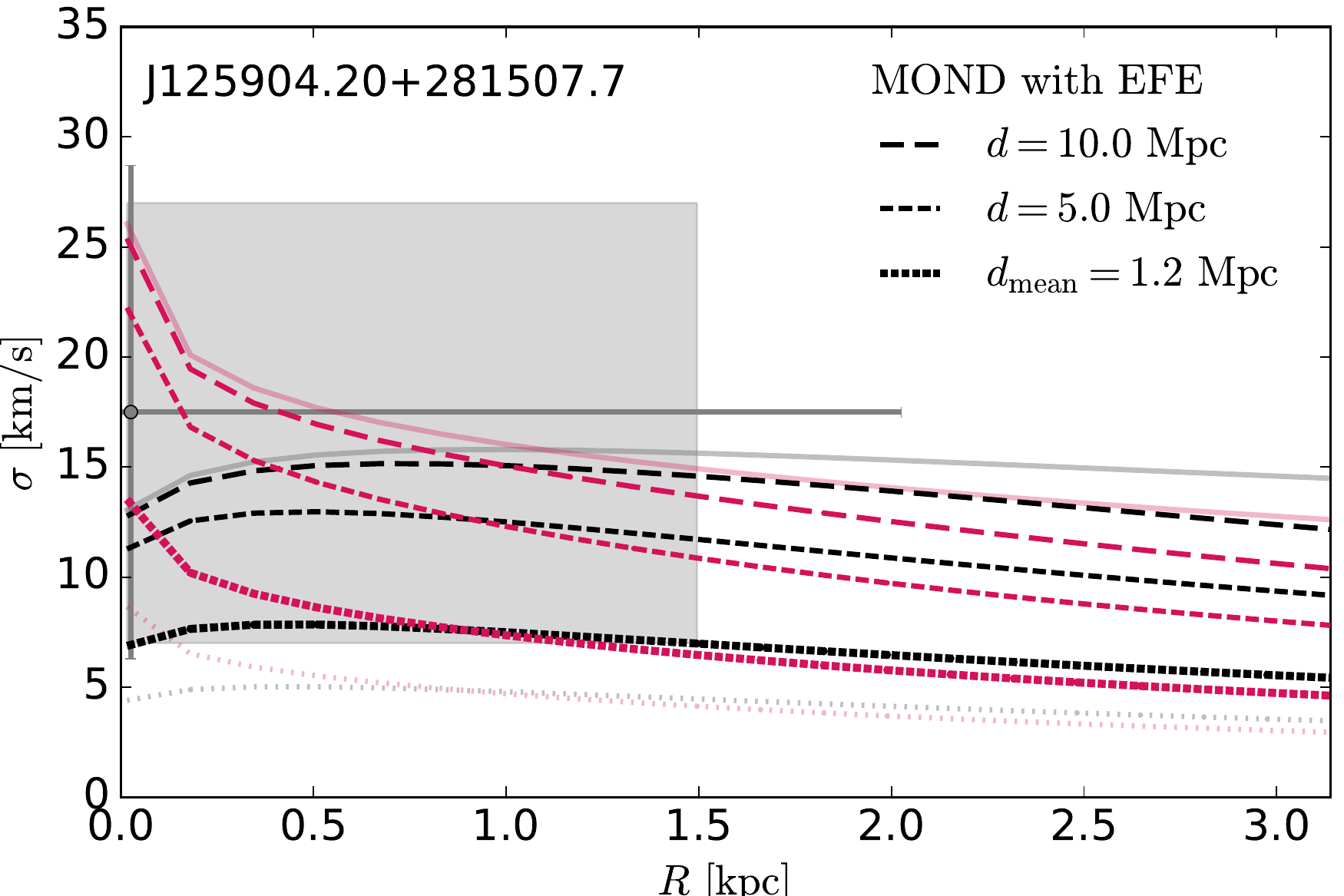}
\hfill
\includegraphics[width=0.46\textwidth,trim={0cm 0cm 0cm 0cm},clip]{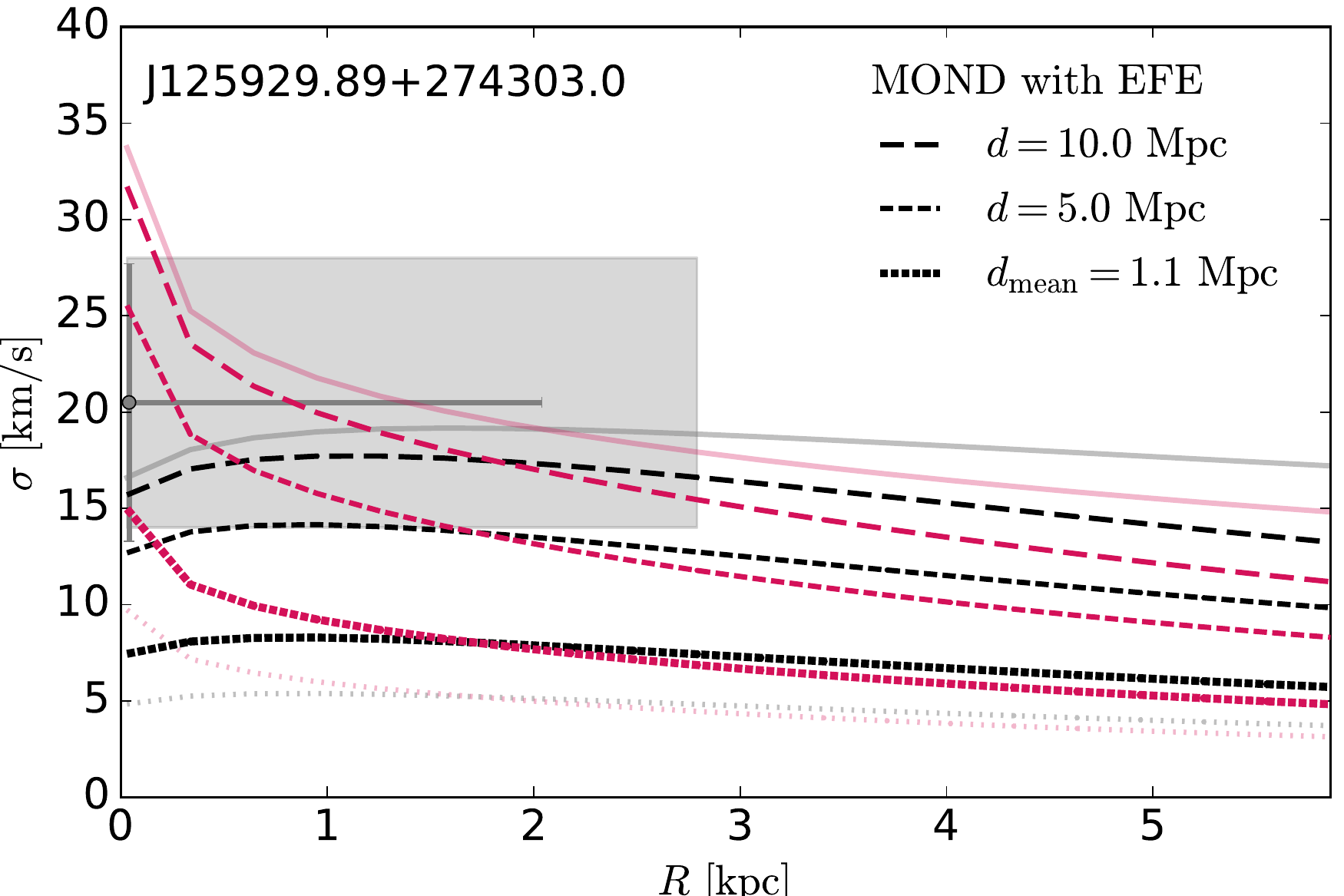}
\includegraphics[width=0.46\textwidth,trim={0cm 0cm 0cm 0cm},clip]{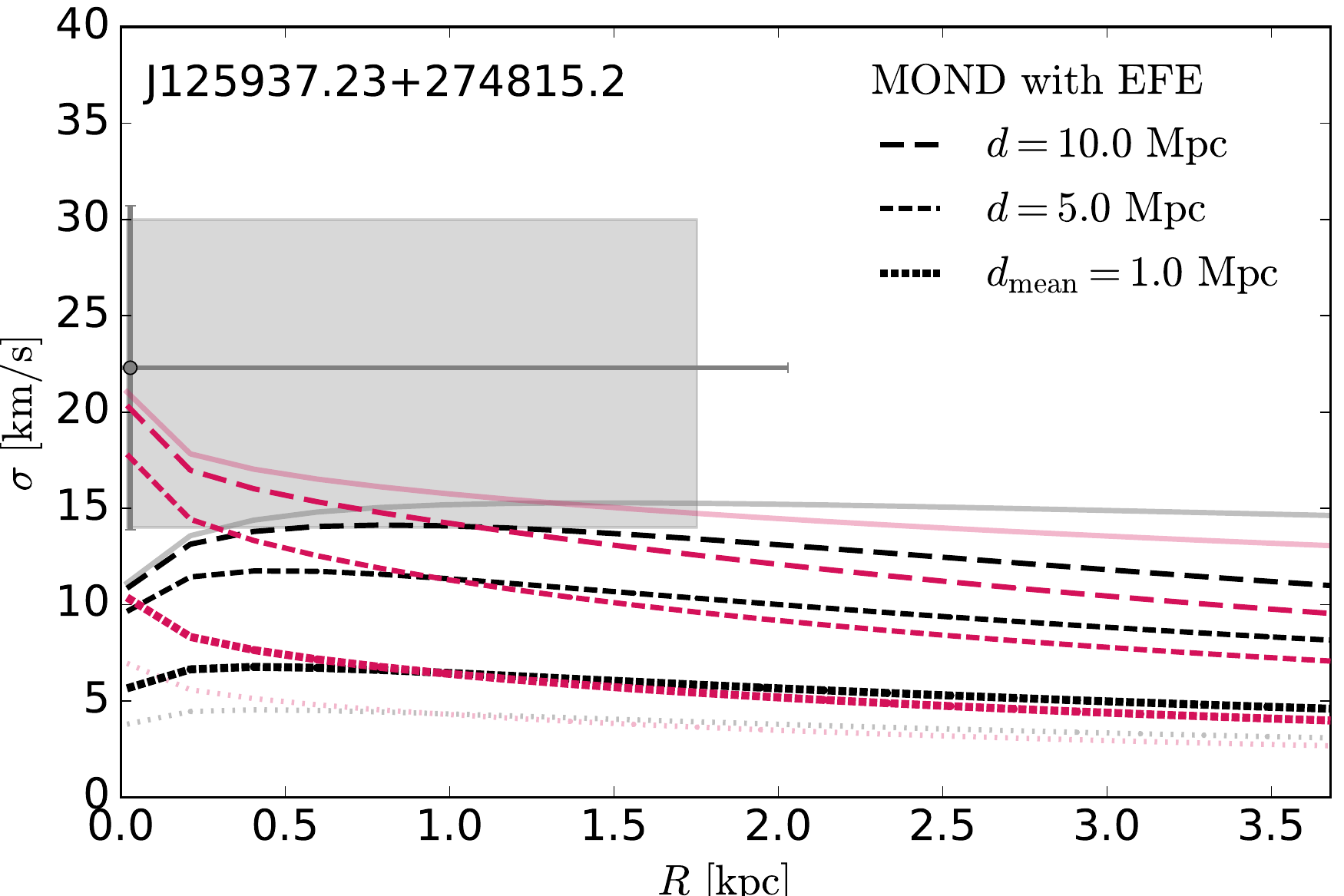}
\hfill
\includegraphics[width=0.46\textwidth,trim={0cm 0cm 0cm 0cm},clip]{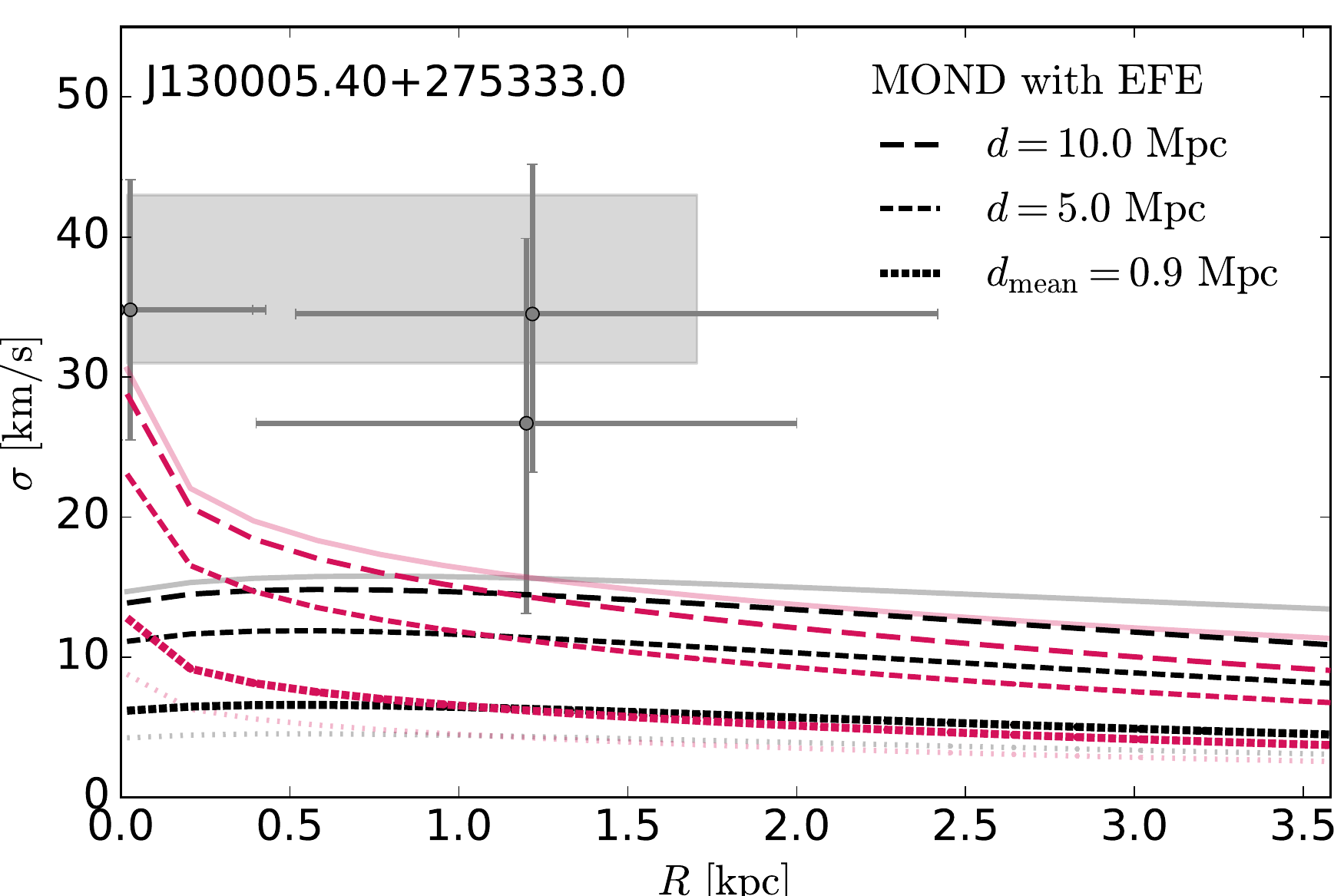}
\vspace{-0.1cm}
\caption{ 
Comparison between the measured line-of-sight velocity dispersion of the sample UDGs (grey error bars and/or shaded area) and the MOND prediction at different distances from the cluster centre. Black lines correspond to a uniform anisotropy parameter $\beta=0$ (isotropic), decreasing magenta lines to $\beta=+0.5$ (radially biased). The plain and dotted light grey (red) lines recall the predicted isotropic (radially-biased) velocity dispersion profiles in the isolated MOND and Newtonian cases, respectively (cf. Fig.~\ref{fig:isoloated}). 
Data points stem from \citet{vanDokkum2019b} for DF44, \citet{VanDokkum2017} for DFX1, and from \citet{Chilingarian2019} for the other UDGs. The shaded area corresponds to the effective velocity dispersion or, when not available (in the case of DFX1), the central stellar velocity dispersion.
}
\label{fig:EFE}
\end{figure*}

\setcounter{figure}{8}
\begin{figure*}
\centering
\includegraphics[width=0.46\textwidth,trim={0cm 0cm 0cm 0cm},clip]{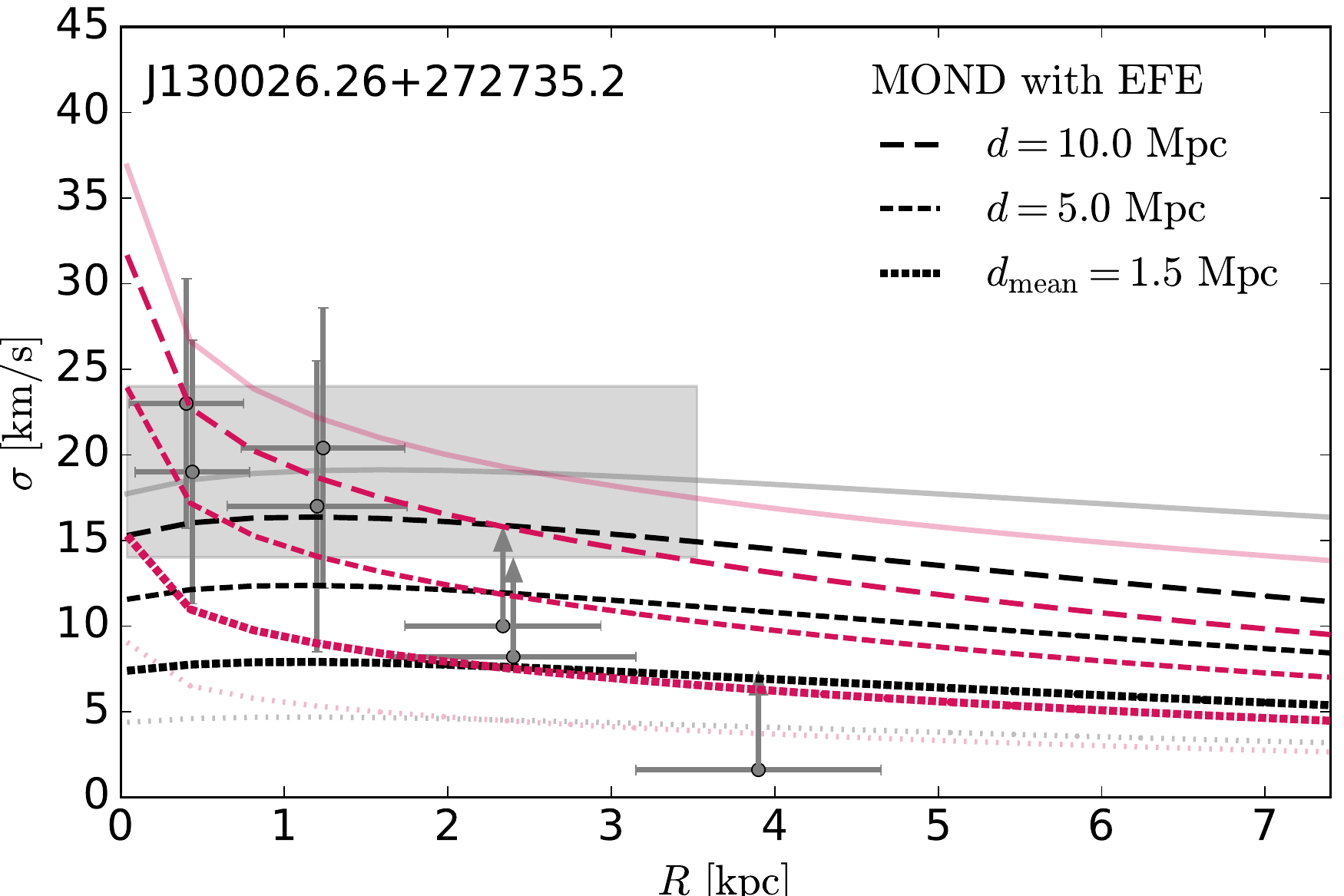}
\hfill
\includegraphics[width=0.46\textwidth,trim={0cm 0cm 0cm 0cm},clip]{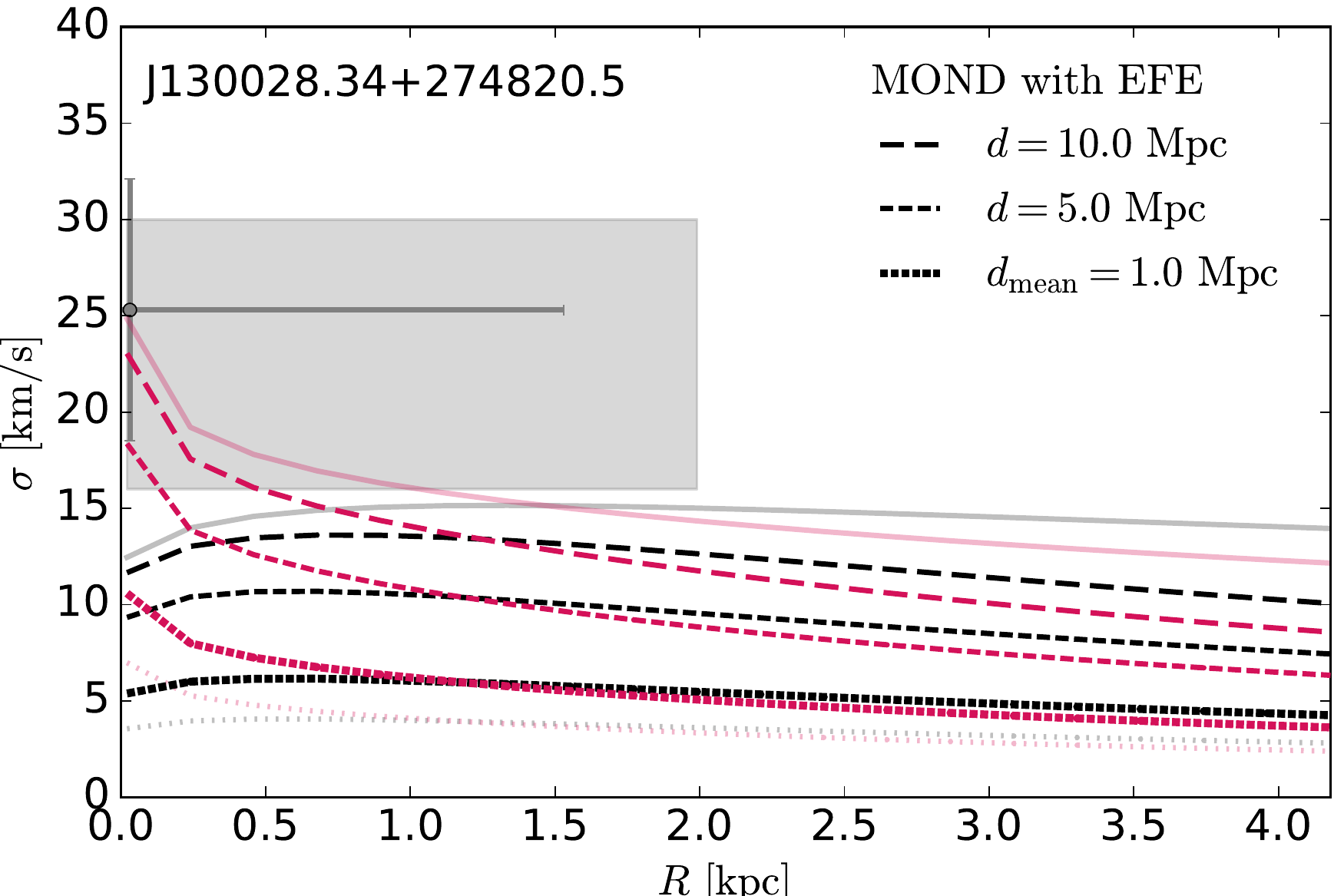}
\includegraphics[width=0.46\textwidth,trim={0cm 0cm 0cm 0cm},clip]{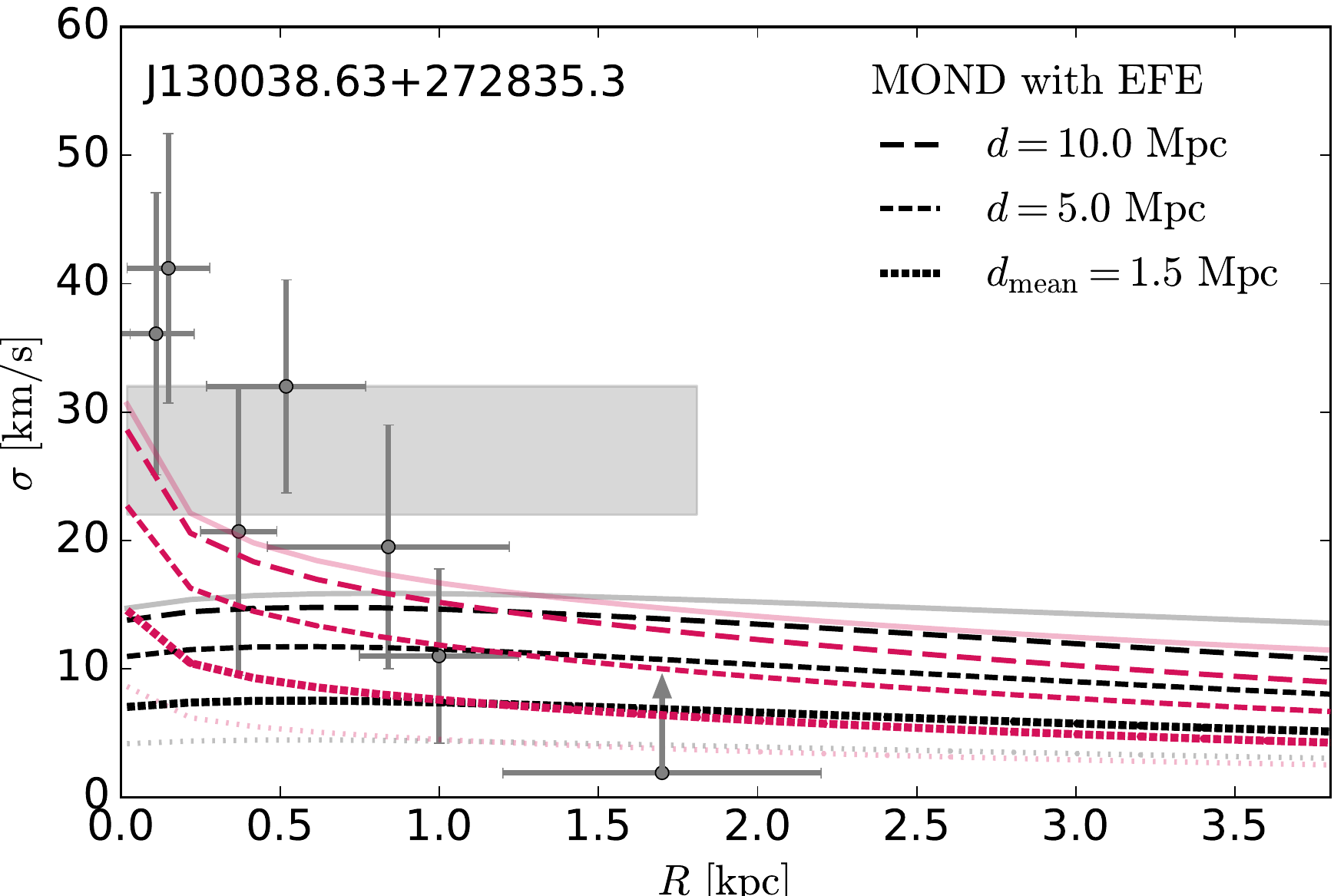}
\hfill \hspace{1cm}
\caption{
Continued. 
}
\label{fig:EFE2}%
\end{figure*}

Figure~\ref{fig:EFE} compares the observed velocity dispersion measurements of each Coma cluster UDG to the MOND predictions with EFE for different distances from the cluster centre, namely the average distance $d_{\rm mean}$ inferred from the Einasto UDG distribution of \citet[][]{vanderBurg2016} (cf. Section \ref{section:coma}), 5 Mpc, and 10 Mpc. 
The EFE pushes the MOND prediction towards the Newtonian case, significantly altering the agreement with the data compared to the isolated MOND prediction: the prediction with EFE is systematically below the measured velocity dispersion for all 11 galaxies of the sample, from about 1$\sigma$ below to more than 3$\sigma$. 
In addition to $\langle g_r\rangle$ from Eq.~(\ref{eq:gr}),  we also considered $g_\parallel$ from Eq.~(60) of \cite{Famaey2012}, recalled in the current Eq.~(\ref{eq:g_parallel}), and $g_\perp$ from Eq.~(\ref{eq:g_perp}). While $g_\perp$ yields similar profiles as $\langle g_r\rangle$, $g_\parallel$ pushes the velocity dispersion profile towards the Newtonian case even further. 
This result confirms the tension already noted by \cite{Bilek2019} and \cite{Haghi2019} for DF44 between observations and the EFE, expanding the sample to 11 Coma cluster UDGs. In particular, Fig.~\ref{fig:distance} has demonstrated that the projected distribution of the considered UDGs would be expected to be very different if they were all further than 10~Mpc from the cluster centre.
As in Section~\ref{section:isolated_profiles}, we note that taking into account the rotation hints for four UDGs from \cite{Chilingarian2019} would typically only cause differences at the level of a km/s in the line-of-sight dispersions. This would not change our conclusions, and it would actually only go into the direction of (slightly) increasing the tension between the data and the MOND models with EFE.
We now discuss the possible implications of this tension.

%-------------------------------------------------------------------

\section{Discussion}
\label{section:discussion}

Assuming sphericity, uniform $\rm M/L$ ratios, a mass model for the Coma cluster, and an accurate analytical QUMOND formula for the EFE,  we find through Jeans modelling that the observed velocity dispersion profiles of all 11 Coma cluster UDGs with velocity dispersion measurements are described well by MOND in isolation but are at odds with the MOND prediction with EFE. We hence do not see any evidence for a violation of the strong equivalence principle in Coma cluster UDGs, contrarily to, for instance, \cite{Chae2020, Chae2021}, for disc galaxies in the field. 
Our work extends that of \cite{Bilek2019} and \cite{Haghi2019}, which is limited to DF44 and makes the result all the more compelling. We recall that the MOND predictions do not involve any free parameter. 

We hereafter state different possible interpretations for the tension between the measurements and the MOND prediction with EFE, constraining either MOND itself or the formation and evolution of UDGs within this theory:
(1) The observed UDGs are further away from the cluster centre than they seem, have fallen inside the cluster relatively recently, and/or are disrupted by tides in the cluster environment. 
(2) They have higher stellar mass-to-light ratios than assumed here.
(3) They are surrounded by additional baryonic dark matter halos.
(4) The EFE varies from one galaxy to another depending on its individual history. 
(5) The characteristic acceleration scale of MOND varies with the environment; it is higher in clusters. 
(6) The cluster environment shuts down the EFE within the parent relativistic theory of MOND. 
Alternatively, MOND being an effective dark matter scaling relation of course also remains a serious possibility; in that context, the fact that cluster UDGs obey the same scaling relation as field spirals, despite their very different environments and likely different formation scenarios, is still particularly intriguing, irrespectively of the underlying theoretical framework.

\subsection{Survivor bias?}
\label{section:survivor}

\subsubsection{Further away from the cluster centre?}
\label{section:further}

The fact that the predictions within MOND in isolation are compatible with the observed stellar velocity dispersion measurements for all 11 Coma cluster UDGs with existing such measurements (Fig.~\ref{fig:isolated2}), without any free parameter, while the MOND predictions with EFE are far away from the data (Fig.~\ref{fig:EFE}), might at first suggest that the UDGs may be further away from the cluster centre than they seem from their projected distance.  
If they were sufficiently close to the cluster centre for a sufficiently long time and at equilibrium, we would expect their velocity dispersions to follow the predictions with EFE, which are close to the Newtonian values and much lower than the measured ones. 
As shown in Fig.~\ref{fig:EFE}, it would require distances $d \gtrsim 10\rm ~Mpc$ for the EFE to be negligible and the predicted velocity dispersions to approach the high measured values. 

As already noted by \cite{Bilek2019} for DF44, the wide field map of UDG candidates around the Coma cluster by \cite{Zaritsky2019} is compatible with UDGs occurring up to at least 15 Mpc from the cluster centre, with overdensities inside the $\sim$3 Mpc virial radius and along large-scale structure features such as cosmic filaments (cf. their Fig. 11). 
We note that \cite{VanDokkum2015} did not detect any UDGs within a projected distance of 300 kpc from the centre of Coma, which was used to put an approximate constraint on the binding mass of UDGs, but the number of objects towards the centre, the extended light profiles of the brighter galaxies, and the intracluster light may have prevented such detections. In particular, \cite{vanderBurg2016} did find UDGs within 300 kpc of the centre. 

It was also argued that the difference in line-of-sight velocity observed by \cite{vanDokkum2015b} between DF44 and the cluster centre ($\Delta v \approx -800 \rm ~km/s$) may indicate that this UDG is in the background of the cluster and hence significantly more distant from the centre than its 1.8 Mpc projected distance. This assumed the UDG to be on a relatively radial orbit towards the cluster centre, which would notably happen if it were falling along a cosmic filament. However, we note that the Newtonian rotation velocity at the projected distance of DF44 is about 1100 km/s given the mass of the Coma cluster ($5.5\times 10^{14}~\rm M_\odot$), such that the velocity difference does not necessarily imply a large distance from the centre. 

Figure~\ref{fig:distance} compared the observed projected distance distribution of Coma cluster UDGs from the \cite{Yagi2016} catalogue (black histogram) with the expectation from a distribution with no UDGs within 10 Mpc and a UDG density decreasing as $1/r^3$ beyond (dotted blue). 
The observed UDG distribution is obviously very different, since it decreases beyond $\sim$1 Mpc and is relatively well fit by an Einasto profile such as that obtained by \cite{vanderBurg2016}. 
Actually, the figure indicates that the \cite{Yagi2016} UDG sample is even more centrally concentrated than the \cite{vanderBurg2016} Einasto profile.
Unless the considered UDGs were not chosen randomly, we therefore conclude that the centrally depleted UDG distribution required to neglect the EFE is unlikely. 

\subsubsection{Recent infall onto the cluster?}
\label{section:infall}

As can be seen in Figs.~\ref{fig:isoloated} and \ref{fig:EFE}, a radially biased anisotropy $\beta=+0.5$ often enables us to increase the velocity dispersion towards the centre and to make it closer to the observed measurements. 
Two \cite{Chilingarian2019} galaxies (J130026.26+272735.2 and J130038.63+272835.3) further have decreasing velocity dispersion profiles consistent with a radially biased anisotropy.
A radially-biased anisotropy indeed leads to a decreasing velocity dispersion profile, while a tangentially biased anisotropy leads to an increasing velocity dispersion profile near the centre.  
DF44 has a generally rising velocity dispersion profile better fit by a tangentially biased anisotropy, as discussed in \cite{Bilek2019} and \cite{Haghi2019b}; however, \cite{vanDokkum2019b} reported a non-Gaussian line profile for DF44 with a positive $h_4$ Gauss-Hermite moment, which instead favours a radial rather than tangential anisotropy. This positive $h_4$ moment indeed reflects the presence of wings in the velocity distribution, which may result from radial orbits \citep[e.g.][]{vanderMarel1993}. Such wings may, however, also stem from deviations from spherical symmetry. 

Radial orbits would be expected if the UDGs are on radial orbits themselves, which could provide an explanation for the absence of EFE. 
While the MOND EFE may contribute to puff-up UDGs as they fall in the cluster environment, it also means that their internal dynamical time increases \citep{Milgrom2015}, such that their observed velocity dispersion at a given time may not reflect the EFE at this time but at an earlier, larger distance from the cluster centre. 
The facts that UDGs may be flowing along cosmic filaments \citep{Zaritsky2019} and that several UDGs around DF44 share similar radial velocities \citep{vanDokkum2019b} are compatible with them falling into the Coma cluster on radial orbits for the first time. 
The cluster orbital time at 1 Mpc of its centre is $\sim$3 Gyr within MOND, while the orbital times at half-mass radius within the UDGs lie between 0.4 and 1.0 Gyr in the deep MOND regime (between 1.3 and 3.3 Gyr in the Newtonian, strong EFE regime) given their de-projected Sérsic mass profiles. 
If the difference may result in a higher velocity dispersion than expected from the EFE at the location of the UDG, it may not be sufficient to explain the fact that measured velocity dispersions are close to the isolated MOND predictions for all sample UDGs. This will need to be confirmed by detailed simulations.
In any case, failing to explain their observed velocity dispersions in this way would not mean that observed UDGs are not significantly biased towards objects that recently fell inside the cluster.

\subsubsection{Disrupted by tides?}
\label{section:tides}

Both a centrally depleted UDG distribution and observed UDGs that only recently fell onto the cluster could be explained by enhanced tides in MOND and rapid tidal disruption: observations would be biased towards not-yet disrupted objects, as in the survivor bias.  
The MOND EFE is indeed expected to decrease the gravitational support of satellite galaxies and hence make them more prone to tidal effects and tidal disruption \cite[e.g. ][]{Brada2000}. 
If the velocity dispersion of the UDG was set before entering the cluster environment in a region where the EFE was negligible, the UDG may become unbound and dissolve when the gravitational support decreases because of the EFE. 
\cite{Bilek2019} accordingly interpreted the observed rising velocity dispersion profile of DF44 as being due to an extra acceleration by tidal forces, which could also explain its relative elongation (with an axis ratio $b/a=0.69$) observed by \cite{vanDokkum2019b}. 
\cite{vanderBurg2016} further reported a significant central deficit of UDGs compared to compact dwarfs in a sample of eight nearby clusters, possibly because of strong dynamical interactions between galaxies in the central cluster region and tidal destruction due to the cluster potential. 
Signs of tidal interactions have been directly observed for some UDGs; for example, 
\cite{Mihos2015}, \cite{Wittmann2017}, and \cite{Bennet2018} found UDGs associated with tidal streams, 
\cite{Merritt2016}, \cite{Toloba2016}, \cite{Venhola2017}, and \cite{Lim2020} reported UDGs with elongated or distorted optical morphologies, 
\cite{Scott2021} observed atomic gas morphology and kinematics of a UDG that are consistent with a recent tidal interaction,  
and \cite{Montes2020} claimed that the small velocity dispersion of the peculiar UDG NGC 1052-DF4 results from tidal disruption. 
For the current UDG sample, the significant elongations reported by \cite{Chilingarian2019} in many of the galaxies ($b/a$ being respectively 
0.69, 
0.62, 
0.65, 
0.56,
0.50, 
0.44, and 
0.42 
%%%%
for 
DF44, 
DFX1, 
J125848.94+281037.1, 
J125904.06+281422.4, 
% J125904.20+281507.7 round
J125929.89+274303.0, 
% J125937.23+274815.2 round
% J130005.40+275333.0 round
J130026.26+272735.2, and 
J130038.63+272835.3)
and the S-shaped image fitting residuals they obtain for J130028.34+274820.5 
may indicate tidal interactions. 

\begin{figure}
\centering
\includegraphics[width=1\linewidth,trim={0cm 0cm 0cm 0cm},clip]{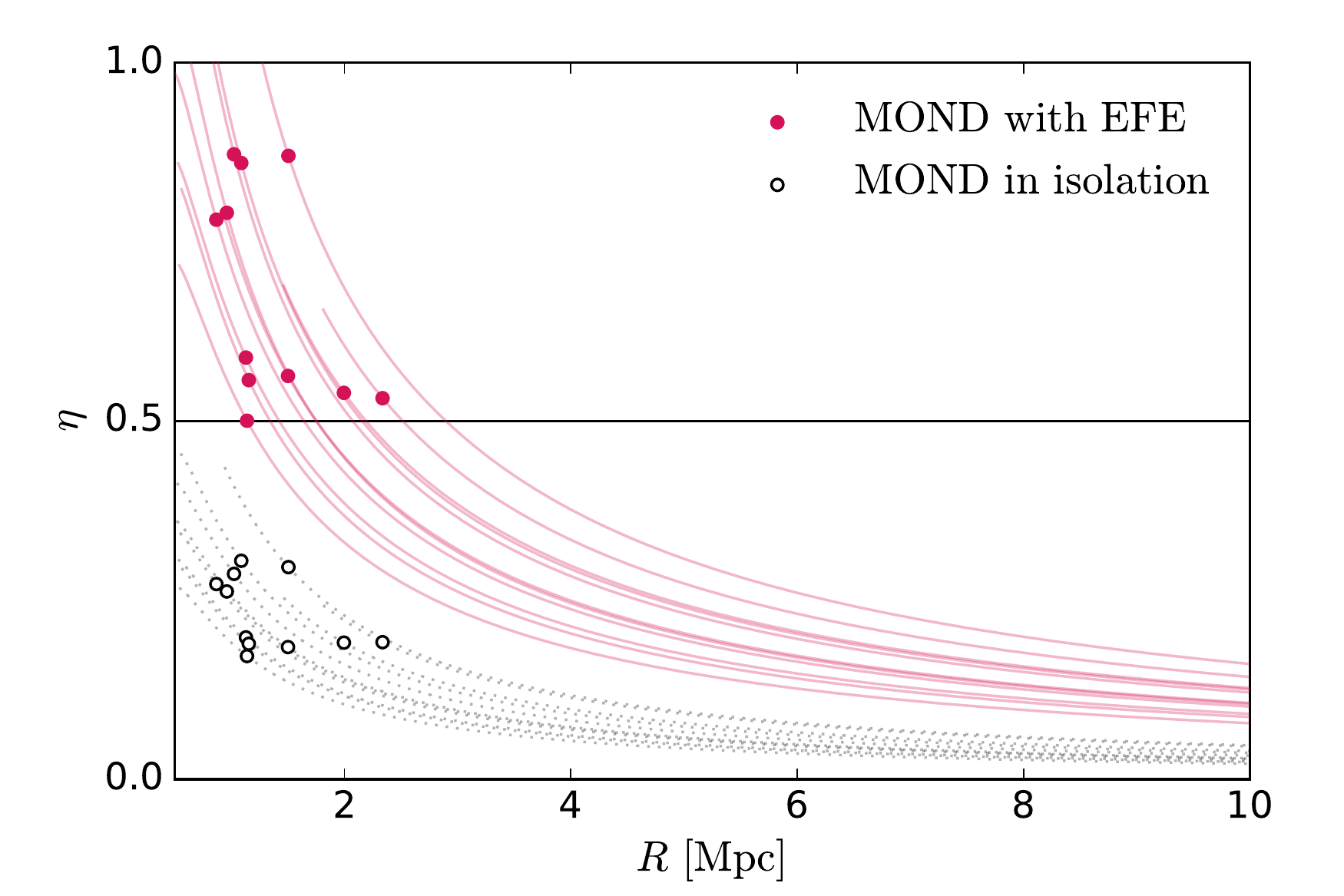}
\caption{
Tidal susceptibility $\eta$ as a function of the distance $R$ from the centre of the Coma cluster for the sample UDGs. The distance is assumed to be larger than the projected distance $d$, i.e., $R>d$. Points correspond to the average distance $d_{\rm mean}$ assuming the \protect\citet{vanderBurg2016} Einasto UDG distribution (cf. Section~\protect\ref{section:coma}): plain magenta for MOND with EFE, open black for MOND in isolation. 
}
\label{fig:tidal_susc}
\end{figure}

To validate the possibility that these Coma cluster UDGs are subjected to tides, we estimate their tidal susceptibility: 
\be
\label{eq:eta}
\eta = \frac{r_{1/2}}{r_2}, 
\ee 
which is the ratio between their half-mass radius $r_{1/2} \approx (4/3) R_e$ and the Roche lobe radius $r_2$. The Roche lobe radius $r_2$ is derived from the radius $r_1$ of the inner Lagrange point using \cite{Zhao2005} and \cite{ZhaoTian2006}, $r_1$ being obtained by numerically solving the equation describing the equilibrium between the UDG internal gravitational force, taking the EFE and the external tidal force into account, as explained in Appendix~\ref{appendix:tidal}.  
Fig.~\ref{fig:tidal_susc} shows $\eta$ as a function of the distance $R$ from the centre of the cluster. 
As highlighted by the plain red data points, the tidal susceptibility at the average distance $d_{\rm mean}$ (cf. Section~\protect\ref{section:coma}) is in the 0.5-1 range for all UDGs, so we expect them to be at least partially affected by tides. 
We further note that the UDGs are most likely not observed at their pericentre, where the tidal susceptibility would be higher, and that 
 the tidal susceptibility at the projected distance is always above one.
The figure also includes the tidal susceptibility of the Coma cluster UDGs without EFE, namely using MOND in isolation. In this case, the tidal susceptibility at the average distance (and even at the projected distance) is always below 0.5. 
 From Fig.~\ref{fig:tidal_susc}, we conclude that UDGs should be affected by tides beyond their half-light radius when taking the MOND EFE into account, which would bias observations towards survivors and/or objects that recently fell onto the cluster and could even deplete the central region of the cluster (although this latter point does not seem to be supported by the observations: cf. Section~\ref{section:further}). 
Since MOND in isolation provides an approximate estimate of the measured velocity dispersion (Section~\ref{section:isolated}), the internal gravitational accelerations of the sample UDGs may be specifically high enough to escape tides, but this requires us to ignore the EFE. 
Alternatively, they may be out-of-equilibrium objects whose stars have been heated by tides, but this would beg the question of why their inflated velocity dispersions would precisely match up with the isolated MOND prediction, which would then appear as a coincidence. Again, future detailed simulations could help to answer these questions more precisely.

\subsection{Higher stellar masses?}
\label{section:higher_ML}

\begin{figure}
\centering
\includegraphics[width=1\linewidth,trim={0cm 0cm 0cm 0cm},clip]{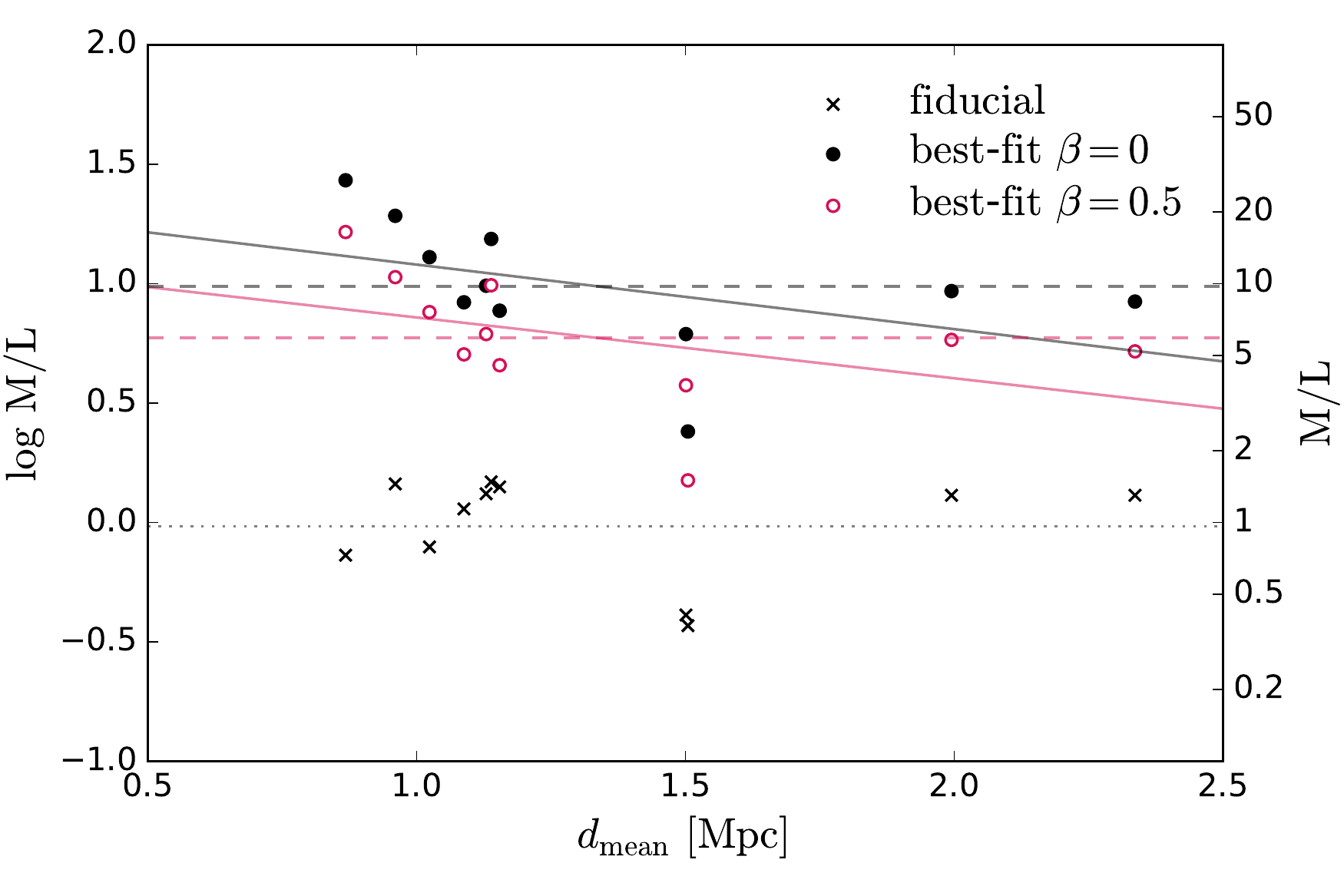}
\caption{
Best-fit $\rm M/L$ ratio matching the effective velocity dispersion $\sigma_{\rm eff}$, assuming MOND with EFE at the average distance $d_{\rm mean}$. Crosses correspond to the fiducial $\rm M/L$ ratio values from Table~\ref{table:sample}, plain black circles to the best-fit values with an anisotropy parameter $\beta=0$, and open magenta circles to the best-fit values with $\beta=0.5$. Plain lines correspond to linear fits, dashed and dotted lines to the averages. The average best-fit $\rm M/L$ ratio is 11.5 with $\beta=0$ and 7.0 with $\beta=+0.5$. The fits with $\beta=+0.5$ are displayed in Fig.~\ref{fig:EFE_ML}. 
}
\label{fig:ML}
\end{figure}

Since UDGs are rather unusual objects compared to other galaxy classes, that is, with Sérsic indices close to one but no disc morphology and pressure support, standard procedures to derive their stellar $\rm M/L$ ratios (such as that based on the \texttt{Pégase.HR} evolutionary synthesis package by \citealp{Leborgne2004} used by \citealp{Chilingarian2019}) may not directly apply.  
In which case, higher stellar $\rm M/L$ ratios than those used here may alleviate the discrepancy between the predicted and measured velocity dispersions. 
In Fig.~\ref{fig:ML}, we show the best-fit stellar $\rm M/L$ ratio that enabled us to recover the effective stellar velocity dispersion $\sigma_{\rm eff}$, assuming MOND with EFE at the average distance $d_{\rm mean}$, both in the isotropic case ($\beta=0$) and in the radially biased case ($\beta=+0.5$). The average best-fit $\rm M/L$ ratio is 11.5 with $\beta=0$, and 7.0 with $\beta=+0.5$. Such values could have been expected from Fig.~\ref{fig:EFE}, where the predicted velocity dispersion profile with EFE generally falls short of the observed measurements by a factor close to $\sqrt{10}\simeq 3$. %
We note that there is a small increasing trend of the best-fit $\rm M/L$ ratio towards the cluster centre, compensating for the enhanced EFE. 
A combination of higher $\rm M/L$ ratio values than fiducial and tidal heating (cf. Section~\ref{section:tides}) could be invoked to explain the high velocity dispersions. In Appendix~C, we explicitly display the radially anisotropic predictions with a stellar mass-to-light ratio $\rm M/L=4$ at the average distance for each galaxy, together with the best-fit $\rm M/L$ curves. 
If stellar $\rm M/L$ ratios up to 4 may be allowed \citep[e.g.][]{Famaey2018}, the particularly high best-fit values may call for a significant amount of baryonic dark matter.

\subsection{Baryonic dark matter?}
\label{section:baryonic_DM}

%Concerning additional baryonic dark matter, 
We recall that although MOND does reduce the mass discrepancy in clusters compared to Newtonian dynamics, clusters within MOND still need to contain an undetected component whose mass is comparable to the observed baryonic mass \citep[e.g.][]{Sanders1999, Sanders2003, Milgrom2008, Angus2008a, Milgrom2015} and is rather concentrated in the central parts \citep{Angus2008a}. If this undetected component were composed of baryons (e.g. molecular hydrogen), this cluster baryonic dark matter would only account for a small fraction ($\sim 15$\%) of the unobserved missing baryons. Moreover, if it is in collisionless form (such as compact molecular hydrogen clouds), it may notably help explain the `bullet cluster' \citep{Clowe2006}, where the peak of the mass distribution deduced from weak lensing lies away from the observed baryons. 
\cite{Milgrom2015} argued that the existence of many cluster UDGs may indeed indicate the presence of baryonic dark matter to stabilise them against tidal disruption. 
\cite{Bilek2019} already interpreted the rising velocity dispersion of DF44 and its large number of globular clusters as possible clues indicating the presence of a surrounding baryonic dark halo. However, they noted that (i) if the UDG were far away from the cluster centre and experiencing little or no EFE, the baryonic halo surrounding it (if any) should be sufficiently extended to have no strong gravitational influence towards the centre of the galaxy, and (ii) if the UDG were, on the contrary, experiencing a substantial EFE, the baryonic halo should conspire to compensate for the decrease from the isolated MOND prediction. 
The two options a priori make the hypothesis of a baryonic dark halo either unnecessary or unappealing, especially as they would have to be generalised to all 11 UDGs of the current sample. 
However, with the (strong) assumption of the baryonic dark matter following the light in all 11 UDGs, a constant dynamical-to-stellar mass ratio of about 10 could be enough to explain the observations, as can also be inferred from Fig.~\ref{fig:ML}. 
The option of cluster baryonic dark matter is obviously difficult to formally exclude, but would nevertheless imply a conspiracy to make it match the isolated MOND prediction for all the galaxies. From a heuristic point of view, this would incidentally decrease the appeal of modified gravity altogether to explain the observed phenomenology in galaxies.

\subsection{Modified inertia?}
\label{section:inertia}

While current MOND theories are of the modified gravity type, only involving a modification of the action of the gravitational field, \citet{Milgrom1994,Milgrom2002,Milgrom2011} has also considered theories of the modified inertia type, where it is the kinetic action (the coupling to matter for relativistic theories) that is modified. For example in the non-relativistic case, this would leave the Poisson equation intact but modify the equation of motion of a test particle into $\vec{A}(\vec{r},t,a_0) = -\vec{\nabla} \Phi (\vec{r}(t))$ where $\vec{A}$ is a function of the trajectory of the particle and $\Phi$ is the potential. In such theories, the MOND effects would depend on the whole trajectory of a system and not just on its instantaneous state. 
As explicitly shown in \citet{Milgrom2011}, it is actually possible to construct such a theory without EFE, or with an EFE which differs significantly from that of modified gravity theories, in which case the isolated MOND behaviour of UDGs might reflect their different history of formation and trajectories (cf. for instance the field spirals of \citet{Chae2020}). 
However, such a framework may slightly hinder the predictive power of MOND if the EFE varies from one galaxy to another depending on its individual history. 
If the EFE were absent altogether, this would also be at odds with the observational evidence and actual need for the EFE in certain places; for instance, in wide binaries (\citealp{Pittordis2019}, \citealp{Banik2019}, although see also \citealp{Hernandez2021}), Andromeda dwarf galaxies \citep{McGaugh2013a,McGaugh2013b}, the crater ultra-faint dwarf \citep{McGaugh2016,Caldwell2017}, or the UDGs NGC 1052-DF2 and NGC 1052-DF4 \citep{Famaey2018,Kroupa2018}.

\subsection{EMOND?}
\label{section:EMOND}

Rather than assuming an additional dark matter component to explain the subsiding mass discrepancy in clusters within MOND, it has been suggested that MOND phenomenology may be the limiting case of a more general relation depending, for example, on the depth of the gravitational potential (or of some scalar field in relativistic versions of the theory). 
In particular, \cite{Zhao2012} proposed an `extended MOND' (EMOND) framework where the critical acceleration $a_0$ is replaced by an acceleration $A_0(\Phi)$ increasing with the depth of the potential well to boost the MOND effect in galaxy clusters. This acceleration would vary from $A_0\approx 1\times 10^{-10}~\rm m~\! s^{-2}$ inside galaxies and the Solar System to $\sim8 \times 10^{-10}~\rm m~\! s^{-2}$ in galaxy clusters \cite[cf. Fig. 1 of][]{Zhao2012}, but it could reach values as high as $\sim100 \times 10^{-10}~\rm m~\! s^{-2}$ in their central regions \citep{Hodson2017a, Hodson2017c}. 
A high critical acceleration in clusters means that the gravitational acceleration would deviate from the Newtonian one at higher accelerations, thus increasing the velocity dispersion in cluster UDGs compared to the MOND predictions with EFE (Section~\ref{section:EFE}), even when the external field of the cluster is strong. 
Although using an unresolved and indirect estimate of the velocity dispersion of Coma cluster UDGs \citep[derived from their effective radius and surface brightness using the fundamental manifold of][]{Zaritsky2006a,Zaritsky2006b,Zaritsky2008},
\cite{Hodson2017c} already studied Coma cluster UDGs in the EMOND framework. By comparing the baryonic mass needed in EMOND to the stellar mass inferred from photometry, they conclude that this framework is consistent with the observations. 

\begin{figure}
\centering
\includegraphics[width=1\linewidth,trim={0cm 0cm 0cm 0cm},clip]{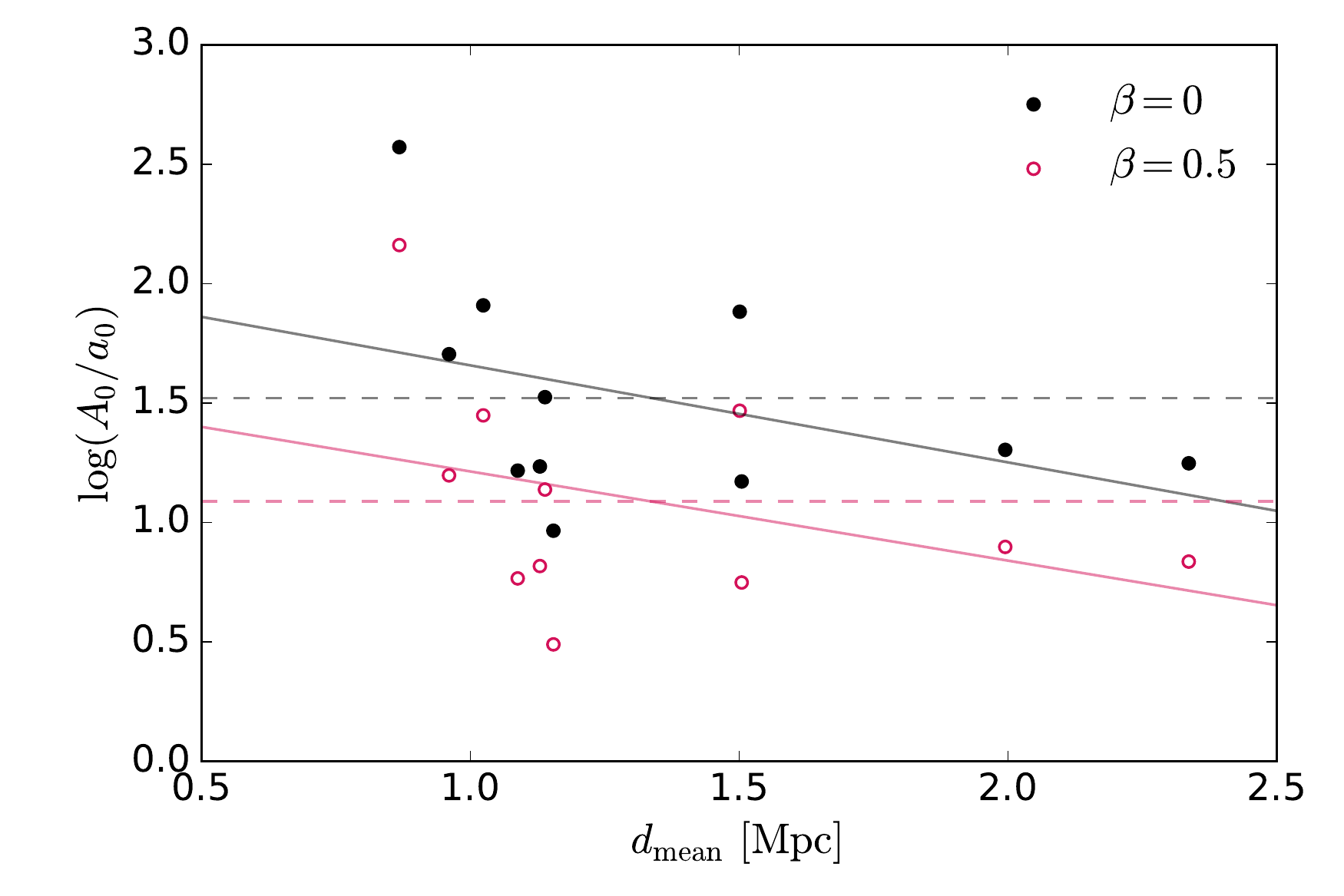}
\caption{
Best-fit acceleration $A_0$ matching the effective velocity dispersion when emulating EMOND at the average distance $d_{\rm mean}$, with the external field derived from the dominant hot gas distribution. Plain black circles correspond to an anisotropy parameter $\beta=0$ and the open magenta circles to $\beta=0.5$. Plain lines correspond to linear fits and dashed lines to the averages. A cluster acceleration $A_0=10 \times a_0$ provides reasonably good fits to the velocity dispersion measurements, especially when some radial anisotropy is allowed (cf. also Fig.~\ref{fig:EMOND_profiles}). 
}
\label{fig:EMOND_A}
\end{figure}

To emulate EMOND, which should not harbour any missing mass in galaxy clusters by construction, we hereafter determine the Newtonian external field from the dominant intracluster hot gas mass distribution, 
\be
\label{eq:Mgas}
M_{\rm gas}(r) = 4\pi \int_0^r \rho_C(r^\prime) {r^\prime}^2 dr^\prime, 
\ee
with $\rho_C$ given by the $\beta$-model of Eq.~(\ref{eq:rho_c}), and replace the critical acceleration $a_0$ by a variable $A_0$. 
As shown in Fig.~\ref{fig:EMOND_A}, the best-fit $A_0$ matching the effective velocity dispersion $\sigma_{\rm eff}$ when including the EFE and assuming the UDGs to be at their average distance $d_{\rm mean}$ is of the order of $10\times a_0,$ as expected by \cite{Zhao2012} in clusters, and decreases with $d_{\rm mean}$ in the isotropic $\beta=0$ case, which follows the increasing need to compensate for the EFE closer to the cluster centre and reach the high measured velocity dispersions.

We note that a slightly higher cluster mass distribution -- to take into account galaxies in addition to the intracluster hot gas -- would increase the EFE, decrease the predicted velocity dispersion, and thus lead to slightly higher best-fit accelerations.
In Fig.~\ref{fig:EMOND_profiles} in the appendix, we compare the measured velocity dispersions to the predicted EMOND profiles with $A_0 = 10\times a_0$, which are obtained without any free parameter. We find that the predictions fit particularly well with the data, especially since we adopted a single fixed value for $A_0$, and notably for DF44 \citep[contrarily to the indirect inference of][]{Hodson2017c}.
Incidentally, the above case would also roughly correspond (modulo the weaker Newtonian external field from the cluster in the EMOND case) to the case of increasing the baryonic mass with a mass-follows-light profile and with a dynamical-to-stellar mass ratio of about 10, as previously discussed in Section \ref{section:higher_ML}.
We also note that galaxy J130005.40+285333.0, for which the EMOND agreement is less convincing, is the closest to the cluster centre, namely that exposed to the strongest tidal forces and residing the deepest in the cluster potential well.

However, the more general problem with EMOND is to account for the displaced peaks of the lensing convergence maps of colliding clusters such as the bullet cluster. While peaks of convergence displaced from the main baryonic peaks are easy to produce, both in MOND and EMOND, producing them outside the baryonic peaks -- and not in between them -- has never been demonstrated to be feasible.

\subsection{Screening the EFE in galaxy clusters?}
\label{section:screening}

We note that the apparent absence of EFE happens precisely within galaxy clusters, where classical MOND fails to explain the overall dynamics of the cluster, and we conjecture here that these two facts might possibly be related. In the case of EMOND, this would be explained by an effective increase of the MOND acceleration constant, not by a screening of the EFE itself. But another possibility is that the EFE is severely damped in galaxy clusters. 

In a theory such as that of \citet{Skordis2020}, the action harbours a free-function, playing the role of the MOND interpolating function, depending both on the spatial gradient squared of the scalar field $|\nabla \varphi|^2$ (with a $3/2$ exponent, characteristic of MOND actions) and on its temporal derivative having a non-zero minimum leading to gravitating `dust'. It is this time-dependent term that allows to reproduce a reasonable angular power spectrum for the CMB, and one could therefore speculate that it may also give rise to additional gravitating `dust' inside galaxy clusters, to explain the residual missing mass of MOND. However, it is not clear that, if the scalar field is dominated by this dust component inside the cluster itself, it would couple to the scalar field within the UDG in the same way as in the fully quasi-static limit. Therefore, one could imagine that, precisely because the residual missing mass in galaxy clusters would be caused by the same scalar field as that creating the MOND effect inside the UDG, the EFE could be effectively screened within clusters. We note that this is especially relevant for any model that would try to explain the residual MOND missing mass in clusters of galaxies, as such an explanation would not work if the residual missing mass is made of additional hot DM such as light sterile neutrinos.

In this context of EFE screening, one could imagine two possibilities: one where the EFE would be solely produced by the baryonic mass of the cluster, and one where it would be almost fully screened, with the UDG living in a MOND bubble, effectively de-correlated from the dynamics of the cluster itself. We tested the first hypothesis by re-doing the analysis in Section~\ref{section:EFE} using only the Coma cluster hot gas mass distribution $M_{\rm gas}$ (Eq.~(\ref{eq:Mgas})) derived from the $\beta$-model of Eq.~(\ref{eq:rho_c}) as a source of EFE, instead of the mass distribution $M_C(r)$ inferred from hydrostatic equilibrium. This mass is about 1~dex below $M_C$ at a distance of 1~Mpc but reaches $M_C$ at 10~Mpc. As a consequence, the resulting velocity dispersions at distances smaller than 10~Mpc are higher than in Fig.~\ref{fig:EFE}, as shown in Fig.~\ref{fig:EFE_gas}. 
However, the difference is not sufficient to significantly alter our conclusion on the mismatch between the observed velocity dispersions and the predictions with EFE at the average $d_{\rm mean}$.  

This means that, to explain our results with the nominal values of the stellar $\rm M/L$ ratios, the EFE should be almost fully screened for the UDGs residing inside clusters. This is actually also the case in some hybrid versions of MOND such as the superfluid DM theory \citep{Khoury2015, Berezhiani2015,Berezhiani2018,Berezhiani2019}. As discussed in detail in Sect.~IX.B of \citet{Berezhiani2018}, the superfluid core would be rather small in galaxy clusters (of the order of a few hundred kpc at most), and no EFE would be expected for cluster UDGs, which is contrary to the case of satellite galaxies orbiting within the superfluid core of their host, where the EFE would be expected to be similar to the MOND case.

%-------------------------------------------------------------------

\section{Conclusion}
\label{section:conclusion}

Given their singularly low internal accelerations, ultra-diffuse galaxies (UDGs) in clusters provide a testing ground for modified Newtonian dynamics (MOND) and its external field effect (EFE) breaking the strong equivalence principle (Section~\ref{section:introduction}). 
In this article, we compared observed velocity dispersion measurements in 11 UDGs of the Coma cluster (Section~\ref{section:observations}) with MOND predictions with and without EFE, hence continuing the work initiated by \cite{Bilek2019} and \cite{Haghi2019b} on DF44. 
Our modelling relied on de-projected Sérsic spherical mass models for the UDGs (Section~\ref{section:method_spherical}), uniform stellar mass-to-light ratios, a mass model stemming from X-ray observations for the Coma cluster (Section~\ref{section:coma}), and quasi-linear MOND \citep[QUMOND,][]{Milgrom2010} for the EFE.
Our work can be summarised as follows: 

\begin{itemize}
    \item We first showed that the Coma cluster UDGs of the sample lie roughly on the same radial acceleration relation (RAR) as spirals in the field, giving a first indication that their dynamical and baryonic masses are similarly correlated (Section \ref{section:RAR}).\\
    \item We then proceeded to compare the velocity dispersion profiles with Jeans modelling in isolated MOND, showing a remarkably good agreement given that there is no free parameter involved. The agreement is even better when introducing a fixed degree of radial anisotropy (Section \ref{section:method_isolated}).\\
    \item To take into account the EFE, we proposed a simple approximate analytical formula for the internal acceleration of a spherical system embedded in a constant external field in the context of QUMOND (Section \ref{section:formula}, Eq.~(\ref{eq:gr})), which we validated using a numerical Poisson solver in QUMOND (Appendix~\ref{section:EFE_numerical}).\\ 
    \item We used this formula to compare the velocity dispersion measurements with MOND models including the EFE, which pushes the predicted velocity dispersion towards the Newtonian case and away from the observations at most plausible distances from the centre of the cluster (Section \ref{section:sigma_EFE}). We noted that all other formulae would generate a stronger EFE and thus increase the tension even more.\\
\end{itemize}

These results therefore disfavour the violation of the strong equivalence principle expected in MOND for the UDGs of the Coma cluster, and imply that the EFE has to be severely damped or screened within these systems. We then discussed several options to explain this:

\begin{itemize}
    \item We considered the option that observed UDGs are biased towards objects further away from the centre of the cluster than it seems, but deemed it unlikely given their projected distance distribution (Section \ref{section:further}).\\
    \item We did not totally exclude the possibility that observed UDGs are falling into the Coma cluster on radial orbits for the first time before being disrupted by tides (Sections \ref{section:infall} and \ref{section:tides}), but we noted that the difference in dynamical times within the cluster at the location of the UDGs and within the UDGs might not be sufficient to explain the fact that measured velocity dispersions are all consistent with the fully isolated MOND prediction.\\
    \item We discussed the option of higher stellar mass-to-light $\rm M/L$ ratios than measured with standard procedures and/or baryonic dark matter located within the UDGs to increase their velocity dispersion and stabilise them against tidal disruption, but noted that it would imply a conspiracy to make it match the isolated MOND prediction for all the sampled UDGs (Sections \ref{section:higher_ML} and \ref{section:baryonic_DM}). A relatively high $\rm M/L\simeq 4$ ratio for all galaxies combined with some tidal heating was nevertheless not excluded.
    \\
    \item We also briefly discussed modified inertia, where the MOND effects would depend on the whole trajectory of a system (Section~\ref{section:inertia}), and in which the EFE could even effectively be absent from most situations. We noted that the latter case of killing the EFE altogether would be in contradiction with the constraints from, for example, wide binaries.\\
    \item We showed that the extended MOND (EMOND) framework where the critical acceleration $a_0$ is replaced by an acceleration $A_0(\Phi)$ depending on the depth of the gravitational potential does provide a good match to the velocity dispersion measurements with $A_0 \sim 10 \times a_0$ (Section \ref{section:EMOND}). We noted, however, that EMOND still struggles to explain the apparently collisionless nature of the residual missing mass in galaxy clusters.\\
    \item Finally, we considered options where the EFE would be screened in galaxy clusters (Section \ref{section:screening}), which is the case within the superfluid DM theory, and could potentially also be the case in theories where the additional degrees of freedom responsible for the MOND behaviour in galaxies would simultaneously be responsible for an additional `dust' component in galaxy clusters \citep{Skordis2020}. 
\end{itemize}

Of course, we note that MOND simply being an effective dark matter scaling relation also remains a serious possibility; however, even in this case, it is still particularly intriguing that UDGs obey the same scaling relations as field spirals despite their very different environments and formation scenarios, irrespectively of the underlying theoretical framework.

%--------------------------------------------------------------------

\begin{acknowledgements}
    We thank the anonymous referee for their uplifting  report.
    The authors would like to thank M. Milgrom, H.S. Zhao, E. Asencio, I. Banik, G. Mamon, R. Errani, F. Jiang, D. Zaritsky, F. Lelli and F. Combes for stimulating discussions and comments. B.F. is grateful for enlightening discussions with C. Skordis and A. Durakovic.
    J.F., B.F., \mbox{P.-A.O.} and R.I. acknowledge funding from the European Research Council (ERC) under the European Union's Horizon 2020 research and innovation programme (grant agreement No. 834148, GreatDigInTheSky). B.F. and R.I. also acknowledge funding from the Agence Nationale de la Recherche (ANR project ANR-18-CE31-0006 and ANR-19-CE31-0017). M.B. acknowledges support from the Polish National Science Centre under the grant 2017/26/D/ST9/00449 and is thankful for the financial support by Cercle Gutenberg. O.M. is grateful to the Swiss National Science Foundation for financial support.
\end{acknowledgements}

%--------------------------------------------------------------------

\bibliographystyle{aa}
\bibliography{Freundlich2021_MOND}

\begin{thebibliography}{206}
\expandafter\ifx\csname natexlab\endcsname\relax\def\natexlab#1{#1}\fi

\bibitem[{{Abraham} \& {van Dokkum}(2014)}]{Abraham2014}
{Abraham}, R.~G. \& {van Dokkum}, P.~G. 2014, \pasp, 126, 55

\bibitem[{{Alabi} {et~al.}(2018){Alabi}, {Ferr{\'e}-Mateu}, {Romanowsky},
  {Brodie}, {Forbes}, {Wasserman}, {Bellstedt}, {Mart{\'\i}n-Navarro},
  {Pandya}, {Stone}, \& {Okabe}}]{Alabi2018}
{Alabi}, A., {Ferr{\'e}-Mateu}, A., {Romanowsky}, A.~J., {et~al.} 2018, \mnras,
  479, 3308

\bibitem[{{Alabi} {et~al.}(2020){Alabi}, {Romanowsky}, {Forbes}, {Brodie}, \&
  {Okabe}}]{Alabi2020}
{Alabi}, A.~B., {Romanowsky}, A.~J., {Forbes}, D.~A., {Brodie}, J.~P., \&
  {Okabe}, N. 2020, \mnras, 496, 3182

\bibitem[{{Amorisco} \& {Loeb}(2016)}]{Amorisco2016}
{Amorisco}, N.~C. \& {Loeb}, A. 2016, \mnras, 459, L51

\bibitem[{{Angus} {et~al.}(2008){Angus}, {Famaey}, \& {Buote}}]{Angus2008a}
{Angus}, G.~W., {Famaey}, B., \& {Buote}, D.~A. 2008, \mnras, 387, 1470

\bibitem[{{Angus} {et~al.}(2010){Angus}, {Famaey}, \& {Diaferio}}]{Angus2010}
{Angus}, G.~W., {Famaey}, B., \& {Diaferio}, A. 2010, \mnras, 402, 395

\bibitem[{{Angus} {et~al.}(2007){Angus}, {Shan}, {Zhao}, \&
  {Famaey}}]{Angus2007}
{Angus}, G.~W., {Shan}, H.~Y., {Zhao}, H.~S., \& {Famaey}, B. 2007, \apjl, 654,
  L13

\bibitem[{{Banik}(2019)}]{Banik2019}
{Banik}, I. 2019, \mnras, 487, 5291

\bibitem[{{Banik} {et~al.}(2020){Banik}, {Thies}, {Famaey}, {Candlish},
  {Kroupa}, \& {Ibata}}]{Banik2020}
{Banik}, I., {Thies}, I., {Famaey}, B., {et~al.} 2020, \apj, 905, 135

\bibitem[{{Banik} \& {Zhao}(2015)}]{Banik2015}
{Banik}, I. \& {Zhao}, H. 2015, arXiv e-prints, arXiv:1509.08457

\bibitem[{{Beasley} {et~al.}(2016){Beasley}, {Romanowsky}, {Pota}, {Navarro},
  {Martinez Delgado}, {Neyer}, \& {Deich}}]{Beasley2016}
{Beasley}, M.~A., {Romanowsky}, A.~J., {Pota}, V., {et~al.} 2016, \apjl, 819,
  L20

\bibitem[{{Beasley} \& {Trujillo}(2016)}]{Beasley2016b}
{Beasley}, M.~A. \& {Trujillo}, I. 2016, \apj, 830, 23

\bibitem[{{Begeman} {et~al.}(1991){Begeman}, {Broeils}, \&
  {Sanders}}]{Begeman1991}
{Begeman}, K.~G., {Broeils}, A.~H., \& {Sanders}, R.~H. 1991, \mnras, 249, 523

\bibitem[{{Bekenstein} \& {Milgrom}(1984)}]{Bekenstein1984}
{Bekenstein}, J. \& {Milgrom}, M. 1984, \apj, 286, 7

\bibitem[{{Bekenstein}(2004)}]{Bekenstein2004}
{Bekenstein}, J.~D. 2004, \prd, 70, 083509

\bibitem[{{Bennet} {et~al.}(2018){Bennet}, {Sand}, {Zaritsky}, {Crnojevi{\'c}},
  {Spekkens}, \& {Karunakaran}}]{Bennet2018}
{Bennet}, P., {Sand}, D.~J., {Zaritsky}, D., {et~al.} 2018, \apjl, 866, L11

\bibitem[{{Berezhiani} {et~al.}(2019){Berezhiani}, {Elder}, \&
  {Khoury}}]{Berezhiani2019}
{Berezhiani}, L., {Elder}, B., \& {Khoury}, J. 2019, \jcap, 2019, 074

\bibitem[{{Berezhiani} {et~al.}(2018){Berezhiani}, {Famaey}, \&
  {Khoury}}]{Berezhiani2018}
{Berezhiani}, L., {Famaey}, B., \& {Khoury}, J. 2018, \jcap, 2018, 021

\bibitem[{{Berezhiani} \& {Khoury}(2015)}]{Berezhiani2015}
{Berezhiani}, L. \& {Khoury}, J. 2015, \prd, 92, 103510

\bibitem[{{Bernard} \& {Blanchet}(2015)}]{Bernard2015}
{Bernard}, L. \& {Blanchet}, L. 2015, \prd, 91, 103536

\bibitem[{{B{\'\i}lek} {et~al.}(2019{\natexlab{a}}){B{\'\i}lek}, {M{\"u}ller},
  \& {Famaey}}]{Bilek2019}
{B{\'\i}lek}, M., {M{\"u}ller}, O., \& {Famaey}, B. 2019{\natexlab{a}}, \aap,
  627, L1

\bibitem[{{B{\'\i}lek} {et~al.}(2019{\natexlab{b}}){B{\'\i}lek},
  {Samurovi{\'c}}, \& {Renaud}}]{Bilek2019a}
{B{\'\i}lek}, M., {Samurovi{\'c}}, S., \& {Renaud}, F. 2019{\natexlab{b}},
  \aap, 625, A32

\bibitem[{{Binney} \& {Mamon}(1982)}]{Binney1982}
{Binney}, J. \& {Mamon}, G.~A. 1982, \mnras, 200, 361

\bibitem[{{Binney} \& {Tremaine}(2008)}]{BinneyTremaine2008}
{Binney}, J. \& {Tremaine}, S. 2008, {Galactic Dynamics: Second Edition}
  (Princeton University Press)

\bibitem[{{Blanchet} \& {Heisenberg}(2015)}]{Blanchet2015}
{Blanchet}, L. \& {Heisenberg}, L. 2015, \jcap, 2015, 026

\bibitem[{{Blanchet} \& {Le Tiec}(2009)}]{Blanchet2009}
{Blanchet}, L. \& {Le Tiec}, A. 2009, \prd, 80, 023524

\bibitem[{{Bosma}(1978)}]{Bosma1978}
{Bosma}, A. 1978, PhD thesis

\bibitem[{{Brada} \& {Milgrom}(2000)}]{Brada2000}
{Brada}, R. \& {Milgrom}, M. 2000, \apj, 541, 556

\bibitem[{{Caldwell} {et~al.}(2017){Caldwell}, {Walker}, {Mateo}, {Olszewski},
  {Koposov}, {Belokurov}, {Torrealba}, {Geringer-Sameth}, \&
  {Johnson}}]{Caldwell2017}
{Caldwell}, N., {Walker}, M.~G., {Mateo}, M., {et~al.} 2017, \apj, 839, 20

\bibitem[{{Carleton} {et~al.}(2019){Carleton}, {Errani}, {Cooper},
  {Kaplinghat}, {Pe{\~n}arrubia}, \& {Guo}}]{Carleton2019}
{Carleton}, T., {Errani}, R., {Cooper}, M., {et~al.} 2019, \mnras, 485, 382

\bibitem[{{Carter} {et~al.}(2008){Carter}, {Goudfrooij}, {Mobasher},
  {Ferguson}, {Puzia}, {Aguerri}, {Balcells}, {Batcheldor}, {Bridges},
  {Davies}, {Erwin}, {Graham}, {Guzm{\'a}n}, {Hammer}, {Hornschemeier},
  {Hoyos}, {Hudson}, {Huxor}, {Jogee}, {Komiyama}, {Lotz}, {Lucey}, {Marzke},
  {Merritt}, {Miller}, {Miller}, {Mouhcine}, {Okamura}, {Peletier},
  {Phillipps}, {Poggianti}, {Sharples}, {Smith}, {Trentham}, {Tully},
  {Valentijn}, \& {Verdoes Kleijn}}]{Carter2008}
{Carter}, D., {Goudfrooij}, P., {Mobasher}, B., {et~al.} 2008, \apjs, 176, 424

\bibitem[{{Chae} {et~al.}(2021){Chae}, {Desmond}, {Lelli}, {McGaugh}, \&
  {Schombert}}]{Chae2021}
{Chae}, K.-H., {Desmond}, H., {Lelli}, F., {McGaugh}, S.~S., \& {Schombert},
  J.~M. 2021, arXiv e-prints, arXiv:2109.04745

\bibitem[{{Chae} {et~al.}(2020){Chae}, {Lelli}, {Desmond}, {McGaugh}, {Li}, \&
  {Schombert}}]{Chae2020}
{Chae}, K.-H., {Lelli}, F., {Desmond}, H., {et~al.} 2020, \apj, 904, 51

\bibitem[{{Chan} {et~al.}(2018){Chan}, {Kere{\v{s}}}, {Wetzel}, {Hopkins},
  {Faucher-Gigu{\`e}re}, {El-Badry}, {Garrison-Kimmel}, \&
  {Boylan-Kolchin}}]{Chan2018}
{Chan}, T.~K., {Kere{\v{s}}}, D., {Wetzel}, A., {et~al.} 2018, \mnras, 478, 906

\bibitem[{{Chilingarian} {et~al.}(2019){Chilingarian}, {Afanasiev}, {Grishin},
  {Fabricant}, \& {Moran}}]{Chilingarian2019}
{Chilingarian}, I.~V., {Afanasiev}, A.~V., {Grishin}, K.~A., {Fabricant}, D.,
  \& {Moran}, S. 2019, \apj, 884, 79

\bibitem[{{Clifton} {et~al.}(2012){Clifton}, {Ferreira}, {Padilla}, \&
  {Skordis}}]{Clifton2012}
{Clifton}, T., {Ferreira}, P.~G., {Padilla}, A., \& {Skordis}, C. 2012,
  \physrep, 513, 1

\bibitem[{{Clowe} {et~al.}(2006){Clowe}, {Brada{\v{c}}}, {Gonzalez},
  {Markevitch}, {Randall}, {Jones}, \& {Zaritsky}}]{Clowe2006}
{Clowe}, D., {Brada{\v{c}}}, M., {Gonzalez}, A.~H., {et~al.} 2006, \apjl, 648,
  L109

\bibitem[{{Colless} \& {Dunn}(1996)}]{Colless1996}
{Colless}, M. \& {Dunn}, A.~M. 1996, \apj, 458, 435

\bibitem[{{Combes}(2014)}]{Combes2014}
{Combes}, F. 2014, \aap, 571, A82

\bibitem[{{Danieli} {et~al.}(2019){Danieli}, {van Dokkum}, {Conroy}, {Abraham},
  \& {Romanowsky}}]{Danieli2019}
{Danieli}, S., {van Dokkum}, P., {Conroy}, C., {Abraham}, R., \& {Romanowsky},
  A.~J. 2019, \apjl, 874, L12

\bibitem[{{de Blok} \& {McGaugh}(1998)}]{deBlok1998}
{de Blok}, W.~J.~G. \& {McGaugh}, S.~S. 1998, \apj, 508, 132

\bibitem[{{Di Cintio} {et~al.}(2017){Di Cintio}, {Brook}, {Dutton},
  {Macci{\`o}}, {Obreja}, \& {Dekel}}]{DiCintio2017}
{Di Cintio}, A., {Brook}, C.~B., {Dutton}, A.~A., {et~al.} 2017, \mnras, 466,
  L1

\bibitem[{{Dutta Chowdhury} {et~al.}(2019){Dutta Chowdhury}, {van den Bosch},
  \& {van Dokkum}}]{DuttaChowdhury2019}
{Dutta Chowdhury}, D., {van den Bosch}, F.~C., \& {van Dokkum}, P. 2019, \apj,
  877, 133

\bibitem[{{Dutta Chowdhury} {et~al.}(2020){Dutta Chowdhury}, {van den Bosch},
  \& {van Dokkum}}]{DuttaChowdhury2020}
{Dutta Chowdhury}, D., {van den Bosch}, F.~C., \& {van Dokkum}, P. 2020, \apj,
  903, 149

\bibitem[{{Einasto}(1965)}]{Einasto1965}
{Einasto}, J. 1965, Trudy Astrofizicheskogo Instituta Alma-Ata, 5, 87

\bibitem[{{Emsellem} {et~al.}(2019){Emsellem}, {van der Burg}, {Fensch},
  {Je{\v{r}}{\'a}bkov{\'a}}, {Zanella}, {Agnello}, {Hilker}, {M{\"u}ller},
  {Rejkuba}, {Duc}, {Durrell}, {Habas}, {Lelli}, {Lim}, {Marleau}, {Peng}, \&
  {S{\'a}nchez-Janssen}}]{Emsellem2019}
{Emsellem}, E., {van der Burg}, R. F.~J., {Fensch}, J., {et~al.} 2019, \aap,
  625, A76

\bibitem[{{Faber} \& {Gallagher}(1979)}]{Faber1979}
{Faber}, S.~M. \& {Gallagher}, J.~S. 1979, \araa, 17, 135

\bibitem[{{Fabricant} {et~al.}(2019){Fabricant}, {Fata}, {Epps}, {Gauron},
  {Mueller}, {Zajac}, {Amato}, {Barberis}, {Bergner}, {Brennan}, {Brown},
  {Chilingarian}, {Geary}, {Kradinov}, {McLeod}, {Smith}, \&
  {Woods}}]{Fabricant2019}
{Fabricant}, D., {Fata}, R., {Epps}, H., {et~al.} 2019, \pasp, 131, 075004

\bibitem[{{Famaey} \& {Binney}(2005)}]{Famaey2005}
{Famaey}, B. \& {Binney}, J. 2005, \mnras, 363, 603

\bibitem[{{Famaey} {et~al.}(2018){Famaey}, {McGaugh}, \&
  {Milgrom}}]{Famaey2018}
{Famaey}, B., {McGaugh}, S., \& {Milgrom}, M. 2018, \mnras, 480, 473

\bibitem[{{Famaey} \& {McGaugh}(2012)}]{Famaey2012}
{Famaey}, B. \& {McGaugh}, S.~S. 2012, Living Reviews in Relativity, 15, 10

\bibitem[{{Fensch} {et~al.}(2019){Fensch}, {van der Burg},
  {Je{\v{r}}{\'a}bkov{\'a}}, {Emsellem}, {Zanella}, {Agnello}, {Hilker},
  {M{\"u}ller}, {Rejkuba}, {Duc}, {Durrell}, {Habas}, {Lim}, {Marleau}, {Peng},
  \& {S{\'a}nchez Janssen}}]{Fensch2019}
{Fensch}, J., {van der Burg}, R. F.~J., {Je{\v{r}}{\'a}bkov{\'a}}, T., {et~al.}
  2019, \aap, 625, A77

\bibitem[{{Ferr{\'e}-Mateu} {et~al.}(2018){Ferr{\'e}-Mateu}, {Alabi}, {Forbes},
  {Romanowsky}, {Brodie}, {Pandya}, {Mart{\'{\i}}n-Navarro}, {Bellstedt},
  {Wasserman}, {Stone}, \& {Okabe}}]{Ferre-Mateu2018}
{Ferr{\'e}-Mateu}, A., {Alabi}, A., {Forbes}, D.~A., {et~al.} 2018, \mnras,
  479, 4891

\bibitem[{{Forbes} {et~al.}(2020){Forbes}, {Alabi}, {Romanowsky}, {Brodie}, \&
  {Arimoto}}]{Forbes2020}
{Forbes}, D.~A., {Alabi}, A., {Romanowsky}, A.~J., {Brodie}, J.~P., \&
  {Arimoto}, N. 2020, \mnras, 492, 4874

\bibitem[{{Forbes} {et~al.}(2021){Forbes}, {Gannon}, {Romanowsky}, {Alabi},
  {Brodie}, {Couch}, \& {Ferr{\'e}-Mateu}}]{Forbes2021}
{Forbes}, D.~A., {Gannon}, J.~S., {Romanowsky}, A.~J., {et~al.} 2021, \mnras,
  500, 1279

\bibitem[{{Fosbury} {et~al.}(1978){Fosbury}, {Mebold}, {Goss}, \&
  {Dopita}}]{Fosbury1978}
{Fosbury}, R.~A.~E., {Mebold}, U., {Goss}, W.~M., \& {Dopita}, M.~A. 1978,
  \mnras, 183, 549

\bibitem[{{Freundlich} {et~al.}(2020{\natexlab{a}}){Freundlich}, {Dekel},
  {Jiang}, {Ishai}, {Cornuault}, {Lapiner}, {Dutton}, \&
  {Macci{\`o}}}]{Freundlich2020a}
{Freundlich}, J., {Dekel}, A., {Jiang}, F., {et~al.} 2020{\natexlab{a}},
  \mnras, 491, 4523

\bibitem[{{Freundlich} {et~al.}(2020{\natexlab{b}}){Freundlich}, {Jiang},
  {Dekel}, {Cornuault}, {Ginzburg}, {Koskas}, {Lapiner}, {Dutton}, \&
  {Macci{\`o}}}]{Freundlich2020b}
{Freundlich}, J., {Jiang}, F., {Dekel}, A., {et~al.} 2020{\natexlab{b}},
  \mnras, 499, 2912

\bibitem[{{Gannon} {et~al.}(2021){Gannon}, {Dullo}, {Forbes}, {Rich},
  {Rom{\'a}n}, {Couch}, {Brodie}, {Ferr{\'e}-Mateu}, {Alabi}, \&
  {Mould}}]{Gannon2021}
{Gannon}, J.~S., {Dullo}, B.~T., {Forbes}, D.~A., {et~al.} 2021, \mnras, 502,
  3144

\bibitem[{{Gannon} {et~al.}(2020){Gannon}, {Forbes}, {Romanowsky},
  {Ferr{\'e}-Mateu}, {Couch}, \& {Brodie}}]{Gannon2020}
{Gannon}, J.~S., {Forbes}, D.~A., {Romanowsky}, A.~J., {et~al.} 2020, \mnras,
  495, 2582

\bibitem[{{Gentile} {et~al.}(2007{\natexlab{a}}){Gentile}, {Famaey}, {Combes},
  {Kroupa}, {Zhao}, \& {Tiret}}]{Gentile2007c}
{Gentile}, G., {Famaey}, B., {Combes}, F., {et~al.} 2007{\natexlab{a}}, \aap,
  472, L25

\bibitem[{{Gentile} {et~al.}(2011){Gentile}, {Famaey}, \& {de
  Blok}}]{Gentile2011}
{Gentile}, G., {Famaey}, B., \& {de Blok}, W.~J.~G. 2011, \aap, 527, A76

\bibitem[{{Gentile} {et~al.}(2007{\natexlab{b}}){Gentile}, {Salucci}, {Klein},
  \& {Granato}}]{Gentile2007a}
{Gentile}, G., {Salucci}, P., {Klein}, U., \& {Granato}, G.~L.
  2007{\natexlab{b}}, \mnras, 375, 199

\bibitem[{{Greco} {et~al.}(2018){Greco}, {Greene}, {Strauss}, {Macarthur},
  {Flowers}, {Goulding}, {Huang}, {Kim}, {Komiyama}, {Leauthaud}, {Leisman},
  {Lupton}, {Sif{\'o}n}, \& {Wang}}]{Greco2018}
{Greco}, J.~P., {Greene}, J.~E., {Strauss}, M.~A., {et~al.} 2018, \apj, 857,
  104

\bibitem[{{Habas} {et~al.}(2020){Habas}, {Marleau}, {Duc}, {Durrell}, {Paudel},
  {Poulain}, {S{\'a}nchez-Janssen}, {Sreejith}, {Ramasawmy}, {Stemock},
  {Leach}, {Cuillandre}, {Gwyn}, {Agnello}, {B{\'\i}lek}, {Fensch},
  {M{\"u}ller}, {Peng}, \& {van der Burg}}]{Habas2020}
{Habas}, R., {Marleau}, F.~R., {Duc}, P.-A., {et~al.} 2020, \mnras, 491, 1901

\bibitem[{{Haghi} {et~al.}(2019{\natexlab{a}}){Haghi}, {Amiri}, {Hasani
  Zonoozi}, {Banik}, {Kroupa}, \& {Haslbauer}}]{Haghi2019b}
{Haghi}, H., {Amiri}, V., {Hasani Zonoozi}, A., {et~al.} 2019{\natexlab{a}},
  \apjl, 884, L25

\bibitem[{{Haghi} {et~al.}(2009){Haghi}, {Baumgardt}, {Kroupa}, {Grebel},
  {Hilker}, \& {Jordi}}]{Haghi2009}
{Haghi}, H., {Baumgardt}, H., {Kroupa}, P., {et~al.} 2009, \mnras, 395, 1549

\bibitem[{{Haghi} {et~al.}(2016){Haghi}, {Bazkiaei}, {Zonoozi}, \&
  {Kroupa}}]{Haghi2016}
{Haghi}, H., {Bazkiaei}, A.~E., {Zonoozi}, A.~H., \& {Kroupa}, P. 2016, \mnras,
  458, 4172

\bibitem[{{Haghi} {et~al.}(2019{\natexlab{b}}){Haghi}, {Kroupa}, {Banik}, {Wu},
  {Zonoozi}, {Javanmardi}, {Ghari}, {M{\"u}ller}, {Dabringhausen}, \&
  {Zhao}}]{Haghi2019}
{Haghi}, H., {Kroupa}, P., {Banik}, I., {et~al.} 2019{\natexlab{b}}, \mnras,
  487, 2441

\bibitem[{{Haslbauer} {et~al.}(2020){Haslbauer}, {Banik}, \&
  {Kroupa}}]{Haslbauer2020}
{Haslbauer}, M., {Banik}, I., \& {Kroupa}, P. 2020, \mnras, 499, 2845

\bibitem[{{Haslbauer} {et~al.}(2019){Haslbauer}, {Banik}, {Kroupa}, \&
  {Grishunin}}]{Haslbauer2019}
{Haslbauer}, M., {Banik}, I., {Kroupa}, P., \& {Grishunin}, K. 2019, \mnras,
  489, 2634

\bibitem[{{Hayashi} \& {Inoue}(2018)}]{Hayashi2018}
{Hayashi}, K. \& {Inoue}, S. 2018, \mnras, 481, L59

\bibitem[{{Hees} {et~al.}(2016){Hees}, {Famaey}, {Angus}, \&
  {Gentile}}]{Hees2016}
{Hees}, A., {Famaey}, B., {Angus}, G.~W., \& {Gentile}, G. 2016, \mnras, 455,
  449

\bibitem[{{Hernandez} {et~al.}(2021){Hernandez}, {Cookson}, \&
  {Cortes}}]{Hernandez2021}
{Hernandez}, X., {Cookson}, S., \& {Cortes}, R.~A.~M. 2021, arXiv e-prints,
  arXiv:2107.14797

\bibitem[{{Hodson} \& {Zhao}(2017{\natexlab{a}})}]{Hodson2017c}
{Hodson}, A.~O. \& {Zhao}, H. 2017{\natexlab{a}}, \aap, 607, A109

\bibitem[{{Hodson} \& {Zhao}(2017{\natexlab{b}})}]{Hodson2017a}
{Hodson}, A.~O. \& {Zhao}, H. 2017{\natexlab{b}}, \aap, 598, A127

\bibitem[{{Hu} {et~al.}(2000){Hu}, {Barkana}, \& {Gruzinov}}]{Hu2000}
{Hu}, W., {Barkana}, R., \& {Gruzinov}, A. 2000, \prl, 85, 1158

\bibitem[{{Impey} {et~al.}(1988){Impey}, {Bothun}, \& {Malin}}]{Impey1988}
{Impey}, C., {Bothun}, G., \& {Malin}, D. 1988, \apj, 330, 634

\bibitem[{{Janowiecki} {et~al.}(2015){Janowiecki}, {Leisman}, {J{\'o}zsa},
  {Salzer}, {Haynes}, {Giovanelli}, {Rhode}, {Cannon}, {Adams}, \&
  {Janesh}}]{Janowiecki2015}
{Janowiecki}, S., {Leisman}, L., {J{\'o}zsa}, G., {et~al.} 2015, \apj, 801, 96

\bibitem[{{Janssens} {et~al.}(2017){Janssens}, {Abraham}, {Brodie}, {Forbes},
  {Romanowsky}, \& {van Dokkum}}]{Janssens2017}
{Janssens}, S., {Abraham}, R., {Brodie}, J., {et~al.} 2017, \apjl, 839, L17

\bibitem[{{Jiang} {et~al.}(2019){Jiang}, {Dekel}, {Freundlich}, {Romanowsky},
  {Dutton}, {Macci{\`o}}, \& {Di Cintio}}]{Jiang2019}
{Jiang}, F., {Dekel}, A., {Freundlich}, J., {et~al.} 2019, \mnras, 487, 5272

\bibitem[{{Kadowaki} {et~al.}(2017){Kadowaki}, {Zaritsky}, \&
  {Donnerstein}}]{Kadowaki2017}
{Kadowaki}, J., {Zaritsky}, D., \& {Donnerstein}, R.~L. 2017, \apjl, 838, L21

\bibitem[{{Karachentsev} {et~al.}(2000){Karachentsev}, {Karachentseva},
  {Suchkov}, \& {Grebel}}]{Karachentsev2000}
{Karachentsev}, I.~D., {Karachentseva}, V.~E., {Suchkov}, A.~A., \& {Grebel},
  E.~K. 2000, \aaps, 145, 415

\bibitem[{{Karunakaran} {et~al.}(2020){Karunakaran}, {Spekkens}, {Zaritsky},
  {Donnerstein}, {Kadowaki}, \& {Dey}}]{Karunakaran2020}
{Karunakaran}, A., {Spekkens}, K., {Zaritsky}, D., {et~al.} 2020, \apj, 902, 39

\bibitem[{{Khoury}(2015)}]{Khoury2015}
{Khoury}, J. 2015, \prd, 91, 024022

\bibitem[{{Koda} {et~al.}(2015){Koda}, {Yagi}, {Yamanoi}, \&
  {Komiyama}}]{Koda2015}
{Koda}, J., {Yagi}, M., {Yamanoi}, H., \& {Komiyama}, Y. 2015, \apjl, 807, L2

\bibitem[{{Kretschmer} {et~al.}(2021){Kretschmer}, {Dekel}, {Freundlich},
  {Lapiner}, {Ceverino}, \& {Primack}}]{Kretschmer2021}
{Kretschmer}, M., {Dekel}, A., {Freundlich}, J., {et~al.} 2021, \mnras, 503,
  5238

\bibitem[{{Kroupa} {et~al.}(2018){Kroupa}, {Haghi}, {Javanmardi}, {Zonoozi},
  {M{\"u}ller}, {Banik}, {Wu}, {Zhao}, \& {Dabringhausen}}]{Kroupa2018}
{Kroupa}, P., {Haghi}, H., {Javanmardi}, B., {et~al.} 2018, \nat, 561, E4

\bibitem[{{Kubo} {et~al.}(2007){Kubo}, {Stebbins}, {Annis}, {Dell'Antonio},
  {Lin}, {Khiabanian}, \& {Frieman}}]{Kubo2007}
{Kubo}, J.~M., {Stebbins}, A., {Annis}, J., {et~al.} 2007, \apj, 671, 1466

\bibitem[{{Laporte} {et~al.}(2019){Laporte}, {Agnello}, \&
  {Navarro}}]{Laporte2019}
{Laporte}, C. F.~P., {Agnello}, A., \& {Navarro}, J.~F. 2019, \mnras, 484, 245

\bibitem[{{Le Borgne} {et~al.}(2004){Le Borgne}, {Rocca-Volmerange},
  {Prugniel}, {Lan{\c{c}}on}, {Fioc}, \& {Soubiran}}]{Leborgne2004}
{Le Borgne}, D., {Rocca-Volmerange}, B., {Prugniel}, P., {et~al.} 2004, \aap,
  425, 881

\bibitem[{{Leisman} {et~al.}(2017){Leisman}, {Haynes}, {Janowiecki},
  {Hallenbeck}, {J{\'o}zsa}, {Giovanelli}, {Adams}, {Bernal Neira}, {Cannon},
  {Janesh}, {Rhode}, \& {Salzer}}]{Leisman2017}
{Leisman}, L., {Haynes}, M.~P., {Janowiecki}, S., {et~al.} 2017, \apj, 842, 133

\bibitem[{{Lelli} {et~al.}(2016{\natexlab{a}}){Lelli}, {McGaugh}, \&
  {Schombert}}]{Lelli2016_SPARC}
{Lelli}, F., {McGaugh}, S.~S., \& {Schombert}, J.~M. 2016{\natexlab{a}}, \aj,
  152, 157

\bibitem[{{Lelli} {et~al.}(2016{\natexlab{b}}){Lelli}, {McGaugh}, \&
  {Schombert}}]{Lelli2016}
{Lelli}, F., {McGaugh}, S.~S., \& {Schombert}, J.~M. 2016{\natexlab{b}}, \apjl,
  816, L14

\bibitem[{{Lelli} {et~al.}(2019){Lelli}, {McGaugh}, {Schombert}, {Desmond}, \&
  {Katz}}]{Lelli2019}
{Lelli}, F., {McGaugh}, S.~S., {Schombert}, J.~M., {Desmond}, H., \& {Katz}, H.
  2019, \mnras, 484, 3267

\bibitem[{{Lelli} {et~al.}(2017){Lelli}, {McGaugh}, {Schombert}, \&
  {Pawlowski}}]{Lelli2017}
{Lelli}, F., {McGaugh}, S.~S., {Schombert}, J.~M., \& {Pawlowski}, M.~S. 2017,
  \apj, 836, 152

\bibitem[{{Li} {et~al.}(2018){Li}, {Lelli}, {McGaugh}, \& {Schombert}}]{Li2018}
{Li}, P., {Lelli}, F., {McGaugh}, S., \& {Schombert}, J. 2018, \aap, 615, A3

\bibitem[{{Lim} {et~al.}(2020){Lim}, {C{\^o}t{\'e}}, {Peng}, {Ferrarese},
  {Roediger}, {Durrell}, {Mihos}, {Wang}, {Gwyn}, {Cuillandre}, {Liu},
  {S{\'a}nchez-Janssen}, {Toloba}, {Sales}, {Guhathakurta}, {Lan{\c{c}}on}, \&
  {Puzia}}]{Lim2020}
{Lim}, S., {C{\^o}t{\'e}}, P., {Peng}, E.~W., {et~al.} 2020, \apj, 899, 69

\bibitem[{{Lim} {et~al.}(2018){Lim}, {Peng}, {C{\^o}t{\'e}}, {Sales}, {den
  Brok}, {Blakeslee}, \& {Guhathakurta}}]{Lim2018}
{Lim}, S., {Peng}, E.~W., {C{\^o}t{\'e}}, P., {et~al.} 2018, \apj, 862, 82

\bibitem[{{Lima Neto} {et~al.}(1999){Lima Neto}, {Gerbal}, \&
  {M{\'a}rquez}}]{LimaNeto1999}
{Lima Neto}, G.~B., {Gerbal}, D., \& {M{\'a}rquez}, I. 1999, \mnras, 309, 481

\bibitem[{{L{\"u}ghausen} {et~al.}(2015){L{\"u}ghausen}, {Famaey}, \&
  {Kroupa}}]{Lughausen2015}
{L{\"u}ghausen}, F., {Famaey}, B., \& {Kroupa}, P. 2015, Canadian Journal of
  Physics, 93, 232

\bibitem[{{L{\"u}ghausen} {et~al.}(2013){L{\"u}ghausen}, {Famaey}, {Kroupa},
  {Angus}, {Combes}, {Gentile}, {Tiret}, \& {Zhao}}]{Lughausen2013}
{L{\"u}ghausen}, F., {Famaey}, B., {Kroupa}, P., {et~al.} 2013, \mnras, 432,
  2846

\bibitem[{{Mamon} \& {{\L}okas}(2005)}]{Mamon2005}
{Mamon}, G.~A. \& {{\L}okas}, E.~L. 2005, \mnras, 363, 705

\bibitem[{{M{\'a}rquez} {et~al.}(2000){M{\'a}rquez}, {Lima Neto}, {Capelato},
  {Durret}, \& {Gerbal}}]{Marquez2000}
{M{\'a}rquez}, I., {Lima Neto}, G.~B., {Capelato}, H., {Durret}, F., \&
  {Gerbal}, D. 2000, \aap, 353, 873

\bibitem[{{Martin} {et~al.}(2018){Martin}, {Collins}, {Longeard}, \&
  {Tollerud}}]{Martin2018}
{Martin}, N.~F., {Collins}, M. L.~M., {Longeard}, N., \& {Tollerud}, E. 2018,
  \apjl, 859, L5

\bibitem[{{Mart{\'{\i}}nez-Delgado} {et~al.}(2016){Mart{\'{\i}}nez-Delgado},
  {L{\"a}sker}, {Sharina}, {Toloba}, {Fliri}, {Beaton}, {Valls-Gabaud},
  {Karachentsev}, {Chonis}, {Grebel}, {Forbes}, {Romanowsky},
  {Gallego-Laborda}, {Teuwen}, {G{\'o}mez-Flechoso}, {Wang}, {Guhathakurta},
  {Kaisin}, \& {Ho}}]{Martinez-Delgado2016}
{Mart{\'{\i}}nez-Delgado}, D., {L{\"a}sker}, R., {Sharina}, M., {et~al.} 2016,
  \aj, 151, 96

\bibitem[{{McGaugh} {et~al.}(2021){McGaugh}, {Lelli}, {Schombert}, {Li},
  {Visgaitis}, {Parker}, \& {Pawlowski}}]{McGaugh2021}
{McGaugh}, S., {Lelli}, F., {Schombert}, J., {et~al.} 2021, arXiv e-prints,
  arXiv:2109.03251

\bibitem[{{McGaugh} \& {Milgrom}(2013{\natexlab{a}})}]{McGaugh2013a}
{McGaugh}, S. \& {Milgrom}, M. 2013{\natexlab{a}}, \apj, 766, 22

\bibitem[{{McGaugh} \& {Milgrom}(2013{\natexlab{b}})}]{McGaugh2013b}
{McGaugh}, S. \& {Milgrom}, M. 2013{\natexlab{b}}, \apj, 775, 139

\bibitem[{{McGaugh}(2005)}]{McGaugh2005}
{McGaugh}, S.~S. 2005, \apj, 632, 859

\bibitem[{{McGaugh}(2008)}]{McGaugh2008}
{McGaugh}, S.~S. 2008, \apj, 683, 137

\bibitem[{{McGaugh}(2012)}]{McGaugh2012}
{McGaugh}, S.~S. 2012, \aj, 143, 40

\bibitem[{{McGaugh}(2016)}]{McGaugh2016_crater}
{McGaugh}, S.~S. 2016, \apjl, 832, L8

\bibitem[{{McGaugh} \& {de Blok}(1998)}]{McGaugh1998}
{McGaugh}, S.~S. \& {de Blok}, W.~J.~G. 1998, \apj, 499, 66

\bibitem[{{McGaugh} {et~al.}(2016){McGaugh}, {Lelli}, \&
  {Schombert}}]{McGaugh2016}
{McGaugh}, S.~S., {Lelli}, F., \& {Schombert}, J.~M. 2016, \prl, 117, 201101

\bibitem[{{McGaugh} {et~al.}(2000){McGaugh}, {Schombert}, {Bothun}, \& {de
  Blok}}]{McGaugh2000}
{McGaugh}, S.~S., {Schombert}, J.~M., {Bothun}, G.~D., \& {de Blok}, W.~J.~G.
  2000, \apjl, 533, L99

\bibitem[{{Merritt} {et~al.}(2016){Merritt}, {van Dokkum}, {Danieli},
  {Abraham}, {Zhang}, {Karachentsev}, \& {Makarova}}]{Merritt2016}
{Merritt}, A., {van Dokkum}, P., {Danieli}, S., {et~al.} 2016, \apj, 833, 168

\bibitem[{{Mihos} {et~al.}(2015){Mihos}, {Durrell}, {Ferrarese}, {Feldmeier},
  {C{\^o}t{\'e}}, {Peng}, {Harding}, {Liu}, {Gwyn}, \&
  {Cuillandre}}]{Mihos2015}
{Mihos}, J.~C., {Durrell}, P.~R., {Ferrarese}, L., {et~al.} 2015, \apjl, 809,
  L21

\bibitem[{{Mihos} {et~al.}(2017){Mihos}, {Harding}, {Feldmeier}, {Rudick},
  {Janowiecki}, {Morrison}, {Slater}, \& {Watkins}}]{Mihos2017}
{Mihos}, J.~C., {Harding}, P., {Feldmeier}, J.~J., {et~al.} 2017, \apj, 834, 16

\bibitem[{{Milgrom}(1983{\natexlab{a}})}]{Milgrom1983b}
{Milgrom}, M. 1983{\natexlab{a}}, \apj, 270, 371

\bibitem[{{Milgrom}(1983{\natexlab{b}})}]{Milgrom1983c}
{Milgrom}, M. 1983{\natexlab{b}}, \apj, 270, 384

\bibitem[{{Milgrom}(1983{\natexlab{c}})}]{Milgrom1983a}
{Milgrom}, M. 1983{\natexlab{c}}, \apj, 270, 365

\bibitem[{{Milgrom}(1984)}]{Milgrom1984}
{Milgrom}, M. 1984, \apj, 287, 571

\bibitem[{{Milgrom}(1986)}]{Milgrom1986b}
{Milgrom}, M. 1986, \apj, 306, 9

\bibitem[{{Milgrom}(1994)}]{Milgrom1994}
{Milgrom}, M. 1994, Annals of Physics, 229, 384

\bibitem[{{Milgrom}(2002)}]{Milgrom2002}
{Milgrom}, M. 2002, \nar, 46, 741

\bibitem[{{Milgrom}(2008)}]{Milgrom2008}
{Milgrom}, M. 2008, \nar, 51, 906

\bibitem[{{Milgrom}(2010)}]{Milgrom2010}
{Milgrom}, M. 2010, \mnras, 403, 886

\bibitem[{{Milgrom}(2011)}]{Milgrom2011}
{Milgrom}, M. 2011, arXiv e-prints, arXiv:1111.1611

\bibitem[{{Milgrom}(2012)}]{Milgrom2012}
{Milgrom}, M. 2012, \prl, 109, 131101

\bibitem[{{Milgrom}(2014)}]{Milgrom2014}
{Milgrom}, M. 2014, \mnras, 437, 2531

\bibitem[{{Milgrom}(2015)}]{Milgrom2015}
{Milgrom}, M. 2015, \mnras, 454, 3810

\bibitem[{{Milgrom}(2016)}]{Milgrom2016}
{Milgrom}, M. 2016, arXiv e-prints, arXiv:1609.06642

\bibitem[{{Montes} {et~al.}(2020){Montes}, {Infante-Sainz}, {Madrigal-Aguado},
  {Rom{\'a}n}, {Monelli}, {Borlaff}, \& {Trujillo}}]{Montes2020}
{Montes}, M., {Infante-Sainz}, R., {Madrigal-Aguado}, A., {et~al.} 2020, \apj,
  904, 114

\bibitem[{{Morrissey} {et~al.}(2012){Morrissey}, {Matuszewski}, {Martin},
  {Moore}, {Adkins}, {Epps}, {Bartos}, {Cabak}, {Cowley}, {Davis}, {Delacroix},
  {Fucik}, {Hilliard}, {James}, {Kaye}, {Lingner}, {Neill}, {Pistor},
  {Phillips}, {Rockosi}, \& {Weber}}]{Morrissey2012}
{Morrissey}, P., {Matuszewski}, M., {Martin}, C., {et~al.} 2012, in Society of
  Photo-Optical Instrumentation Engineers (SPIE) Conference Series, Vol. 8446,
  Ground-based and Airborne Instrumentation for Astronomy IV, ed. I.~S.
  {McLean}, S.~K. {Ramsay}, \& H.~{Takami}, 844613

\bibitem[{{Morrissey} {et~al.}(2018){Morrissey}, {Matuszewski}, {Martin},
  {Neill}, {Epps}, {Fucik}, {Weber}, {Darvish}, {Adkins}, {Allen}, {Bartos},
  {Belicki}, {Cabak}, {Callahan}, {Cowley}, {Crabill}, {Deich}, {Delecroix},
  {Doppman}, {Hilyard}, {James}, {Kaye}, {Kokorowski}, {Kwok}, {Lanclos},
  {Milner}, {Moore}, {O'Sullivan}, {Parihar}, {Park}, {Phillips}, {Rizzi},
  {Rockosi}, {Rodriguez}, {Salaun}, {Seaman}, {Sheikh}, {Weiss}, \&
  {Zarzaca}}]{Morrissey2018}
{Morrissey}, P., {Matuszewski}, M., {Martin}, D.~C., {et~al.} 2018, \apj, 864,
  93

\bibitem[{{Mu{\~n}oz} {et~al.}(2015){Mu{\~n}oz}, {Eigenthaler}, {Puzia},
  {Taylor}, {Ordenes-Brice{\~n}o}, {Alamo-Mart{\'{\i}}nez}, {Ribbeck},
  {{\'A}ngel}, {Capaccioli}, {C{\^o}t{\'e}}, {Ferrarese}, {Galaz}, {Hempel},
  {Hilker}, {Jord{\'a}n}, {Lan{\c c}on}, {Mieske}, {Paolillo}, {Richtler},
  {S{\'a}nchez-Janssen}, \& {Zhang}}]{Munoz2015}
{Mu{\~n}oz}, R.~P., {Eigenthaler}, P., {Puzia}, T.~H., {et~al.} 2015, \apjl,
  813, L15

\bibitem[{{M{\"u}ller} {et~al.}(2021){M{\"u}ller}, {Durrell}, {Marleau}, {Duc},
  {Lim}, {Posti}, {Agnello}, {S{\'a}nchez-Janssen}, {Poulain}, {Habas},
  {Emsellem}, {Paudel}, {van der Burg}, \& {Fensch}}]{Muller2021}
{M{\"u}ller}, O., {Durrell}, P.~R., {Marleau}, F.~R., {et~al.} 2021, arXiv
  e-prints, arXiv:2101.10659

\bibitem[{{M{\"u}ller} {et~al.}(2019){M{\"u}ller}, {Famaey}, \&
  {Zhao}}]{Muller2019a}
{M{\"u}ller}, O., {Famaey}, B., \& {Zhao}, H. 2019, \aap, 623, A36

\bibitem[{{M{\"u}ller} {et~al.}(2018){M{\"u}ller}, {Jerjen}, \&
  {Binggeli}}]{Muller2018}
{M{\"u}ller}, O., {Jerjen}, H., \& {Binggeli}, B. 2018, \aap, 615, A105

\bibitem[{{M{\"u}ller} {et~al.}(2020){M{\"u}ller}, {Marleau}, {Duc}, {Habas},
  {Fensch}, {Emsellem}, {Poulain}, {Lim}, {Agnello}, {Durrell}, {Paudel},
  {S{\'a}nchez-Janssen}, \& {van der Burg}}]{Muller2020}
{M{\"u}ller}, O., {Marleau}, F.~R., {Duc}, P.-A., {et~al.} 2020, \aap, 640,
  A106

\bibitem[{{Natarajan} \& {Zhao}(2008)}]{Natarajan2008}
{Natarajan}, P. \& {Zhao}, H. 2008, \mnras, 389, 250

\bibitem[{{Nusser}(2018)}]{Nusser2018}
{Nusser}, A. 2018, \apjl, 863, L17

\bibitem[{{Nusser}(2019)}]{Nusser2019}
{Nusser}, A. 2019, \mnras, 484, 510

\bibitem[{{Oria} {et~al.}(2021){Oria}, {Famaey}, {Thomas}, {Ibata},
  {Freundlich}, {Posti}, {Korsaga}, {Monari}, {M{\"u}ller}, {Libeskind}, \&
  {Pawlowski}}]{Oria2021}
{Oria}, P.~A., {Famaey}, B., {Thomas}, G.~F., {et~al.} 2021, arXiv e-prints,
  arXiv:2109.10160

\bibitem[{{Pandya} {et~al.}(2018){Pandya}, {Romanowsky}, {Laine}, {Brodie},
  {Johnson}, {Glaccum}, {Villaume}, {Cuillandre}, {Gwyn}, {Krick}, {Lasker},
  {Mart{\'{\i}}n-Navarro}, {Martinez-Delgado}, \& {van Dokkum}}]{Pandya2018}
{Pandya}, V., {Romanowsky}, A.~J., {Laine}, S., {et~al.} 2018, \apj, 858, 29

\bibitem[{{Papastergis} {et~al.}(2017){Papastergis}, {Adams}, \&
  {Romanowsky}}]{Papastergis2017}
{Papastergis}, E., {Adams}, E.~A.~K., \& {Romanowsky}, A.~J. 2017, \aap, 601,
  L10

\bibitem[{{Pawlowski} \& {McGaugh}(2014)}]{Pawlowski2014}
{Pawlowski}, M.~S. \& {McGaugh}, S.~S. 2014, \mnras, 440, 908

\bibitem[{{Pawlowski} {et~al.}(2015){Pawlowski}, {McGaugh}, \&
  {Jerjen}}]{Pawlowski2015}
{Pawlowski}, M.~S., {McGaugh}, S.~S., \& {Jerjen}, H. 2015, \mnras, 453, 1047

\bibitem[{{Peng} \& {Lim}(2016)}]{Peng2016}
{Peng}, E.~W. \& {Lim}, S. 2016, \apjl, 822, L31

\bibitem[{{Pittordis} \& {Sutherland}(2019)}]{Pittordis2019}
{Pittordis}, C. \& {Sutherland}, W. 2019, \mnras, 488, 4740

\bibitem[{{Pointecouteau} \& {Silk}(2005)}]{Pointecouteau2005}
{Pointecouteau}, E. \& {Silk}, J. 2005, \mnras, 364, 654

\bibitem[{{Prole} {et~al.}(2019){Prole}, {van der Burg}, {Hilker}, \&
  {Davies}}]{Prole2019}
{Prole}, D.~J., {van der Burg}, R.~F.~J., {Hilker}, M., \& {Davies}, J.~I.
  2019, \mnras, 488, 2143

\bibitem[{{Reiprich}(2001)}]{Reiprich2001}
{Reiprich}, T.~H. 2001, PhD thesis, Max-Planck-Institut f{\"u}r
  extraterrestrische Physik, P.O. Box 1312, Garching bei M{\"u}nchen, Germany

\bibitem[{{Rom{\'a}n} \& {Trujillo}(2017)}]{Roman2017}
{Rom{\'a}n}, J. \& {Trujillo}, I. 2017, \mnras, 468, 703

\bibitem[{{Rong} {et~al.}(2017){Rong}, {Guo}, {Gao}, {Liao}, {Xie}, {Puzia},
  {Sun}, \& {Pan}}]{Rong2017}
{Rong}, Y., {Guo}, Q., {Gao}, L., {et~al.} 2017, \mnras, 470, 4231

\bibitem[{{Roshan} {et~al.}(2021){Roshan}, {Banik}, {Ghafourian}, {Thies},
  {Famaey}, {Asencio}, \& {Kroupa}}]{Roshan2021}
{Roshan}, M., {Banik}, I., {Ghafourian}, N., {et~al.} 2021, \mnras, 503, 2833

\bibitem[{{Rubin} {et~al.}(1978){Rubin}, {Ford}, \& {Thonnard}}]{Rubin1978}
{Rubin}, V.~C., {Ford}, W.~K., J., \& {Thonnard}, N. 1978, \apjl, 225, L107

\bibitem[{{Ruiz-Lara} {et~al.}(2018){Ruiz-Lara}, {Beasley},
  {Falc{\'o}n-Barroso}, {Rom{\'a}n}, {Pinna}, {Brook}, {Di Cintio},
  {Mart{\'\i}n-Navarro}, {Trujillo}, \& {Vazdekis}}]{Ruiz-Lara2018}
{Ruiz-Lara}, T., {Beasley}, M.~A., {Falc{\'o}n-Barroso}, J., {et~al.} 2018,
  \mnras, 478, 2034

\bibitem[{{Sales} {et~al.}(2020){Sales}, {Navarro}, {Pe{\~n}afiel}, {Peng},
  {Lim}, \& {Hernquist}}]{Sales2020}
{Sales}, L.~V., {Navarro}, J.~F., {Pe{\~n}afiel}, L., {et~al.} 2020, \mnras,
  494, 1848

\bibitem[{{Sandage} \& {Binggeli}(1984)}]{Sandage1984}
{Sandage}, A. \& {Binggeli}, B. 1984, \aj, 89, 919

\bibitem[{{Sanders}(1996)}]{Sanders1996}
{Sanders}, R.~H. 1996, \apj, 473, 117

\bibitem[{{Sanders}(1999)}]{Sanders1999}
{Sanders}, R.~H. 1999, \apjl, 512, L23

\bibitem[{{Sanders}(2003)}]{Sanders2003}
{Sanders}, R.~H. 2003, \mnras, 342, 901

\bibitem[{{Sanders}(2019)}]{Sanders2019}
{Sanders}, R.~H. 2019, \mnras, 485, 513

\bibitem[{{Sanders} \& {McGaugh}(2002)}]{Sanders2002}
{Sanders}, R.~H. \& {McGaugh}, S.~S. 2002, \araa, 40, 263

\bibitem[{{Sanders} \& {Noordermeer}(2007)}]{Sanders2007}
{Sanders}, R.~H. \& {Noordermeer}, E. 2007, \mnras, 379, 702

\bibitem[{{Sanders} \& {Verheijen}(1998)}]{Sanders1998}
{Sanders}, R.~H. \& {Verheijen}, M.~A.~W. 1998, \apj, 503, 97

\bibitem[{{Scott} {et~al.}(2021){Scott}, {Sengupta}, {Lagos}, {Chung}, \&
  {Wong}}]{Scott2021}
{Scott}, T.~C., {Sengupta}, C., {Lagos}, P., {Chung}, A., \& {Wong}, O.~I.
  2021, \mnras, 503, 3953

\bibitem[{{Shi} {et~al.}(2017){Shi}, {Zheng}, {Zhao}, {Pan}, {Li}, {Zou},
  {Zhou}, {Guo}, {An}, \& {Li}}]{Shi2017}
{Shi}, D.~D., {Zheng}, X.~Z., {Zhao}, H.~B., {et~al.} 2017, \apj, 846, 26

\bibitem[{{Skordis} \& {Z{\l}o{\'s}nik}(2019)}]{Skordis2019}
{Skordis}, C. \& {Z{\l}o{\'s}nik}, T. 2019, \prd, 100, 104013

\bibitem[{{Skordis} \& {Z{\l}o{\'s}nik}(2020)}]{Skordis2020}
{Skordis}, C. \& {Z{\l}o{\'s}nik}, T. 2020, arXiv e-prints, arXiv:2007.00082

\bibitem[{{Swaters} {et~al.}(2010){Swaters}, {Sanders}, \&
  {McGaugh}}]{Swaters2010}
{Swaters}, R.~A., {Sanders}, R.~H., \& {McGaugh}, S.~S. 2010, \apj, 718, 380

\bibitem[{{Thomas} {et~al.}(2018){Thomas}, {Famaey}, {Ibata}, {Renaud},
  {Martin}, \& {Kroupa}}]{Thomas2018}
{Thomas}, G.~F., {Famaey}, B., {Ibata}, R., {et~al.} 2018, \aap, 609, A44

\bibitem[{{Tiret} \& {Combes}(2007)}]{Tiret2007}
{Tiret}, O. \& {Combes}, F. 2007, \aap, 464, 517

\bibitem[{{Tiret} \& {Combes}(2008)}]{Tiret2008}
{Tiret}, O. \& {Combes}, F. 2008, \aap, 483, 719

\bibitem[{{Toloba} {et~al.}(2018){Toloba}, {Lim}, {Peng}, {Sales},
  {Guhathakurta}, {Mihos}, {C{\^o}t{\'e}}, {Boselli}, {Cuillandre},
  {Ferrarese}, {Gwyn}, {Lan{\c{c}}on}, {Mu{\~n}oz}, \& {Puzia}}]{Toloba2018}
{Toloba}, E., {Lim}, S., {Peng}, E., {et~al.} 2018, \apjl, 856, L31

\bibitem[{{Toloba} {et~al.}(2016){Toloba}, {Sand}, {Spekkens}, {Crnojevi{\'c}},
  {Simon}, {Guhathakurta}, {Strader}, {Caldwell}, {McLeod}, \&
  {Seth}}]{Toloba2016}
{Toloba}, E., {Sand}, D.~J., {Spekkens}, K., {et~al.} 2016, \apjl, 816, L5

\bibitem[{{Trujillo} {et~al.}(2019){Trujillo}, {Beasley}, {Borlaff},
  {Carrasco}, {Di Cintio}, {Filho}, {Monelli}, {Montes}, {Rom{\'a}n},
  {Ruiz-Lara}, {S{\'a}nchez Almeida}, {Valls-Gabaud}, \&
  {Vazdekis}}]{Trujillo2019}
{Trujillo}, I., {Beasley}, M.~A., {Borlaff}, A., {et~al.} 2019, \mnras, 486,
  1192

\bibitem[{{van der Burg} {et~al.}(2016){van der Burg}, {Muzzin}, \&
  {Hoekstra}}]{vanderBurg2016}
{van der Burg}, R. F.~J., {Muzzin}, A., \& {Hoekstra}, H. 2016, \aap, 590, A20

\bibitem[{{van der Marel} \& {Franx}(1993)}]{vanderMarel1993}
{van der Marel}, R.~P. \& {Franx}, M. 1993, \apj, 407, 525

\bibitem[{{van Dokkum} {et~al.}(2016){van Dokkum}, {Abraham}, {Brodie},
  {Conroy}, {Danieli}, {Merritt}, {Mowla}, {Romanowsky}, \&
  {Zhang}}]{vanDokkum2016}
{van Dokkum}, P., {Abraham}, R., {Brodie}, J., {et~al.} 2016, \apjl, 828, L6

\bibitem[{{van Dokkum} {et~al.}(2017){van Dokkum}, {Abraham}, {Romanowsky},
  {Brodie}, {Conroy}, {Danieli}, {Lokhorst}, {Merritt}, {Mowla}, \&
  {Zhang}}]{VanDokkum2017}
{van Dokkum}, P., {Abraham}, R., {Romanowsky}, A.~J., {et~al.} 2017, \apjl,
  844, L11

\bibitem[{{van Dokkum} {et~al.}(2019{\natexlab{a}}){van Dokkum}, {Danieli},
  {Abraham}, {Conroy}, \& {Romanowsky}}]{VanDokkum2019a}
{van Dokkum}, P., {Danieli}, S., {Abraham}, R., {Conroy}, C., \& {Romanowsky},
  A.~J. 2019{\natexlab{a}}, \apjl, 874, L5

\bibitem[{{van Dokkum} {et~al.}(2018){van Dokkum}, {Danieli}, {Cohen},
  {Merritt}, {Romanowsky}, {Abraham}, {Brodie}, {Conroy}, {Lokhorst}, {Mowla},
  {O'Sullivan}, \& {Zhang}}]{VanDokkum2018a}
{van Dokkum}, P., {Danieli}, S., {Cohen}, Y., {et~al.} 2018, \nat, 555, 629

\bibitem[{{van Dokkum} {et~al.}(2019{\natexlab{b}}){van Dokkum}, {Wasserman},
  {Danieli}, {Abraham}, {Brodie}, {Conroy}, {Forbes}, {Martin}, {Matuszewski},
  {Romanowsky}, \& {Villaume}}]{vanDokkum2019b}
{van Dokkum}, P., {Wasserman}, A., {Danieli}, S., {et~al.} 2019{\natexlab{b}},
  \apj, 880, 91

\bibitem[{{van Dokkum} {et~al.}(2015{\natexlab{a}}){van Dokkum}, {Abraham},
  {Merritt}, {Zhang}, {Geha}, \& {Conroy}}]{VanDokkum2015}
{van Dokkum}, P.~G., {Abraham}, R., {Merritt}, A., {et~al.} 2015{\natexlab{a}},
  \apjl, 798, L45

\bibitem[{{van Dokkum} {et~al.}(2015{\natexlab{b}}){van Dokkum}, {Romanowsky},
  {Abraham}, {Brodie}, {Conroy}, {Geha}, {Merritt}, {Villaume}, \&
  {Zhang}}]{vanDokkum2015b}
{van Dokkum}, P.~G., {Romanowsky}, A.~J., {Abraham}, R., {et~al.}
  2015{\natexlab{b}}, \apjl, 804, L26

\bibitem[{{Venhola} {et~al.}(2017){Venhola}, {Peletier}, {Laurikainen}, {Salo},
  {Lisker}, {Iodice}, {Capaccioli}, {Verdois Kleijn}, {Valentijn}, {Mieske},
  {Hilker}, {Wittmann}, {van de Ven}, {Grado}, {Spavone}, {Cantiello},
  {Napolitano}, {Paolillo}, \& {Falc{\'o}n-Barroso}}]{Venhola2017}
{Venhola}, A., {Peletier}, R., {Laurikainen}, E., {et~al.} 2017, \aap, 608,
  A142

\bibitem[{{Villaume} {et~al.}(2021){Villaume}, {Romanowsky}, {Brodie}, {van
  Dokkum}, {Conroy}, {Forbes}, {Danieli}, {Martin}, \&
  {Matuszewski}}]{Villaume2021}
{Villaume}, A., {Romanowsky}, A.~J., {Brodie}, J., {et~al.} 2021, arXiv
  e-prints, arXiv:2101.02220

\bibitem[{{Wasserman} {et~al.}(2018){Wasserman}, {Romanowsky}, {Brodie}, {van
  Dokkum}, {Conroy}, {Abraham}, {Cohen}, \& {Danieli}}]{Wasserman2018}
{Wasserman}, A., {Romanowsky}, A.~J., {Brodie}, J., {et~al.} 2018, \apjl, 863,
  L15

\bibitem[{{Wasserman} {et~al.}(2019){Wasserman}, {van Dokkum}, {Romanowsky},
  {Brodie}, {Danieli}, {Forbes}, {Abraham}, {Martin}, {Matuszewski},
  {Villaume}, {Tamanas}, \& {Profumo}}]{Wasserman2019}
{Wasserman}, A., {van Dokkum}, P., {Romanowsky}, A.~J., {et~al.} 2019, \apj,
  885, 155

\bibitem[{{Wittmann} {et~al.}(2017){Wittmann}, {Lisker}, {Ambachew Tilahun},
  {Grebel}, {Conselice}, {Penny}, {Janz}, {Gallagher}, {Kotulla}, \&
  {McCormac}}]{Wittmann2017}
{Wittmann}, C., {Lisker}, T., {Ambachew Tilahun}, L., {et~al.} 2017, \mnras,
  470, 1512

\bibitem[{{Wolf} {et~al.}(2010){Wolf}, {Martinez}, {Bullock}, {Kaplinghat},
  {Geha}, {Mu{\~n}oz}, {Simon}, \& {Avedo}}]{Wolf2010}
{Wolf}, J., {Martinez}, G.~D., {Bullock}, J.~S., {et~al.} 2010, \mnras, 406,
  1220

\bibitem[{{Wu} \& {Kroupa}(2015)}]{Wu2015}
{Wu}, X. \& {Kroupa}, P. 2015, \mnras, 446, 330

\bibitem[{{Wu} {et~al.}(2007){Wu}, {Zhao}, {Famaey}, {Gentile}, {Tiret},
  {Combes}, {Angus}, \& {Robin}}]{Wu2007}
{Wu}, X., {Zhao}, H., {Famaey}, B., {et~al.} 2007, \apjl, 665, L101

\bibitem[{{Yagi} {et~al.}(2016){Yagi}, {Koda}, {Komiyama}, \&
  {Yamanoi}}]{Yagi2016}
{Yagi}, M., {Koda}, J., {Komiyama}, Y., \& {Yamanoi}, H. 2016, \apjs, 225, 11

\bibitem[{{Yozin} \& {Bekki}(2015)}]{Yozin2015}
{Yozin}, C. \& {Bekki}, K. 2015, \mnras, 452, 937

\bibitem[{{Zaritsky} {et~al.}(2019){Zaritsky}, {Donnerstein}, {Dey},
  {Kadowaki}, {Zhang}, {Karunakaran}, {Mart{\'\i}nez-Delgado}, {Rahman}, \&
  {Spekkens}}]{Zaritsky2019}
{Zaritsky}, D., {Donnerstein}, R., {Dey}, A., {et~al.} 2019, \apjs, 240, 1

\bibitem[{{Zaritsky} {et~al.}(2006{\natexlab{a}}){Zaritsky}, {Gonzalez}, \&
  {Zabludoff}}]{Zaritsky2006b}
{Zaritsky}, D., {Gonzalez}, A.~H., \& {Zabludoff}, A.~I. 2006{\natexlab{a}},
  \apjl, 642, L37

\bibitem[{{Zaritsky} {et~al.}(2006{\natexlab{b}}){Zaritsky}, {Gonzalez}, \&
  {Zabludoff}}]{Zaritsky2006a}
{Zaritsky}, D., {Gonzalez}, A.~H., \& {Zabludoff}, A.~I. 2006{\natexlab{b}},
  \apj, 638, 725

\bibitem[{{Zaritsky} {et~al.}(2008){Zaritsky}, {Zabludoff}, \&
  {Gonzalez}}]{Zaritsky2008}
{Zaritsky}, D., {Zabludoff}, A.~I., \& {Gonzalez}, A.~H. 2008, in Astronomical
  Society of the Pacific Conference Series, Vol. 396, Formation and Evolution
  of Galaxy Disks, ed. J.~G. {Funes} \& E.~M. {Corsini}, 381

\bibitem[{{Zhao} \& {Famaey}(2012)}]{Zhao2012}
{Zhao}, H. \& {Famaey}, B. 2012, \prd, 86, 067301

\bibitem[{{Zhao} \& {Tian}(2006)}]{ZhaoTian2006}
{Zhao}, H. \& {Tian}, L. 2006, \aap, 450, 1005

\bibitem[{{Zhao}(2005)}]{Zhao2005}
{Zhao}, H.~S. 2005, \aap, 444, L25

\bibitem[{{Zhao} \& {Famaey}(2006)}]{Zhao2006}
{Zhao}, H.~S. \& {Famaey}, B. 2006, \apjl, 638, L9

\end{thebibliography}

\appendix

%--------------------------------------------------------------------

\section{Numerical test of the EFE formula}
\label{section:EFE_numerical}

In order to test the accuracy of Eq.~(\ref{eq:gr}), we compared the acceleration it yields to the QUMOND gravitational acceleration obtained by numerical computation on a given test case. For this purpose, we considered a UDG modelled as a Plummer sphere of baryonic mass $M_0=3.9\times 10^{8}~\rm M_\odot$ and characteristic radius $R_P=3.6 \rm ~kpc$ (i.e. a half-mass radius of 4.7 kpc) located at distances $d= 0.5$, 1, 2, 5, 10, and 20 Mpc from a point mass $M_C(d)$ representing the centre of the Coma cluster. The mass and radius of the UDG are comparable to those of DF44 (cf. Table~\ref{table:sample}).
The MOND potential $\Phi$ was obtained by numerical integration of the QUMOND Poisson equation: 
\begin{equation}
\label{eq:Poisson_PDM}
\Delta \Phi = 4\pi G(\rho+\rho_{\rm PDM}) 
,\end{equation}
where $\rho$ is the baryonic density and $\rho_{\rm PDM}$ the phantom dark matter (PDM) density \citep[e.g.][]{Milgrom1986b,Milgrom2010}.
The PDM is the theoretical matter that would be needed to obtain MOND effects with Newtonian gravity; its density can be expressed by combining the Newtonian Poisson equation, 
\begin{equation}
\label{eq:Poisson_N}
\Delta \Phi_N = 4\pi G \rho
,\end{equation}
with the QUMOND generalised Poisson equation, 
\begin{equation}
\label{eq:Poisson_QUMOND}
\Delta \Phi = \nabla\cdot\left[\nu\left(\frac{\left | \nabla\Phi_N \right |}{a_0}\right)\nabla\Phi_N\right], 
\end{equation}
yielding
\begin{equation}
\label{eq:rho_PDM}
\rho_{\rm PDM} = \frac{1}{4\pi G} \nabla\cdot\left[\left(\nu\left(\frac{\left | \nabla\Phi_N \right |}{a_0}\right)-1\right)\nabla\Phi_N\right]
,\end{equation}
where $\Phi_N$ is the Newtonian potential. 
This formula was discretised on a grid pattern using finite differences and computed numerically, as, for example, in \cite{Lughausen2015} and \cite{Oria2021}.

\begin{figure*}
\centering
\includegraphics[width=0.48\linewidth,trim={0cm 0.4cm 0cm 0.4cm},clip]{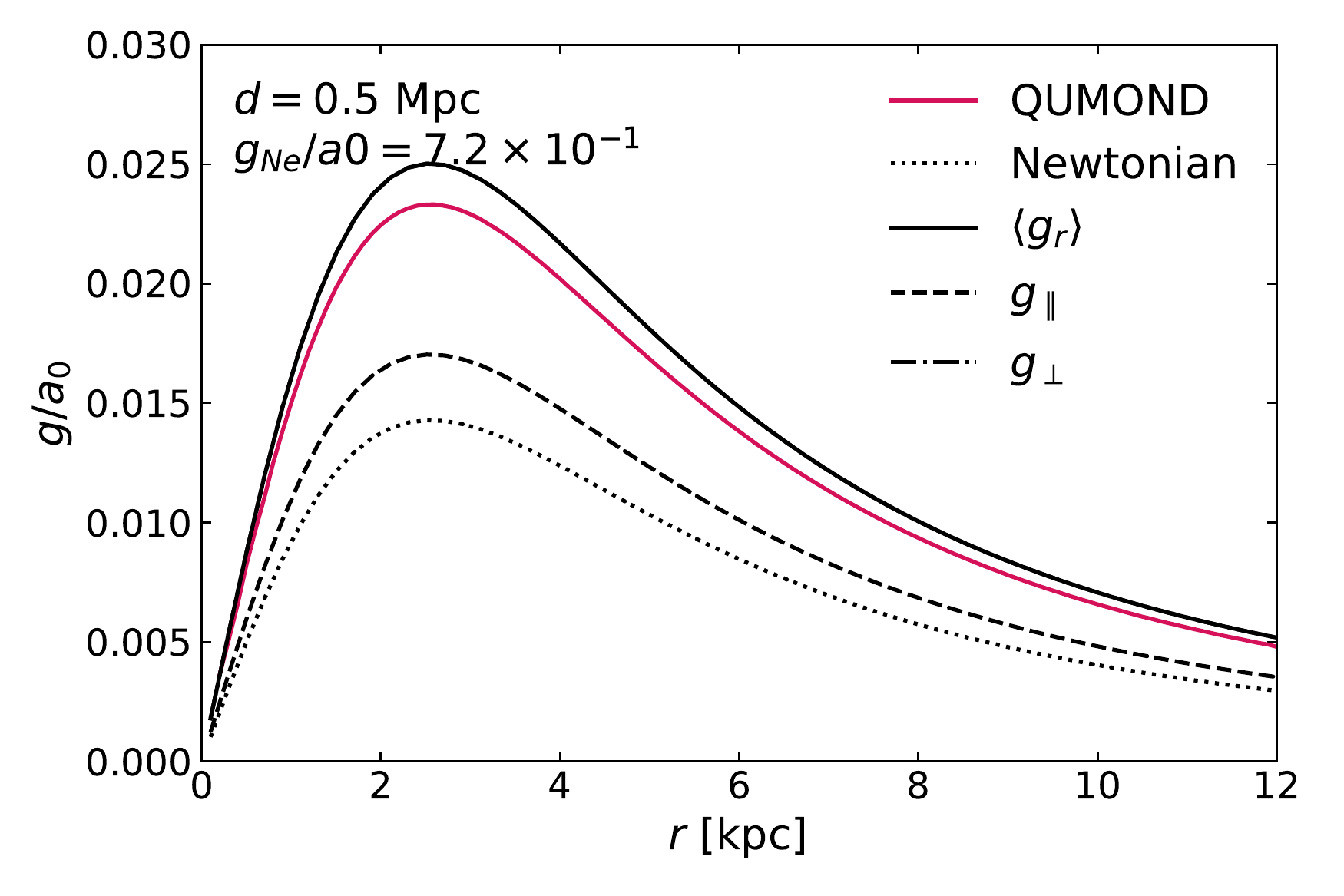}
\includegraphics[width=0.48\linewidth,trim={0cm 0.4cm 0cm 0.4cm},clip]{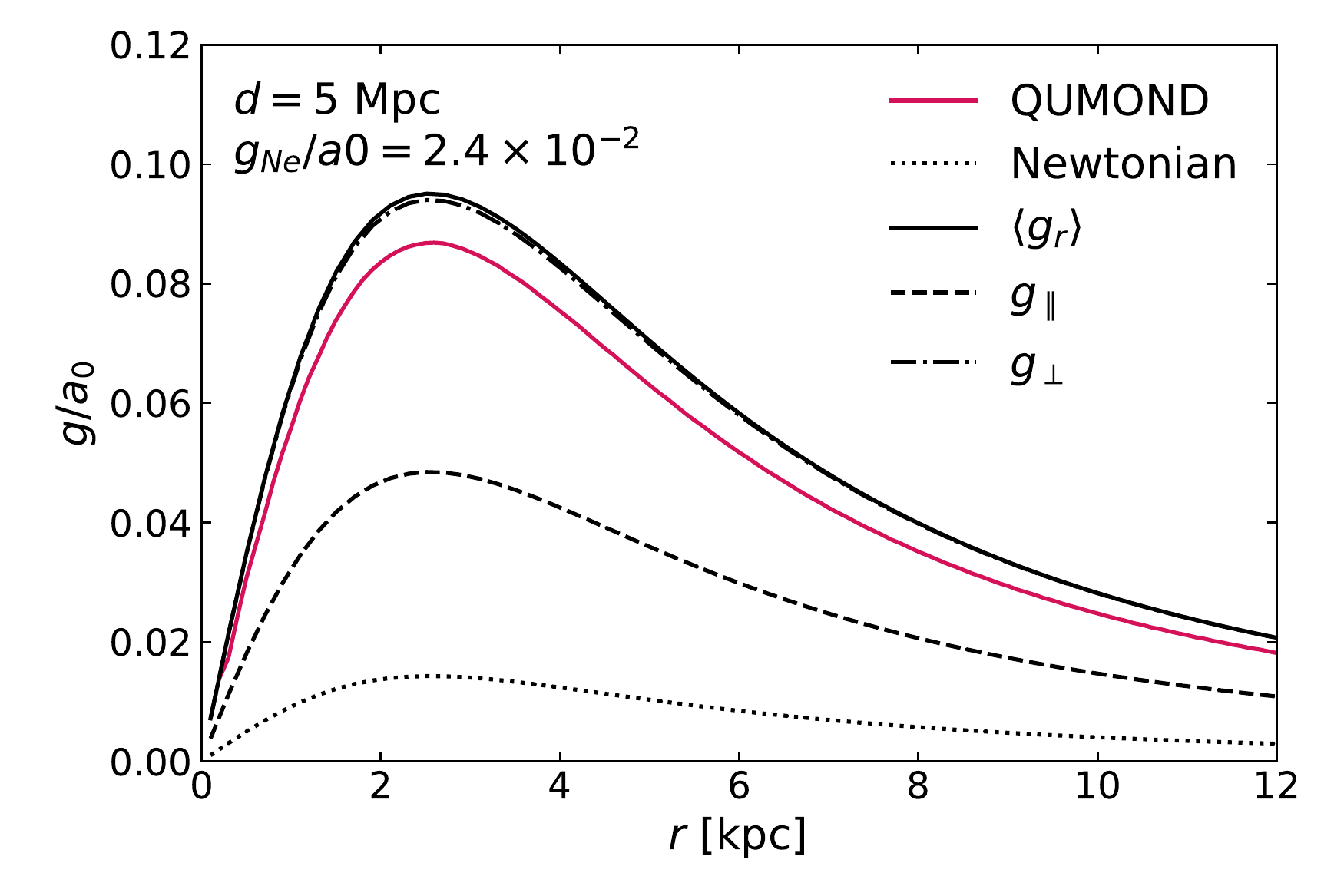}

\includegraphics[width=0.48\linewidth,trim={0cm 0.4cm 0cm 0.4cm},clip]{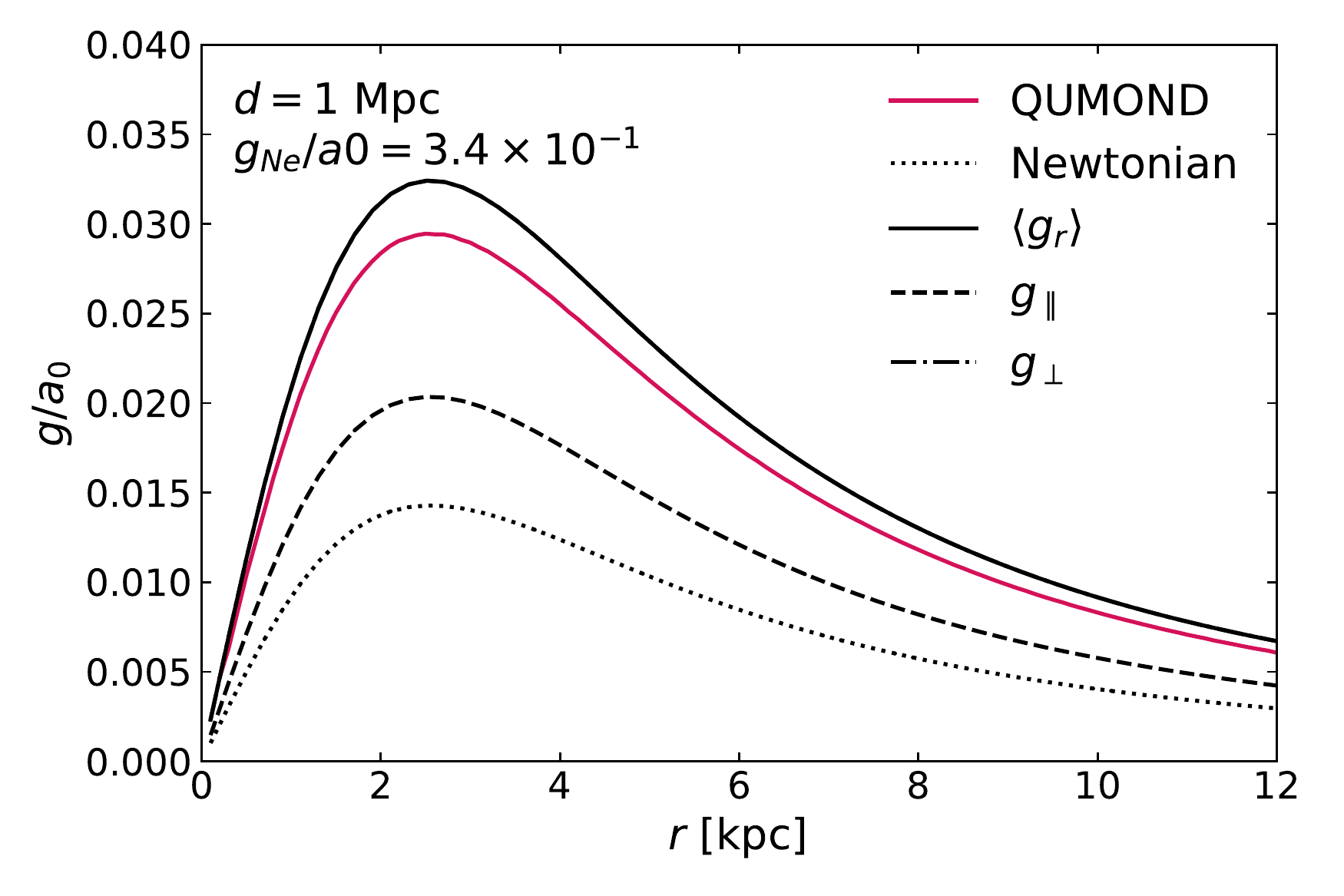}
\includegraphics[width=0.48\linewidth,trim={0cm 0.4cm 0cm 0.4cm},clip]{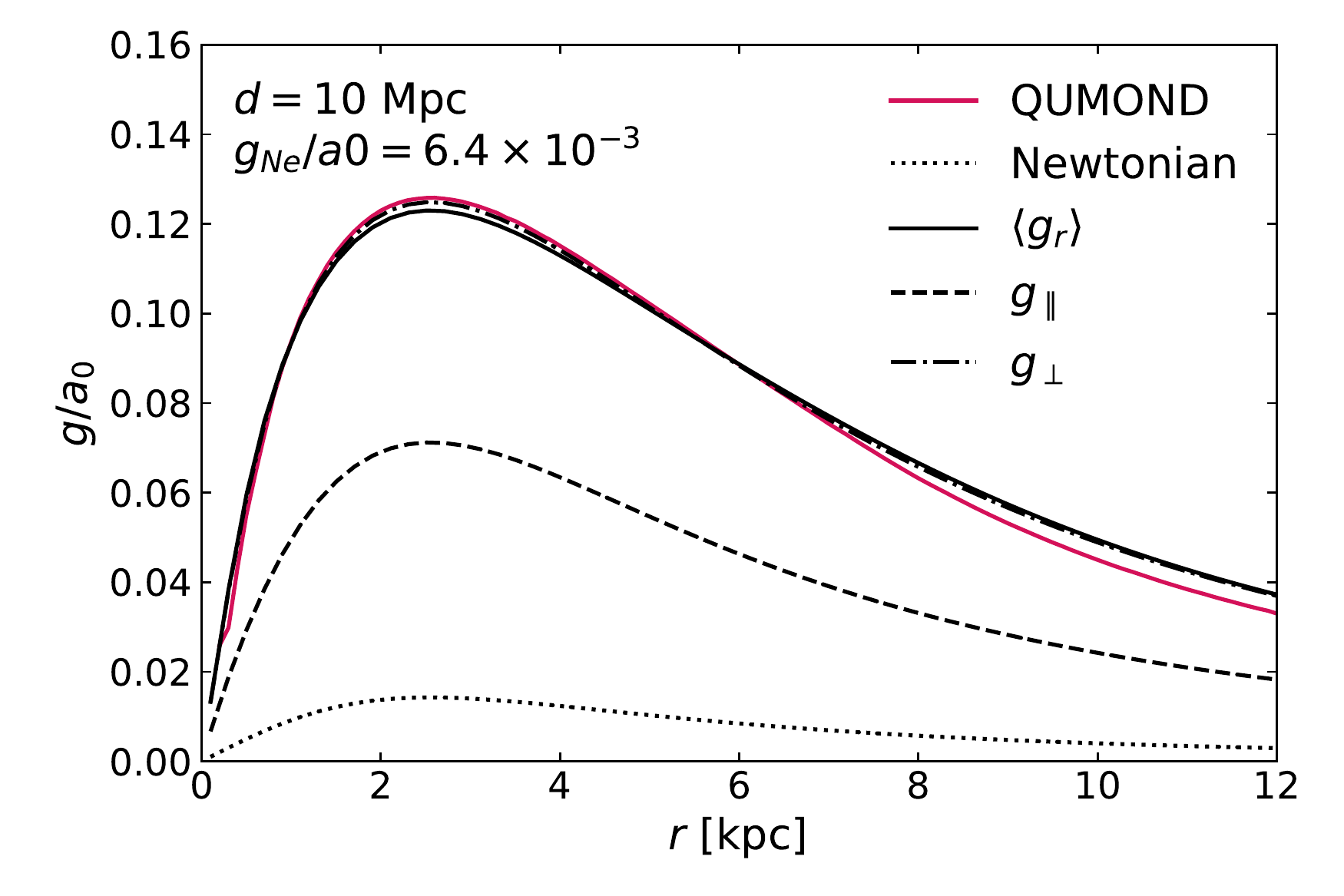}

\includegraphics[width=0.48\linewidth,trim={0cm 0.4cm 0cm 0.4cm},clip]{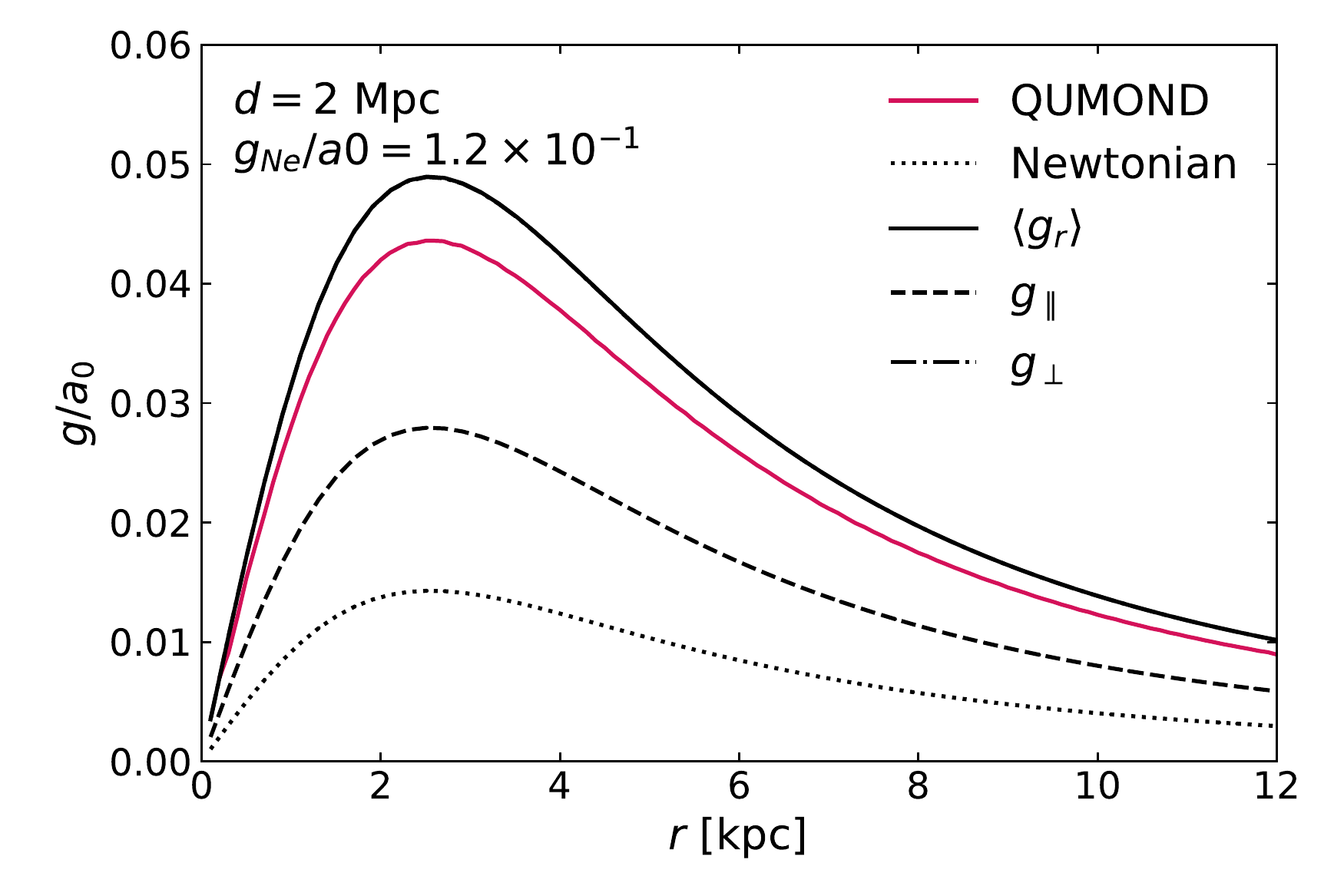}
\includegraphics[width=0.48\linewidth,trim={0cm 0.4cm 0cm 0.4cm},clip]{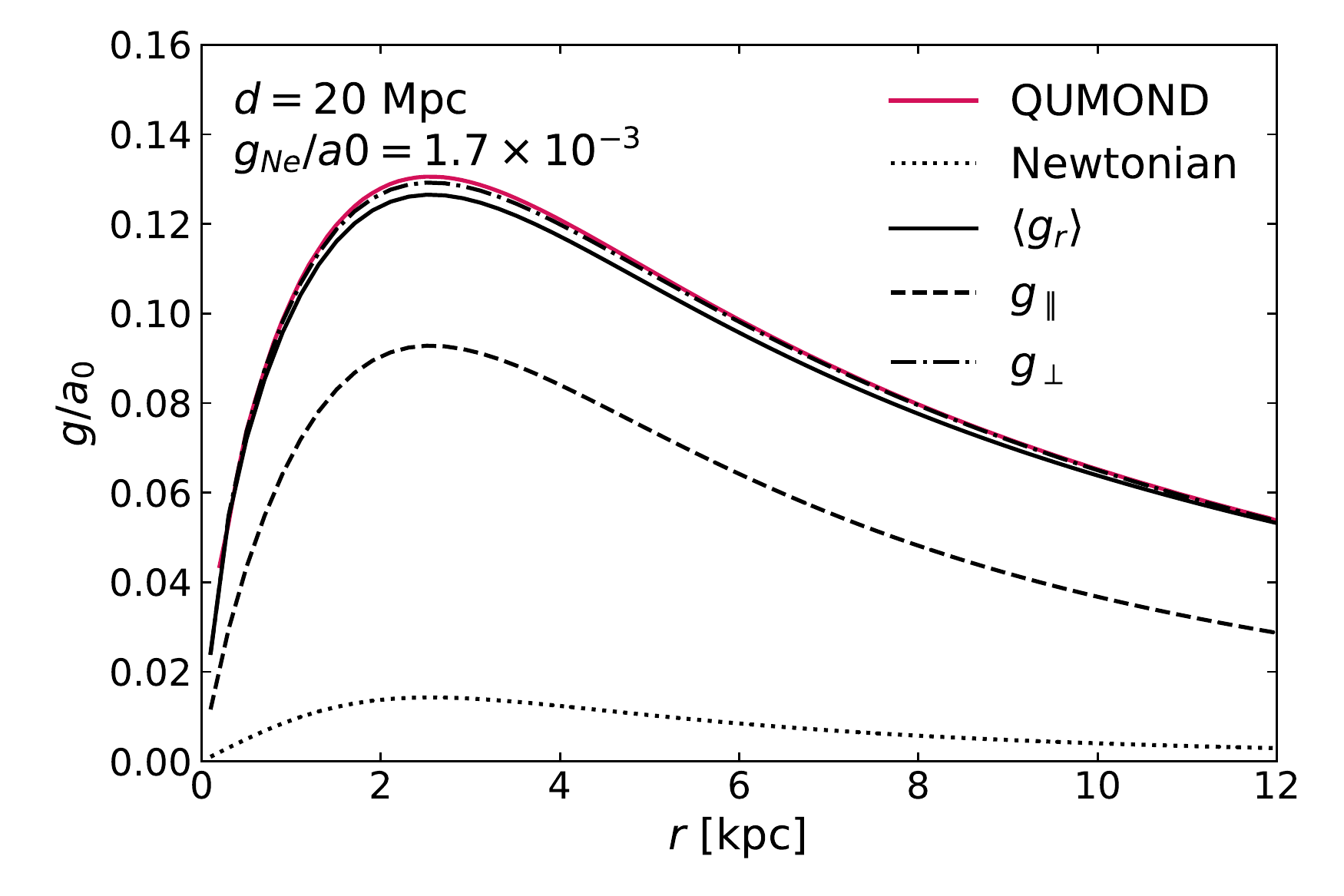}
\caption{Comparison between the MOND acceleration with EFE resulting from a QUMOND numerical integration and the analytical expressions $\langle g_r\rangle$ from Eq.~(\ref{eq:gr}), $g_\parallel$ from Eq. (60) of \protect\citet{Famaey2012}, and $g_\perp$ from Eq.~(\ref{eq:g_perp}), considering a Plummer sphere ($M_0=3.9\times 10^{8}~\rm M_\odot$, $R_P=3.6 \rm ~kpc$) at different distances $d$ from a point mass equal to that of the Coma cluster at that distance given the mass model presented in Section~\ref{section:coma}.
The plots show that $\langle g_r\rangle \approx g_\perp$ provides a good approximation of the MOND acceleration with EFE. 
}
\label{fig:EFE_formula_g}%
\end{figure*}

Figure~\ref{fig:EFE_formula_g} compares the resulting average QUMOND acceleration profile $\langle g_r\rangle$ with the analytic expression from Eq.~(\ref{eq:gr}) for different distances from the point mass modelling the Coma cluster. For comparison, we also considered $g_\parallel$ corresponding to the case where $\vec{g}$ and $\vec{g_e}$ are aligned \citep[Eq. (60) of ][or Eq.~(\ref{eq:g_parallel}) above]{Famaey2012}, and $g_\perp$ corresponding to the case where the two fields are perpendicular (Eq.~(\ref{eq:g_perp}) above).
The points where $g_N=a_0/100$ in each profile (namely, at radii $r=1.1$ and 5.2 kpc, cf. the dotted curve in the upper left panel of Fig.~\ref{fig:EFE_formula_g}) are reported in Fig.~\ref{fig:EFE_formula_ggn}, which highlights how QUMOND and the different analytic expressions transition from the deep-MOND regime ($g=\nu(g_N/a_0)g_N$) to the EFE-dominated Newtonian regime ($g=g_N$). 
The different plots confirm that $\langle g_r\rangle \approx g_\perp$ and show that $\langle g_r\rangle$ fits the numerically computed acceleration much better than $g_\parallel$, which systematically overestimates the EFE and thus underestimates the acceleration field.

%--------------------------------------------------------------------

\section{Tidal susceptibility}
\label{appendix:tidal}

We estimated the tidal susceptibility of a UDG as the ratio $\eta = r_{1/2}/r_2$
between its de-projected half-mass radius $r_{1/2}\approx (4/3) R_e$ and the radius $r_2$ of the Roche lobe perpendicularly to the axis linking the UDG and the centre of the cluster. This latter radius $r_2$ is the minimum radius associated with the teardrop-shaped Roche lobe bounded by the gravitational equipotential where the external tidal field from the Coma cluster balances the internal field $g$ of the UDG in the frame of reference of the UDG \citep[e.g. Fig. 1 of ][]{Zhao2005}. The apex of the Roche lobe pointing towards the cluster centre is the inner Lagrange point $\rm L_1$, situated at radius $r_1>r_2$ from the UDG centre.

We determined the radius of the inner Lagrange point $r_1$ for a UDG at a distance $R$ from the cluster centre by numerically solving the force equilibrium equation: 
\be
\label{eq:gr1}
|g(r_1)| = r_1 \left.\frac{d g_e}{dR}\right|_{R}, 
\ee
where the left-hand term corresponds to the internal gravitational force and the right-hand term to the tidal force, involving the spatial derivative of the external field at $R$. We note that $r_1$ is not defined when this derivative is negative, since that would correspond to a compressive rather than disruptive tidal force. %
The internal acceleration $g$ assumes a spherical mass profile for the UDG (cf. Section~\ref{section:method_spherical}) and takes into account the EFE (cf. Section~\ref{section:formula}). 
The radial derivative of the spherically symmetric external Newtonian acceleration $g_{Ne}(R)=-GM_C(R)/R^2$ is
\be
\label{eq:dgnedr}
\frac{{\rm d} g_{Ne}}{{\rm d}R} = - \frac{G}{R^2} \frac{d M_C}{dR} + \frac{2 GM_C}{R^3}, 
\ee 
and that of the corresponding MOND acceleration {$g_e=\nu(|g_{Ne}/a_0|) g_{Ne}$},
\be
\label{eq:dgedr}
\frac{{\rm d} g_{e}}{{\rm d}R} = \frac{d g_{Ne}}{dR} \times \left[ \nu\left(\left|\frac{g_{Ne}}{a_0}\right|\right) + \left.\frac{d\nu}{dy}\right|_{\left| \frac{g_{Ne}}{a_0}\right|} \times \left| \frac{g_{Ne}}{a_0}\right| \right]. 
\ee

For the mass model of the Coma cluster derived by \cite{Sanders2003} and presented in Section \ref{section:coma}, Eq.~(\ref{eq:rho_c}) for $\rho_C$ yields
\be
M_{NC} (R) = 3 \beta_C M_0 \frac{X^3}{1+X^2}
,\ee 
and
\be
a_C (R) = \frac{3 \beta_C G M_0}{r_C^2} \frac{X}{1+X^2}
,\ee 
with $M_0= kT_C r_C/G\mu m_p$ and $X=R/r_C$. The derivatives 
\be
\frac{{\rm d} M_{NC}}{{\rm d}R} = \frac{3 \beta_C M_0}{r_C} X^2 \frac{3+X^2}{(1+X^2)^2}
\ee 
and
\be
\frac{{\rm d} a_{C}}{{\rm d}R} = \frac{3 \beta_C G M_0}{r_C^3}  \frac{1-X^2}{(1+X^2)^2}
\ee 
enable us to calculate
\be
\frac{{\rm d} M_{C}}{{\rm d}R} = \frac{a_C}{a_C+a_0} \frac{{\rm d} M_{NC}}{{\rm d}R} 
+ \frac{a_0}{(a_C+a_0)^2}
\frac{{\rm d} a_{C}}{{\rm d}R}
M_{NC}.
\ee
This expression stemming from Eq.~(\ref{eq:MC}) can be plugged into Eqs.~(\ref{eq:dgnedr}) and (\ref{eq:dgedr}) to retrieve ${\rm d}g_e/{\rm d}R$, and subsequently to obtain the radius $r_1$ of the inner Lagrange point by solving Eq.~(\ref{eq:gr1}) numerically.

Approximating the satellite as a point mass, \cite{Zhao2005} and \cite{ZhaoTian2006} determined
\be
\label{eq:r2r1}
\frac{r_2}{r_1} = \frac{2}{3\sqrt{\Delta_1}}
,\ee 
with 
\be
\label{eq:Delta1}
\Delta_1 \equiv \frac{{\rm d} \ln g_{Ne}}{{\rm d}\ln g_e} = \left(1+\frac{1}{\nu\left(\left|\frac{g_{Ne}}{a_0}\right|\right)} \left.\frac{d\nu}{dy}\right|_{\left| \frac{g_{Ne}}{a_0}\right|} \times \left| \frac{g_{Ne}}{a_0}\right| \right)^{-1}, 
\ee 
from which we obtained $r_2$. Eq.~(\ref{eq:Delta1}) yields $\Delta_1=2$ in the weak acceleration regime of MOND ($g_{Ne}\ll a_0$), $\Delta_1=1$ in the strong acceleration regime ($g_{Ne}\gg a_0$), and we obtained, on average, an intermediate $\Delta \approx 1.6$ for the Coma cluster UDGs at their average distance $d_{\rm mean}$. 
The ratio $r_2/r_1$ is thus equal to 0.471 in the weak acceleration regime, to 0.667 in the strong acceleration regime, and somewhere in between for the Coma cluster UDGs. 
Although Eq.~(\ref{eq:r2r1}) may not be exact for a given satellite mass distribution $m(r)$, it does give the right order of magnitude for the tidal susceptibility and converges to the exact value when $r_{1/2}\ll r_2$ (i.e. $\eta\ll 1$).

We note that \cite{Zhao2005} and \cite{ZhaoTian2006} also provided an expression for the radius of the inner Lagrange point for a point mass $m$ situated at a distance $R$ from the centre of a spherical mass distribution $M(R)$, 
\begin{equation}
r_1 = R \left( \frac{m}{\zeta_1 M}\right)^{1/3}~{\rm with}~~\zeta_1 = - \frac{{\rm d} \ln \Omega^2}{{\rm d}\ln R}, 
\end{equation}
where $\Omega(R)=\sqrt{g_e(R)/R}$ is the angular frequency for a circular orbit at radius $R$. Given Eq.~(\ref{eq:r2r1}), this yields $r_2=0.374 R (m/M)^{1/3}$ in the weak acceleration regime ($\zeta_1=2$, $\Delta_1=2$) and $r_2=0.462 (m/M)^{1/3}$ in the strong acceleration regime ($\zeta_1=3$, $\Delta_1=1$). In the present work, we instead used Eq.~(\ref{eq:gr1}) to determine $r_1$, taking into account both the spatial extent of the UDGs and the EFE.

%--------------------------------------------------------------------

\section{Higher stellar $\rm M/L$ ratio}
\label{appendix:ML}

In Fig.~\ref{fig:EFE_ML}, we show the MOND predicted velocity dispersion with EFE at the average distance $d_{\rm mean}$ inferred from the Einasto UDG distribution of \cite{vanderBurg2016} (cf. Section \ref{section:coma}) for different $\rm M/L$ ratio values, in the case of a radially biased anisotropy $\beta=+0.5$. The best-fit values recovering the luminosity-weighted effective stellar velocity dispersion $\sigma_{\rm eff}$ range from 1.5 to 16.5, with an average of 7.0.

\begin{figure*}
\centering
\includegraphics[width=0.46\textwidth,trim={0cm 0cm 0cm 0cm},clip]{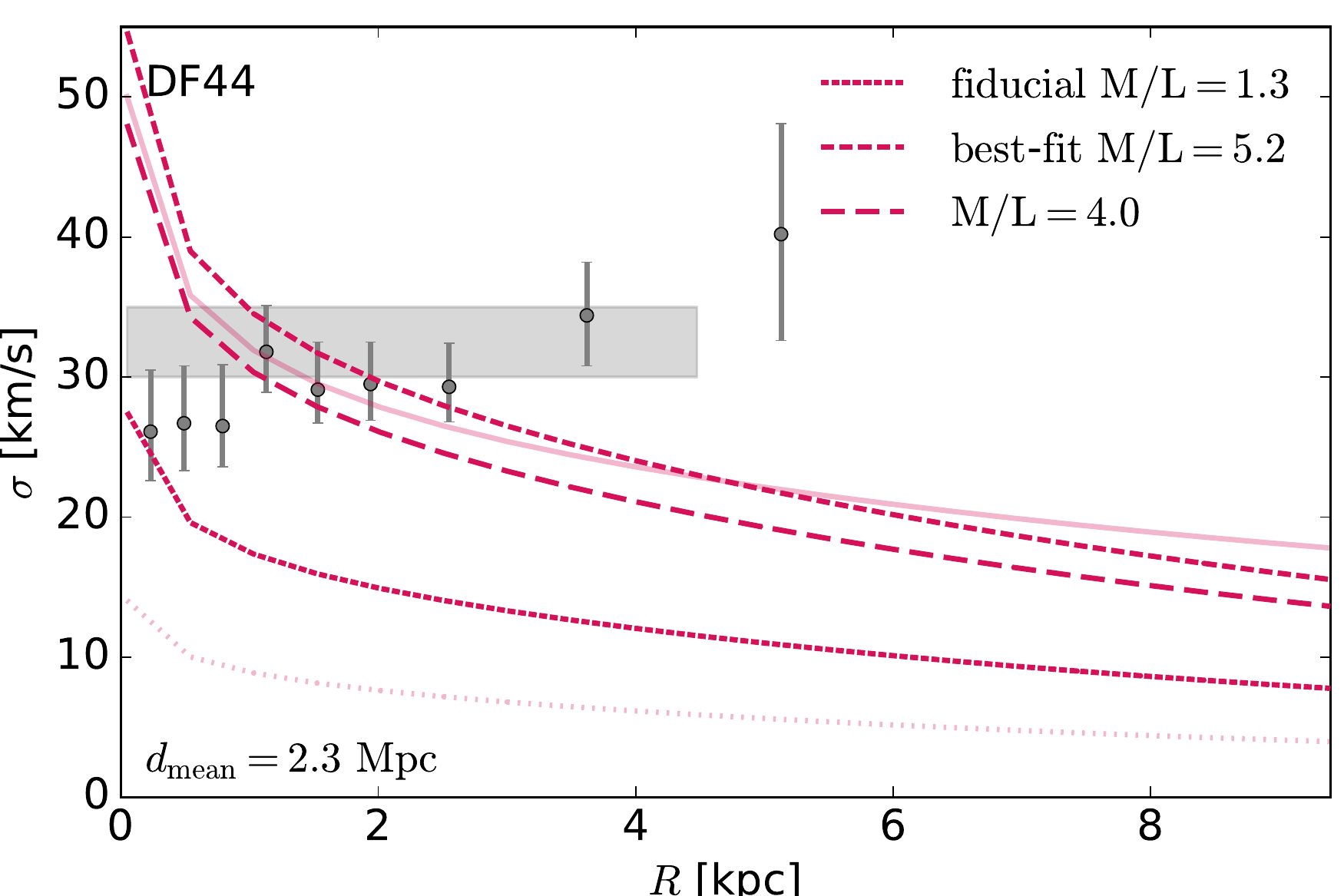}
\hfill
\includegraphics[width=0.46\textwidth,trim={0cm 0cm 0cm 0cm},clip]{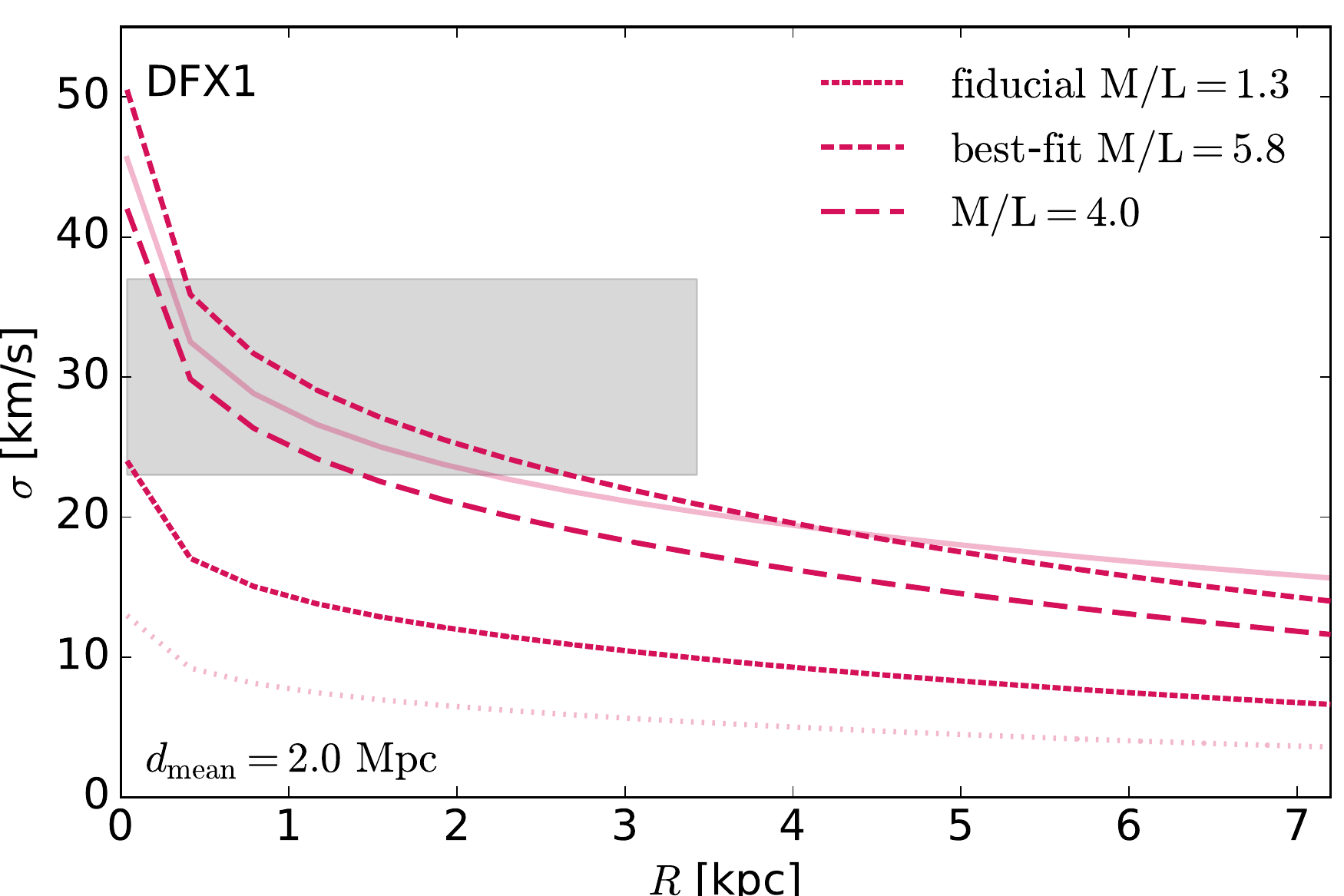}
\includegraphics[width=0.46\textwidth,trim={0cm 0cm 0cm 0cm},clip]{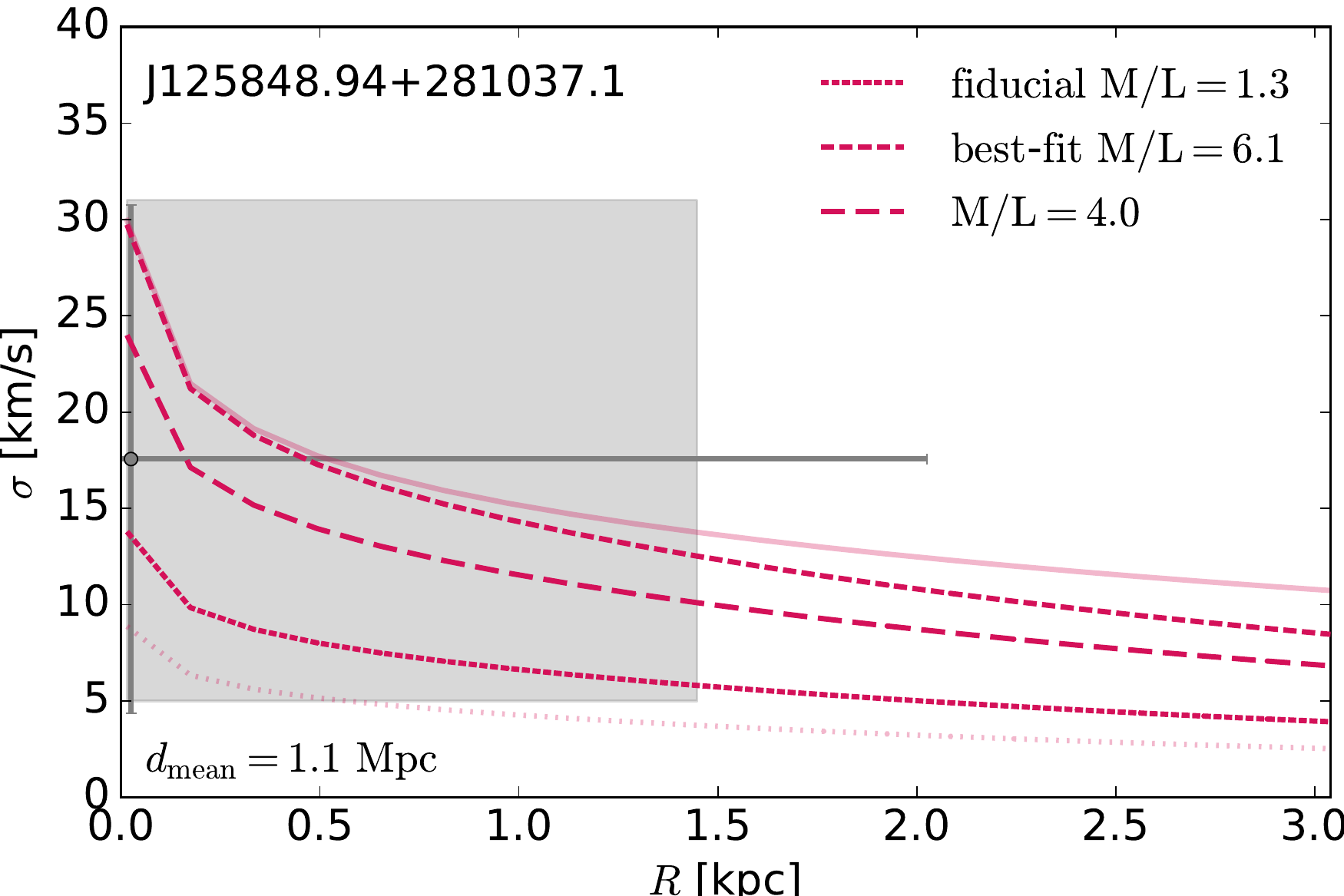}
\hfill
\includegraphics[width=0.46\textwidth,trim={0cm 0cm 0cm 0cm},clip]{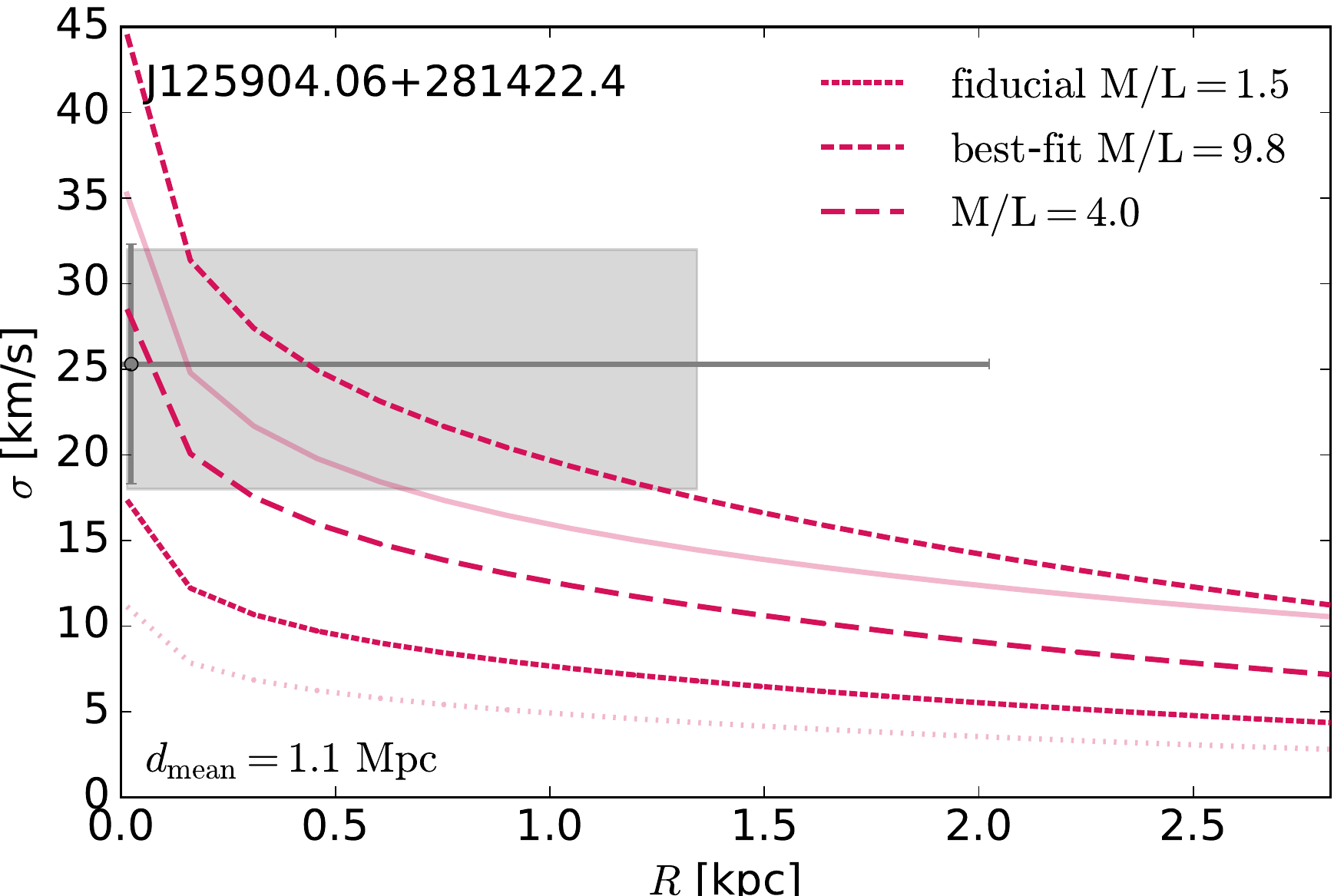}
\includegraphics[width=0.46\textwidth,trim={0cm 0cm 0cm 0cm},clip]{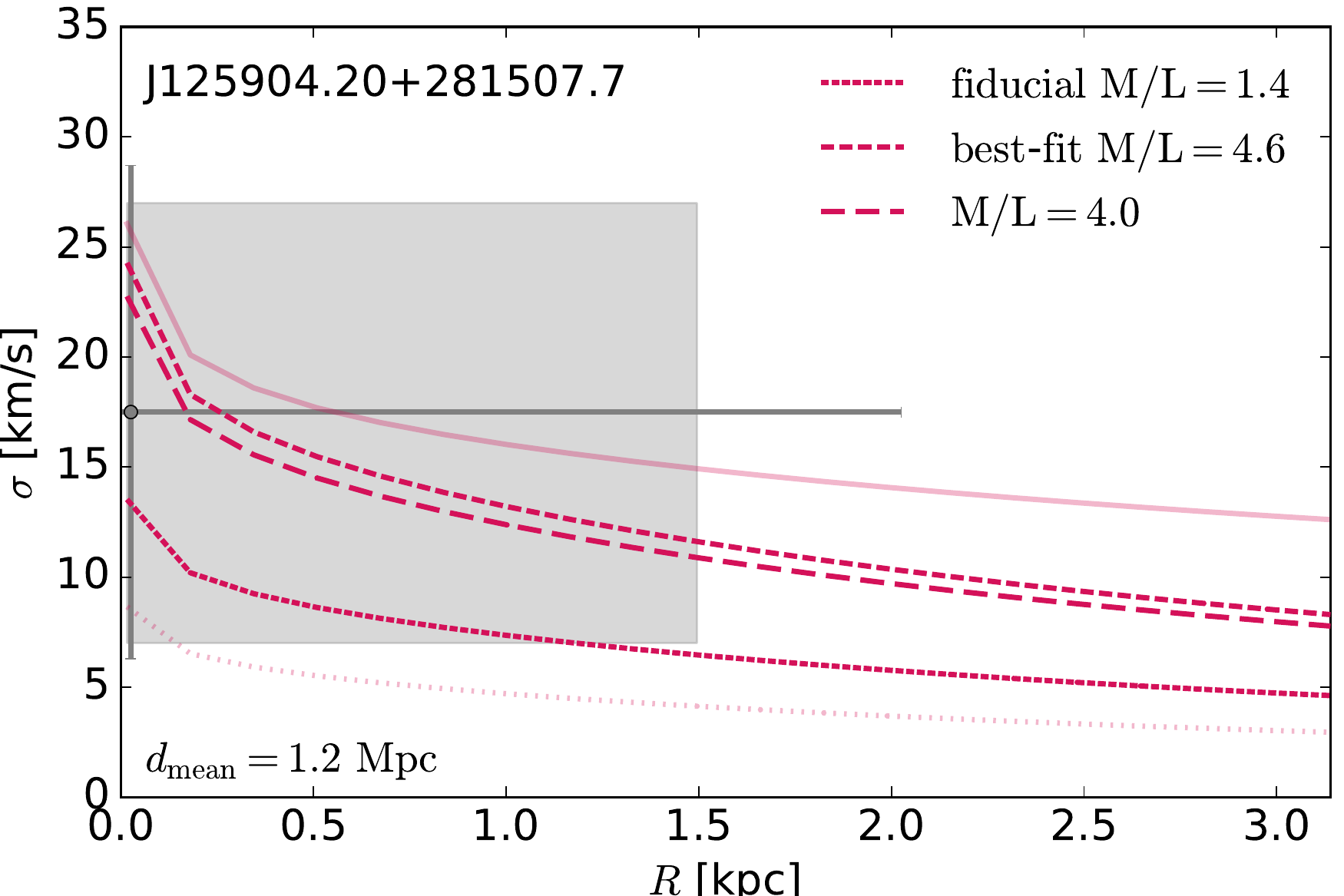}
\hfill
\includegraphics[width=0.46\textwidth,trim={0cm 0cm 0cm 0cm},clip]{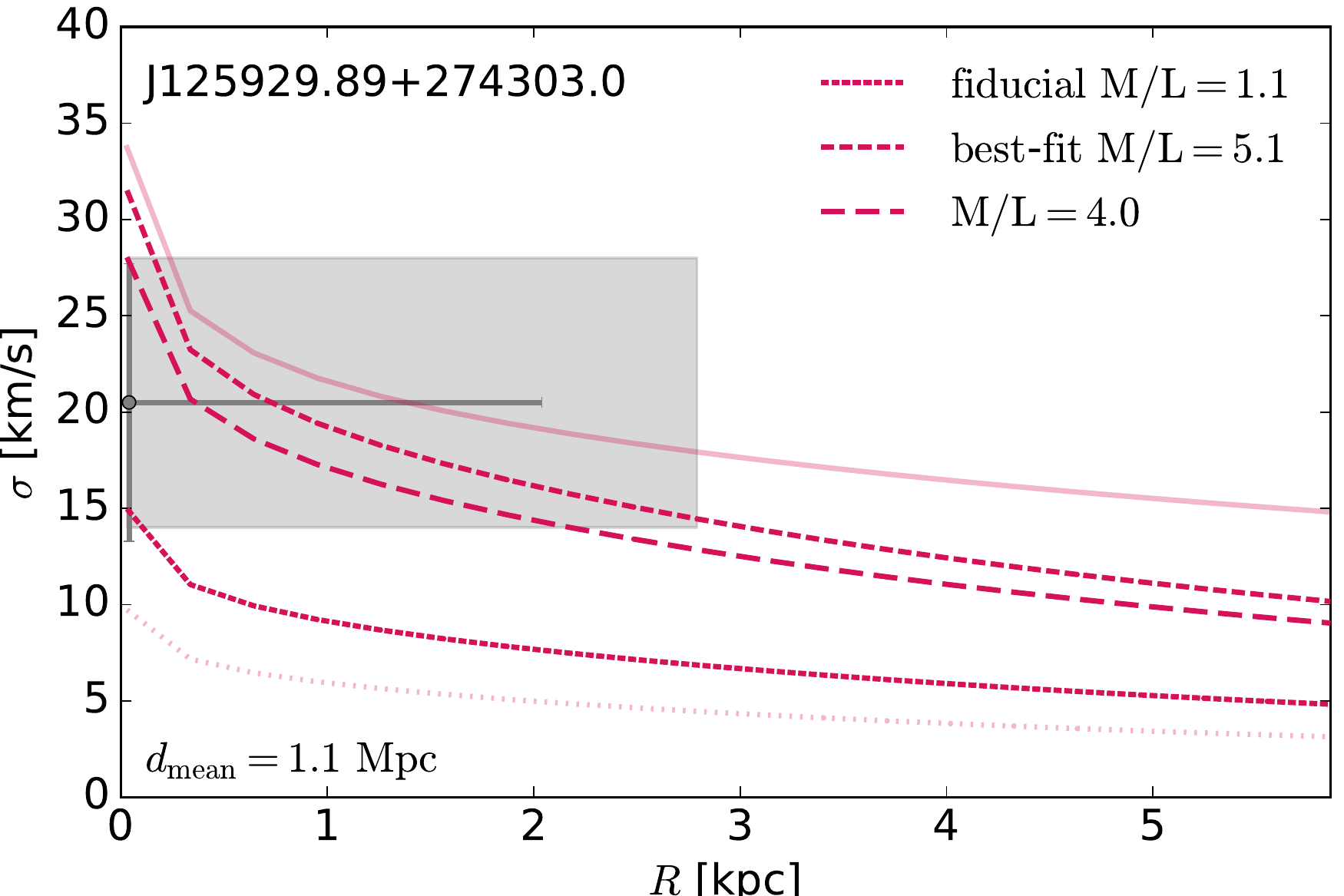}
\includegraphics[width=0.46\textwidth,trim={0cm 0cm 0cm 0cm},clip]{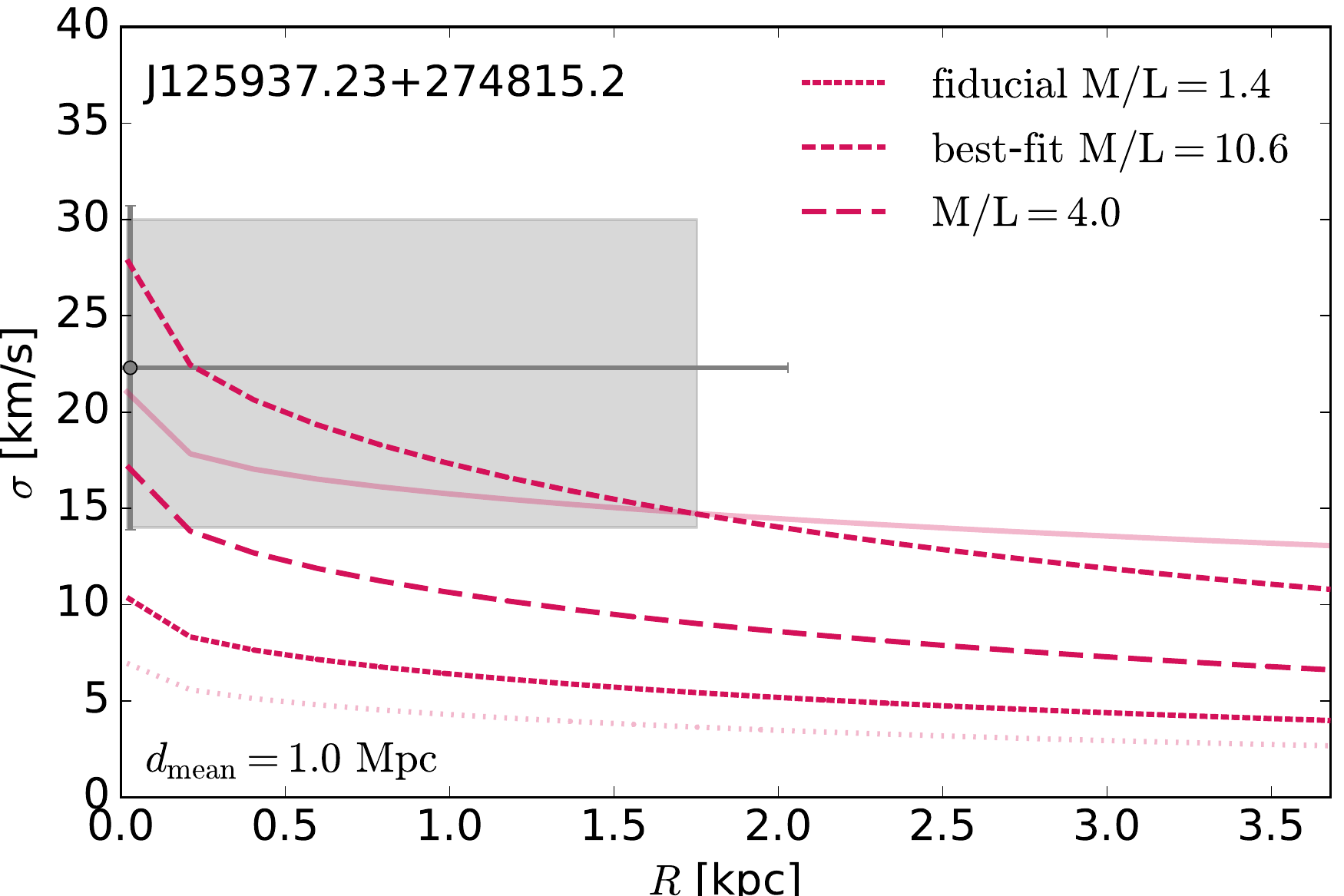}
\hfill
\includegraphics[width=0.46\textwidth,trim={0cm 0cm 0cm 0cm},clip]{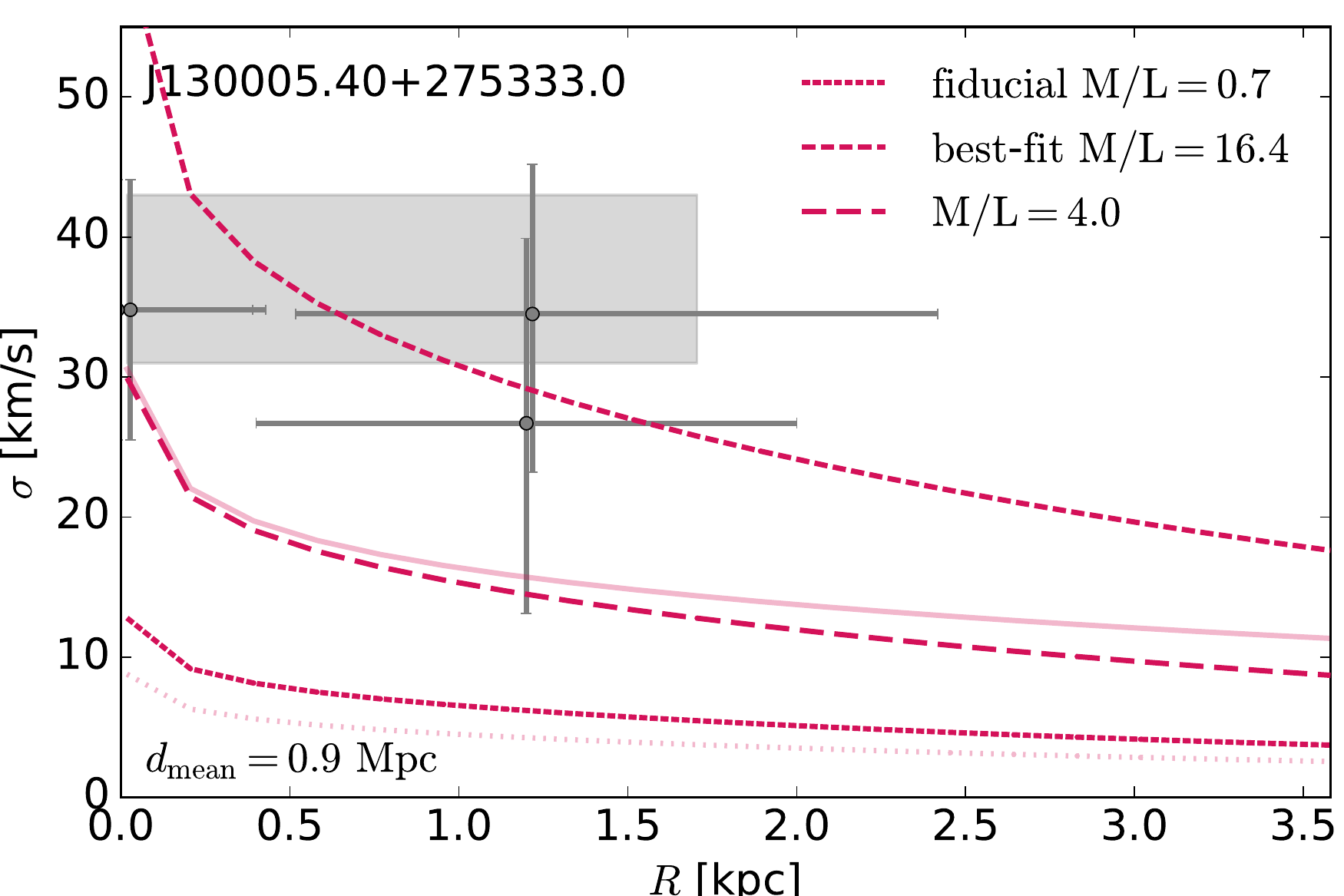}
\caption{ 
Comparison between the measured line-of-sight velocity dispersion of the sample UDGs (grey error bars and/or shaded area) and the MOND prediction with EFE at the average distance $d_{\rm mean}$ for different values of the stellar mass-to-light ratio $\rm M/L$, assuming a radially biased anisotropy $\beta=+0.5$. The fiducial $\rm M/L$ ratio from Table~\ref{table:sample}, the best-fit $\rm M/L$ ratio recovering the effective velocity dispersion $\sigma_{\rm eff}$, and the large bust still acceptable $\rm M/L=4 $ are compared. As in Fig.~\ref{fig:EFE}, the plain and dotted light red curves recall the predicted radially biased velocity dispersion profiles in the isolated MOND and Newtonian cases, respectively (cf. Fig.~\ref{fig:isoloated}). 
}
\label{fig:EFE_ML}
\end{figure*}

\setcounter{figure}{0}
\begin{figure*}
\centering
\includegraphics[width=0.46\textwidth,trim={0cm 0cm 0cm 0cm},clip]{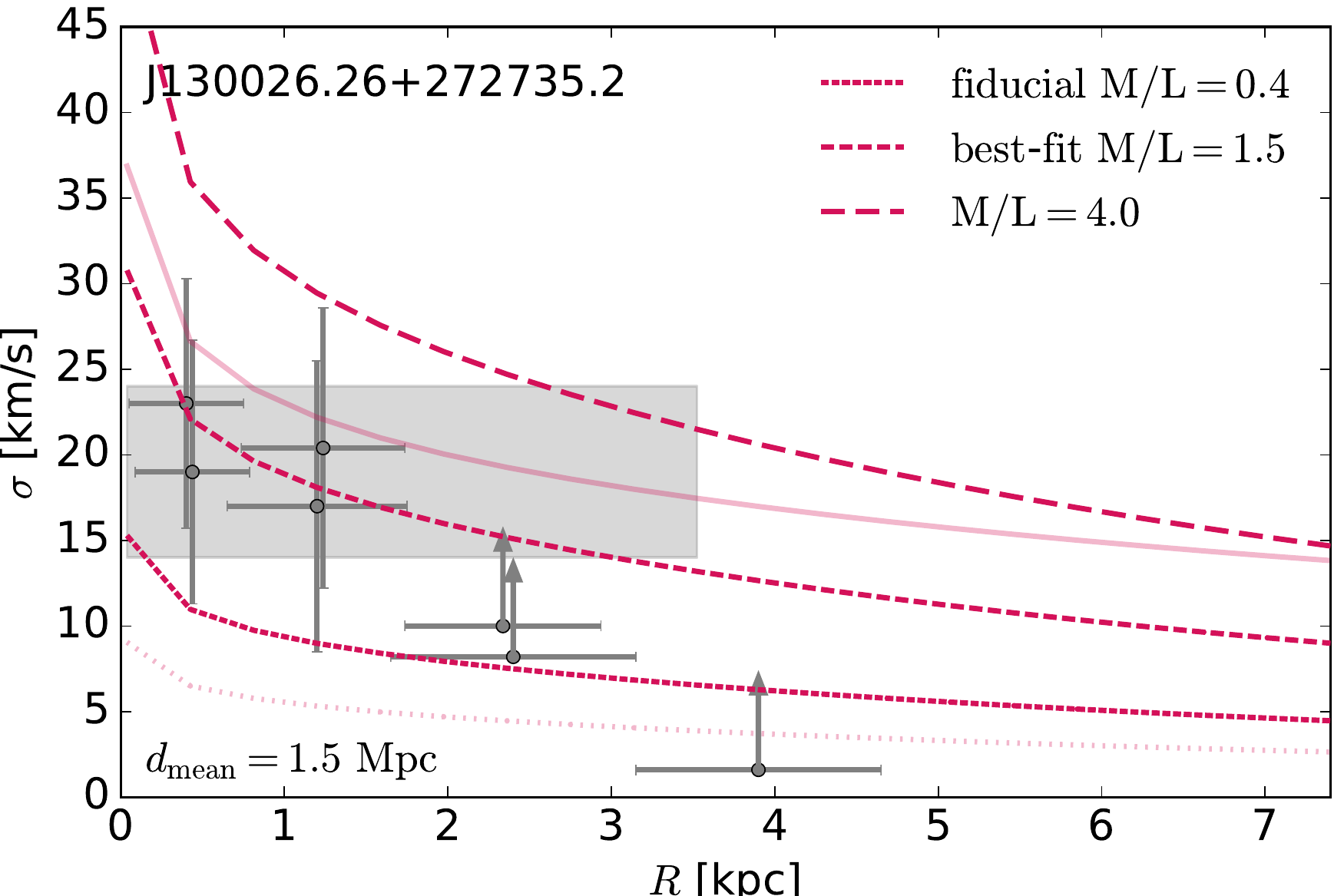}
\hfill
\includegraphics[width=0.46\textwidth,trim={0cm 0cm 0cm 0cm},clip]{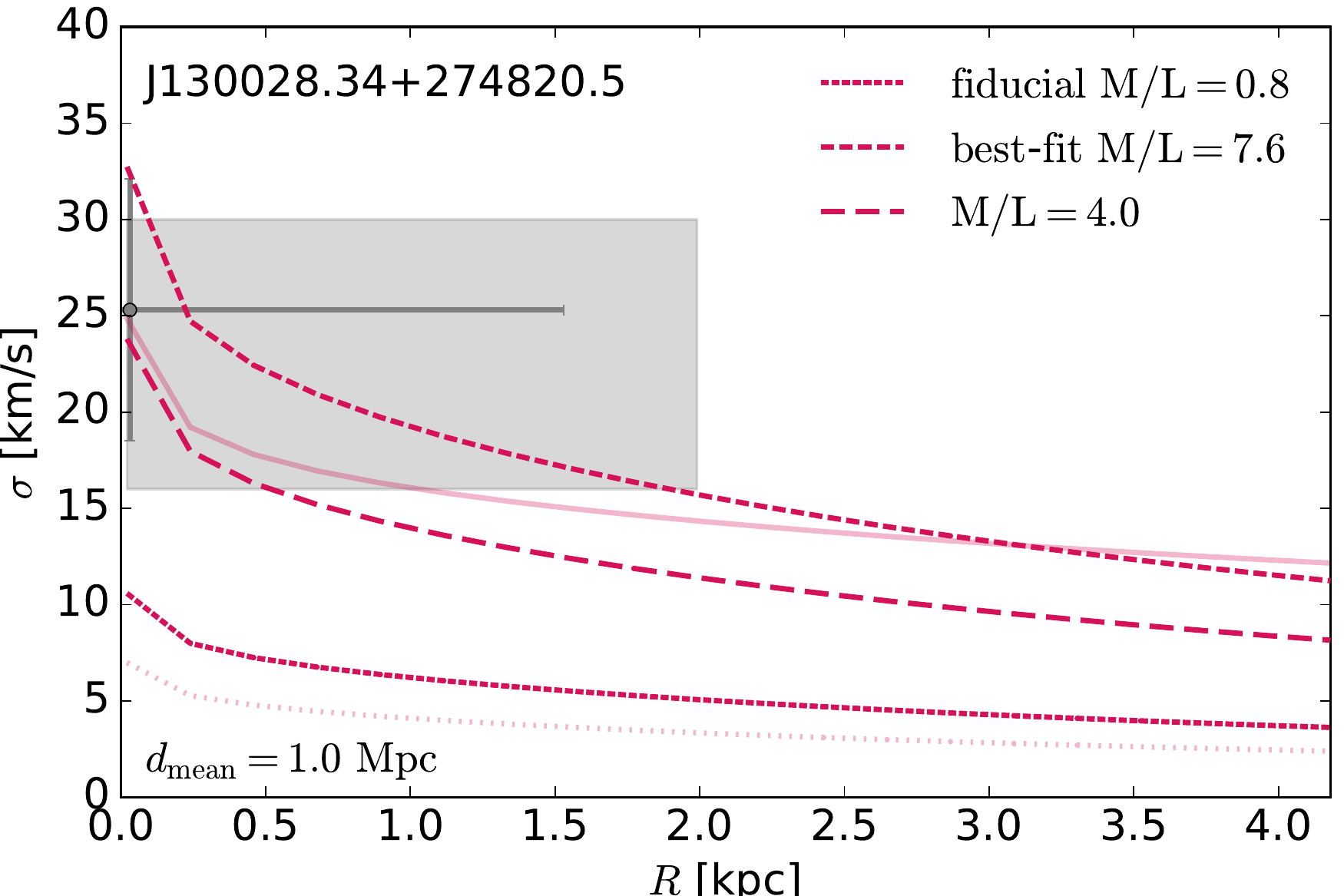}
\includegraphics[width=0.46\textwidth,trim={0cm 0cm 0cm 0cm},clip]{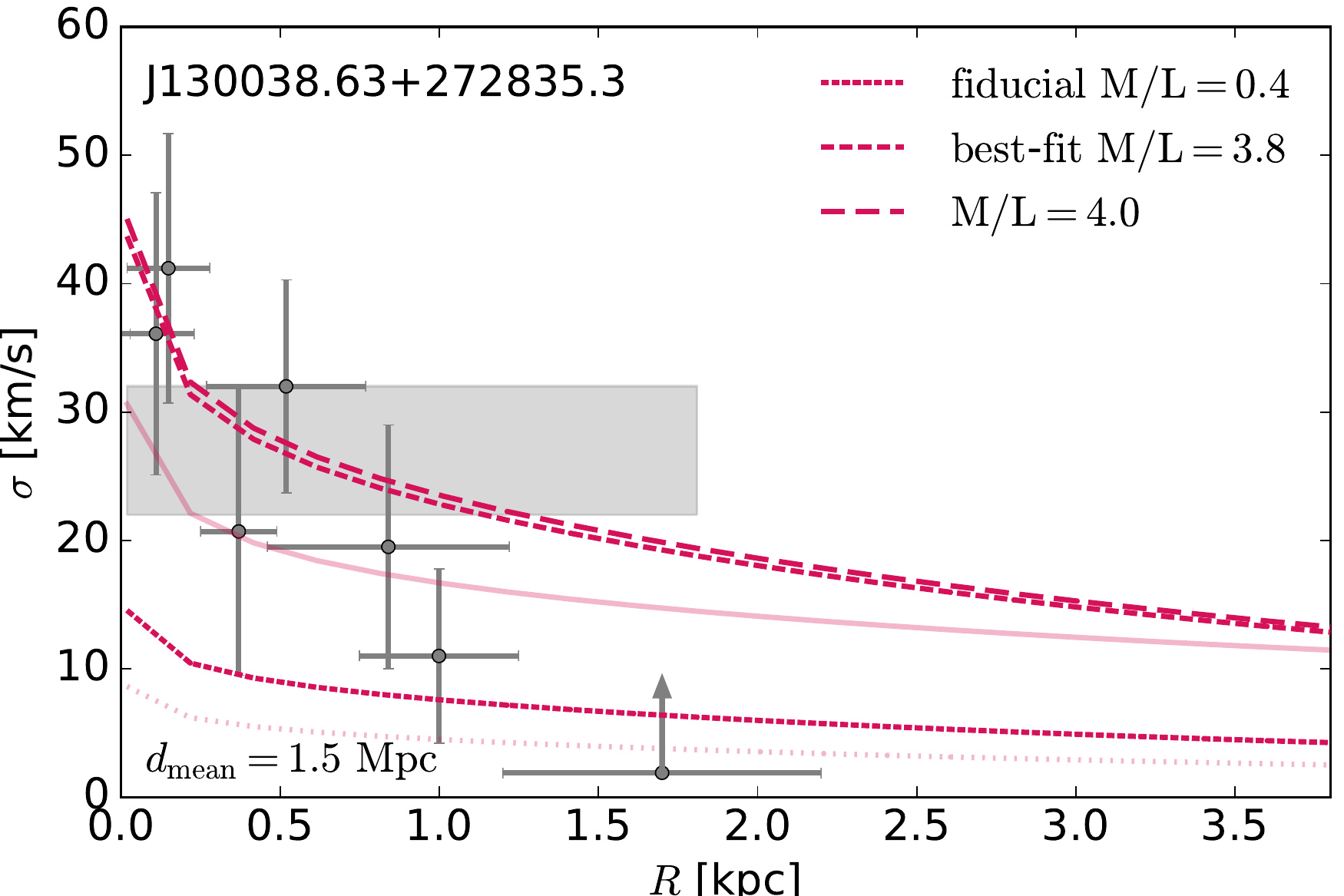}
\hfill \hspace{1cm}
\caption{
Continued. 
}
\label{fig:EFE_ML3}%
\end{figure*}

%--------------------------------------------------------------------

\section{EMOND profiles}

\begin{figure*}
\centering
\includegraphics[width=0.46\textwidth,trim={0cm 0cm 0cm 0cm},clip]{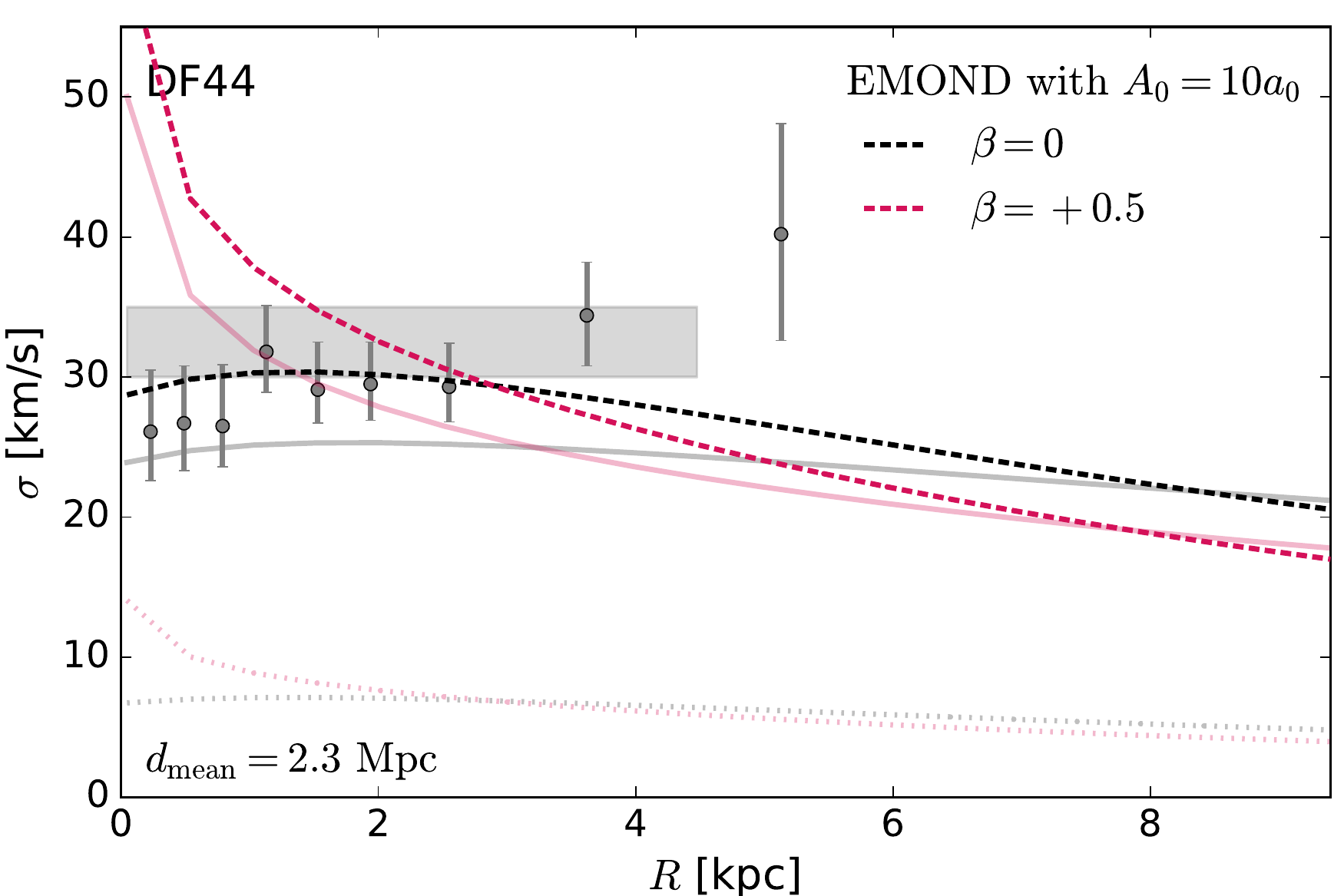}
\hfill
\includegraphics[width=0.46\textwidth,trim={0cm 0cm 0cm 0cm},clip]{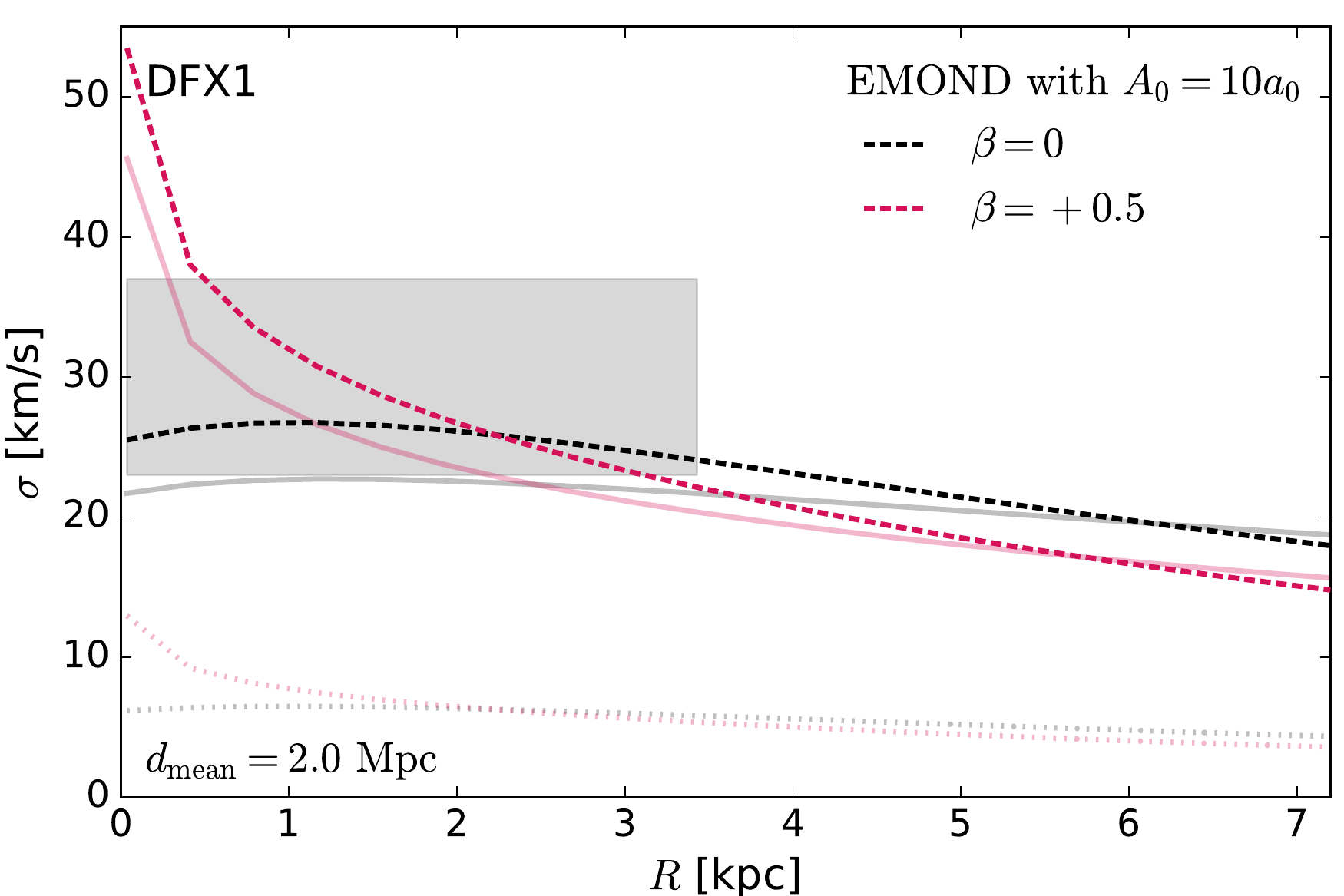}
\includegraphics[width=0.46\textwidth,trim={0cm 0cm 0cm 0cm},clip]{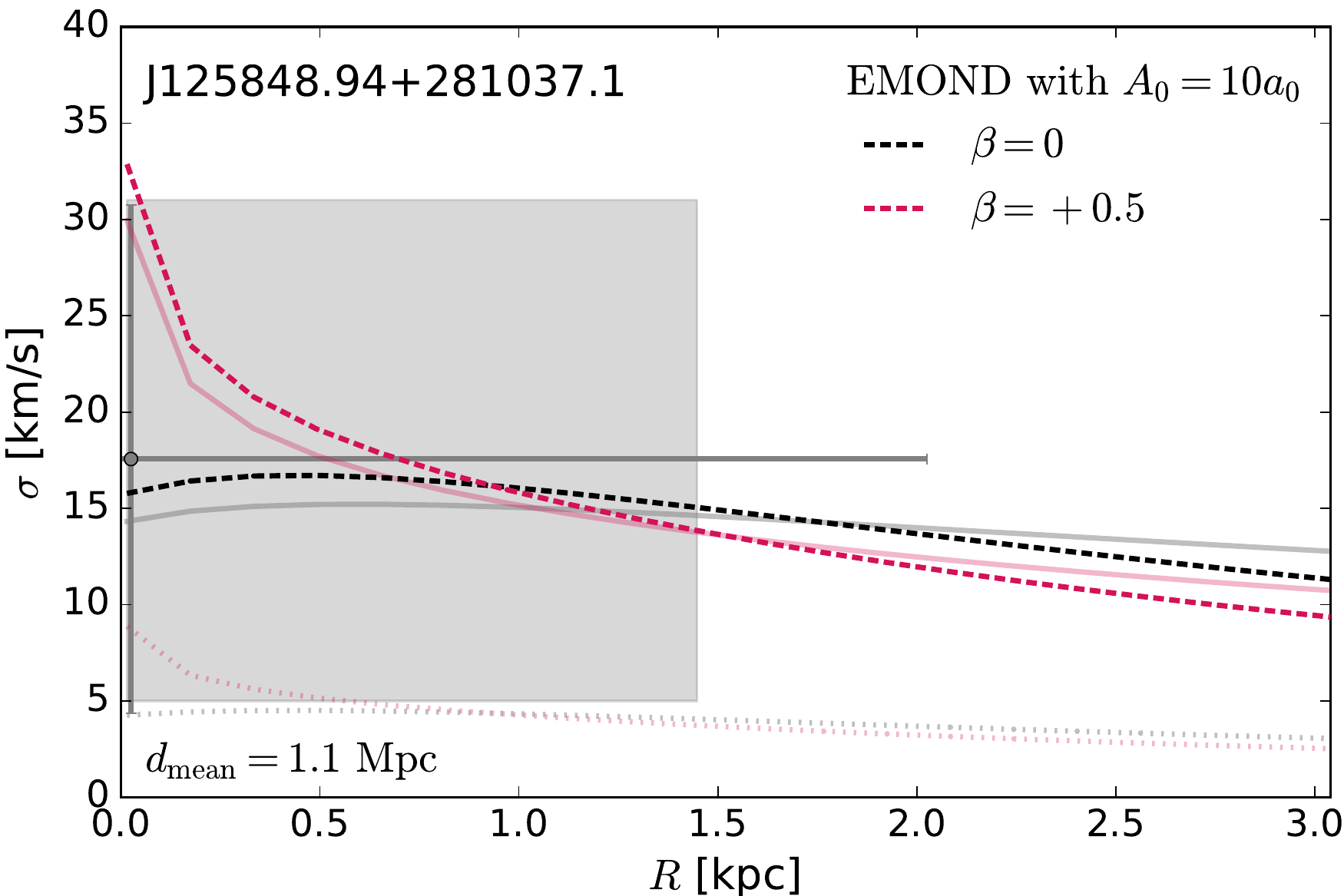}
\hfill
\includegraphics[width=0.46\textwidth,trim={0cm 0cm 0cm 0cm},clip]{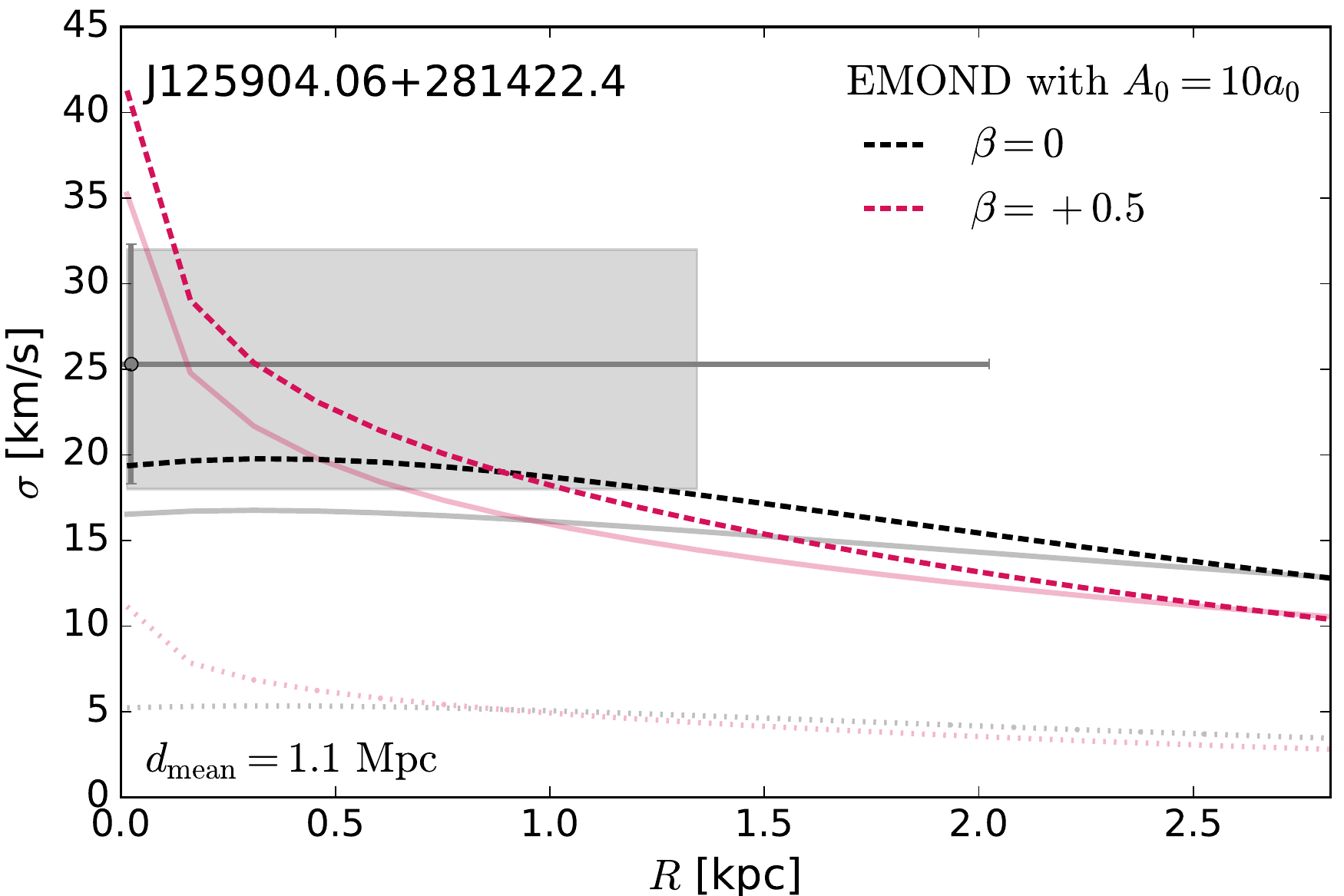}
\includegraphics[width=0.46\textwidth,trim={0cm 0cm 0cm 0cm},clip]{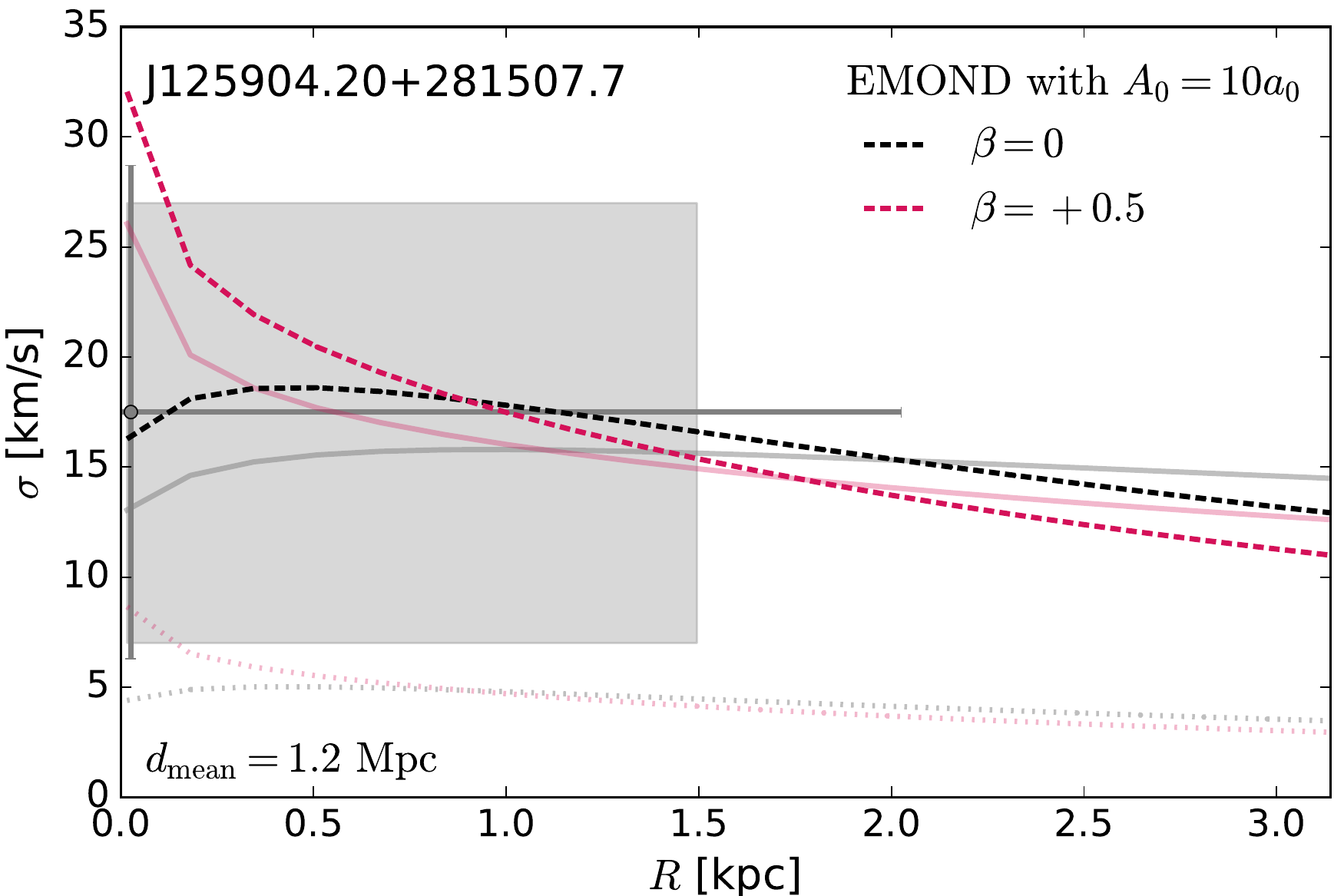}
\hfill
\includegraphics[width=0.46\textwidth,trim={0cm 0cm 0cm 0cm},clip]{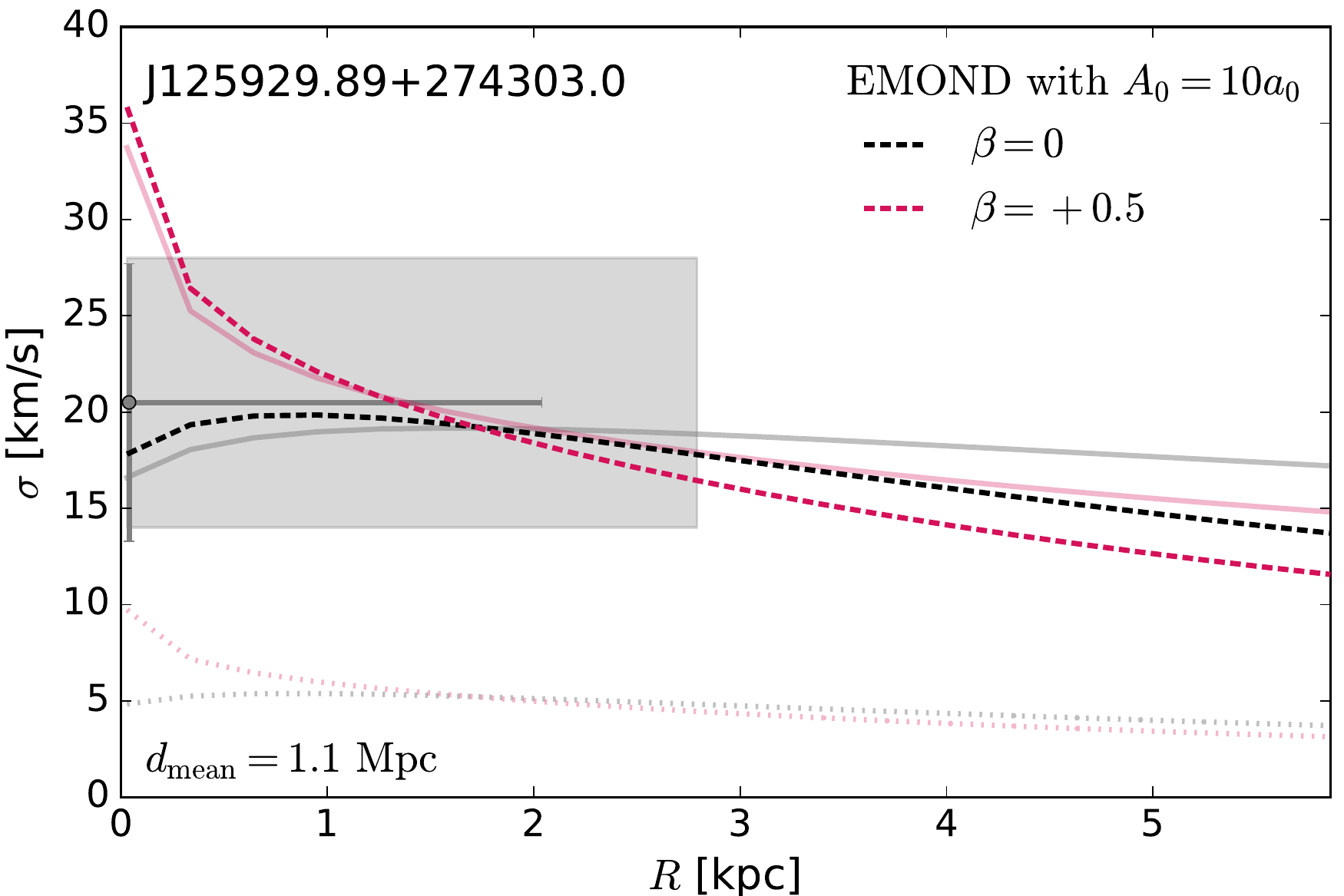}
\includegraphics[width=0.46\textwidth,trim={0cm 0cm 0cm 0cm},clip]{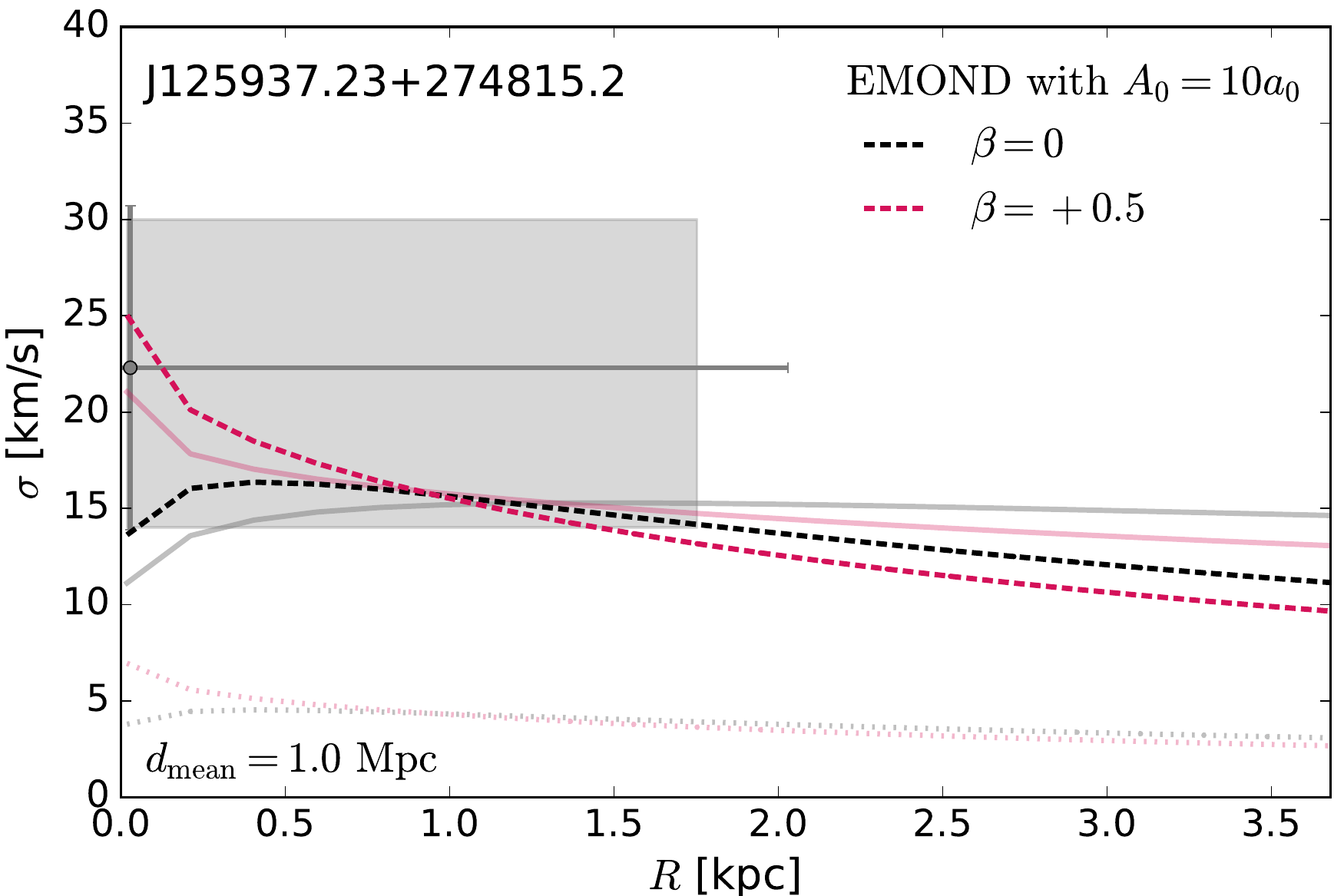}
\hfill
\includegraphics[width=0.46\textwidth,trim={0cm 0cm 0cm 0cm},clip]{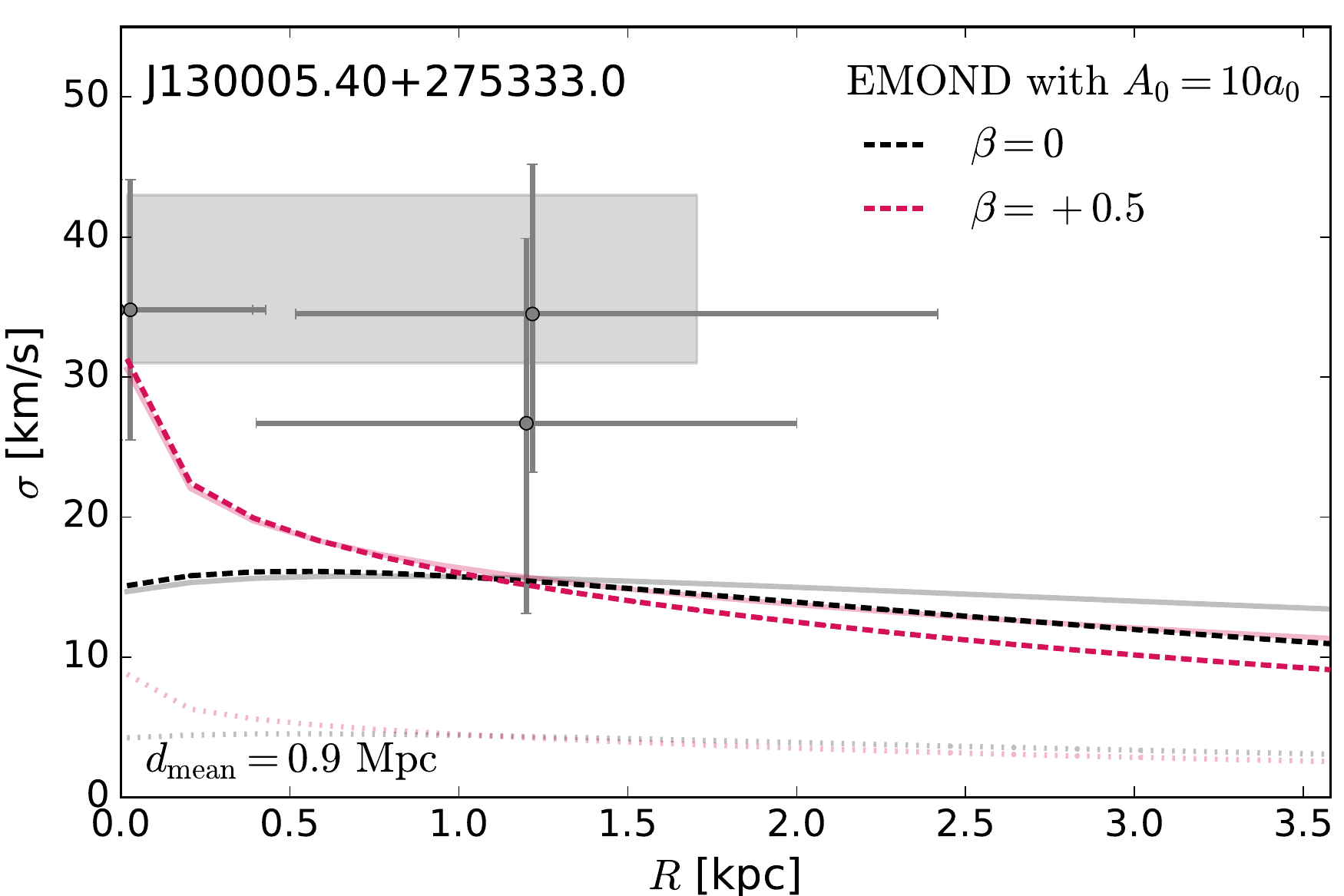}
\caption{ 
Comparison between the measured line-of-sight velocity dispersion of the sample UDGs, as in Fig.~\ref{fig:EFE}, to the predicted velocity dispersion at the average distance $d_{\rm mean}$ obtained by emulating EMOND with $A_0=10\times a_0$, shown as dashed black ($\beta=0$) and magenta ($\beta=+0.5$) lines. 
}
\label{fig:EMOND_profiles}
\end{figure*}

\setcounter{figure}{0}
\begin{figure*}
\centering
\includegraphics[width=0.46\textwidth,trim={0cm 0cm 0cm 0cm},clip]{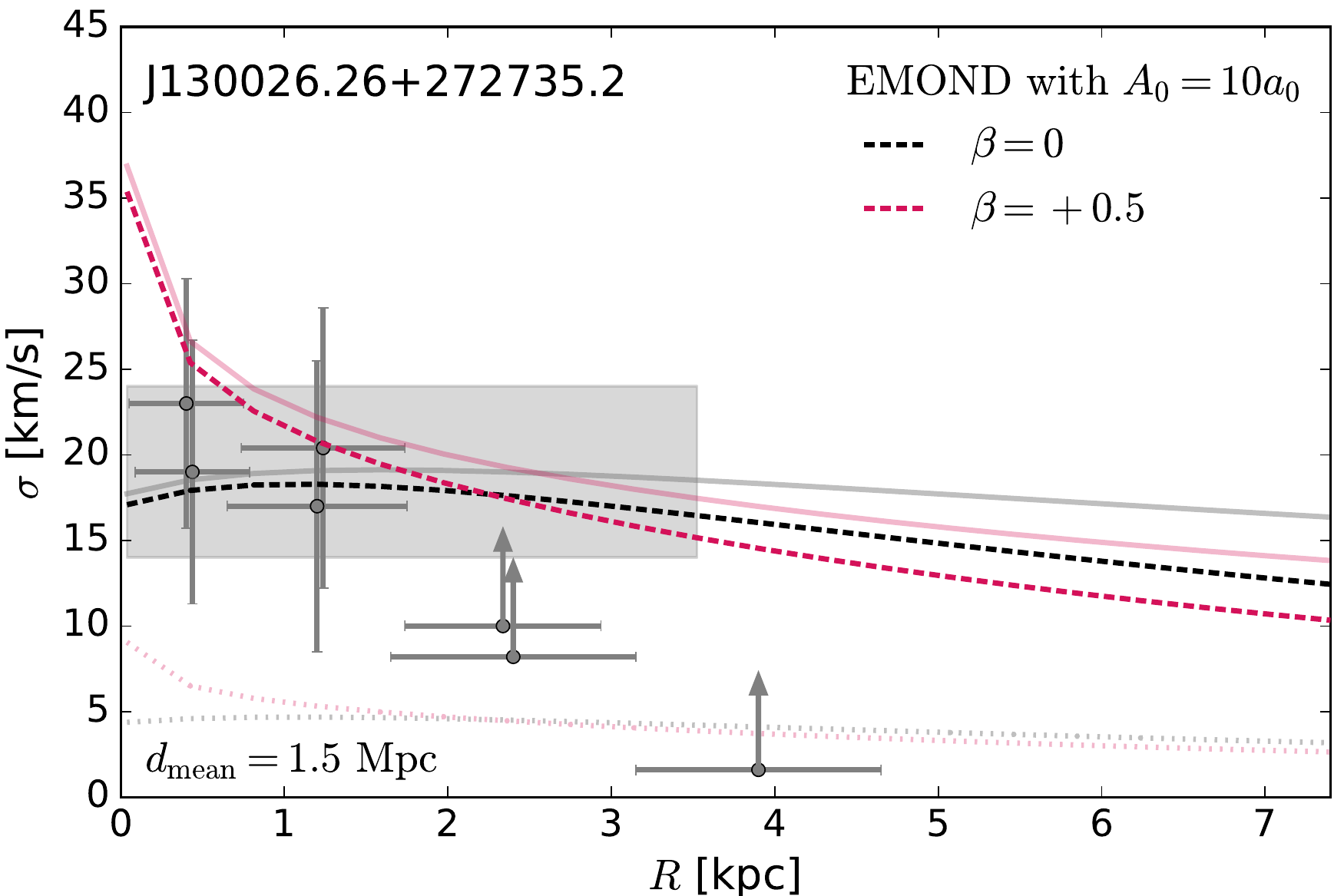}
\hfill
\includegraphics[width=0.46\textwidth,trim={0cm 0cm 0cm 0cm},clip]{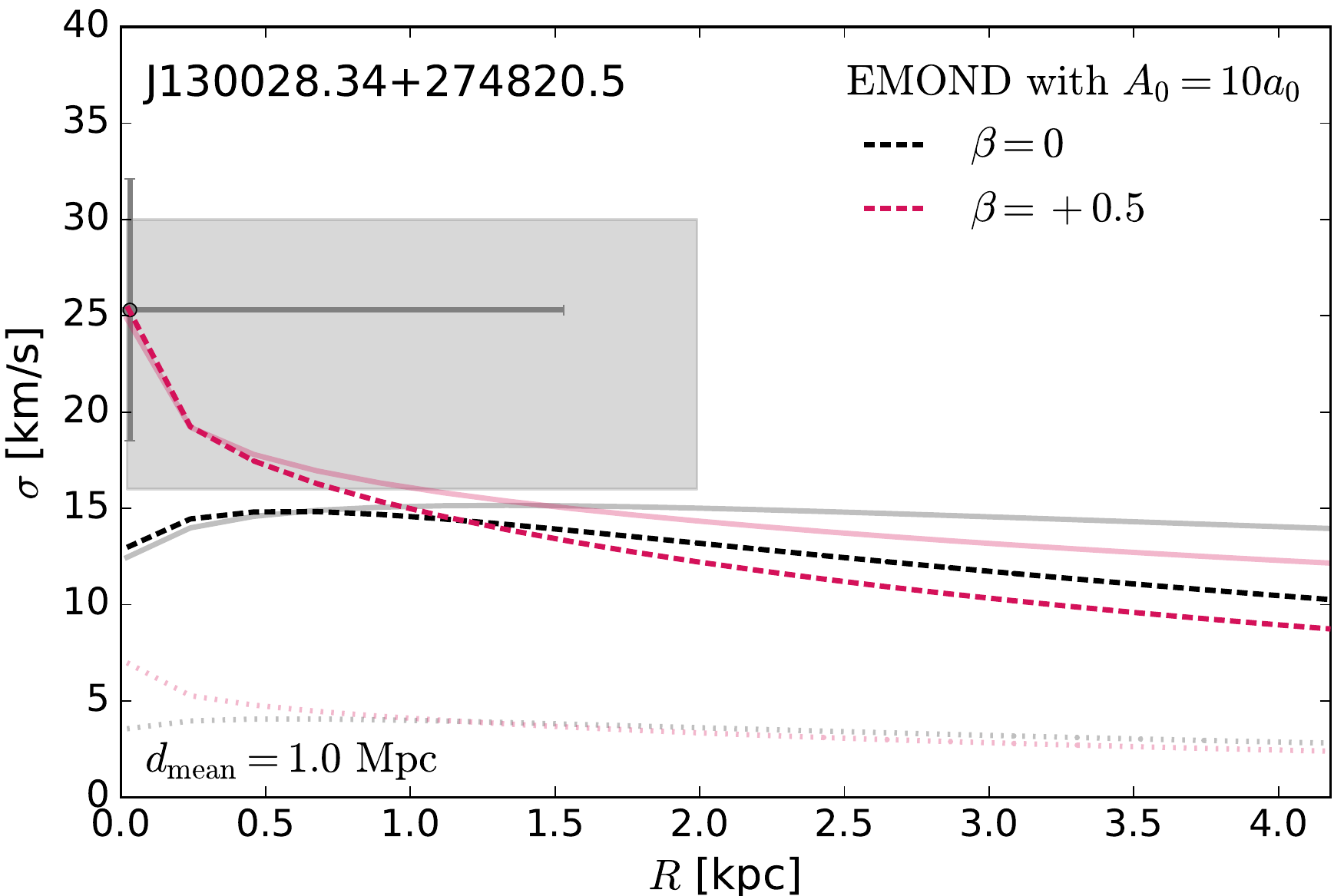}
\includegraphics[width=0.46\textwidth,trim={0cm 0cm 0cm 0cm},clip]{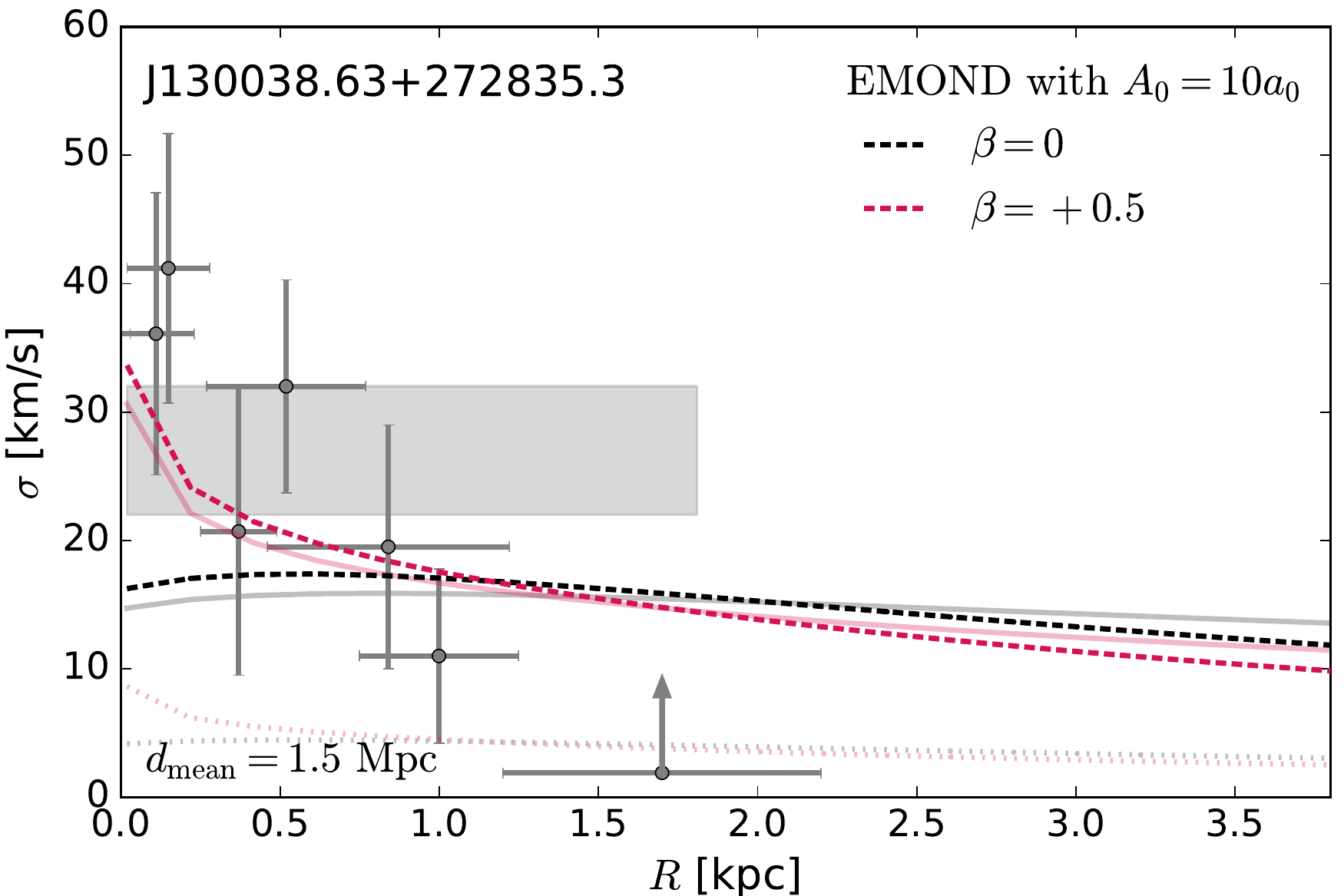}
\hfill \hspace{1cm}
\caption{
Continued. 
}
\label{fig:EMOND_profiles2}%
\end{figure*}

In Fig.~\ref{fig:EMOND_profiles}, we show the velocity dispersion profiles predicted when emulating EMOND with a critical acceleration $A_0=10\times a_0$ at the average distance $d_{\rm mean}$ inferred from the Einasto UDG distribution of \cite{vanderBurg2016} (cf. Section \ref{section:coma}). The external field was determined from the dominant hot gas mass $M_{\rm gas}$ of Eq.~(\ref{eq:Mgas}) and its effect using Eq.~(\ref{eq:gr}). 
EMOND with $A_0=10\times a_0$ provides a good match to the velocity dispersion measurements. 
Galaxy J130005.40+285333.0, for which the agreement is less convincing, is the closest to the cluster centre; hence, it is exposed to the strongest tidal forces and resides the deepest in the potential well of the cluster.

%--------------------------------------------------------------------

\section{Baryonic EFE profiles}
\label{appendix:EFE_bar}

\begin{figure*}
\centering
\includegraphics[width=0.46\textwidth,trim={0cm 0cm 0cm 0cm},clip]{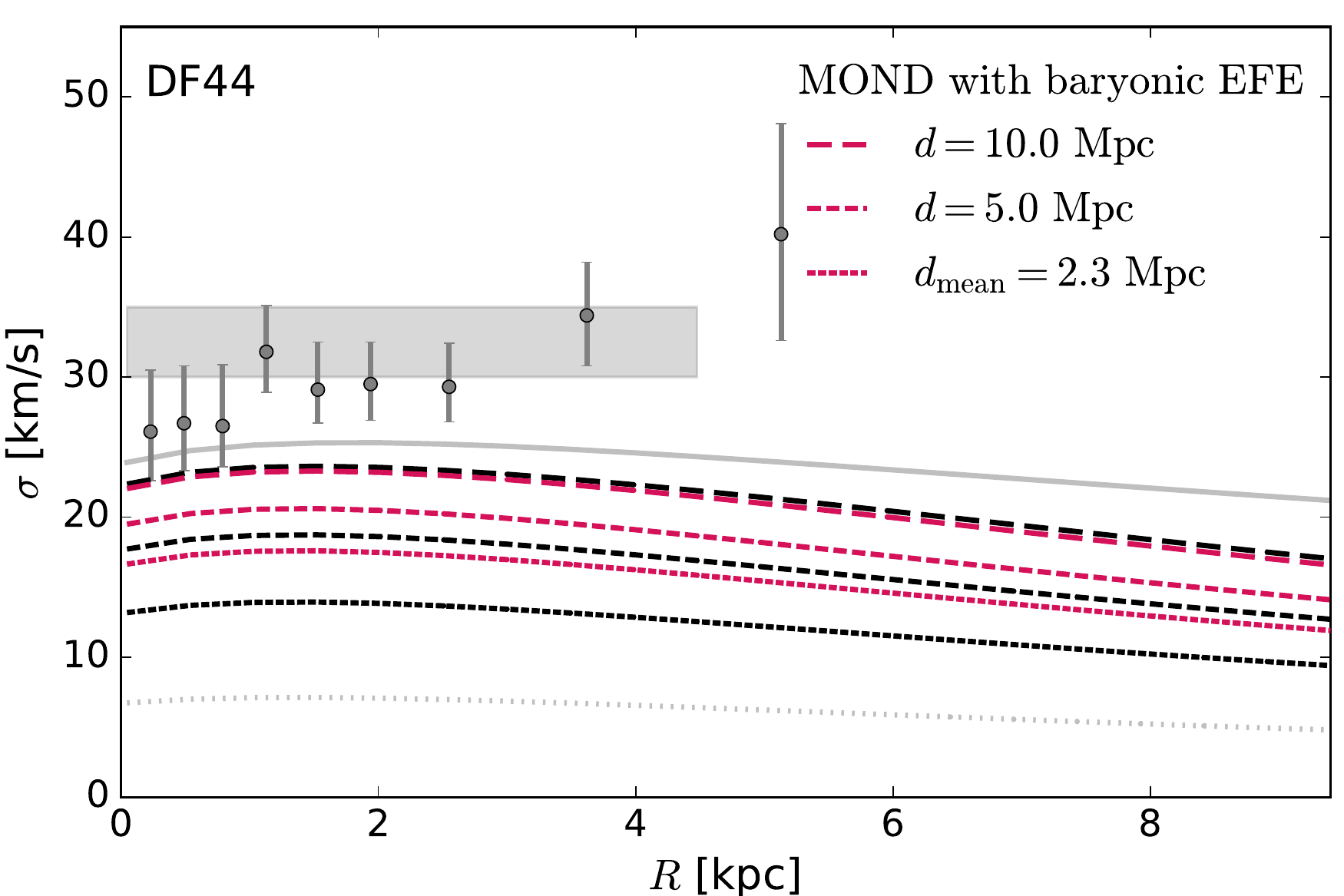}
\hfill
\includegraphics[width=0.46\textwidth,trim={0cm 0cm 0cm 0cm},clip]{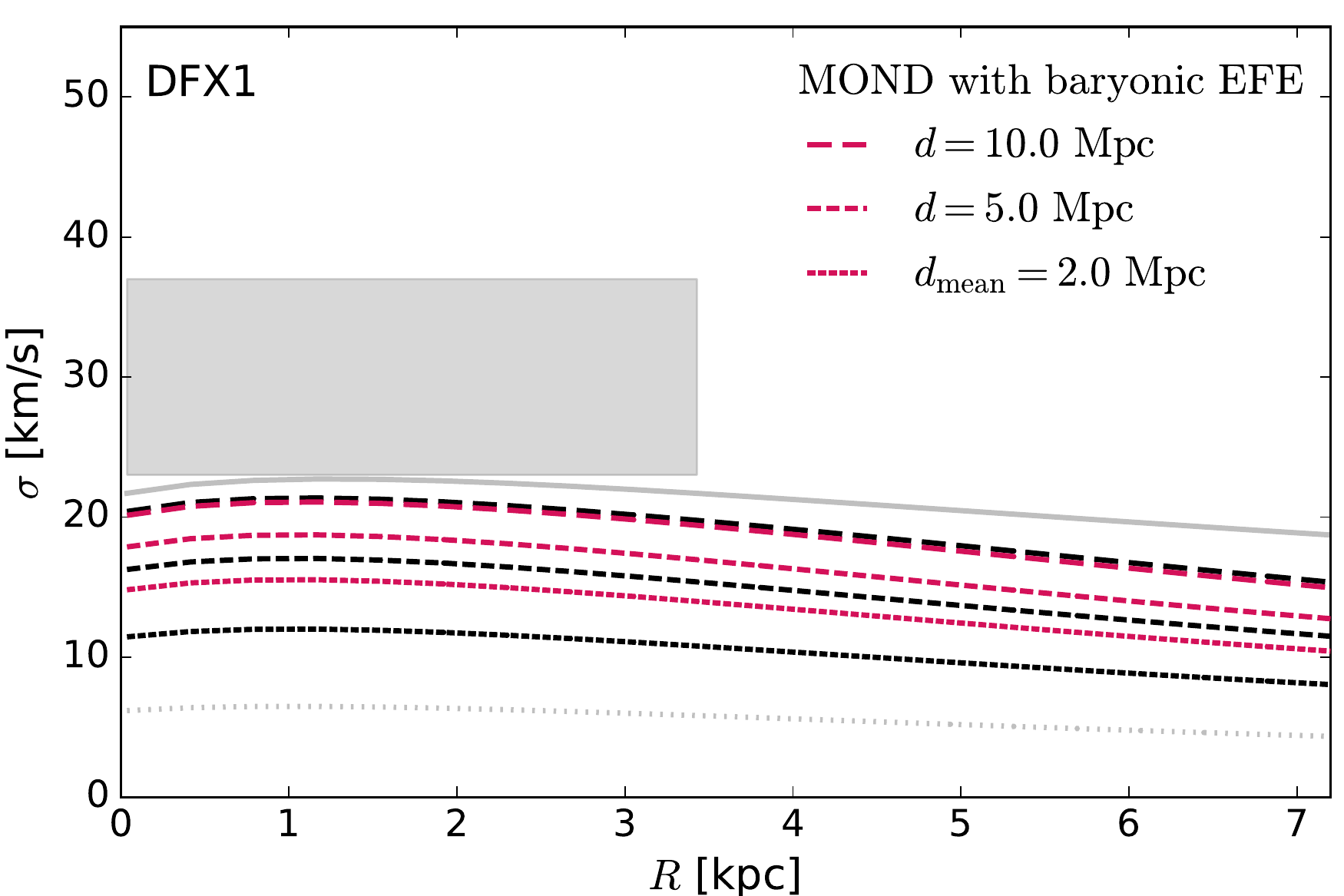}
\includegraphics[width=0.46\textwidth,trim={0cm 0cm 0cm 0cm},clip]{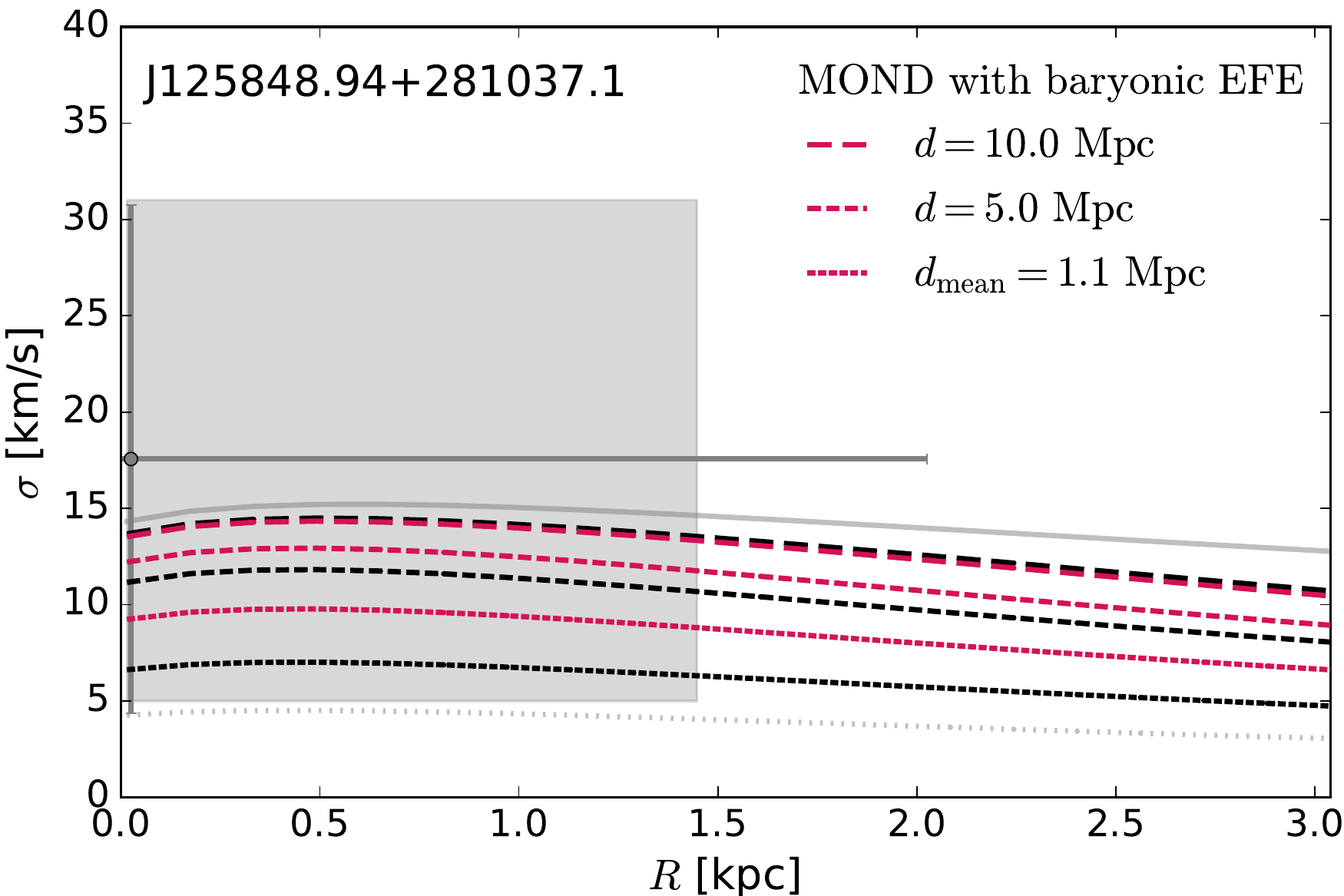}
\hfill
\includegraphics[width=0.46\textwidth,trim={0cm 0cm 0cm 0cm},clip]{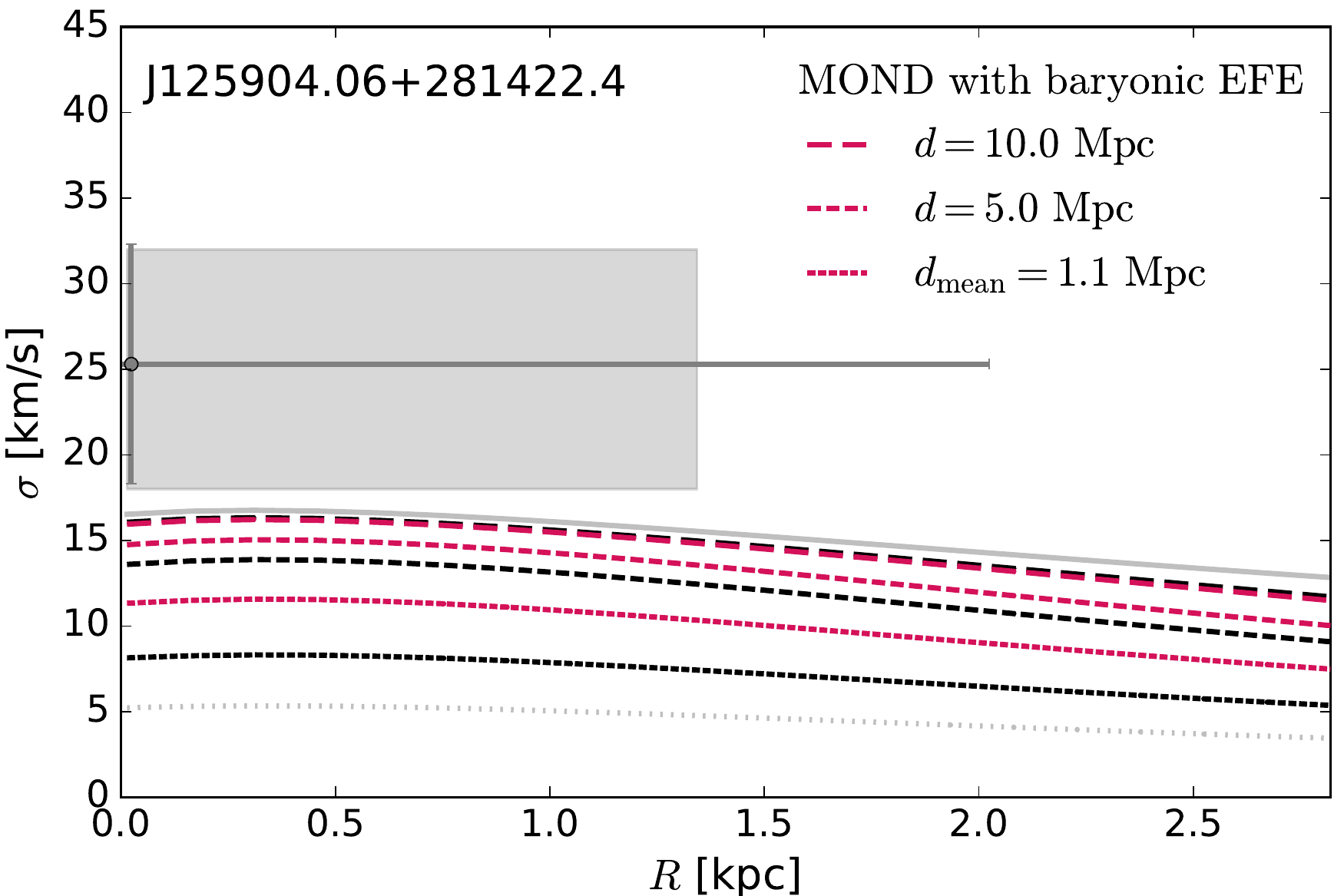}
\includegraphics[width=0.46\textwidth,trim={0cm 0cm 0cm 0cm},clip]{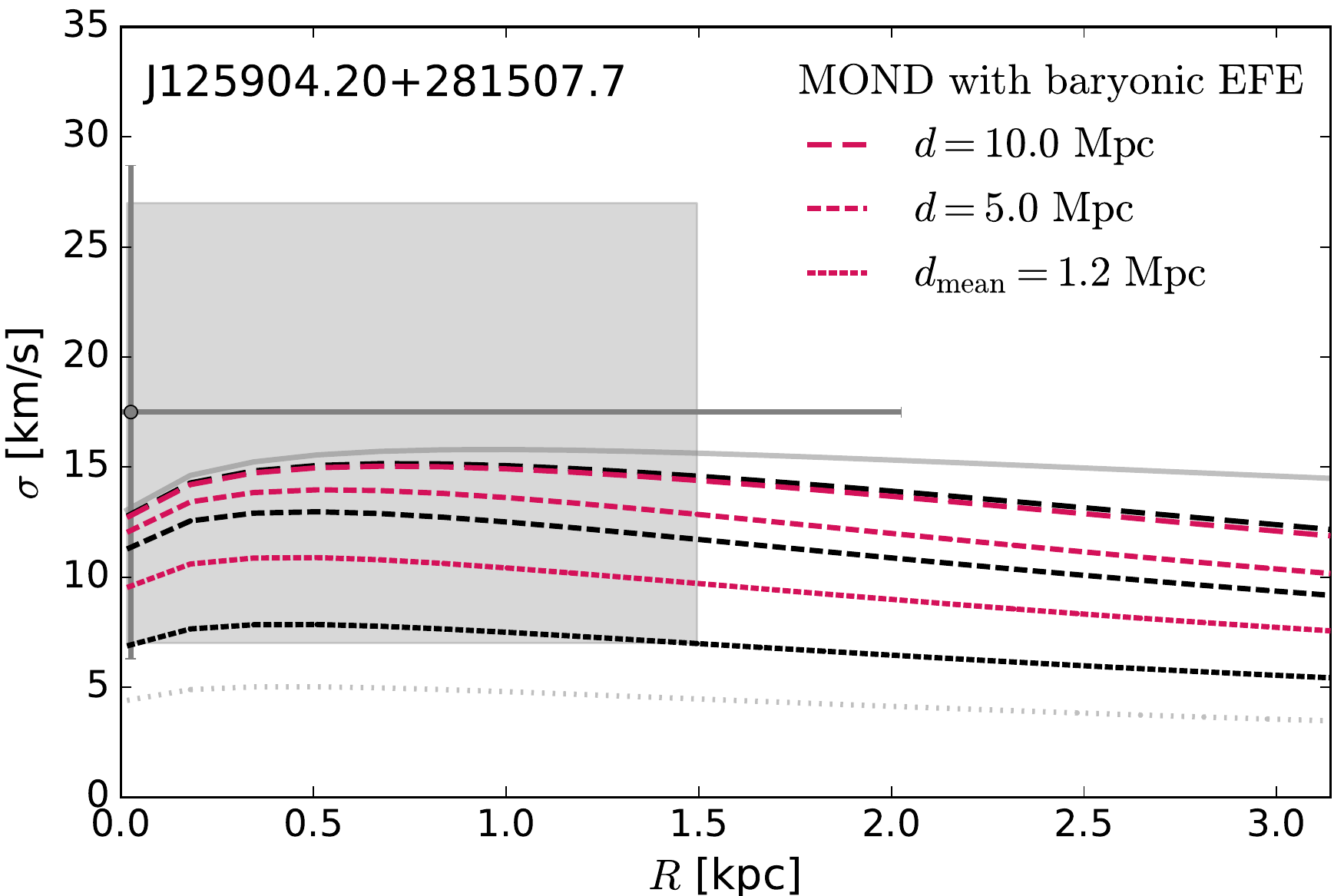}
\hfill
\includegraphics[width=0.46\textwidth,trim={0cm 0cm 0cm 0cm},clip]{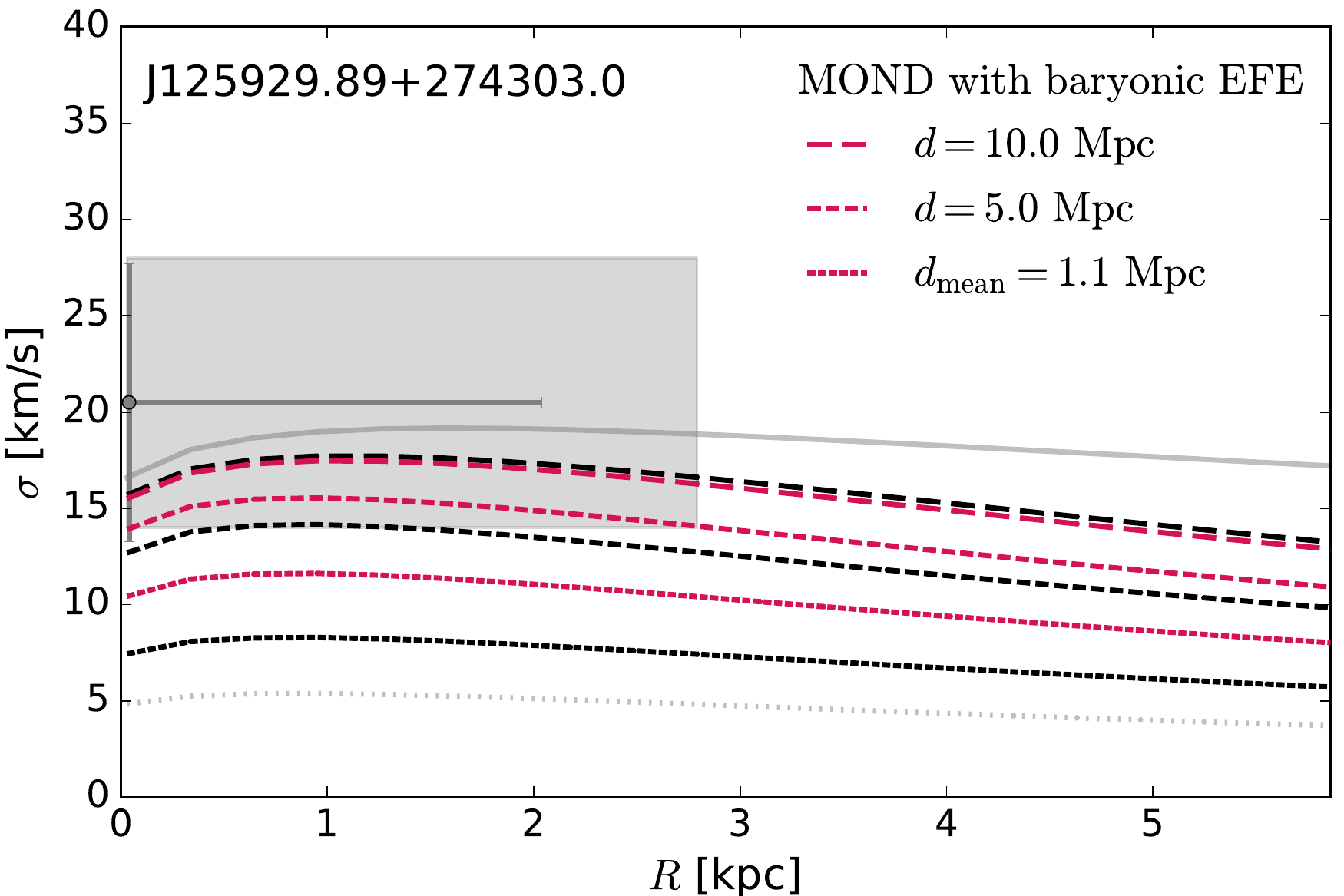}
\includegraphics[width=0.46\textwidth,trim={0cm 0cm 0cm 0cm},clip]{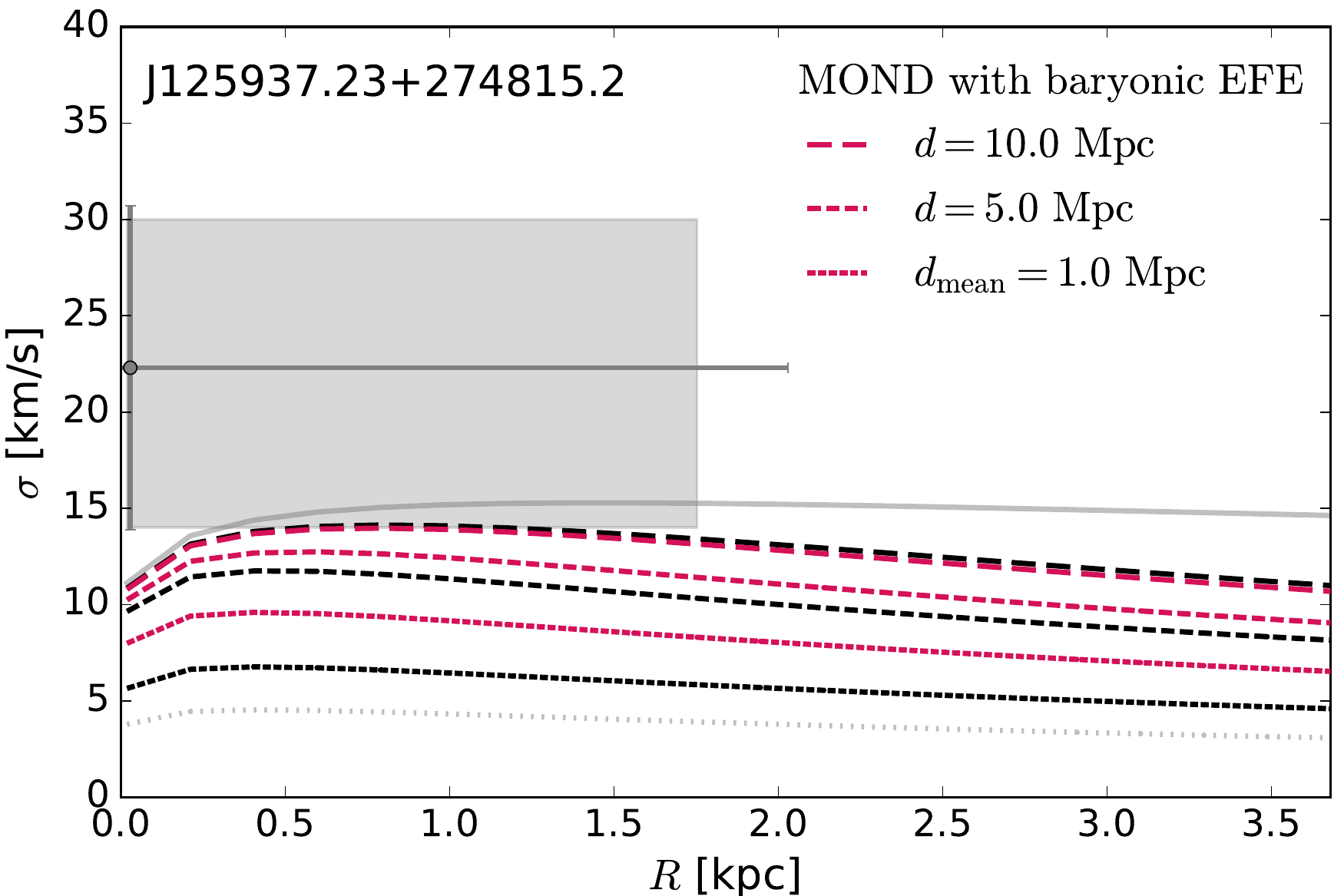}
\hfill
\includegraphics[width=0.46\textwidth,trim={0cm 0cm 0cm 0cm},clip]{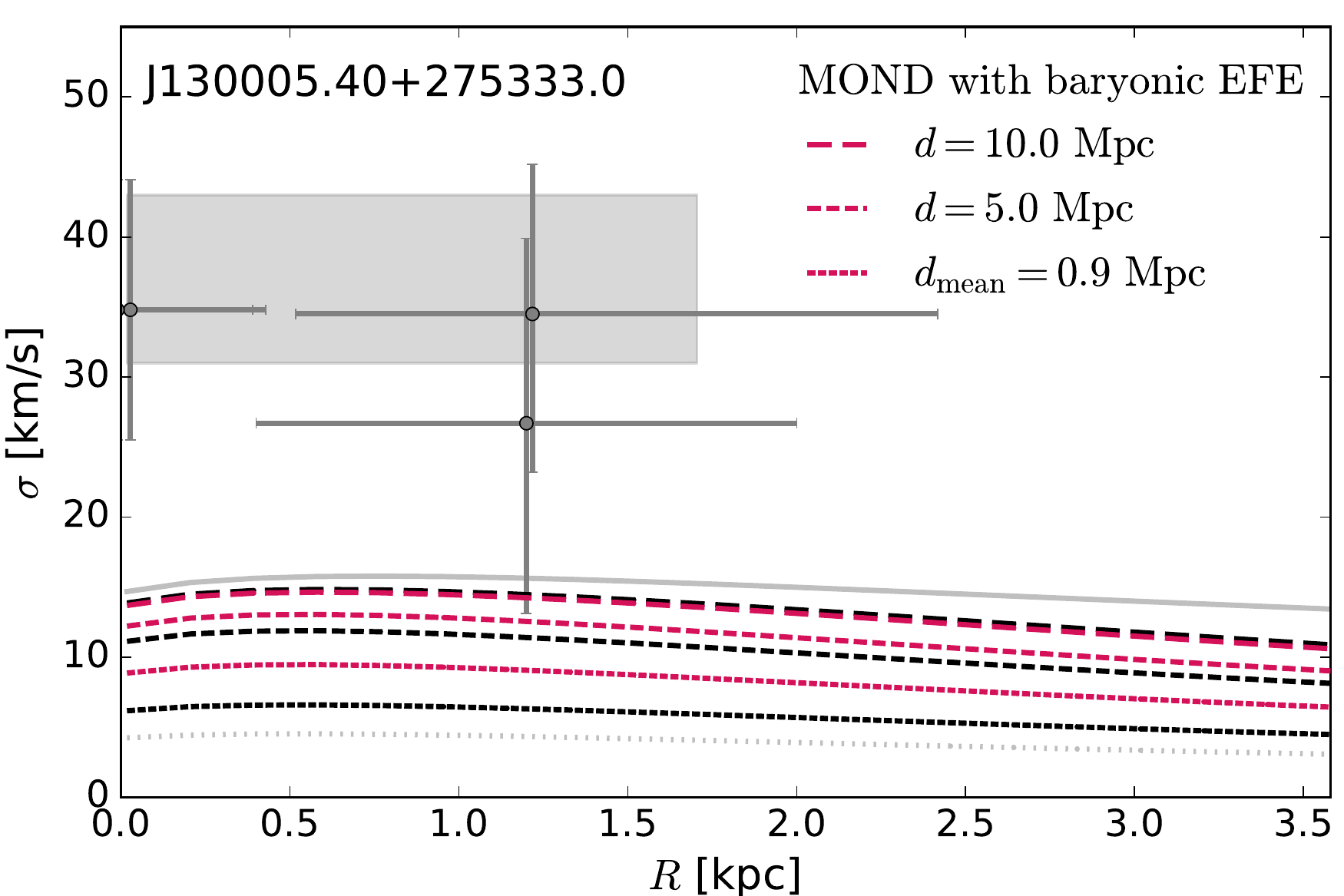}
\caption{ 
Similar to Fig.~\ref{fig:EFE}, with the EFE calculated solely from the intracluster hot gas distribution $M_{\rm gas}$ (magenta dashed lines) instead of the mass distribution $M_C$ inferred from hydrostatic equilibrium (black dashed lines, similar to Fig.~\ref{fig:EFE}). Only the case of a uniform anisotropy parameter $\beta=0$ is shown.
}
\label{fig:EFE_gas}
\end{figure*}

\setcounter{figure}{0}
\begin{figure*}
\centering
\includegraphics[width=0.46\textwidth,trim={0cm 0cm 0cm 0cm},clip]{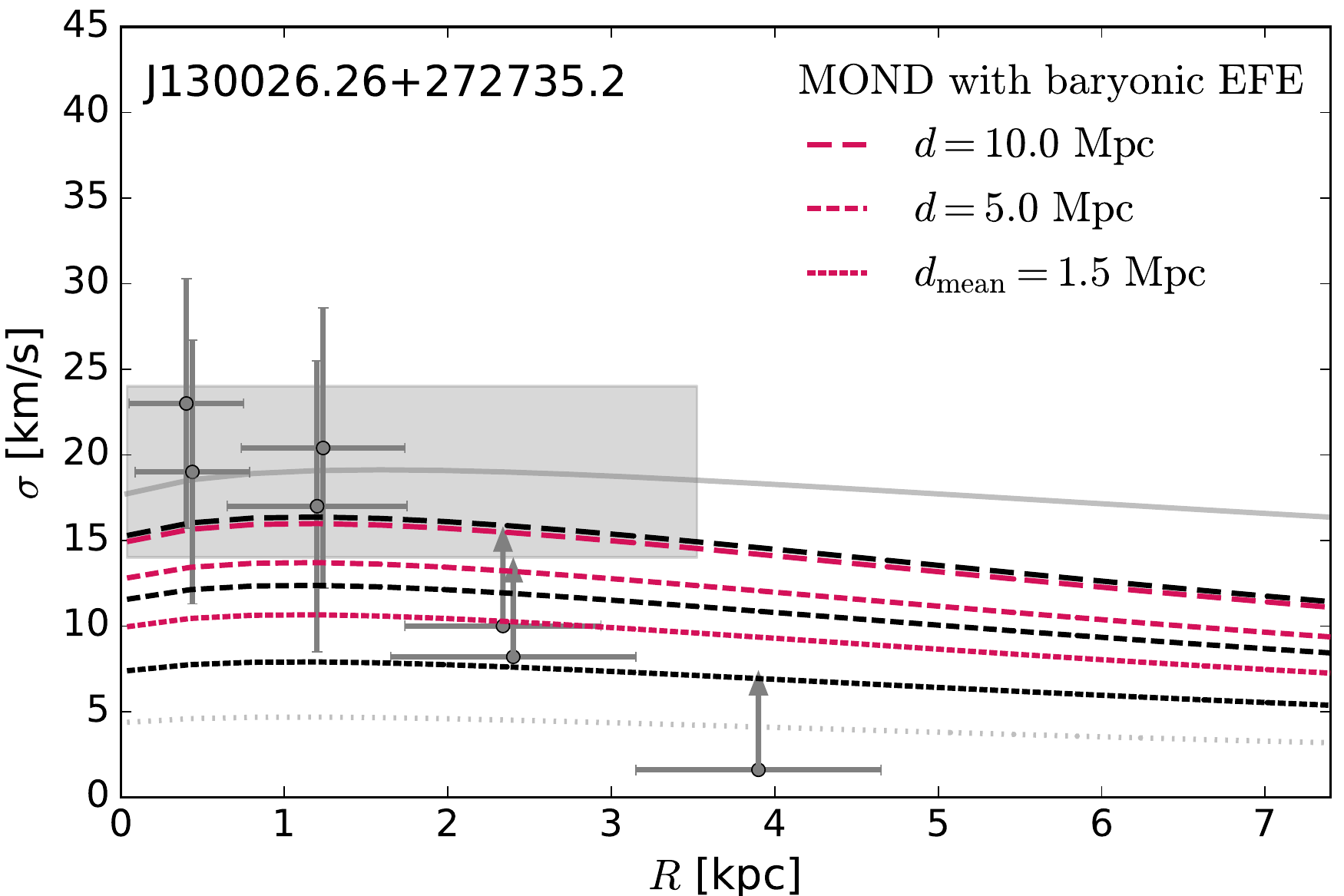}
\hfill
\includegraphics[width=0.46\textwidth,trim={0cm 0cm 0cm 0cm},clip]{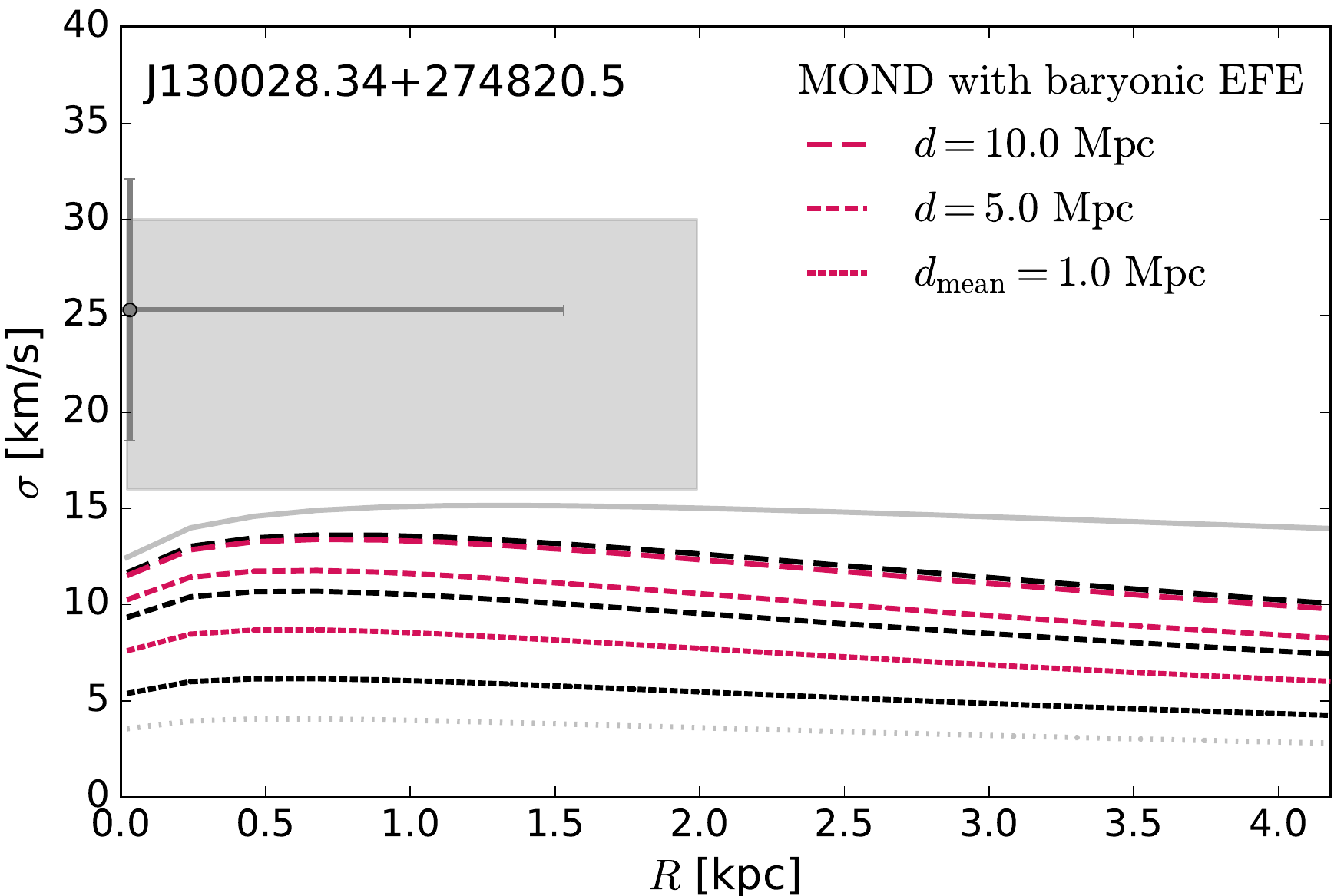}
\includegraphics[width=0.46\textwidth,trim={0cm 0cm 0cm 0cm},clip]{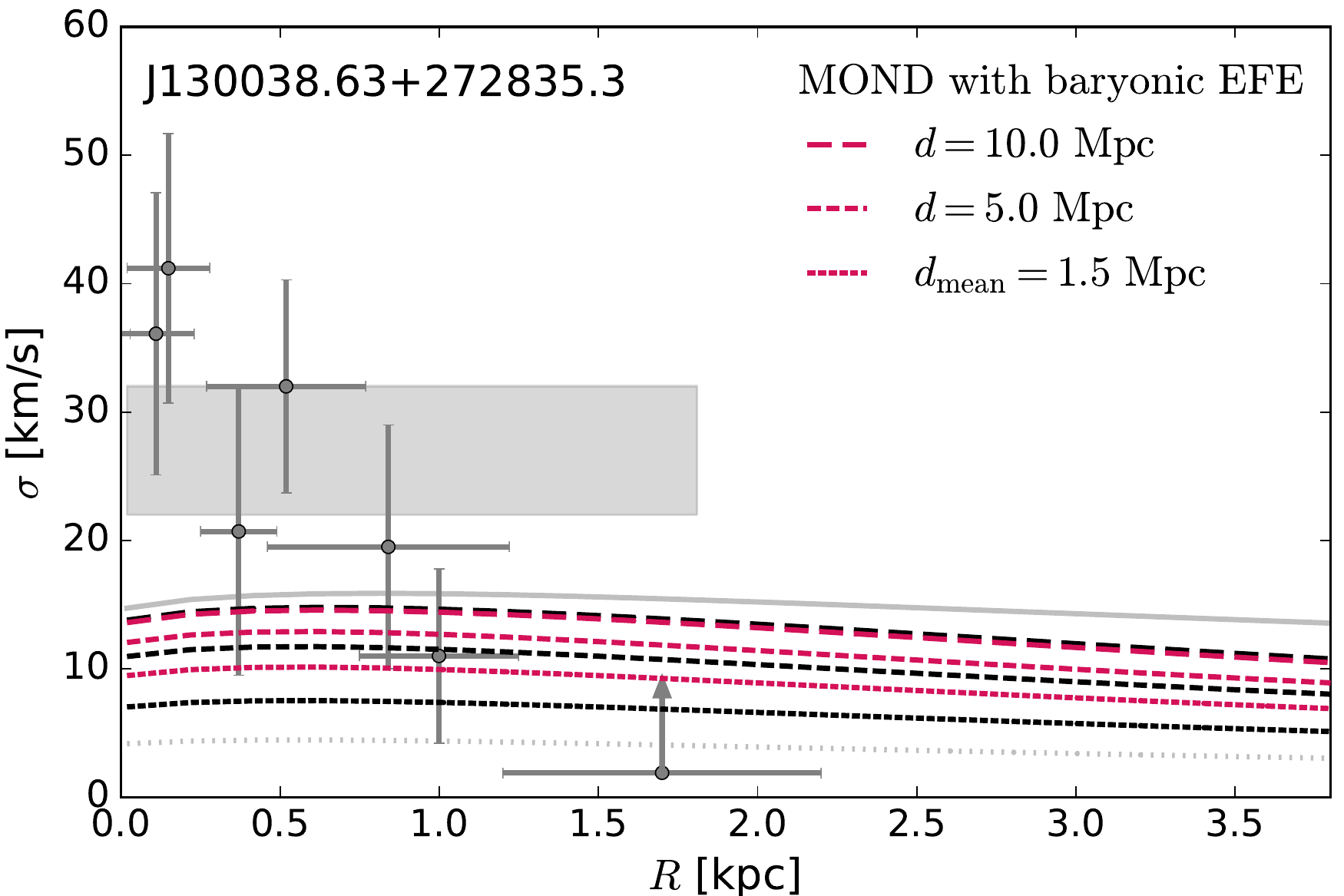}
\hfill \hspace{1cm}
\caption{
Continued. 
}
\label{fig:EFE_gas2}%
\end{figure*}

In Fig.~\ref{fig:EFE_gas}, we show the MOND predicted velocity dispersion when the EFE is calculated solely from the intracluster hot gas distribution $M_{\rm gas}$ (Eq.~(\ref{eq:Mgas})) instead of $M_C$ from hydrostatic equilibrium (Eq.~(\ref{eq:MC})). Although $M_{\rm gas}$ is 1 dex below $M_C$ at 1 Mpc, the difference is not sufficient to reach the high measured velocity dispersions.

%--------------------------------------------------------------------

\end{document}